%% file: tesi.tex
\documentclass[a4paper,11pt,twoside]{book}

\usepackage{graphicx}
\usepackage[english]{babel}
\usepackage{amsmath,amssymb}
\usepackage{natbib}
\usepackage[utf8x]{inputenc}
\usepackage{times}
\usepackage{makeidx}
\usepackage[dvips]{epsfig}
\usepackage[dvips]{graphicx}
\usepackage{textcomp}
\usepackage{longtable}
\usepackage{pdfpages}
\usepackage{import}
\usepackage[titletoc,title]{appendix} % for A,B,ecc appendix
\usepackage{verbatim}
\usepackage{ulem}
\usepackage{lscape}
\usepackage{multirow}
\usepackage{hhline}
\usepackage[T1]{fontenc}
\usepackage{subfigure}
\usepackage{deluxetable}
\usepackage{amsfonts}
\usepackage{url}

\def\ltsim{\raise 2pt \hbox {$<$} \kern-1.1em \lower 4pt \hbox {$\sim$}}
\def\ltapprox{\raise 2pt \hbox {$<$} \kern-1.1em \lower 5pt \hbox {$\approx$}}
\def\gtsim{\raise 2pt \hbox {$>$} \kern-1.1em \lower 4pt \hbox {$\sim$}}
\def\gtapprox{\raise 2pt \hbox {$>$} \kern-1.1em \lower 5pt \hbox {$\approx$}}

\def\lapp{\ifmmode\stackrel{<}{_{\sim}}\else$\stackrel{<}{_{\sim}}$\fi}
\def\gapp{\ifmmode\stackrel{>}{_{\sim}}\else$\stackrel{<}{_{\sim}}$\fi}

\def\eps@scaling{.95}
\def\epsscale#1{\gdef\eps@scaling{#1}}
\def\plotone#1{\centering \leavevmode
    \epsfxsize=\eps@scaling\columnwidth \epsfbox{#1}}
\def\plottwo#1#2{\centering \leavevmode
    \epsfxsize=.45\columnwidth \hbox{\epsfbox{#1}} \hfil
    \epsfxsize=.45\columnwidth \epsfbox{#2}}

\def\Msun{M_{\odot}}

\def\ltsima{$\; \buildrel < \over \sim \;$}
\def\lsim{\lower.5ex\hbox{\ltsima}}

\usepackage[top=4cm, bottom=3cm, left=3.5cm, right=3cm]{geometry}
\usepackage{fancyhdr}
\newcommand\arcsec{\mbox{$^{\prime\prime}$}}%
\newcommand\arcmin{\mbox{$^\prime$}}%

\def\aj{AJ}%
\def\araa{ARA\&A}%
\def\apj{ApJ}%
\def\apjl{ApJL}%
\def\apjs{ApJS}%
\def\aap{A\&A}%
\def\aapr{A\&A~Rev.}%
\def\aaps{A\&AS}%
\def\mnras{MNRAS}%
\def\pasp{PASP}%
\def\pasj{PASJ}%
\def\solphys{Sol.~Phys.}%
\def\ssr{Space~Sci.~Rev.}%
\def\nat{Nature}%
\def\aplett{Astrophys.~Lett.}%
\def\memsai{Mem.~Soc.~Astron.~Italiana}%
\def\physrep{Phys.~Rep.}%
\def\procspie{Proc.~SPIE}%
\begin{document}

%\begin{comment}

  %crea l'ambiente per il frontespizio
  \thispagestyle{empty}   \hsize 145 mm
  \topmargin 0mm
  \enlargethispage*{3cm} %aumenta la lunghezza della pagina di 3cm
%-------------------------------------------------------------------------%
\begin{center}

    {\Large \textbf{Alma Mater Studiorum}}\\\vspace{5mm}
    {\Large \textbf{Universit\`{a} degli Studi di Bologna}}\\
    \vspace{1mm}
    \rule{140mm}{0.2mm}\\
    \rule[2ex]{140mm}{0.5mm}
    \vspace{5mm}
    \large{DIPARTIMENTO DI FISICA E ASTRONOMIA\\
           Dottorato di ricerca in Astronomia\\
            Ciclo XXVIII\\

           \vspace{2.5cm}}

                  {\huge\textbf{COSMIC-LAB: Unexpected Results from\\
                  \vspace{5mm}
                  High-resolution Spectra of AGB Stars
                  \vspace{5mm}
                  in Globular Clusters}}\\

\end{center}

  \vspace{2.5  cm}

  \hspace{-6mm}
  \parbox[t]{50mm}{
    \begin{flushleft}
      {\large Dottorando: \vspace{1mm}\\
        \textbf{Emilio Lapenna}}\\
    \end{flushleft}
  } \ \          % lascia uno spazio tra \ e \ NON LASCIARE RIGHE VUOTE
  \hspace{-5mm}
  \parbox[t]{90mm}{
    \vspace{0cm}
    \begin{flushright}
      {\large Relatore:\\ \vspace{1mm}         \textbf{Chiar.mo Prof. Francesco R. Ferraro}\\
        \vspace{0.3 cm}
        Co-Relatori: \\ \vspace{1mm}
        \textbf{Dr. Alessio Mucciarelli} \\
        \textbf{Dr. Livia Origlia} \\
	\textbf{Chiar.ma Prof. Barbara Lanzoni} \\
        %\textbf{Prof. Barbara Lanzoni} \\
        %\vspace{0.3 cm}
        %Co--Relatrice: \\ \vspace{1mm}
        %\textbf{Prof.sa Barbara Lanzoni} \\
        \vspace{1 cm}
        Coordinatore: \\ \vspace{2mm}
        \textbf{Chiar.mo Prof. Lauro Moscardini}} \\
    \end{flushright}
  } \\

\begin{center}

    \vspace{1.5cm}
      {\large Esame finale anno 2015}\\
    \rule{140mm}{0.2mm}\\
    \rule[2ex]{140mm}{0.5mm}\\

       {\small         Settore Concorsuale: 02/C1 -- Astronomia, Astrofisica, Fisica della Terra e dei Pianeti\\
     Settore Scientifico-Disciplinare: FIS/05 -- Astronomia e Astrofisica\\}

          \end{center}

 \hsize 140 mm

\clearpage{\pagestyle{empty}\cleardoublepage}
%\end{comment}

%\includepdf{frontespizio.pdf}

%\includepdf[pages={1}]{frontespizio.pdf}

%\newpage
%\mbox{ }
%\thispagestyle{empty}

\newpage
\mbox{ }
\thispagestyle{empty}

\vspace{6cm}
\begin{flushright}
{\sl Alla Mia Famiglia}\\
\vspace{2mm}
\end{flushright}

%%\newpage
%%\mbox{ }
%%\thispagestyle{empty}
%%\newpage
%%\includepdf{abstract}

%\newpage
%\mbox{ }
%\thispagestyle{empty}

\clearpage{\pagestyle{empty}\cleardoublepage}

\newpage
\mbox{ }
\thispagestyle{empty}

\vspace{4cm}
\begin{flushright}
{\sl ``L'astronomia costringe l'anima a guardare oltre e ci conduce da un mondo ad un altro.''}\\
   \vspace{2mm}
               {Platone\ \ \  }
\\
\end{flushright}

\clearpage{\pagestyle{empty}\cleardoublepage}

%\begin{flushright}
%{\sl ``Non puoi aspettarti di vedere al primo sguardo. Osservare è per certi versi un'arte che bisogna apprendere.''}\\
%   \vspace{2mm}
%               {William Herschel\ \ \  }
%\\
%\end{flushright}

%%\newpage
%%\mbox{ }
%%\thispagestyle{empty} 
%%\newpage
%%\includepdf{abstract}

%\newpage
%\mbox{ }
%\thispagestyle{empty}

\baselineskip 4ex

\newcommand{\ltae}{\raisebox{-0.6ex}{$\,\stackrel
{\raisebox{-.2ex}{$\textstyle <$}}{\sim}\,$}}
\newcommand{\gtae}{\raisebox{-0.6ex}{$\,\stackrel
{\raisebox{-.2ex}{$\textstyle >$}}{\sim}\,$}}

\baselineskip 4ex

% questo numera l' in numeri romani: i,ii,iii,iv ecc.

%\tableofcontents
%\clearpage{\pagestyle{empty}\cleardoublepage}

%\listoftables
%\listoffigures

%\newpage

% questo numera le pagine della tesi in arabico: 1,2,3, ecc.

%\pagenumbering{arabic}
% gli "\input" includono i relativi files.tex che contengono i vari
% capitoli della tesi, che si trovano nel directory [lauNN.tesi]
%(bisogna togliere il "%" in testa alla riga)

%\input{stud$disk:capitolouno.tex}
%\input{stud$disk:[lauNN.tesi]CAP2.tex}
%\addcontentsline{toc}{chapter}{Introduzione}
%\input {prem.tex}
%\input{primo.tex}

%\cleardoublepage
%\renewcommand{\thepage}{  }
\newpage

\pagenumbering{roman}

\markboth{\sc \ }{\sc Contents}
\tableofcontents
\clearpage{\pagestyle{empty}\cleardoublepage}

%\listoffigures
%\listoftables
%\clearpage{\pagestyle{empty}\cleardoublepage}

%%%%%\clearpage
%%%%%\markboth{ACRONYMS}{ACRONYMS}
%%%%%\newpage
%%%%%\input{listacr.tex}
%%%%%\clearpage{\pagestyle{empty}\cleardoublepage}

%\newpage
%\mbox{ }
%\thispagestyle{empty}
%\clearpage{\pagestyle{empty}\cleardoublepage}

\setcounter{page}{1}

\renewcommand{\thepage}{\arabic{page}}

\clearpage
%\markboth{ABSTRACT}{ABSTRACT}
%\addcontentsline{toc}{chapter}{Abstract}
%\newpage
%\input{}
%\clearpage{\pagestyle{empty}\cleardoublepage}

\markboth{Introduction}{Introduction}
\addcontentsline{toc}{chapter}{Introduction}

\input{intro/intro.tex}

\clearpage{\pagestyle{empty}\cleardoublepage}

%\newpage
\input{c1a/c1a.tex}

\clearpage{\pagestyle{empty}\cleardoublepage}

%\newpage
\input{c2a/c2b.tex}

\clearpage{\pagestyle{empty}\cleardoublepage}

%\newpage
\input{c3a/c3b.tex}

\clearpage{\pagestyle{empty}\cleardoublepage}

%\newpage
\input{c4/ms_4.tex}

\clearpage{\pagestyle{empty}\cleardoublepage}

%\newpage
\input{c5/ms_5.tex}

\clearpage{\pagestyle{empty}\cleardoublepage}

%\newpage
\input{c6/ms_6.tex}

\clearpage{\pagestyle{empty}\cleardoublepage}

%\newpage
\input{c7/ms_7.tex}
\clearpage{\pagestyle{empty}\cleardoublepage}

%\newpage
\input{c8/ms_8.tex}

\clearpage{\pagestyle{empty}\cleardoublepage}

\addcontentsline{toc}{chapter}{Conclusions}
\newpage
\markboth{CONCLUSIONS}{CONCLUSIONS}
\input{concl/conclusions.tex}

\clearpage{\pagestyle{empty}\cleardoublepage}

\addcontentsline{toc}{chapter}{Appendix}
\newpage
\markboth{APPENDIX}{APPENDIX}

\input{ap1/ap_1.tex}

\clearpage{\pagestyle{empty}\cleardoublepage}

\input{ap2/ap_2.tex}

\clearpage{\pagestyle{empty}\cleardoublepage}

% \newpage
% \mbox{ }
% \thispagestyle{empty}
% \clearpage{\pagestyle{empty}\cleardoublepage}
% \include{thanx1}

%\input{CAPITOLI/fronte.tex}

%\input{copertina.tex}
%\input{"path completo del file"]capitolo_10.tex}

\clearpage{\pagestyle{empty}\cleardoublepage}
\clearpage{\pagestyle{empty}\cleardoublepage}

\clearpage{\pagestyle{empty}\cleardoublepage}

\newpage
\markboth{ACKNOWLEDGEMENTS}{ACKNOWLEDGEMENTS}
\input{acknowl/acknowledgements.tex}

%\addcontentsline{toc}{chapter}{Appendice}
%\nocite{*}
%\normalsize{%
%\bibliographystyle{plain}
%\bibliography{/CAPITOLI/biblio.tex}
%}
%\input {app.tex}
%\newpage
%\addcontentsline{toc}{chapter}{Conclusioni e lavori futuri}
%\input {concl.tex}
%\addcontentsline{toc}{chapter}{Appendice}
%\input {app.tex}
%\addcontentsline{toc}{chapter}{Bibliografia}
%\input{biblio.tex}
%\addcontentsline{toc}{chapter}{Ringraziamenti}
%\input{r.tex}

%\input{../CAPITOLI/curriculum.tex}

%\input{../CAPITOLI/riassunto.tex}

% fine di tutto

\end{document}

%% file: intro/intro.tex
% PREFACE

\chapter*{Introduction}

Globular clusters (GCs) are among the most interesting stellar systems and have been
targeted by several studies from the beginning of the astrophysical research.
These systems are thought to be the first stellar aggregates formed
in the very early epochs of the Galaxy formation.

For decades, the GCs have been thought to be the best example of simple stellar population (SSP).
This terminology was suggested to highlight the overall property of a population
in which stars show similar characteristics in terms of age and metallicity.
In fact, a SSP is assumed to be originated in a single star formation episode from
a cloud with an homogeneous chemical composition.
In this way, all the stars share the same age, since they were born at the same time,
and the same metallicity, since they were formed from the same material.
Moreover, since GCs typically are older than $\sim$ 10 Gyr, they can be considered as
living ``relics'' of the first stellar aggregates formed and they can be used as tracers of
the chemical enrichment history of the host galaxies.

Today, the huge progress that has been accomplished thanks to the rising number of
photometric and spectroscopic facilities demonstrated, however, that the SSP concept should be revised.
In fact, important chemical anomalies have been detected in GCs,
the main ones consisting in the so-called ``anticorrelations'' among light-elements and
in peculiar behaviours observed for iron-peak and neutron-capture elements.

This observational evidence suggests that GCs are complex systems,
which during their lifetimes undergo a series of processes able to deeply affect
the global characteristics.
In this sense, GCs can be considered, at least to a first approximation,
as ``closed'' systems, which mainly suffered auto-enrichment processes
with a negligible interaction with the Inter Galactic Medium.
In this way, the chemical imprint of different polluters takes a
fundamental role in defining the chemical characteristics of the descendants.
The study of the chemical composition of stars in GCs is thus fundamental
not only to test their chemical models and to unveil the nature of the polluters,
but also to better constrain the nucleosynthesis of the elements.

In this context, my thesis is focused on AGB stars, which have an important role
in defining the characteristics of GCs for several reasons:
(1) the AGB stars dominate the integrated light of stellar populations of intermediate ages ($t<$ 2 Gyr),
(2) they are an important nucleosynthesis site for the chemical element formed through $proton$- and $neutron$-capture chains and
(3) they are thought to be the main polluters in the self-enrichment processes during GC lifetimes.
However, due to their short evolutive timescales, AGB stars are numerically the smallest population in GCs,
and nowadays only a few studies have investigated their chemical characteristics.
Moreover, a few recent results \citep{ivans01,beccari06,campbell13} have awoken
the interest for these stars and highlighted the need of new and comprehensive characterizations.

This thesis is aimed at chemically clarifying this poorly studied evolutionary phase.
The thesis presents the analysis of a large sample of high-resolution spectra of AGB stars
in GCs acquired at the Very Large Telescope (ESO) and at the MPG-2.2m telescope (ESO).
The results are quite unexpected and they are contributing to a new understanding of GC chemistry.
The work is part of the project Cosmic-Lab, a five year research program funded by the European Research council.

The work is organized as follows: 
Chapter~\ref{c1} presents an introduction to the nucleosynthesis sites and channels from which
the main chemical elements form. Chapter~\ref{c2} is focused on the main properties of GCs,
their stellar populations and the main formation scenarios suggested so far.
Chapter~\ref{c3} is devoted to illustrate the known properties of AGB, in light of a few recent results from the literature.
Chapter~\ref{c4} reports on the spectroscopic analysis of a sample of 24 AGB stars belonging to
the GC 47Tucanae ([Fe/H] = $-0.7$ dex) which has shown evidence of important non-local thermodynamic equilibrium (NLTE) effects.
Chapter~\ref{c5} and~\ref{c6} present the discussion of the iron content of two GCs (namely NGC3201 and M22),
demonstrating that (at odds with previous claims) they show no intrinsic metallicity spreads.
Chapter~\ref{c7} presents the high-resolution spectroscopic analysis of a sample of 19 giant stars in M62
which has revealed that the same NLTE mechanism discovered in 47Tucanae affects also the titanium lines.
Chapter~\ref{c8} describes how the ionization balance between chemical abundances derived
from neutral and ionized elements can be used as a powerful weighing device and presents the discovery of
an anomalously heavy (1.4 M$_{\odot}$) star in a sample of photometrically indistinguishable giants
of 47Tucanae (this objects most likely is an evolved blue straggler star).

Finally in the Appendices \ref{a1} and \ref{a2} two side-product works are discussed.
They deal with the characterization of the performances of two new-generation spectrographs:
as GIANO and KMOS.

%% file: c1a/c1a.tex
% CAPITOLO 1

\chapter{Elemental Abundances and Chemical Evolution}

\label{c1}

The chemical enrichment history of galaxies is driven by the nucleosynthesis occurring in many generations of stars.
Indeed, stars are the most important nucleosynthetic site in which the chemical elements heavier than He are build up with
different processes on different timescales. The processes that form the elements are strictly linked
to the physics of stellar interiors and their evolution over time.

Generally speaking, according to their mass, stars are able to synthesize different
elements through different channels and to release a fraction of them (the so-called yields) 
at different epochs from the star formation onset, 
not only at the end of their lifetimes (e.g. as Supernovae) but also during their evolution
(e.g. through wind activity from AGB stars, fast rotating massive stars and Wolf-Rayet stars).
When these yields are released they are mixed in the Inter Stellar Medium (ISM),
and subsequent stellar generations can form from this pre-enriched material.
The total yield of a given element depends on the mass of the corresponding metal ejected by the stars and on the
relative frequency of stars of different masses born in a stellar generation (the so-called Initial Mass Function, IMF).
Another chemical evolution parameter is the star formation rate (SFR),
which is commonly assumed to be proportional to the star formation efficiency and
to some power of the normalized gas surface mass density.

Hence, chemical abundance ratios are powerful diagnostics of the IMF
and SFR parameters of stellar systems, also flagging the timescales of
chemical evolution. In particular, investigating the chemical
composition of the oldest stars is especially important, because these
are fossils of the earliest epoch of the galaxy chemical evolution
history. The main elements that can be observed and studied in stars
can be divided in three main families based on their different formation
processes and nucleosynthetic sites: $\alpha$ and other light elements,
iron-peak elements, and neutron-capture elements.
In the following sections we will briefly examine some of their characteristics and peculiarities.

\section{$\alpha$ and other light elements}

\begin{figure}[h]
\centering
\includegraphics[width=0.98\textwidth]{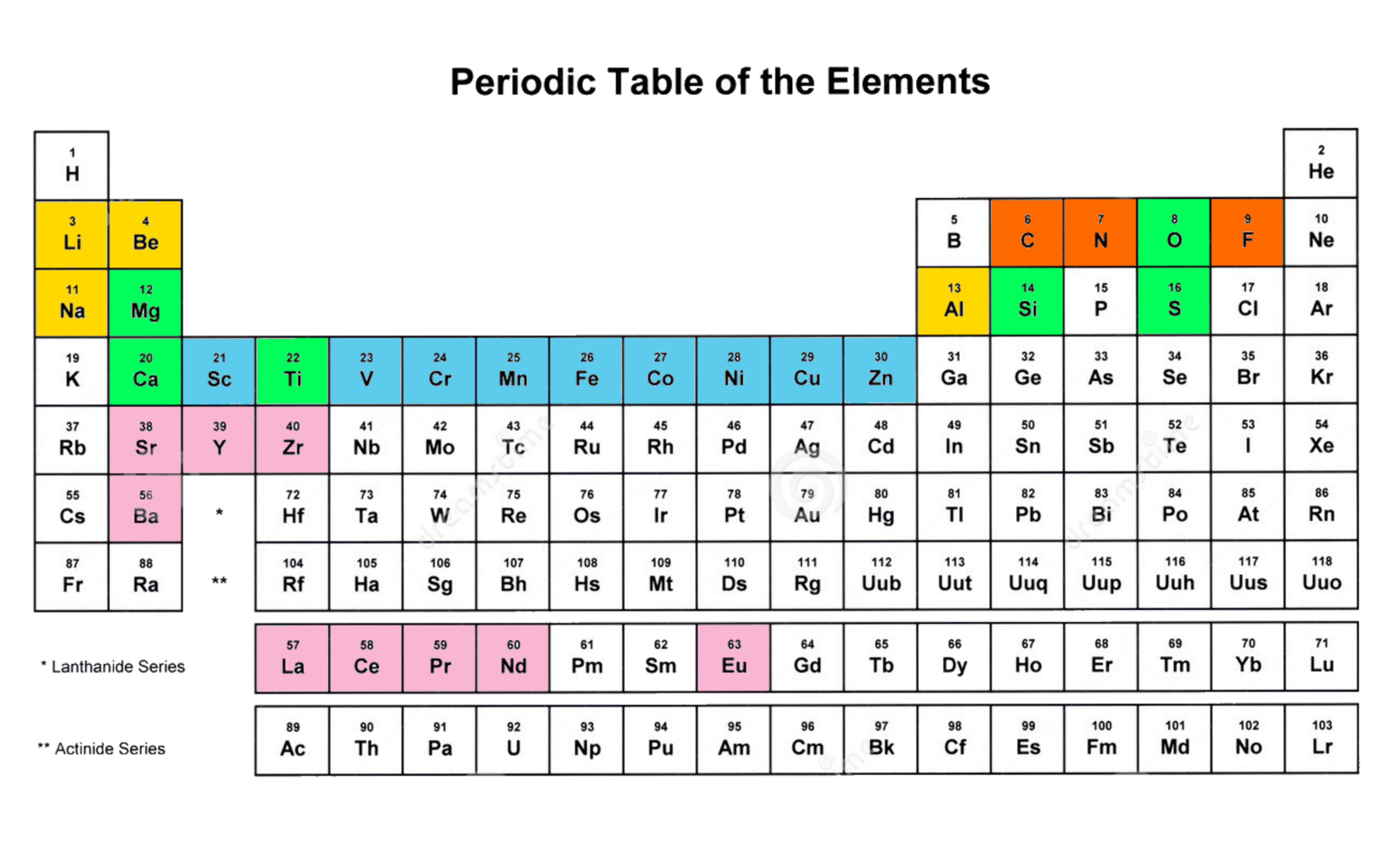}
\caption{The periodic table of elements. The colors highlight the different classes:
Green = $\alpha$-elements, Yellow and Red = light-elements, Blue = Iron-Peak elements
and Pink = $s$- and $r$-process elements. Only the main and well studied
elements are highlighted.}
\label{fig2}
\end{figure}

The name ``$\alpha$-elements'' is due to the fact that these chemical species are formed
through $\alpha$-capture processes on seed nuclei. 
The $\alpha$ particle consists in a nucleus of Helium made of two protons and two neutrons.
The main elements belonging to this family,
i.e. {\bf O, Ne, Mg, Si, S, Ca,} and {\bf Ti} (see green boxes in Figure~\ref{fig2}), 
are progressively built up starting from the burning of He and C \citep[see][]{woosley95}.

% FIGURE 1
%

\begin{figure}[h]
\centering
\includegraphics[width=0.80\textwidth]{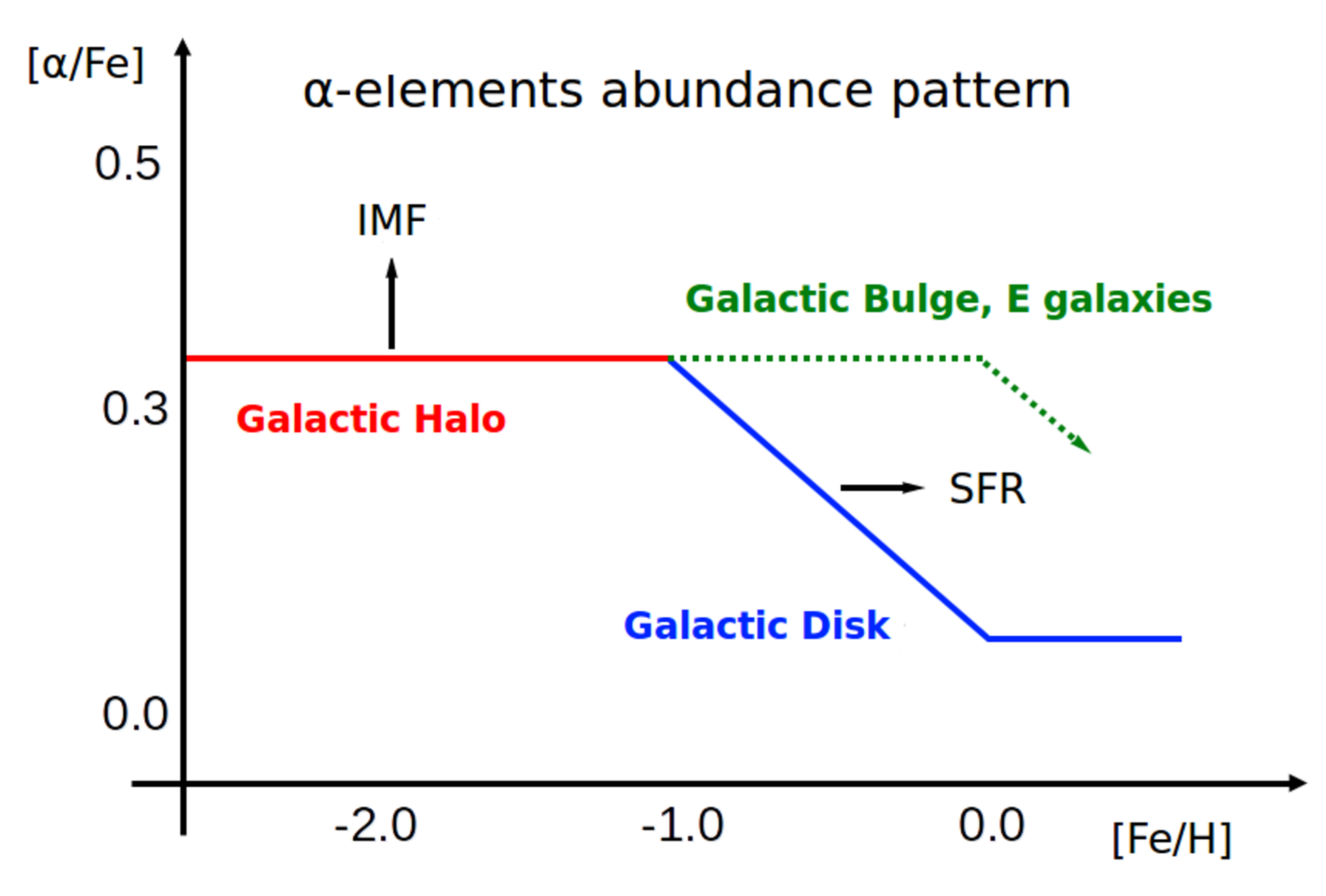}
\caption{The effect of the change of both IMF and SFR on the $\alpha$ abundance
pattern observed in our Galaxy.}
\label{fig1}
\end{figure}

The stellar evolution theory indicates that $\alpha$-elements are mostly synthesized in massive stars.
Indeed, thanks to the very high temperatures reached in the interior of such stars,
the $\alpha$-capture process is able to convert the seed nuclei into heavier elements,
with a copious production of the other even-Z elements in between.
The elements produced by such massive stars are released into the ISM
at the end of the stellar life, during SN explosions.
From the very early investigations of
\cite{aller60} and \cite{wallerstein62} it was found that in
Galactic metal-poor stars the [$\alpha$/Fe] abundance ratio is overabundant
with respect to the solar value. The [$\alpha$/Fe]-[Fe/H] trend in our Galaxy
shows two different regimes (see Figure~\ref{fig1}): for $-1.0 <$ [Fe/H] $< 0.0$ dex,
the [$\alpha$/Fe] ratio increases as the metallicity decreases, reaching a factor
of 2-3 above solar (i.e., [$\alpha$/Fe] $\simeq$ 0.3 dex) at [Fe/H] $\simeq -1$ dex,
while for [Fe/H] $< -1$ dex the [$\alpha$/Fe] ratio remains almost constant.

% FIGURE 2
%

\cite{tinsley79} suggested that the [$\alpha$/Fe] trend is due to the time delay
between the explosions of Type II Supernovae (SNII), which produce
$\alpha$-elements and little amount of iron-peak elements \citep[e.g.][]{arnett78,woosley95},
and Type Ia Supernovae (SNIa), which provide mostly iron-peak elements and little $\alpha$-elements
\citep[e.g.][]{nomoto84,thielemann86}. Thus, after the delay for the onset
of SNIa, the [$\alpha$/Fe] ratio decreases from the value set by the SNII
ejecta. Accordingly, the position of the knee in the [$\alpha$/Fe]-[Fe/H] trend
flags the metallicity reached by the system at the epoch when SNIa
start to dominate. In turn, this depends on the SFR: if it is high, then the
gas reaches larger [Fe/H] before the ejecta of SNIa are fully
mixed into ISM, and the position of the knee in the diagram will be at a
higher [Fe/H] (see black arrow in Figure~\ref{fig1}). On the other hand, a change
of the IMF determines an increase or decrease of the number of high-mass stars formed,
which explode as SNII; this corresponds to different
amount of $\alpha$ elements produced and, therefore, to a different value of
the ``plateau'' in the [$\alpha$/Fe]-[Fe/H] diagram (see black arrow in Figure~\ref{fig1}).
Hence, the study of the [$\alpha$/Fe] trend (and of the chemical
composition of stars, in general) in stellar populations is very
important, since it traces the modes and timescales of the chemical
enrichment process.

Other important light elements such as {\bf He, Li, Be, C, N, F, Na, Al, P,} and {\bf K}
(yellow and red boxes in Figure~\ref{fig2}) may have a fundamental role in defining
the properties of stellar populations in specific evolutionary stages.

Helium is the most abundant among the few chemical elements produced
directly during the primordial nucleosynthesis of the Big Bang
($^{3}$He and $^{4}$He). The most recent determination of the primordial
He mass fraction provides a value Y=0.254$\pm$0.003 \citep{izotov13}.
Also, $^{4}$He is produced in stars with masses larger than $\sim$0.08
$M_{\odot}$ through the hydrogen-burning chains.

Lithium is produced mainly during the Big Bang nucleosynthesis,
in the form of $^{7}$Li. In metal-poor ([Fe/H]$<$-1.5 dex) halo dwarf
stars the surface abundance of lithium turns out to be constant
regardless of the metallicity and the temperature, with a value
A(Li)$\sim$2.2 \citep[the so-called Spite-Plateau,][]{spite82}.
Additionally, $^{7}$Li can be produced in AGB stars (with mass
between 5 and 8 $M_{\odot}$) through the Cameron-Fowler mechanism
and perhaps in novae and SNII \citep[see][]{romano01}.

The only stable isotope of beryllium, $^{9}$Be, is a pure product of cosmic-ray
spallation of heavy nuclei (mostly CNO) in the interstellar medium
\citep{reeves70} with negligible/null contribution from the Big Bang
and stellar nucleosynthesis. Recently, the measurement of the Be abundance
has been proposed as a ``clock'' to date the oldest stars \citep{pasquini04}.

Carbon and nitrogen, together with oxygen, are the most abundant metals in the Universe.
During the Post main Sequence evolution, the stellar surface abundances of carbon and nitrogen can be significantly altered by 
the dredge-up of material processed by the CNO cycle, with a resulting 
[C/Fe] depletion and a corresponding [N/Fe] enhancement, while O abundance remains almost unaffected. 
Additional extra-mixing processes are also invoked to explain some extreme [C/Fe] depletions and [N/Fe] enhancements, 
although the precise physical mechanisms responsible for their occurrence are still unknown and/or debated.
Such modifications of the {\it ab initio} C and N abundances induced by stellar evolution complicate
the interpretation of the observed trends in different stellar populations and galactic environments.

From a nucleosynthetic perspective, fluorine is a very interesting element, and its cosmic origin is still to be understood.
Three main production mechanisms for $^{19}$F, the only stable isotope of fluorine, have been proposed. 
1) neutrino nucleosynthesis in SNII. The core collapse of a massive star,
following a SNII explosion, leads to a prodigious neutrino flux.
In spite of the small cross sections, the large amount of neutrinos gives rise to a significant
spallation of $^{20}$Ne to $^{19}$F \citep{woosley88} in the overlying (neon-rich) shells of the core.
2) Thermal-pulsing asymptotic giant branch stars. The production of fluorine starts from $^{14}$N burning
\citep{forestini92,jorissen92,gallino10,abia11,kobayashi11},
then it is transported up to the surface by the third dredge-up. Fluorine production in AGB stars is expected
to be accompanied by the slow neutron-capture nucleosynthesis (the $s$-process).
It has been demonstrated observationally that AGB stars produce fluorine,
see for example \cite{jorissen92} and \cite{abia11}.
3) Wolf-Rayet stars. \cite{meynet93} and \cite{meynet96,meynet00} suggested that Wolf-Rayet (W-R)
stars might contribute to the Galactic fluorine budget.
$^{19}$F is produced in the convective cores of W-R stars during the core He-burning phase. 
Using a semi-analytic multizone chemical evolution model,
\cite{renda04} showed for the first time the impact of the AGB and W-R star
contributions to the Galactic chemical evolution of fluorine.
They showed that the production was dominant in the early Universe
and that the contribution of AGB stars successively grows.
Based on old yields and nonrotating models, they also showed that the contribution
of W-R stars is significant for solar and supersolar metallicities,
increasing the [F/O] ratio by a factor of two at solar metallicities.
Their conclusion was that all three production sites are needed to explain
the Galactic chemical evolution of fluorine for a range of metallicities.
The abundance of fluorine in stars is difficult to measure because of
a paucity of suitable spectral lines and a systematic,
massive study of its abundance in different stellar populations and environment is still lacking. 
Highly ionized F V and F VI lines in the UV have been used by \cite{werner05} in extremely hot post-AGB stars,
and a handful of F I lines between 6800-7800 $\rm \mathring{A}$ have been used in extreme helium stars
and R Coronae Borealis stars \citep{pandey06,pandey08}.
All other studies have used the HF molecular lines in the K-band and mostly the HF line at 23358 $\rm \mathring{A}$.

Other important odd-Z elements are Na and Al, which have been studied in several astrophysical
environments. Both in field and cluster stars the Na and Al abundances showed a clear star-to-star dispersion.
From the analysis of a sample of field stars in the interval $-3 <$ [Fe/H] $< -1$ dex, 
\cite{pilachowski96} found a small deficiency of [Na/Fe] with values around $-0.2/-0.3$ dex.
\cite{reddy06} found a hint for an increase of [Na/Fe] with increasing [Fe/H] in the metallicity range
$-1.0 <$ [Fe/H] $< -0.6$ dex, followed by a decrease towards solar values for $-0.6 <$ [Fe/H] $< 0$ dex.
On the contrary, in globular cluster (GCs) stars a very high dispersion was observed with
abundances spanning a range between slightly subsolar [Na/Fe] up to 1.0 dex.
A similar behaviour was observed for Al in field stars, which show an increase
of the [Al/Fe] abundances as [Fe/H] increases, the trend is not as strong for Na.
However, at very low metallicities ([Fe/H] $< -2.8$ dex), the relation [Al/Fe] versus [Fe/H]
seems to become rather flat \citep{andrievsky08}.
On the contrary, in GCs the [Al/Fe] behaviour well resembles that of [Na/Fe],
with abundances spanning up to $1.0$ dex.
So the Na and Al abundances in GCs appear to be correlated,
and they are correlated with N enhancements and O depletions.
These abundance patterns have been interpreted either as evidence of internal nucleosynthesis
and mixing operating in individual stars or, alternatively, as characteristic of a dispersion
in the composition of the material out of which the stars formed
\citep[see][and references therein for details]{kraft94,shetrone96a,shetrone96b,kraft97}.
Currently, the formation channels of the Na and Al remain controversial.
Na should likely form by means of hydrostatic carbon and hydrogen burning through the NeNa cycle, 
while Al should form by means of hydrostatic carbon and neon burning and during hydrogen burning through the MgAl chain.
In the NeNa cycle the $^{20}$Ne is progressively converted in $^{23}$Na by several proton captures,
while in the MgAl chains the $^{24}$Mg is finally converted in $^{26}$Al
\citep{denisenkov89,langer93,cavallo96,prantzos07,straniero13}.
However, several other formation channels have been proposed so far,
thus it is difficult to use Na and Al as probes of Galactic chemical evolution
until their nucleosynthesis is better constrained.

Phosphorus has a single stable isotope $^{31}$P, and its most likely sites of production
are O and Ne burning shells in the late stages of the evolution of massive stars, which end up as SNII.
The production mechanism probably occurs via neutron capture,
as it is for the parent nuclei $^{29}$Si and $^{30}$Si.
According to \cite{woosley95},
there is no significant P production during the explosive phases.
Recently, \cite{caffau11} analyzed the high excitation IR PI lines
at 1051-1068 nm in a sample of twenty Galactic stars,
finding a systematic increase of [P/Fe] for decreasing [Fe/H];
[P/Fe] is close to zero for solar metallicity stars.
\cite{cescutti12} compared the observed results to a model of the chemical evolution of P in the Milky Way,
adopting different sets of yields.
They conclude that P is formed mainly in massive stars (core-collapse SNe) and that the yields
of P available in the literature are all too low and have to be artificially increased by a factor
of 3 to satisfactory reproduce the observed data, including the solar photospheric P.
From their best model they predict a ``plateau'' at [P/Fe] $\sim$ +0.5 dex in
the metal poor metallicity range ($-3.0 <$ [Fe/H] $< -1.0$ dex), if normal yields from SNII are adopted,
and at [P/Fe] = $+0.2$ to $+0.3$ dex if hypernova yields are assumed.
However, previous chemical evolution models for P \citep{fenner05,kobayashi06}
suggested a flat trend in the range $-2.0 <$ [Fe/H] $< 0.0$ dex.

Potassium is mainly produced by a combination of hydrostatic oxygen 
shell burning and explosive oxygen burning in proportions that vary 
depending on the stellar mass. Measurements of [K/Fe] in the Milky Way 
suggest an enhanced value in metal-poor stars and a decrease toward 
the solar value for stars more metal-rich than $-0.5$ dex
\citep{reddy03,zhang06,andrievsky10}.

\section{Iron-peak elements}

The elements with a nucleosythesis closely linked to that of iron are usually
tagged as ``iron-peak elements''. This family includes {\bf Sc, V, Cr, Mn, Fe, Co, Ni, Cu,} and {\bf Zn}
(blue boxes in Figure~\ref{fig2}).
These elements are mostly produced during the explosive nucleosynthesis associated to SNIa,
but some contributions from the weak $s$-processes in massive stars (e.g. for Cu) and from
SNII have been also proposed (e.g. for Fe, Cu and Zn).

Iron is probably the best known chemical element and it has been extensively studied in almost all astrophysical contexts.
This is due to the huge number of available atomic transitions over the whole spectral range,
and at all the metallicities, which normally allows one to obtain a very precise estimate of the iron abundance.
Moreover, as the majority (but not all) of iron-peak elements, iron has only
two main formation channels: SNII and SNIa.
After the collapse of the core and the subsequent explosion, the SNII are able to
release in the ISM a few tenths of solar masses of iron.
Conversely, SNIa mainly release iron and other iron-peak elements.
Thus, there are no nuclear reactions able to alter the iron abundance of a star.
Because of this, all abundance ratios are usually expressed in terms of iron content. 

Up to now, only few studies have investigated the behaviour of Scandium and Vanadium in our Galaxy.
Although \cite{zhao90a,zhao90b} have found that the [Sc/Fe] is slightly supersolar 
in metal-poor dwarfs, several other studies found no evidence for a deviation from
[Sc/Fe] = 0.0 in field stars \citep{peterson90,gratton91,mcwilliam95b}.
The same applies for V for which abundances fully compatible with a solar [V/Fe] $\sim$ 0.0 dex have been
measured by \cite{gratton91}.

The case of Manganese is quite different.
From [Fe/H] = $0.0$ to $-1.0$ dex, the [Mn/Fe] ratios are sub-solar in a manner that mirrors
the $\alpha$-element overabundances, and in the interval [Fe/H] = $-1.0$ to $-2.5$ dex,
[Mn/Fe] remains roughly constant around $-0.4$ dex.
Thus the trend of [Mn/Fe] with [Fe/H] is similar, but opposite, to that of [$\alpha$/Fe].
Moreover, \cite{mcwilliam95a} discovered that below [Fe/H] $\sim -2.5$ dex,
the [Mn/Fe] ratio decreases steadily with decreasing [Fe/H], suggesting that
Mn is among those elements whose yields depend on the metallicity of the parent stars.
This is also supported by the results of \cite{mcwilliam03}
who compared the [Mn/Fe] versus [Fe/H] relation in the Galactic bulge,
in the solar neighbourhood and in the Sagittarius dwarf spheroidal galaxy
concluding that the Mn is produced by both SNIa and SNII in a metallicity-dependent way.
Very similar to that of [Mn/Fe] is the behaviour of [Cr/Fe], which is found to rise as the metallicity
increase. Also for this element a metallicity-dependent yield by both SN types has been proposed
although the main production should occur during incomplete explosive Si burning
\citep{woosley95,limongi03}.

[Ni/Fe] is usually found to be close to the solar value at different metallicities,
suggesting that the origin of Ni is strictly linked to that of Fe from both SN types.

[Co/Fe] is found to increase as metallicity decreases, in particular,
below [Fe/H] $\sim -2.5$ dex \cite{mcwilliam95b} found a steep rise of [Co/Fe] up to $1$ dex.
This likely suggests that the main contribution in the production of Co comes from SNII.
However, the explosion energies may have a big impact on the amount of Co produced as suggested
by \cite{umeda05}.

Cu and Zn abundances are difficult to obtain due to the paucity of available atomic transitions.
In our Galaxy \cite{sneden88} and \cite{sneden91} studied the abundances
of Cu and Zn as a function of metallicity and discovered that [Cu/Fe] decreases
linearly with declining metallicity, while Zn is roughly constant at [Zn/Fe] = $0.0$ dex
for all metallicities \citep[see also][]{bensby05,nissen07}.
However, [Zn/Fe] rises steeply to $\sim$ +0.5 at the lowest metallicities \citep{cayrel04}.
\cite{sneden91} suggested that the nucleosynthesis of Cu may be predominantly due to ``weak $s$-process''
in the cores of massive stars, with a small contribution from explosive burning in SNII.
The $s$-process is one of the possible channels through which the neutron-capture process occur
(see the next section for details).
However, \cite{matteucci93} suggested that the main production of Cu and Zn occurs in SNIa.
A more recent comprehensive study by \cite{romano07} have shown that Cu should be mainly
synthesized in massive stars, during core-helium and carbon-shell hydrostatic burnings,
as well as in explosive complete Ne burning \citep[e.g.][]{woosley95,limongi03}.

\section{Neutron-capture elements}

The ``neutron-capture'' elements are characterized by a proton number Z larger than 30
(pink boxes in Figure~\ref{fig2}) and form through subsequent neutron captures on a seed nucleus,
in general an iron-peak element.
These processes, if compared to the radiative decay timescales of the unstable nuclei,
can be slower or faster, thus resulting in the slow ($s$-process) or rapid ($r$-process)
neutron captures.
The $s$-process occurs in two different branches: the $weak$-$s$ process is responsible for the formation
of nuclei with an atomic number 29 $<$ Z $<$ 40 and the $main$-$s$ process can produce heavier nuclei up to Z = 84.
The $s$-process captures through the $weak$ channel mainly occur in massive stars
(M $\gtrsim$ 13 $\Msun$) while the $main$ channel dominates the nucleosynthesis during
the thermal pulse stage of low-mass (1-3 M$_{\odot}$) AGB stars at neutron densities of 10$^{7}$-10$^{9}$ cm$^{-3}$
\citep[e.g. see][]{busso95,lambert95}.
Here the main source of neutrons is expected be the $^{13}C(\alpha,n)^{16}O$ reaction.
The neutron capture process can also happen in a very fast way, thus resulting in the so called $r$-process.
This makes use of the large number of neutrons available and occurs in unstable
neutron-rich and very radioactive regions \citep{sneden03,cowan04,arnould07}.
The most plausible sites for this process are the core-collapse SNe during the explosion,
in which a density of up to $\sim$10$^{25}$ neutrons cm$^{-3}$ can be achieved.
The typical $s$-process elements observable in the stars
are {\bf Sr, Y, Zr, Ba,} and {\bf La}, while the main chemical element produced through $r$-process is {\bf Eu}.
Generally speaking, the behaviour of such groups in the Galaxy can be considered roughly specular.
At low metallicities the amount of $r$-elements in stars is clearly supersolar, reaching values up to
$+1.5$ dex for [Eu/Fe] \citep[see][]{barklem05}.
With the increase of [Fe/H] these abundances decline reaching a [Eu/Fe] $\sim$ 0.0 dex at solar metallicity.
This behaviour closely resembles that of $\alpha$-elements
thus suggesting that $r$-elements may have a common origin mainly due to the contribution of SNII.
On the other hand, the production of $s$-elements may occur in a metallicity-dependent way, mainly due to 
neutron-captures processes during the thermal pulses of AGB stars.
In fact, at low metallicities a large number of neutrons is available for a limited amount of seed nuclei,
in general iron-peak elements.
This would result in a larger number of captures, thus favoring the production
of $s$-elements belonging to the group of Ba and La (A $>$ 130).
With the increase of metallicity, i.e. of the available seed nuclei, the number of neutron captures per atom
decreases, thus favoring the formation of $s$-elements belonging to the group of Sr, Y, and Zr (A $\sim$ 90).
In our Galaxy the [$s$/Fe] ratio generally increases with metallicity thanks to the rising contribution of
AGB stars, and the proportion between heavy-$s$ and light-$s$ progressively changes.
In any case, it is important to underline that it is difficult to tightly constrain the chemical evolution history
of neutron capture elements for two main reasons: (1) for almost all of these elements a high dispersion of
the abundances is observed, regardless of the metallicity (this makes difficult a clean detection of any possible trend 
able to constrain the evolution), and (2) ``pure'' $s$ or $r$-element do not exist,
in the sense that secondary channels have a non-negligible contribution to the formation of each of these elements
\citep[see][]{arlandini99}.

%%%%%%%%%%%%%%%%%%%%%%%%%%%%%%%%%%%%%%%%%%%%%%%%%%%%%%%%%%%%%%

%% file: c2a/c2b.tex
% CAPITOLO 2

\chapter{Globular Clusters: Evolutionary Sequences and Chemical Composition}

\label{c2}

GCs are roughly spherical ensambles of stars, which are tightly bound by gravity.
Their name comes from the Latin word {\it globulus} which means ``small sphere''.
Our Galaxy hosts more than $\sim$ 150 GCs, spanning a wide range of metallicities (over $\sim$ 2 dex),
but a small range of ages: they are mostly older than 10 Gyrs.

\section{Evolutionary sequences}

During their lifetimes, stars evolve through different stages characterized by different
thermonuclear reactions and processes, that, in turn, determine the physical conditions in their atmospheres.
These have fundamental implications on the global parameters of the stars like
luminosity and surface temperature.

\begin{figure}[h]
\includegraphics[width=0.99\textwidth]{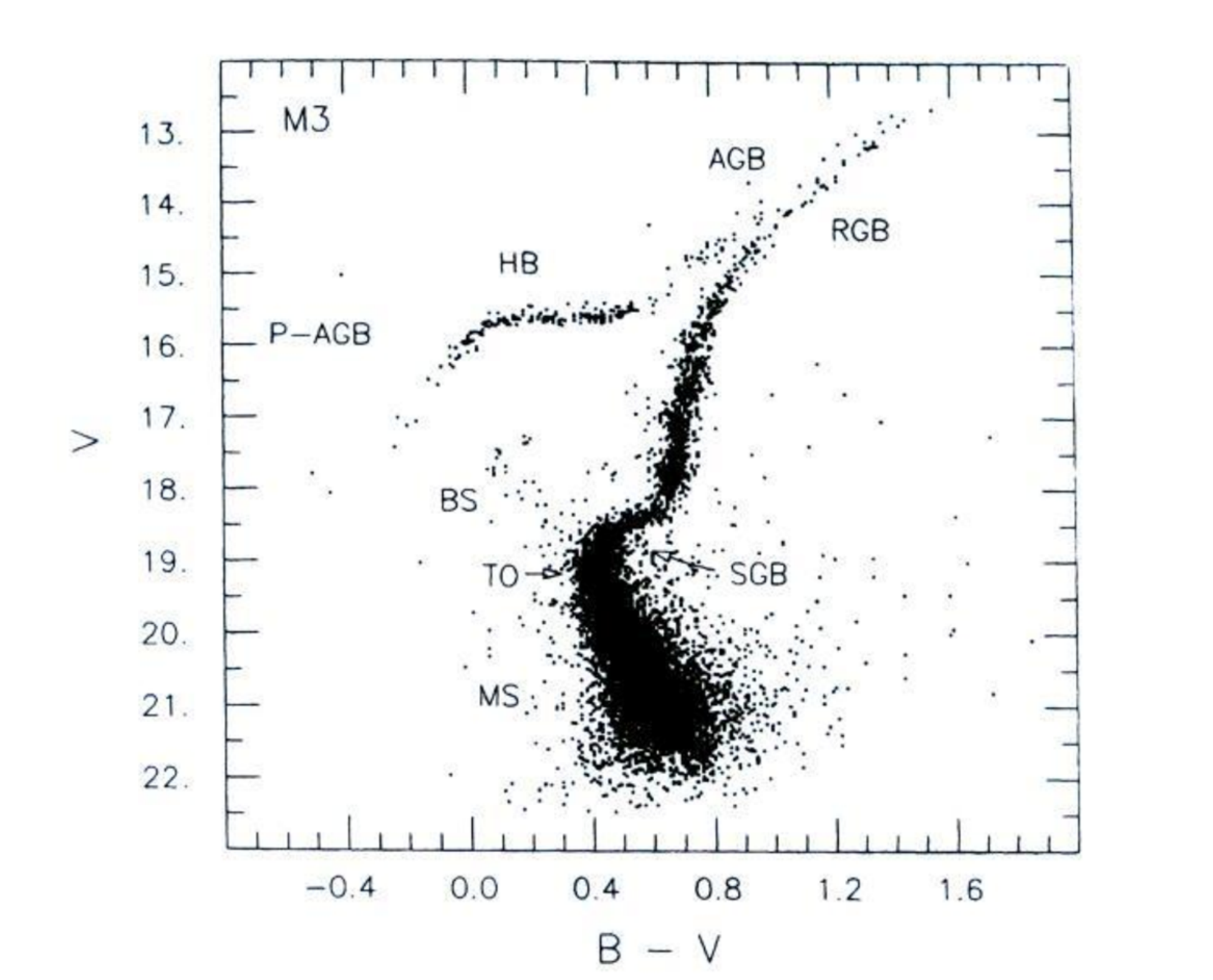}
\caption{The CMD of the GC M3 from \cite{buonanno86}.
The main evolutive stages of the stars are also higlighted.}
\label{fig3}
\end{figure}

GSs are old ($>$ 10 Gyr), hence they host stars with masses $\lesssim$ 1 M$_{\odot}$.
The main stages of the stellar evolution which have been targeted by an invaluable number of
studies in the last decades are Main Sequence, Sub Giant Branch, Red Giant Branch,
Horizontal Branch and Asymptotic Giant Branch.
In the following we will summarize the main properties of each evolutionary sequence as
observed in the color magnitude diagram of GCs (see Figure~\ref{fig3}).

% FIGURE 3
%

\begin{itemize}

\item Main Sequence (MS) -
The MS is the evolutionary stage where stars spend most of their lifetime (more than 90$\%$).
This stage is characterized by the thermonuclear burning of hydrogen in the core,
through which hydrogen is converted in helium.
This can happen with two different reactions: the ``proton-proton chain'', which mainly occurs
in stars with a mass lower than 1.3 M$_{\odot}$ where a temperature larger than $\sim$ 10$^{7}$ K
is reached in the core, and the so-called ``CNO-cycle'' which occurs in stars with a M $>$ 1.3 M$_{\odot}$
with a temperature greater than $\sim$ 15 $\times$ 10$^{7}$ K.
For low mass stars this stage is very long (a few Gyr) since the time spent in MS is an inverse function of the stellar mass
(this is because, in order to sustain the stellar structure, high-mass stars burn hydrogen at a very high rate, while low-mass
stars can burn their fuel reservoir at much lower rates).
In any case, each star, sooner or later, will finish its hydrogen reservoir in the core.
When this happens the thermonuclear reactions in the core end.

\item Sub Giant Branch (SGB) -
At the end of the MS, when the core thermonuclear reactions are going to end,
hydrogen is ignited in a thick shell surrounding the stellar nucleus,
where the temperature are high enough for this burning to occur.
As time goes on, the shell moves outwards, becoming progressively thinner.
This phase lasts from a few tens to several hundreds Myr,
depending on the mass and metallicity of the star,
and it is characterized by an important variation in the surface temperature,
while the luminosity does not change appreciably.

\item Red Giant Branch (RGB) -
After the SGB phase, a phase begins in which the temperature
does not change much, while the luminosity increases significantly.
The RGB phase lasts only a small fraction ($\sim$ 10$\%$) of the MS time.
During this phase, the hydrogen burning shell ignited during the SGB phase reachs a stable rate
and progressively determines an increase of the star luminosity.
In this phase stars are partially convective.
The progressive penetration of the convection produces the first dredge-up (FDU),
in which the outer convection zone progressively engulfs deeper regions,
dredging for the first time to the surface matter which has been partially
processed through H-burning during the MS. The chemical elements affected by
the FDU are essentially C, N, Li, and He. Both He and N abundances increase, whilst Li and C decreases.
However, the FDU is not the only important event that occurs during the RGB,
in fact, when the incoming H-burning shell crosses the chemical discontinuity left
in the deep layers by the convective envelope after the FDU, the stellar luminosity temporarily
decreases, and then stars to increase again.
The macroscopic effect of these events produce the so-called RGB-Bump,
which appears as a ``local'' (at a given magnitude) excess of stars in the RGB luminosity function of GCs
\citep[see e.g.][]{fusipecci90,ferraro00,valenti04}.
It is important to recall also that, during the RGB, an important mass-loss phase occurs and
can remove up to 15-20$\%$ (i.e. 0.1-0.2 M$_{\odot}$) of the mass held at the turn-off.
The RGB phase ends when helium in the stellar core is ignited with a thermal runaway,
the so-called ``Helium-flash'', which releases a huge quantity of energy.
This energy, however, is not released as luminosity but it is used to remove the degeneracy
of the helium core. From this moment the star is able to burn helium into carbon through
the triple-$\alpha$ process and it moves toward the next evolutive stage,
i.e. the HB phase.

\item Horizontal Branch (HB) -
The HB phase begins when the He burning process starts into the non-degenerate core.
In this phase, which lasts roughly $\sim$ 1$\%$ of the time spent in the MS,
stars are powered by helium burning in the core and by hydrogen burning
in a shell surrounding the core.
Each star begins its He-burning phase starting from a ``reference'' sequence in the CMD:
the so-called Zero Age Horizontal Branch (ZAHB).
It potentially extends from $\sim$ 4000 up to 40000 K.
The position occupied by the star on the ZAHB is determined by its surface temperature
and it depends primarily on the star mass.
Generally speaking, more massive stars tend to occupy the red part of the HB,
while less massive stars lie in the bluest part.
The mass with which stars reach the HB is, of course, a function of the global parameters
of the stellar population, like age and metallicity, but it also depends on the mass-loss occurred during the RGB phase.
While the metallicity is identified as the dominant parameter
determining the HB morphology, several studies in the GCs of our Galaxy suggest
that additional parameters are required \citep[see e.g.][]{ferraro97b,ferraro98}. 
This is the so called ``second parameter problem'' and several options have been proposed,
e.g. Helium content, mass-loss, etc.
The next evolutive stage of stars with masses larger than 0.55-0.60 M$_{\odot}$ is the AGB phase,
this is due to the fact that these stars have at least 0.1 M$_{\odot}$ to ignite the hydrogen and helium
burning in the layers surrounding the nucleus of carbon now off.
On the contrary, the less massive HB stars, mostly located in the bluest part the HB,
evolve directly through the White Dwarf (WD) cooling sequence.

\item Asymptotic Giant Branch (AGB) -
At the end of the HB phase, when thermonuclear reactions in the core stop,
the He burning shifts in a shell surrounding the C-O core.
This flags the beginning of the AGB phase, which is characterized by a slightly decrease
of the surface temperature while the luminosity sensibly increases.
The AGB phase is the last evolutive stage of intermediate and low-mass stars before they end their lives
as WD and is a very rapid phase which typically lasts a few percent of the time spend in the HB, i.e. some Myr.
The AGB phase is usually divided in two main stages: the ``early'' AGB (E-AGB) and the ``thermal pulses'' AGB (TP-AGB).
The E-AGB begins when the He burning shell becomes active, the energy released pushes outwards the
overlying H burning shell, which expands and became inactive due to the temperature drop.
However, soon after its ignition the He burning shell loses efficiency and switch off.
When this happen the contraction of the stellar atmosphere and the subsequent increase of temperature
make the H burning shell active.
The temporary stop of the He burning shell flags the beginning of AGB phase,
in which the He and H shells become alternately active.
Only in AGB stars with progenitor masses $>$ 1.2-1.3 M$_{\odot}$ (which are all evolved
as WDs or died in Galactic GCs) the TP phase is accompanied by additional dredge-up processes.

\end{itemize}

\section{Overall chemistry}

GCs have been traditionally assumed to be SSPs, i.e. made of single, coeval and chemically homogeneous stars.

However, soon after the very first spectroscopic observations in the late seventies, 
it became clear that the chemistry of GCs is more complex.
Thanks to the huge technological development of spectroscopic facilities with
high spectral resolution and multi-object capabilities in the last two decades, 
large samples of giant stars in GCs have been observed and
abundances of several chemical elements have been obtained.
These measurements indicate that different chemical elements show different abundance spreads.
In particular, these can exceed a few tenths of dex in the cases of C, N, O, Na, Mg and Al, 
suggesting that some self-enrichment process should have occurred 
during the first stages of the GC evolution, on typical timescales of a few hundreds Myrs.
Such a behaviour is not peculiar to our Galactic environment
but it has been also found in the GCs of M31, Large Magellanic Cloud (LMC)
and Fornax dwarf galaxy \citep{colucci09,johnson06,mucciarelli09,letarte06}.

\subsection{Metallicity distribution}

At the very early stage of the study of GC systems one of the main issues was
the derivation of a proper metallicity scale.
The first attempt to derive homogeneous metallicities for GCs
was performed by \cite{zinn84} who observed 60 Galactic clusters obtaining
values of [Fe/H] ranging from $-2.58$ to $+0.24$ dex.
This scale was refined by \cite{carretta97} using for the first time medium-resolution
spectra collected with the CASPEC spectrograph mounted at the 3.6m ESO telescope and based
on 160 stars observed in 24 Galactic GCs \citep[see also][]{origlia97}.
They have found no hints of intrinsic iron dispersion in the studied sample.
\cite{kraft03} derived a new metallicity scale by using FeII lines in a sample of 149 stars in 11 GCs.
They argued that a reliable derivation of [Fe/H] should be based
on FeII lines instead of FeI in order to avoid possible effects related
to Local Thermodynamic Equilibrum (LTE) departures, which mainly affect the abundances
derived from neutral lines. Also in this case, the authors found no metallicity spread in the analysed sample.
The most recent attempt to refine the previous metallicity scale for GCs was performed by
\cite{carretta09b}, who have used medium- and high-resolution spectra collected with
FLAMES@VLT to study a sample of $\sim$ 2000 stars in 19 Galactic GCs.
They have not found any noticeable iron spread in the studied sample,
the intrinsic scatter being less than 0.05 dex for the majority of GCs.
Only for some of the most massive GCs they found some evidence of iron spread and of a mild correlation
with the cluster mass.

However, there are at least two major exceptions where a wide metallicity distribution has been found.

\begin{itemize}
% 1
\item $\omega$ Centauri $-$ This object is the most massive (M $>$ 10$^{6}$ M$_{\odot}$)
and most luminous (M$_{V}$ $< -10$ mag) stellar system of the Milky Way.
It has a complex color-magnitude morphology, which has been interpreted as due to
the presence of several stellar populations \citep{lee99,pancino00,ferraro04,bedin04,sollima05,villanova07,
pancino11a,pancino11b}. Nowadays, from the analysis of the SGB morphology,
which can help in the identification of different components, at least 6 different stellar
populations have been clearly detected \citep{villanova14}.
The metallicity distribution of the system is very wide, ranging from a [Fe/H] $\sim -2.0$ to
$\sim -0.2$ dex. Of course, such a wide mettalicity distribution coupled with the existence of several stellar
populations with different properties makes $\omega$ Centauri an ``anomalous GC''.
In fact, due to some similarities with the satellites of our Galaxy it was suggested that
$\omega$ Centauri could be the remnant of a dwarf spheroidal accreted by the Milky Way in the past.
However, this complex stellar system shows several peculiarities which may challenge also this
conclusion. In fact, \cite{origlia03} and \cite{gratton11} have measured enhanced $\alpha$-element abundances
in the most metal-rich population of the cluser, which seems incompatible with the deficiency of
$\alpha$-elements usually observed in the metal-rich stars of dwarf spheroidals.

% 2
\item Terzan 5 $-$ This system was originally catalogued as a GC and is located in the Galactic Bulge
in a very extincted region where the $E(B-V)$ ranges from 2.15 to 2.82 mag \citep{massari12}.
For this reason, accurate photometric measurements have been possible only by means of infrared
observations in J, H and K bands. \cite{ferraro09} by using
the Multi-Conjugate Adaptive Optics Demonstrator (MAD) mounted at the VLT have detected the presence
of two distinct Red Clumps (RC) belonging to two different populations.
Subsequent spectroscopic observations have demonstrated that these two populations have very different
metal content: an average [Fe/H] = $-0.25 \pm 0.07$ dex for the metal-poor population and
[Fe/H] = $+0.27 \pm 0.04$ dex for the metal-rich one \citep{origlia11}.
Moreover, Terzan 5 does not show the typical anticorrelations observed in the majority of GCs.
Recently, also another very metal-poor component with an average [Fe/H] = $-0.8$ dex
has been detected \citep{origlia13,massari14a,massari14b}, thus leading to an iron distribution wider than 1 dex.
The chemical patterns observed in Terzan 5, when interpreted in the context of a self-enrichment scenario,
require that the proto-Terzan5 system was originally more massive in order to retain
the high-velocity SN ejecta \citep{ferraro09,lanzoni10}.
For these reasons, Terzan 5 is no longer considered as a genuine GC but it was suggested to be
the remnant of one of the pristine building-blocks from which the Galactic Bulge was born \citep{immeli04}.
\end{itemize}

% Minor systems
In a few other GCs some iron spreads, at a level of a few tenths of a de, have been also claimed.

\begin{itemize}

% 3
\item M22 $-$ This is a metal-poor GC which is suspected to have an intrinsic iron spread.
This suspect was originally based on the large colour spread of the RGB in the CMD \citep{monaco04}.
However, M22 resides in a region deeply affected by differential reddening which makes
difficult to properly assess whether the observed colour broadening is due
or not to an intrinsic metal spread \citep{monaco04}.
From a spectroscopic point of view, divergent results have been obtained, all being based, however,
on small samples \citep{cohen81,pilachowski82,gratton82}.
Recently, \cite{marino09,marino11b} have analysed high-resolution spectra of 35 giant stars finding a $\sim$0.4 dex
wide [Fe/H] distribution.
Also, they highlighted that M22 is composed by two groups of stars,
characterized by different metallicity, CNO and $s$-process element abundances:
the first group with an average iron abundance of [Fe/H] = $-1.82$ dex,
[(C+N+O)/Fe] = $+0.28$ dex and solar-scaled [s/Fe] abundance ratios,
while the second group has [Fe/H] = $-1.67$ dex,
[(C+N+O)/Fe] = $+0.41$ dex and an enhancement of $+0.3/+0.4$ dex for the [s/Fe] ratios.
As a reasonable explanation the authors have suggested that in the past M22
was massive enough to retain the ejecta of SNe and AGB stars,
being these stars the main site of $s$-process element production.

% 4
\item M54 $-$ This GC is a peculiar system which is located in the center of
the Sagittarius dwarf spheroidal, the remnant of a galaxy that is going
to be disrupted by the interaction with the Milky Way \citep{ibata94,bellazzini99}.
M54 has a complex CMD morphology, which is the result of the presence of two main components:
the metal-poor stars associated with the Sagittarius dwarf galaxy,
and those belonging to the GC \citep{siegel07}.
From the analysis of 76 RGB stars associated to the original population of the cluster,
\cite{carretta10b} have detected a small iron spread of $\sim$ 0.19 dex and the presence of
the Na-O anticorrelation. A comparable, but slightly smaller, spread of $\sim$ 0.15 dex was also
found by \cite{mucciarelli14} from the anaysis of a sample of 51 stars belonging to the cluster.
In any case, the presence of Na-O anticorrelation is thought to be a genuine characteristic of GCs,
which tend to exclude the idea that M54 is the remnant of the nucleus of the Sagittarius dwarf galaxy.
This idea was supported also by the results of \cite{bellazzini08}, who suggested that the current position
of M54 is due to the decay of the original orbit due to dynamical friction.
A possible speculative explanation has been proposed by \cite{carretta10b} who
suggested that M54 is an anomalous cluster who have experienced a prolonged period of
star formation occurred at high rate if compared to normal GCs.

% 5
\item NGC3201 $-$ Among the anomalous GC candidates, NGC3201 is a controversial case
because different analyses provide conflicting results about its level of iron homogeneity.
\cite{gonzalez98} first analysed a sample of CTIO high-resolution spectra of 18 cluster stars,
finding a large iron variations ($\Delta$[Fe/H] $\sim$ 0.4 dex).
Further analyses by \cite{carretta09c} and \cite{munoz13}
based on high-resolution, high signal-to-noise ratio
(S/N) spectra (FLAMES@VLT and MIKE@Magellan, respectively),
do not highlight similar spreads, ruling out large star-to-star variations.
However, both studies are based on small star samples (13 and 8 respectively),
so they may suffer from low statistics.
On the other hand, \cite{simmerer13} analyzed UVES@FLAMES and MIKE@Magellan high-resolution spectra of 24 giant stars,
revealing a metallicity distribution as large as 0.4 dex (not explainable within the uncertainties)
and with an evident metal-poor tail (5 out of 24 stars).
This iron spread, qualitatively similar to that observed in M22 by \cite{marino09,marino11b},
would make NGC3201 the least massive GC ($\sim$ 1.1 $\times$ 10$^{5}$ $\Msun$; \citealt{mclaughlin05}) with
evidence of SN ejecta retention.

% 6
\item M2 $-$ This massive system is characterized by a complex CMD morphology
which flags the presence of multiple RGB and SGB components \citep{piotto12,milone15}.
In particular, M2 is known to host a second, redder, RGB interpreted as due to
the presence of a population with different C and N abundances \citep{lardo12}.
\cite{lardo12} have also confirmed that two C-enhanced stars previously detected by \cite{smith90}
belong to this peculiar RGB.
In a subsequent analysis, \cite{lardo13} have also detected a peculiar pattern for
the $s$-process elements of the second RGB, in particular for what concerns Sr and Ba.
Finally, \cite{yong14} have found that M2 hosts at least three populations with different
average metallicities ([Fe/H] $\sim$ $-1.7$, $-1.5$ and $-1.0$ dex) and
the star-to-star variations are not limited to iron but are also
detected for both $s$- and $r$-process elements.
Hence, also for M2, a firm conclusion about its star formation history
and evolution cannot be firmly drawn.

% 7
\item NGC5286 $-$ This cluster has been recently investigated by \cite{marino15} who have analysed
a sample of 62 RGB stars by using FLAMES@VLT data. They highlighted that NGC5286 well
resembles the case of M22, in which two populations with different [Fe/H] and $s$-process elements
are present. The iron-poor component resulted to be also poor in $s$-process elements,
while the iron-rich component is also $s$-process rich. Moreover, they concluded that
both the components show important iron star-to-star variations, which globally tags
NGC5286 as an anomalous GC with a non-negligible metallicity dispersion.
Accordingly, \cite{marino15} have suggested that this object was probably more massive in the past,
and thus able to retain SNe ejecta, or even that it could be the remnant of the nucleus
of a dwarf galaxy tidally disrupted by the interactions with the Milky Way.

\end{itemize}

\subsection{Pre-enrichment}

Lithium, together with hydrogen and helium, is a product of the Big Bang nucleosynthesis.
Hence, the first generation of stars in GCs are expected to have
a Li abundance close to the cosmological value, while in the subsequent stellar generations
we may expect to find only Li-poor or Li-free stars.
This is due to the fact that Li burns at $\sim$ 2.5 $\times$ 10$^{6}$ K,
a temperature that is roughly one order of magnitude lower than any other
characteristic temperature ($>$ 10$^{7}$ K) of thermonuclear reactions occurring in stars.
However, observations have revealed a more complex scenario: 
three Na-rich stars with low Li abundance have been detected in NGC6397 \citep{lind09},
while most of the observed stars display a uniform Li abundance (compatible with the cosmological value)
but a large range of Na abundances. 
M4 displays a very small intrinsic Li dispersion,
without correlation between O and Li abundances \citep{mucciarelli11} 
and with a weak Li–Na anticorrelation \citep{monaco12}.

The study of the Helium content of stars in GCs is crucial for a number of fundamental aspects.
First of all, the He content in GC stars is thought to be a good tracer of
the primordial He abundance because these are among the first 
stars formed in the Universe and the mixing episodes occurring
during their evolution only marginally affect their surface He
abundance \citep{sweigart97}.

However, there are a number of direct and indirect pieces of evidence that
different sub-populations in GCs can have different He abundance.
For example, a different He content has been 
invoked as one of the possible ``second parameters''
\citep[together with age, CNO/Fe ratio, stellar density; see e.g.,][]{dotter10,gratton10,dalessandro13},
to explain the observed distribution of stars along the HB,
with the overall metallicity being the first parameter. 
Stars on the bluest side of the HB may have a higher He content.
Recently, a differential analysis performed by \cite{pasquini11}
on two giants in NGC2808 with different Na content indicates that the Na-rich
star is also He enriched at odds with the Na-poor one.
\cite{villanova09,villanova12} derived He, Na and O abundances for HB
stars in NGC6752 and M4, respectively, finding that the stars
along the reddest part of the HB of NGC6752 have a standard
He content, as well as Na and O abundances compatible with
the first generation, while the stars in the bluest part of the HB
of M4 are slightly He-enhanced (by $\sim$ 0.05), with Na and O
abundance ratios compatible with the second stellar generation.
In a similar way, \cite{marino14} found clear evidence
of He enhancement (by $\sim$ 0.09) among the bluest HB stars in
NGC2808 that are also all Na-rich.

SNII are likely the first, main polluters of the Intra-Cluster Medium (ICM) after the
Big Bang nucleosynthesis from which stars in GCs formed.
These objects originate from the core collapse of massive stars, which have very short lifetimes.
Hence, SNII can be responsible for a rapid pre-enrichment of gas with $\alpha$-elements and some iron,
thus explaining why to date no GCs with [Fe/H] $\leq -2.5$ dex have been found and
why the [$\alpha$/Fe] abundance ratios are normally enhanced by a factor of 2-3.

\begin{figure}[!h]
\centering
\includegraphics[width=0.80\textwidth]{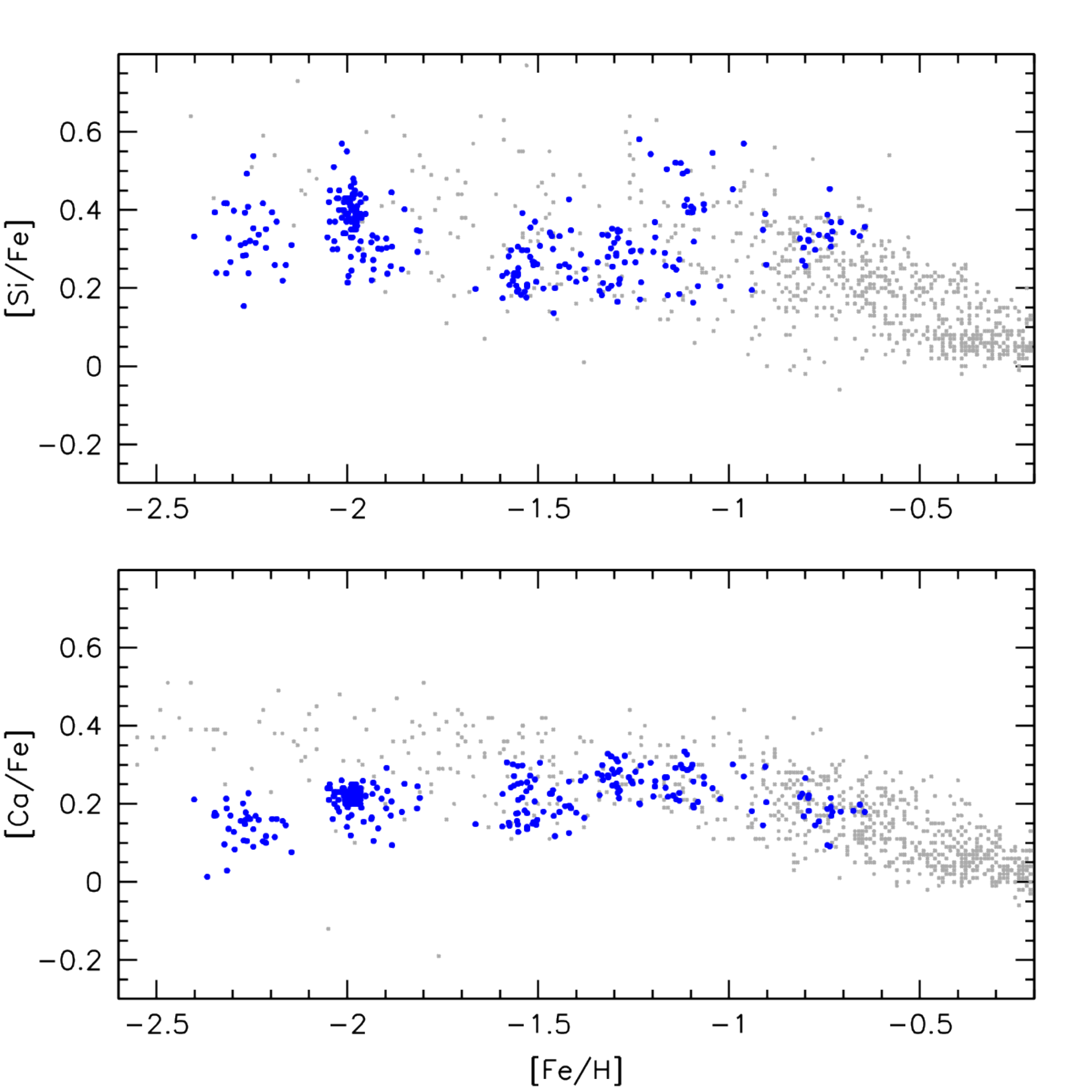}
\caption{The behaviour of [Si/Fe] and [Ca/Fe] as a function of [Fe/H] (large blue dots)
measured in several GCs by \cite{carretta09a}.
Values measured in field halo and disk stars (light gray, small dots) by
\citet{edvardsson93,fulbright00,bensby03,gratton03,reddy03,reddy06}
are also plotted for comparison.}
\label{fig4}
\end{figure}

The $\alpha$-element tipically measured in GC giants are O, Mg, Si, Ca and Ti.
While O and Mg can show significant star-to-star variations (see Section~\ref{var}), 
Si, Ca and Ti do not.
The average values of [Si/Fe] and [Ca/Fe] in GCs at different metallicities (see Figure~\ref{fig4})
well match the typical values observed in the halo stars \citep{gratton04}.
This suggests that the nucleosynthesis processes occurred in GCs were not able to alter
the abundance of these elements, which probably reflects the original composition
of the material from which they have formed.

% FIGURE 4
%

\subsection{Star-to-star variations and signatures of self-enrichment}
\label{var}

Historically, the first example of chemical star-to-star variations within a GC has been observed in
the strengths of molecular bands of CH, CN and NH \citep{kraft79,freeman81,smith84,smith87}
and in the depths of lines of OI, NaI, MgI and AlI \citep{pilachowski83,smith91,gratton01}.
One of the first hints of the presence of anticorrelated variations between CH and CN abundances
have been detected in RGB star samples of several GCs \citep{norris81,norris81b,norris82,smith82}.
Recently, C and N abundance measurements have been also obtained from low resolution spectroscopy
of blue-optical CN-CH molecular bands \citep[see e.g.][]{alvesbrito08,martell08,pancino10}
providing evidence of CN bimodality and CN-CH anticorrelation.
Such a behaviour cannot be univocally explained by standard nucleosynthesis, which converts both carbon and oxygen into nitrogen 
through the CNO cycle and brings the processed material to the star surface by means 
of the first dredge-up mixing process \citep{iben65}, occurring soon after the star leaves the MS.

\begin{figure}[!h]
\centering
\includegraphics[width=0.68\textwidth]{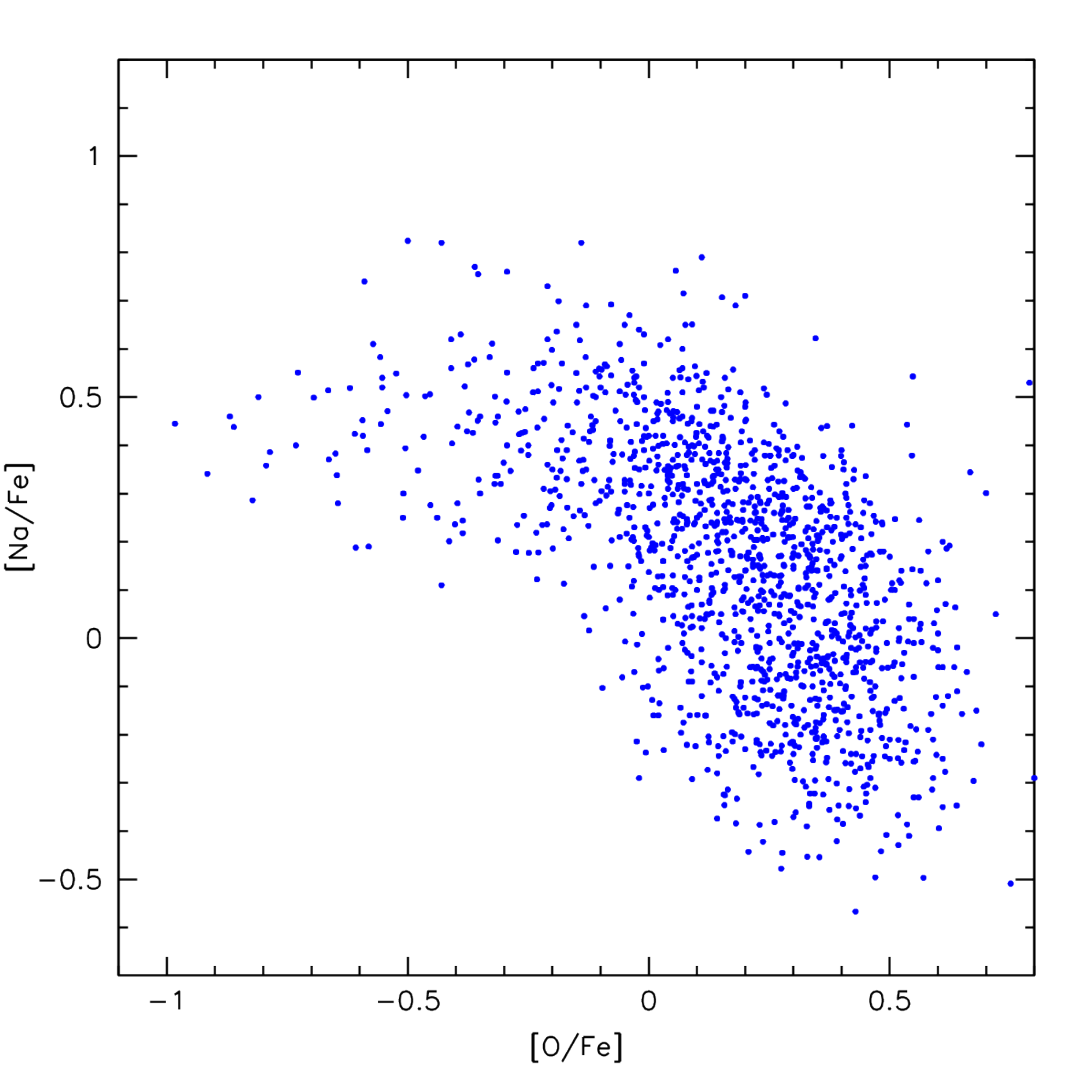}
\includegraphics[width=0.68\textwidth]{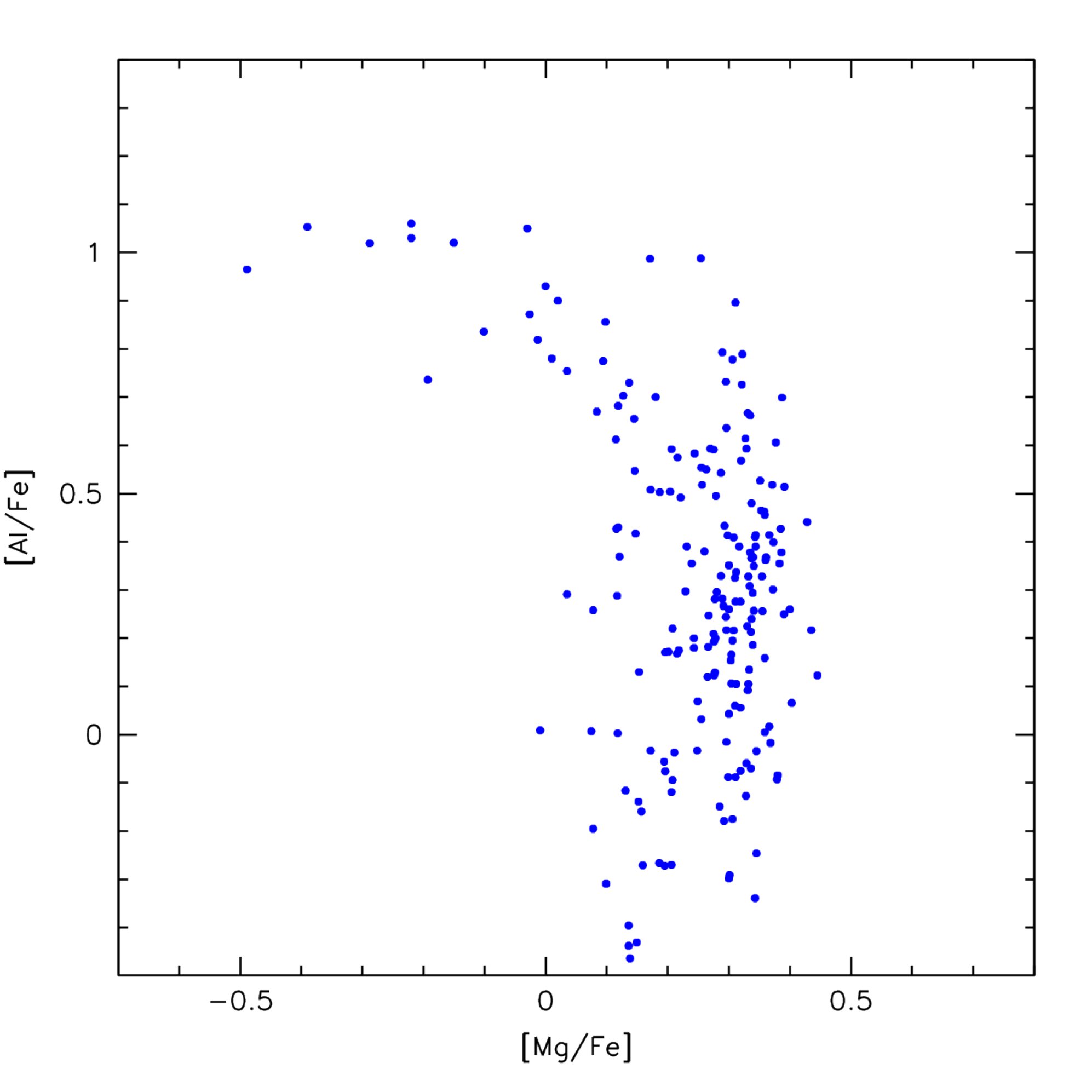}
\caption{The Na-O and Mg-Al anticorrelations observed in GCs, from data collected with FLAMES and UVES
spectrographs by \cite{carretta04,carretta06,carretta09a,carretta09b,carretta14}.}
\label{fig5}
\end{figure}

\begin{figure}[!h]
\centering
\includegraphics[width=0.85\textwidth]{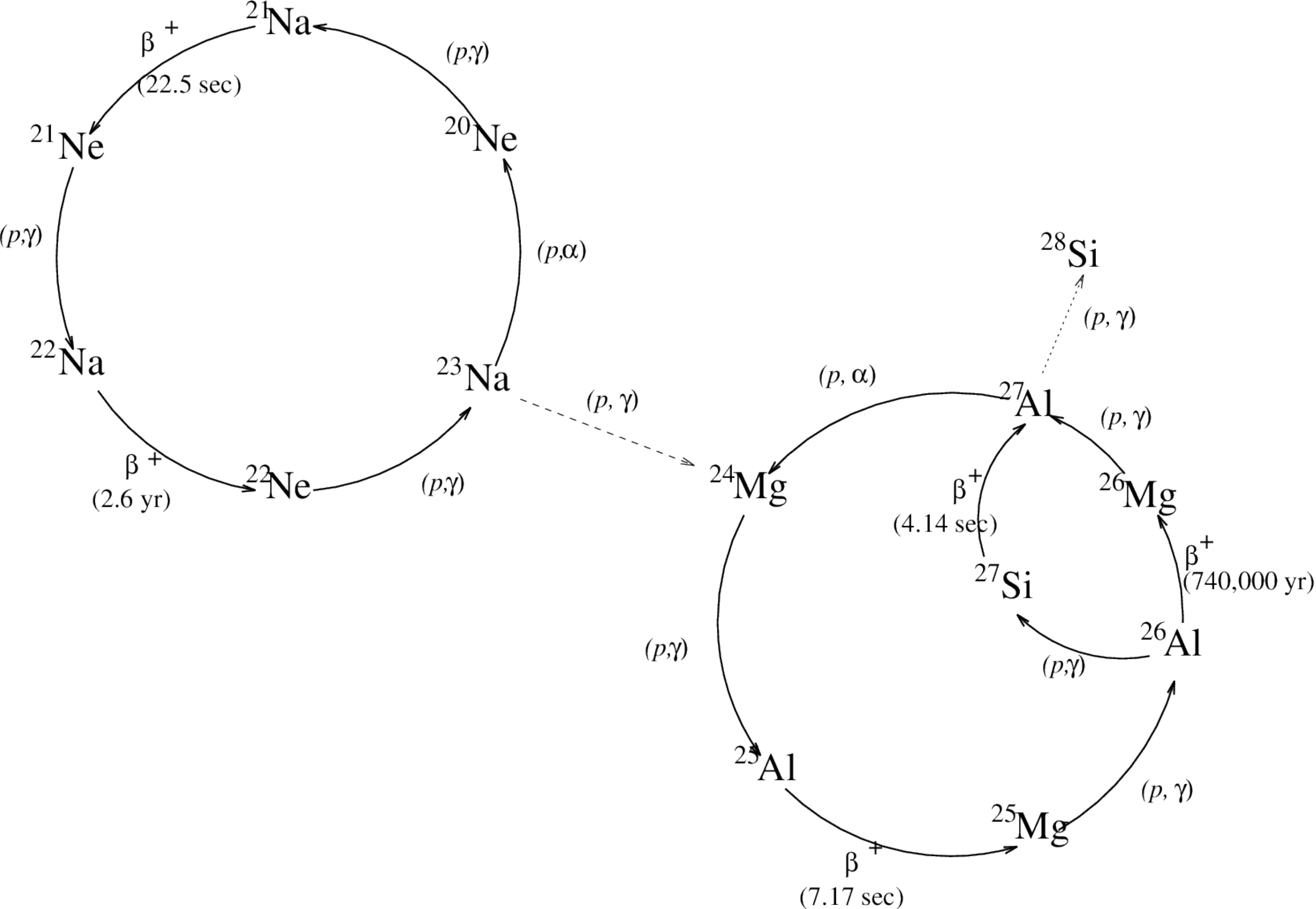}
\caption{Schematic view of the NeNa and MgAl cycles \citep[from][]{cavallo96}.
The labels near the arrows show the reactions occurring in each step.}
\label{fig5b}
\end{figure}

From the very first samples of RGB stars analysed by using high-resolution spectra it was
also found that those depleted in carbon and enhanced in nitrogen are also depleted in oxygen
and magnesium \citep{cottrell81}, while they are enhanced in sodium
and aluminium \citep{sneden92,carretta10}.
The presence of Na-O and Mg-Al anticorrelations (see Figure~\ref{fig5}) seems to suggest a common origin
strictly linked to the formation channels of these elements in the stellar interiors,
where a series of proton-captures happen thanks to the high-temperatures reached.
In Figure~\ref{fig5b} the main chains of the NeNa and MgAl cycles are shown.

% FIGURE 5a
%

In the NeNa cycle, $^{20}$Ne is progressively converted in $^{23}$Na by several proton captures
while in the MgAl chains, $^{24}$Mg is finally converted in $^{26}$Al \citep{denisenkov89,langer93,cavallo96,prantzos07,straniero13}.
These two cycles, however, require temperatures of about 3$\times$10$^{7}$ K and $\sim$ 7$\times$10$^{7}$ K
to take place, i.e. temperature much higher than that at which the CNO-cycle occurs in low mass stars \citep{langer97}.
Interestingly, these anticorrelations have been detected in the majority of Galactic GCs \citep{carretta09a},
althought these systems host low mass stars only.
It is thus impossible to justify the presence of chemical anomalies by assuming that
they are the result of nuclear reactions occurred in these stars,
and some sort of external pollution is unavoidably required.

% FIGURE 5b
%

Thus, the presence of anticorrelations, involving elements from carbon to aluminium,
is likely the result of some pre-enrichment of the gas from which stars originated.
This of course has raised new questions about the nature of the polluters.

One of the first attempt to explain the presence of anticorrelations claimed some
enrichment from SNII. However, this scenario faces two main problems:
(a) the total energy released in a SNII explosion is of the order of $\sim$ 10$^{51}$ ergs \citep{salaris-cassisi},
which is comparable to the typical GC binding energy; hence, the materials ejected by the
star can easily escape the GC potential well and (b) just before the explosion, a SNII 
has an onion-like structure with layers rich of C, O, Mg, Si, Ca but poor in Na and Al \citep{salaris-cassisi},
so the contribution of these stars does not favor the development of anticorrelations.
In the latest years a number of different polluters of the ICM have been suggested: 
intermediate mass AGB stars \citep{ventura01,fenner04,ventura05}, fast rotating massive stars (FRMS) \citep{decressin07},
massive interacting binaries \citep{demink09} and novae \citep{smith96,maccarone12}.
The ejecta of these stars have the main advantage that they are able to
remain in the potential well of GCs and pollute the ICM with different chemical imprints on
different timescales.
However, the complex chemical patterns observed to date in GCs, which often also differ from
cluster to cluster, are likely the result of the interplay between different
polluters \citep{marcolini09,bastian13}.

Among the light element abundances measured in GC giants there are also those of fluorine and potassium, 
whose trends and overall nucleosynthesis are however still debated.

Fluorine is one of the most elusive species measurable in stars, but
it has a particular importance since it is extremely sensitive to the physical conditions
within the stellar atmospheres.
Very few and sparse abundance estimates in GCs are available in literature:
\cite{cunha13} measured F in red giants in the Large Magellanic Cloud and $\omega$ Centauri
finding that the F/O ratio declines as the O abundance decreases and
the two giants observed in $\omega$ Centauri have particularly low F/O values.
They argue that these results are consistent with most
F production coming from either neutrino nucleosynthesis or
W-R stars rather than from AGB stars in these systems.
On the contrary, the abundance of fluorine is found to correlate with oxygen and
anticorrelate with the sodium and aluminium abundances in 7 giants of M4 \citep{smith05}
and in 6 giants NGC6721 \citep{yong08b}.
The same correlations have been found from the analysis of six cool giant stars in M22 by \cite{dorazi13},
however they also found that the F content positively correlates with two different
sub-groups of $s$-process rich and $s$-process poor.
The comparison with theoretical models has suggested that AGB stars of intermediate mass
may be responsible for this chemical pattern.

Very few abundance measurements of potassium are available in GCs.
This element is thought to be formed during hydrostatic oxygen burning in massive stars,
so a behaviour not so different from those of $\alpha$-elements is expected.
However, \cite{mucciarelli12}, from the analysis of a sample of 49 giant stars
belonging to the GC NGC2419, have detected a clear anticorrelation between K and Mg.
The analysis of a sample of 119 stars in the GC NGC2808 by \cite{mucciarelli15c}
highlighted the same trend. This is not expected since, in the framework of SNII yields,
K abundances are expected to correlate with Mg ones.
\cite{mucciarelli15c} have suggested that the behaviour of K could be the result of the
self-enrichment processes occurred in GCs with the contribution of different polluters.

\begin{figure}[!h]
\centering
\includegraphics[width=0.70\textwidth]{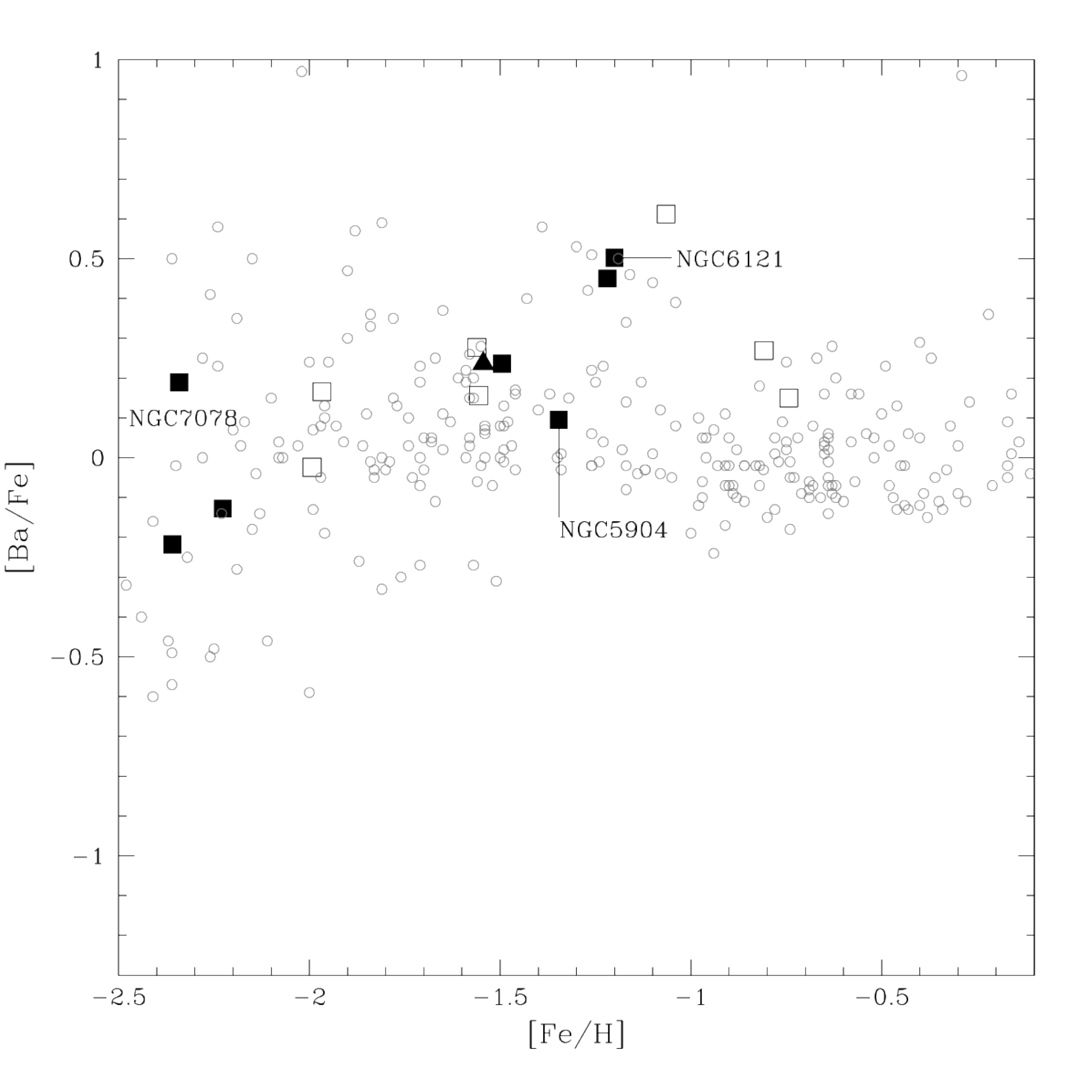}
\caption{The behaviour of [Ba/Fe] as a function of [Fe/H]
measured in several GCs (large symbols) by \cite{dorazi10}. Empty and filled
squares are for disk/bulge and inner halo GCs, while the filled triangle is for the
only outer halo cluster NGC1904. The small gray dots mark the values measured in field stars.}
\label{fig6}
\end{figure}

The abundances of neutron-capture elements measured so far in GCs also show some
peculiarities that change from cluster to cluster.
Regarding the $s$-process elements, the most deeply investigated element is Ba.
This is due to the fact that some strong Ba transitions are available in the
blue/optical spectrum and they can be observed and measured also in low-SNR spectra.
In GCs, the largest survey of Ba abundances has been performed by \cite{dorazi10},
who have measured Ba abundances in more than 1200 individual stars in 15 Galactic GCs (see Figure~\ref{fig6}).
The studied GCs cover a range in metallicity of up to $\sim$ 1.6 dex,
from [Fe/H] = $-2.3$ dex (NGC7099) up to [Fe/H] = $-0.7$ dex (47Tuc).
They have found that in each GC the [Ba/Fe] ratios show a dispersion of the order of 0.15-0.3 dex,
compatible with the measurement errors, so no clear hint of intrinsic star-to-star variations
in the Ba content has been found.
Slightly different is the comparison with the average [Ba/Fe] of clusters at similar metallicities.
For the GCs with [Fe/H] $< -2.0$ dex \cite{dorazi10} have found a solar [Ba/Fe] values,
while for those with [Fe/H] $> -1.7$ dex they have found [Ba/Fe] values which increase from $\sim +0.2$ dex
up to $+0.7$ dex (see their Figure 1).
No clusters present a correlation between Na and Ba abundances,
indicating that there is no significant contribution from low-mass AGB stars
to the intracluster pollution.
The same behaviour has been observed also for other two $s$-process elements like Sr and Y,
although not so striking.
\cite{james04}, from the analysis of a sample of MS and SGB stars belonging to the GCs
NGC6397, NGC6752 and 47Tuc have detected a mild increasing trend of [Sr/Fe] as [Fe/H] increase,
while for [Y/Fe] they have found values roughly compatible with the solar value.
The authors have concluded that the observed behaviour for light- and heavy-elements in GCs
cannot be explained with a self-pollution scenario.

% FIGURE 6
%

It is interesting to note that for the majority of the GCs in which an iron spread has been found
also some peculiar behaviours of the $s$-process abundances have been observed.
In $\omega$ Centauri there is a positive correlation between [Fe/H] and neutron-capture
elements, suggesting that $s$-processes have been increasingly efficient during the evolution of the system
(see \citealt{johnson10,marino12}).
For M22, \cite{marino09,marino11b,marino12} have highlighted the presence of two populations with a difference
in [Fe/H] of $\sim$ 0.15 dex which show also a difference in $s$-process elements.
Moreover, these two population seems to follow their own Na-O anticorrelation.
The same behaviour has been observed in the GC NGC1851 where the $s$-process elements appear to be
correlated with Na and Al abundances \citep{yong08}.
For these two clusters, there are also hints of an overabundance of CNO sum
for the observed stars \citep{marino12,yong09}.
This is a strong evidence pointing toward a chemical imprint left by AGB stars.
These stars seems to be a natural explanation for these peculiar chemical patterns since they can account for
the presence of the Na-O anticorrelation and they are the main source of $s$-process elements
during thermal pulses. Furthermore, AGB stars are able to release CNO-processed materials,
thus altering the CNO sum in subsequent stellar populations.

Among the $r$-process elements, Eu is the one mostly measured in GC giants, however,
the widely used feature is the weak line at 6645 $\rm \mathring{A}$, which requires
high-quality and high-SNR spectra.
This makes the Eu measurements quite difficult and only sparse derivations are available.
\cite{shetrone96a} has derived Eu abundances in several Galactic GCs covering a metallicity
range from roughly [Fe/H] $\sim -2.2$ up to $-0.8$ dex.
The author found an average [Eu/Fe] which remains almost constant at about $+0.4$ dex.
Moreover, within each cluster, a very small dispersion of [Eu/Fe] abundances has been measured.
The lack of correlation with O abundances measured in the same stars suggests that,
if SNII are the source of both elements, one of the two elements should be produced in a
metallicity-independent way.
The small dispersion of [Eu/Fe] detected in the majority of GCs, however, is not always observed.
An important exception is the GC M15 in which [Eu/Fe] ranges over 0.5 dex
at constant [Fe/H] \citep{sneden97}, which may challenge the idea that Eu is produced exclusively by $r$-processes.

\subsection{Open questions}

The observed light-element abundance anomalies in GCs are currently interpreted as signatures
of the presence of at least two stellar generations.

The first generation (FG) stars formed in the very early epoch of the GC formation while
the second generation (SG) stars formed after some hundreds Myr and the chemical imprint of each stellar generation
takes a fundamental role in defining the chemical properties of the descendants.

However, there are still several open questions:
\textbf{(1)} how many star formation episodes actually occurred in GCs? Which is the mechanism that triggered these events?
\textbf{(2)} Which mechanism is responsible of the pollution of the gas from which the second stellar generation formed?
How the gas polluted by the FG stars could remain bound in the system despite the shallow potential well?
\textbf{(3)} Considering that, nowadays, GCs are almost devoid of gas, which mechanism has been able to remove the residual gas after
the star formation episodes?
\textbf{(4)} In several GCs the number of FG stars is roughly equal to the number of SG ones,
while the FG component is expected to be numerically more consistent; thus which was the fate of the missing FG stars?

Several authors attempted to answer these questions by suggesting plausible scenarios based on
both theoretical and observational arguments.
In the following we will briefly summarize them:

\begin{itemize}

\item \citeauthor{dercole08} scenario -
By means of hydrodynamical and N-body simulations \cite{dercole08} have suggested
that GCs were at least 10-20 times more massive in the past.
The FG stars, which have formed from a pristine gas mainly enriched by SNII,
release in the ICM yields with chemical patterns associated to super-AGB and
massive AGB nucleosynthesis \citep{pumo08,ventura09}.
In this way, the SG component mainly forms in the center of the cluster where these materials have been settled.
\cite{dercole08} also suggested that, according to this, the stars that escape the cluster during its lifetime
belong to the FG component. This idea can explain why the SG stars are more centrally concentrated and
why nowadays the proportion between FG and SG is roughly the same. After the SG formation, the first stars
which explode as SNIa are responsible of the quenching of the star formation (SF), removing most of the
residual gas from the cluster. After some time, this gas may flow again in the central region driven by cooling
mechanisms and mix with low-velocity AGB ejecta triggering a new SF episode, which should produce
subsequent stellar generations. These new stellar generations may have a very peculiar chemical composition
which should explain, at least in part, the presence ``extreme'' stars in the NaO and MgAl planes.

\item \citeauthor{conroy11} scenario -
In this scenario the GCs are assumed to have a initial mass similar to the current one.
The FG stars are assumed to form from material pre-enriched in metals (i.e. with the present-day metallicity)
and the SF episode ends when the first SNII explode and sweep out the gas from the system.
After this phase, the ICM of the cluster is polluted by the AGB ejecta which can mix with the materials
accreted from the environment surrounding the GC. This freshly available gas can now take part in the
SG star formation.
Also this SF episode produces a number of SNII and subsequent SNIa, which are able to remove
the residual gas from the cluster, thus quenching any subsequent SF event.
This scenario has the main advantage that it can naturally explain the presence of multiple stellar
populations and the absence of residual gas in the clusters.
However, \cite{dercole11} have highlighted that it does not account for
the high helium content observed in some clusters (which it cannot be released by FG stars only).

\item \citeauthor{carretta10c} scenario -
According to this scenario, the GCs that we observe today are the remnants of more massive structures,
which have been tidally disrupted by the interactions with the Galaxy.
The FG stars formed from a gas pre-enriched by SNII and
they can pollute the ICM with typical imprint of massive-AGB and FRMS.
The ICM pre-enriched in this way is mainly located in the center of the systems,
where the SG stars form. After some Myr also the more massive
SG stars end their lives as core-collapse SNe and quench the SF processes by removing
the residual gas from the cluster. During their evolution, GCs are expected
to loose the majority of the original mass and all dark matter component,
and also the primordial population of stars is almost completely lost.
Indeed, the scenario suggested by \cite{carretta10c} seems to explain why metal-poor stars,
with [Fe/H] $< -2.5$ dex, have not been found in GCs:
they supposed to be lost and currently form the halo population.

\item \citeauthor{valcarce11} scenario -
This scenario suggests that the main actor in defining the presence of star-to-star abundance scatter
and the presence of multiple stellar populations is the mass of the \textit{precursor},
i.e. the mass of the cloud that formed the system.
The most massive precursors are able to retain the ejecta
of massive stars and core-collapse SNe, thus they can explain both chemical anomalies
and the presence of multiple stellar populations. The intermediate-mass precursors
can only retain the gas ejected by massive stars through stellar winds, while they are
unable to keep the ejecta of core-collapse SNe.
The systems formed by small precursors are able to retain only the slow wind of massive stars
and low-mass AGBs. In any case, an important mass-loss during the system lives
needed to account for the lack of a fraction of FG stars. At the same time,
after the formation of the FG component, the pristine gas located in the outskirts of the system
begin to fall in the central region, thus triggering a new SF episode.
This scenario is thus able to explain the presence of multiple stellar populations and
the different behaviour of anticorrelations. In fact it is suggested that the extreme high-Na/low-O
stars are formed with materials processed by high-mass stars while the normal high-Na/low-O stars
have been formed from massive- and intermediate-AGB ejecta.

\item \citeauthor{decressin10} scenario -
This scenario is mainly based on the enrichment due to FRMS \citep{decressin07}.
They postulated that the initial mass segregation determines the formation of massive stars
in the very central part of the system, and since SF occurs locally around the massive stars,
also the SG component is concentrated in the center of the system.
So the mass-loss, which preferentially take place in the most external regions of the system,
will reduce the number of FG stars.
However, even this loss is not able to balance the ratio between FG and SG stars, which was
observationally estimated in several GCs and resulted to be near 50:50.
Hence, an additional mechanism is needed to efficiently expel a non negligible fraction of FG stars.
\cite{decressin10} have suggested that the gas expelled by the very first SNe
is able to lower the GC potential well and make easier the expulsion of the most external FG stars.

\end{itemize}

%% file: c3a/c3b.tex
% CAPITOLO 3

\chapter{Chemical abundances of AGB stars in globular clusters}

\label{c3}

In principle, MS stars would be the ideal tracers of the chemical enrichment history of GCs
since they did not undergo mixing processes and hence all their surface abundances
fully reflect the initial composition of the gas from which they formed.
However, being much fainter than evolved stars, they can be observed
mainly at low-medium resolution in the closest GCs.

Hence, up to now, most of the chemical abundance measures in GC stars have been obtained
by observing RGB and Red Clump (RC) stars, which are luminous enough to be studied
with the current generation of spectrographs with medium-high resolutions at 8-10mt class telescopes.
The spectra of these stars have many metal lines sampling all the most
important chemical species.

Among the bright evolved stars in GCs there are those evolving along the AGB, which can also provide crucial
constraints on the surface abundance change of chemical species like CNO and $s$-process elements due to mixing
processes in the stellar interiors during the post-MS evolutionary stages.

However, due to their short lifetimes, the number of AGB stars observable in an old GC is small,
about a factor of 4-5 smaller than the number of RGB stars of comparable luminosity.
Additionally, high-quality photometry and large color baselines are needed to
properly separate genuine AGB from RGB stars.
Hence, despite their luminosity, massive and systematic chemical studies of AGB stars in GCs are still lacking.
Accurate determinations of their atmospheric parameters and abundances based on
high-resolution spectroscopy are available for a few GCs only, namely
M4 \citep{ivans99}, M5 \citep{ivans01,koch10} and 47Tuc
\citep[hereafter 47Tuc,][]{wylie06,koch08,worley09}, and they are all based on small
samples (no more than 6 AGB stars in each of the quoted clusters).
A few other studies have been focused on the determinations of C, N and Na abundances.
\cite{mallia78} and \cite{norris81} have used medium- and high-resolution spectra to study
RGB and AGB star samples in the GCs 47Tuc, NGC6397, NGC6656 and NGC6752 finding
the presence of CN-strong and CN-weak stars.
The same investigation has been performed by \cite{briley93} on a sample of 24 giant stars in M55,
finding only one star compatible with CN enhancement.
Finally, \cite{campbell13} and \cite{johnson15} have investigated the behaviour of Na abundances
in NGC6752 and 47Tuc, by using medium-resolution FLAMES spectra.

However, three studies in particular have recently provided quite unexpected results, which opened
new questions about the physics of AGB stellar atmospheres and their properties in GCs.
In the following we will briefly examine them.

\section{NLTE effects in AGB stars}
% NLTE

\cite{ivans01} performed a detailed analysis of a sample of 36 AGB and RGB stars belonging to the GC M5
by using high-resolution spectra taken with HIRES@KECK spectrograph.
In the analysis they adopted the full spectroscopic approach to compute the abundance of chemical species,
implying that the surface gravity has been constrained by imposing ionization equilibrium of iron lines.
They found that the average [Fe/H] of AGB stars is $\sim$ 0.2 dex lower than that computed from RGB stars.
After several checks on the adopted procedure they found that, if photometric gravities are assumed,
the average [FeII/H] of AGB stars is in perfect agreement with that derived from RGB stars,
while the average [FeI/H] of AGB stars remains lower than that derived for RGB stars.
This discrepancy cannot be explained by measurement uncertainties or an incorrect derivation of the atmospheric parameters,
and it was not observed in RGB stars belonging to the same cluster and analysed in a homogeneous way.
The authors interpreted this evidence in terms of possible NLTE (Non Local Thermodynamic Equilibrium) effects
in the atmospheres of the AGB stars, since these effects mainly affect the neutral lines while leaving unaltered the ionized ones.

\section{The lack of SG AGB stars in GCs}
% LACK OF SG AGB

% FIGURE 7
%
\begin{figure}[h]
\centering
\includegraphics[width=0.70\textwidth]{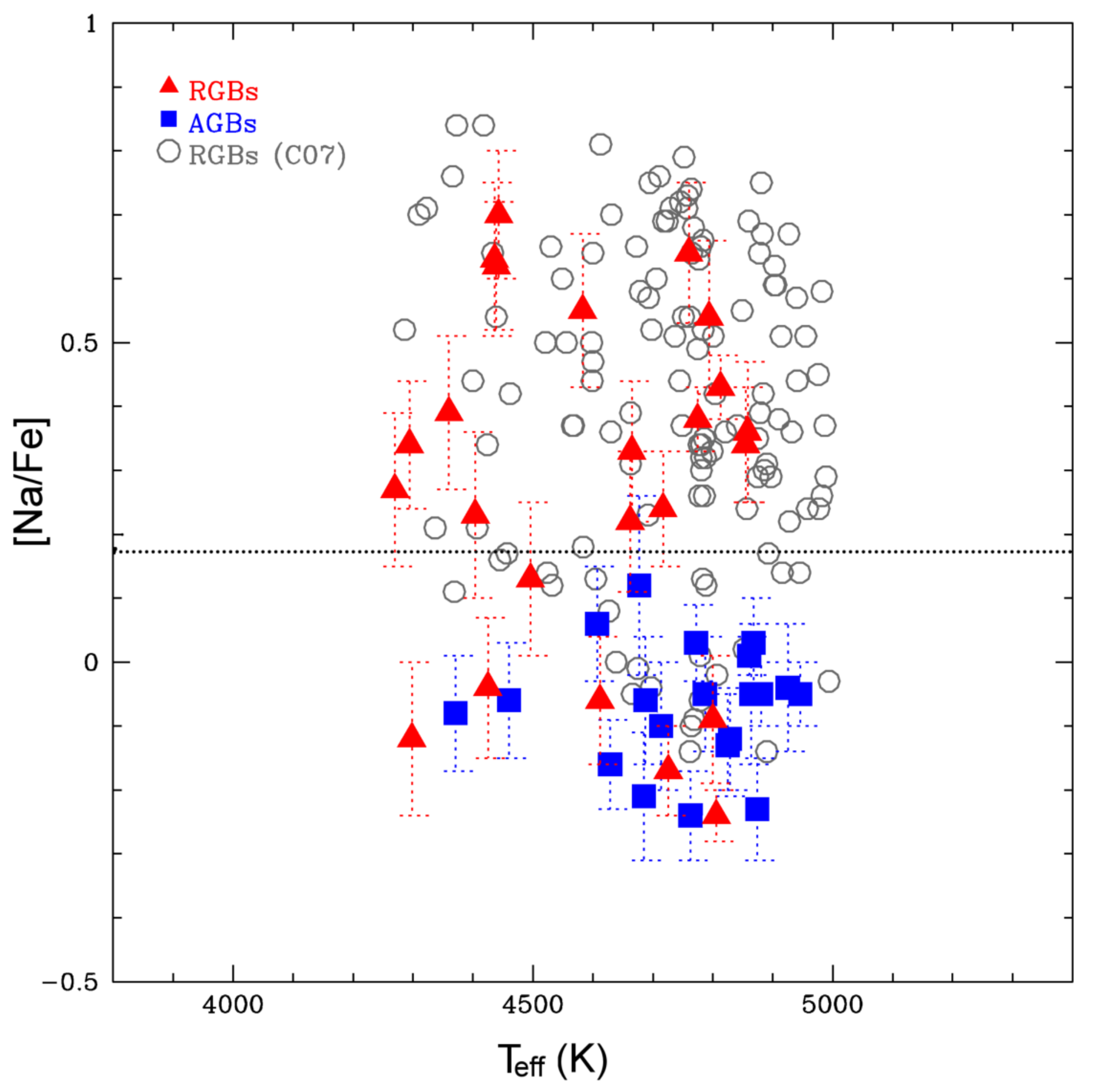}
\caption{The behaviour of [Na/Fe] as a function of T$_{eff}$ for RGB (red triangles, 24 stars)
and AGB stars (blue squares, 20 stars) found by \cite{campbell13}.
For comparison, the RGB stars analysed by \cite{carretta07} are showed as large empty dots.}
\label{fig7}
\end{figure}

As discussed in Secion~\ref{var}, the Na abundances of the stars in GCs is
usually used as a proxy to distinguish the different stellar populations within the same cluster.
Recently, \cite{campbell13} have analysed a sample of 20 AGB and 24 RGB stars belonging to the
GC NGC6752 observed with the FLAMES@VLT spectrograph.
They have found that the RGB stars show the presence of both Na-rich and Na-poor stars,
spanning a range fully compatible with previous determinations available in literature
(see Figure~\ref{fig7}). However, all the AGB stars in their sample resulted to be
Na-poor, with [Na/Fe] values compatible with those observed in FG stars.
\cite{campbell13} have concluded that all SG stars fail to ascend the AGB.
This result is even more challenging by considering that the second-generation population
contains the majority (70$\%$) of the stars in NGC6752,
hence the presence of Na-rich SG AGB stars is expected.
The authors have also critically discussed some possible explanations;
however, up to now it is still unclear why and how the the SG stars of NGC6752 can skip the AGB phase.

\section{AGB overconcentration in the core of 47Tuc}
% EBSS IN 47TUC

% FIGURE 8
%
\begin{figure}[!h]
\centering
\includegraphics[width=0.70\textwidth]{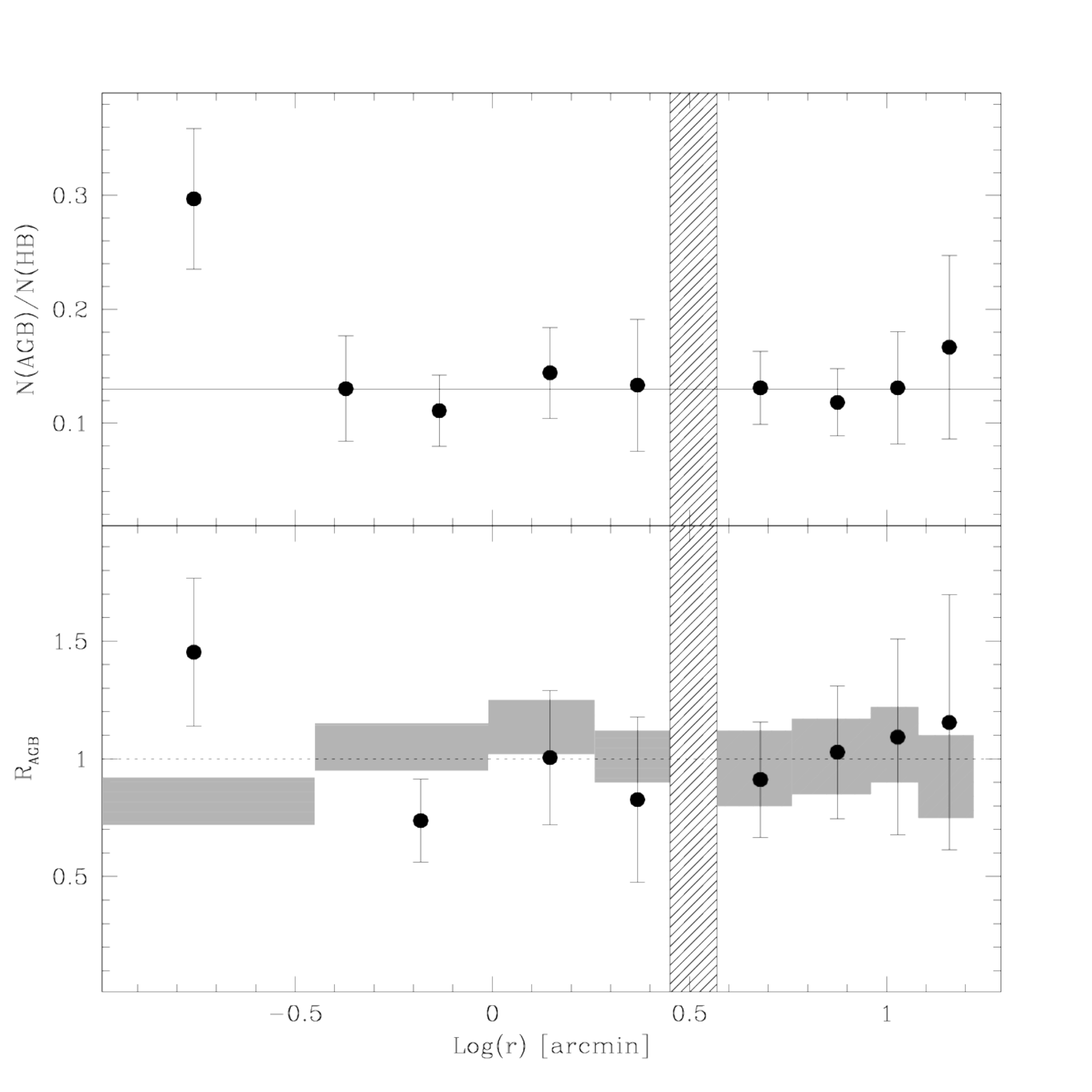}
\caption{Relative frequency of AGB-to-HB stars (upper panel)
and double-normalized specific frequency of AGB stars (lower panel)
as a function of the projected distance from the center of the GC 47Tuc
by \cite{beccari06}. In the innermost radial bin the number of AGB stars is roughly
twice the average value (black horizontal line).}
\label{fig8}
\end{figure}

\cite{beccari06} have computed the radial distribution
of RGB, HB, bright-HB (hereafter bHB, defined as stars slightly brighter than in normal HB stars)
and AGB stars in the GC 47Tuc by using several photometric images obtained with
the ACS camera mounted on board the HST.
They have detected an ``overconcentration'' of bHB and AGB stars toward
the cluster central region, with respect to RGB and HB stars.
In particular, for the AGB sample a number count excess of about 30$\%$ has been found within
the innermost 21 arcsec (see the innermost point in Figure~\ref{fig8}),
while in the outer cluster regions, the AGB stars follow the same radial distribution of the other populations.
This overconcentration cannot be justified by invoking the standard stellar
evolution theory and cannot be explained by dynamical processes (as mass segregation)
acting on genuine AGB stars.
This evidence (overabundance and central segregation) suggests the presence among
the AGB stars of an extrapopulation of massive objects, probably related to the evolution of binary systems
(for instance, like evolved Blue Straggler Stars: BSSs).
Thus, AGB stars in the innermost region of 47Tuc promise to be an ideal sample where evolved BSS could be detected.

\subsection{Blue straggler stars in GCs}
% BSS IN GCS

\begin{figure}[!h]
\centering
\includegraphics[width=0.70\textwidth]{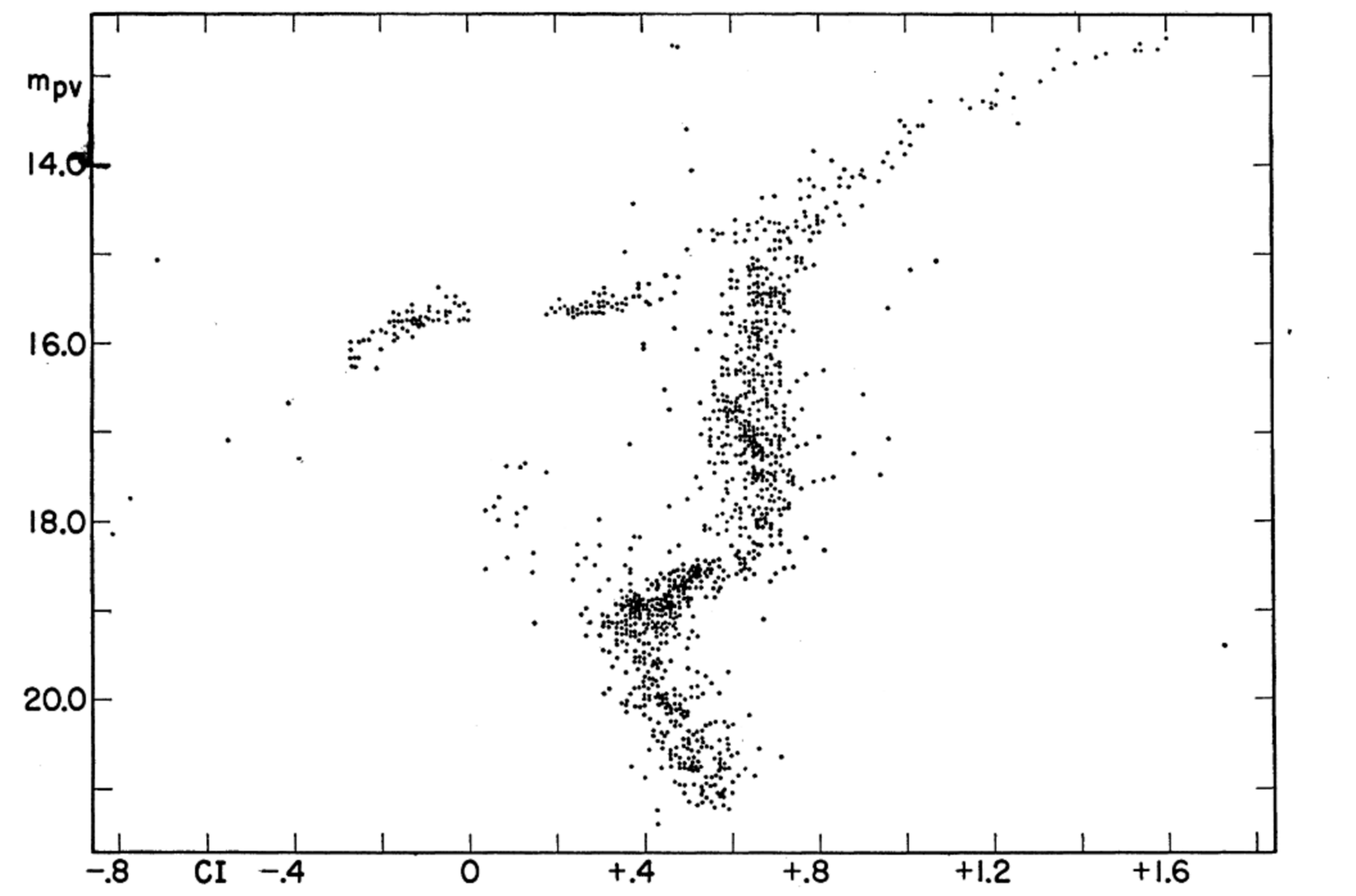}
\caption{CMD of the GC M3 taken from \cite{sandage53},
where the BSS stars have been identified for the first time.}
\label{fig9}
\end{figure}

BSSs are commonly defined as stars brighter and
bluer than the MS turn-off point in the CMD of the host stellar cluster.
Firstly identified by Allan Sandage in the GC M3 (see Figure~\ref{fig9}),
they are thought to be central hydrogen-burning stars, more massive than the MS
stars \citep{shara97,gilliland98,fiorentino14}.
In GCs, BSSs have typical masses in the range 1.2-1.6 M$_{\odot}$ \citep{ferraro06,tian06,lanzoni07,sills09}.
Two main formation channels have been proposed:
mass-transfer in binary systems \citep{mccrea64} and direct stellar collisions \citep{hills76}.
Since BSSs are more massive than normal cluster stars, they suffer from the effect of dynamical friction,
which makes them progressively sinking towards the cluster center \citep{mapelli06,lanzoni07,alessandrini14},
and for this reason they have been found to be powerful probes of the internal dynamical evolution
of the host cluster \citep{ferraro12,ferraro15,miocchi15}.

% FIGURE 9
%

While BSSs can be easily identified during their core hydrogen-burning phase,
they are photometrically indistinguishable from their low-mass sisters
in advanced stages of the subsequent evolution.
This is the reason why, although BSSs have been routinely observed for decades
in GCs \citep{ferraro03,piotto04,leigh07},
in several open clusters \citep{geller11,gosnell14}
and also in dwarf galaxies \citep{mapelli09,monelli12}, only a few
identifications of evolved BSSs have been obtained so far.
Three candidates have been recently identified from asteroseismology studies
in two open clusters (NGC6791 and NGC6819; \citealp{brogaard12,corsaro12})
and with an estimated mass in the range 1.2-1.5 M$_{\odot}$.
Only a candidate evolved BSS is known in GCs: the anomalous cepheid V19 in NGC5466
\citep{zinn82,mccharty97} with an estimated mass of 1.6 M$_{\odot}$.
Photometric criteria have been suggested to optimize the
search for candidate evolved BSSs: for instance \citet{renzinifusi88}
and \citet{fusi92} suggested to look at a region of the CMD between
the horizontal branch (HB) and the base of the asymptotic giant branch
(AGB, the so-called AGB-clump; \citealp{ferraro99a}), where evolved
BSSs experiencing the core helium burning phase (predicted to be
brighter than that of {\it canonical}, lower mass, HB stars) are expected to lie.
Following this prescription, evidence suggesting the presence of evolved BSSs ``contaminating''
the genuine HB-AGB cluster population has been found in M3 \citep{ferraro97},
M80 \citep{ferraro99b} and 47Tuc \citep{beccari06}.

\section{A few words on the NLTE effect}

The interpretation of observed spectra of stars in terms of fundamental
stellar properties is a key problem in astrophysics.
For FGK-type stars, the radiative transfer models are often computed using
the assumption of Local Thermodynamic Equilibrium (LTE).
As often happens in astrophysics research, the reason for adopting LTE
is that it substantially simplifies the calculation of number densities of atoms and molecules.
However, by definition, this is only an ``assumption'', since the atmospheres of the stars
are better described as a series of different layers in which the temperature and
gas density change continuously.
Hence, in the following we consider a simple one-dimensional hydrostatic model of a stellar atmosphere,
in which temperature and density are linked to the depth and the radiation is transported outward.
At each depth point, we can assume that matter particles (ions, atoms, electrons, molecules)
are in LTE with each other.
This equilibrium is established by intra-particle collisions, i.e. the energy distribution
of matter depends on the collisions between particles.
The LTE condition is usually expressed as J$_{\nu}$ = B$_{\nu}$,
where J$_{\nu}$ is the radiation field component while B$_{\nu}$ is the Planck function.
In the dense part of the stellar atmosphere the collision rate is very high and the photon mean free path
$l_{\lambda}$ is smaller than the scale over which the physical variables (temperature, pressure) change.
For this reason the radiation and the particles can be assumed to be in equilibrium (guaranteed by collisions).
The departure from LTE condition (NLTE) happens when $l_{\lambda}$ becomes larger than the scale height of the material.
Thus, as photons diffuse outward, their decoupling from matter increases,
and the radiation field becomes ``non-local'' (i.e. suddenly changes even between close points), anisotropic,
and strongly non-Planckian, i.e. J$_{\nu}$ $\neq$ B$_{\nu}$.

The same occurs in the outermost layers where density decreases and collisions become progressively
less frequent. Accordingly, the collisions are not able to thermalize the matter any more,
while the radiative excitations/de-excitations become important.
In giant stars it is usually possible to observe neutral and single ionized lines.
By definition, the neutral lines are formed in the outer part of the stellar atmosphere,
where the energetics is not high enough to ionize the atoms.
On the contrary, the single ionized lines are formed much deeper inside,
were the higher energetics is able to ionize the atoms.
This is the reason why the NLTE phenomenon mainly affects the neutral lines,
while it leaves quite unaltered the single ionized lines formed deep inside the atmosphere.
However, also other mechanisms contribute to the departures from LTE conditions
\citep[see][for an extended discussion]{bergemann14}.

In the last decades, the treatment of NLTE in stellar atmospheres has made huge progress,
and several NLTE correction grids for the abundances of different elements are available
(see e.g. the INSPECT project\footnote{http://www.inspect-stars.com/}).
However, deriving proper NLTE corrections is a quite hard task, because of two main reasons:
(1) the development of reliable atoms and molecules models, which should take into account
all the possible transitions from all the energetic levels and
(2) the treatment of the collision rate with neutral hydrogen atoms (the $S_{H}$ parameter).
In particular, the $S_{H}$ coefficient is usually computed with the Drawin's formula \citep{drawin69},
which laboratory measurements and quantum mechanical calculations
indicate that it overestimates the rate coefficient for optically allowed transitions
by one to seven orders of magnitude.
Therefore, various approaches have been adopted to empirically constrain the $S_{H}$ parameter,
however, up to now, a fully consistent treatment of the collision rates with hydrogen atoms is still lacking.
In any case, big progresses have already been made for some species like Fe, Na and Li \citep[see e.g.][]{thevenin99,lind09,lind11}.

%% file: c4/ms_4.tex
% CHAPTER 4

\chapter{Non Local Thermodynamic Equilibrium Effects on Asymptotic Giant Branch Stars in 47Tucanae}

\label{c4}

{\bf Published in Lapenna et al. 2014, ApJ, 797, 124L}

{\it We present the iron abundance of 24 asymptotic giant branch (AGB)
stars members of the globular cluster 47 Tucanae, obtained with
high-resolution spectra collected with the FEROS spectrograph at the
MPG/ESO-2.2m Telescope.  We find that the iron abundances derived from neutral
lines (with mean value [FeI/H] = $-0.94 \pm 0.01$, $\sigma= 0.08$ dex)
are systematically lower than those derived from single ionized lines
([FeII/H] = $-0.83 \pm 0.01$, $\sigma= 0.05$ dex). Only the latter are
in agreement with those obtained for a sample of red giant branch
(RGB) cluster stars, for which FeI and FeII lines provide the same
iron abundance.  This finding suggests that Non Local Thermodynamical
Equilibrium (NLTE) effects driven by overionization mechanisms are
present in the atmosphere of AGB stars and significantly affect FeI
lines, while leaving FeII features unaltered. On the other hand, the
very good ionization equilibrium found for RGB stars indicates that these NLTE
effects may depend on the evolutionary stage.  We discuss the impact
of this finding both on the chemical analysis of AGB stars, and on the
search for evolved blue stragglers.}

% FIGURE 1
%
\begin{figure}[h]
\centering
\includegraphics[trim=0cm 0cm 0cm 0cm,clip=true,scale=.40,angle=270]{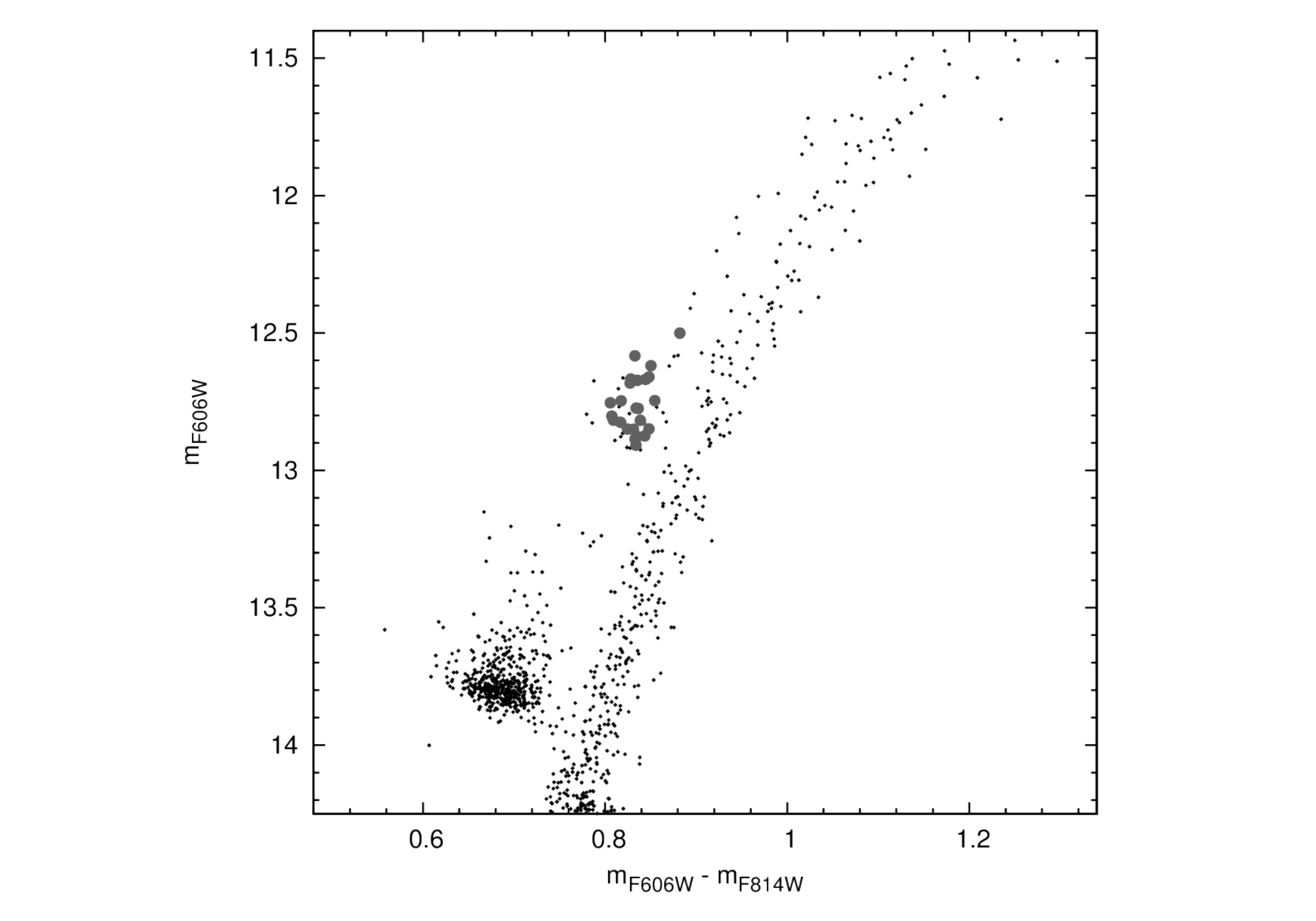}
\includegraphics[trim=0cm 0cm 0cm 0cm,clip=true,scale=.40,angle=270]{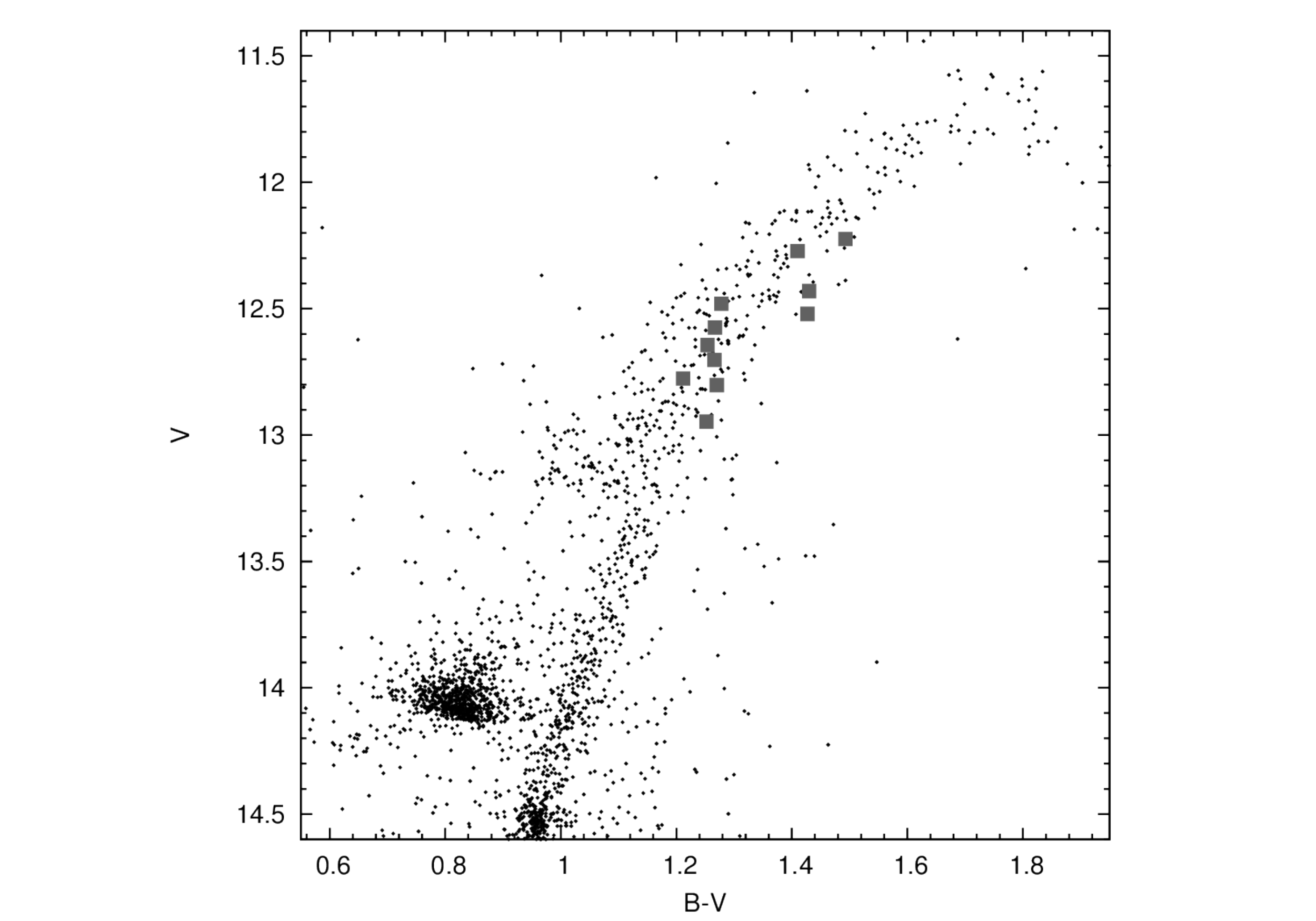}
\caption{Top panel: $(m_{\rm F606W},m_{\rm F606W}-m_{\rm F814W})$ color magnitude diagram of 47 Tuc from
\cite{beccari06}. The large solid circles mark the 24 targets studied in the present work.
Bottom panel: ($V,V-I$) color magnitude diagram of 47 Tuc obtained from WFI data by \cite{ferraro04b}.
The large solid squares mark the 11 RGB stars used for comparison.}
\label{c4_cmd}
\end{figure}

\section{Observations}
\label{c4_obs}

We acquired high resolution spectra of 24 AGB stars in 47 Tuc (Program
ID 090.D-0153, PI: Lanzoni) by using the Fibre-fed Extended Range
Optical Spectrograph \citep[FEROS;][]{kaufer99} mounted at the
MPG/ESO-2.2m telescope.  The spectra cover a wavelength range between
$\lambda \sim 3500~\mathring{\rm A}$ and $\lambda \sim 9200
~\mathring{\rm A}$, with a spectral resolution of $\sim 48000$.
FEROS allows to allocate simultaneously two fibers at a relative distance of
$2.9\arcmin$, one on the source and the other on the sky.  The targets
have been selected from the photometric catalog of \cite{beccari06},
within $\sim 100\arcsec$ from the cluster center. In the
color-magnitude diagram (CMD) of 47 Tuc they are located in the AGB
clump (corresponding to the beginning of this evolutionary phase;
\citealp{ferraro99a}), at $m_{\rm F606W} \sim 12.8$ and
$(m_{\rm F606W}-m_{\rm F814W})\simeq 0.8$ (see top panel of Figure~\ref{c4_cmd}).
Only isolated stars have been selected, in order to avoid
contamination of the spectra from close objects of larger or comparable
luminosity.  The identification number, coordinates and magnitudes of
each target are listed in Table~\ref{c4_tab1}.

For each target a single exposure of $\sim$30-40 min has been
acquired, reaching signal-to-noise ratios S/N$\geq 70$ per pixel.
The data reduction was performed by using the ESO FEROS pipeline,
including bias subtraction, flat fielding, wavelength calibration
by using a Th-Ar-Ne reference lamp, spectrum extraction and final
merging and rebinning of the orders.  Since the background level of
the sky is negligible ($<1\%$) compared to the brightness of the
observed targets, we did not perform the sky subtraction from the
final spectra in order to preserve its maximum quality.  We accurately
checked that the lack of sky subtraction has no impact on the derived
abundances, by comparing the equivalent widths (EWs) measured 
for some spectra with and without the sky subtraction.

%%%%%%%%%%%%%%%%%%%%%%%%%%%%%%%%%%%%%%%%%%%%%%%%%%%%%%%%%%

\section{Analysis}
\label{c4_analysis}

\subsection{Radial velocities}
\label{c4_vrad}

The radial velocity of each target has been obtained by means of
the code DAOSPEC \citep{stetson08}, measuring the positions of more than 300
metallic lines. The accuracy of the wavelength calibration has been
checked by measuring telluric absorptions and oxygen sky lines,
finding no significant zero-point offsets. Uncertainties have
been computed as the dispersion of the measured radial velocities
divided by the square root of the number of used lines, and they
turned out to be smaller than 0.04 km s$^{-1}$.
Heliocentric corrections obtained with the IRAF task RVCORRECT have
been adopted.  The heliocentric radial velocities for all the targets
are listed in Table~\ref{c4_tab1}. They range between $\sim -41.5$ and
$\sim +9.5$ km s$^{-1}$, with mean value of $-17.6 \pm 2.3$ km
s$^{-1}$ and dispersion $\sigma = 11.5$ km s$^{-1}$.  These values are
in good agreement with previous determinations of the systemic radial
velocity of 47 Tuc \citep[see e.g.][]{mayor83,meylan91,gebhardt95,carretta04b,alvesbrito05,koch08,lane10}.
All the targets have been considered as members of the
cluster, according to their radial velocities and distance from the
cluster center.

%------------------------------------------------------------------------ CHEMICAL ANALYSIS

\subsection{Chemical analysis}
\label{c4_chem}

The chemical abundances have been derived by using the package
GALA\footnote{http://www.cosmic-lab.eu/gala/gala.php}
\citep{mucciarelli13} which matches the measured and the theoretical
equivalent widths \citep[see][for a detailed description of this
method]{castelli05}.  The model atmospheres have been computed by
using the ATLAS9 code, under the assumption of plane-parallel
geometry, local thermodynamical equilibrium (LTE) and no overshooting
in the computation of the convective flux.  We adopted the last
release of the opacity distribution functions from \cite{castelli04},
assuming a global metallicity of [M/H] = $-1$ dex with [$\alpha$/Fe]$ = +0.4$ dex
for the model atmospheres.

The effective temperatures ($T_{\rm eff}$) and surface gravities
(log~$g$) of the targets have been derived photometrically, by
projecting the position of each star in the CMD onto the isochrone
best fitting the main evolutionary sequences of 47 Tuc. The isochrone
has been extracted from the BaSTI database \citep{pietrinferni06}
assuming an age of 12 Gyr, metallicity Z $= 0.008$ and $\alpha$-enhanced
chemical mixture. We adopted a distance modulus $(m-M)_V = 13.32$ mag
and a color excess $E(B-V) = 0.04$ mag \citep{ferraro99a}.  Microturbulent
velocities ($v_{\rm turb}$) have been derived by requiring that no
trends exist between FeI abundances and the reduced EWs, defined as
$\log(EW/\lambda)$. The adopted atmospheric parameters are listed in
Table~\ref{c4_tab2}.

Only absorption lines that are predicted to be unblended at the FEROS
resolution have been included in our analysis. The line selection has
been performed through a careful inspection of synthetic spectra
calculated with the code SYNTHE \citep{sbordone05} assuming the
typical atmospheric parameters of our targets and the typical
metallicity of 47 Tuc.
We considered only transitions with accurate theoretical/laboratory
atomic data taken from the last version of the Kurucz/Castelli
compilation.\footnote{http://wwwuser.oat.ts.astro.it/castelli/linelists.html}
The EWs have been obtained with DAOSPEC \citep{stetson08}, iteratively
launched by means of the package
4DAO\footnote{http://www.cosmic-lab.eu/4dao/4dao.php}\citep{mucciarelli13b}
that allows an analysis cascade of a large sample of stellar spectra
and a visual inspection of the Gaussian fit obtained for all the
investigated lines.  Due to the extreme crowding of spectral lines in
the region between $\lambda \sim 3800~\mathring{\rm A}$ and $\lambda
\sim 4500~\mathring{\rm A}$, and to the presence of several absorption
telluric line bands beyond $\sim 6800~\mathring{\rm A}$, we restricted
the analysis to the spectral range between $\sim 4500~\mathring{\rm A}$
and $\sim 6800~\mathring{\rm A}$.  In order to avoid too weak or
saturated features, we considered only lines with reduced EWs between
$-5.6$ and $-4.7$ (these correspond to EW = 11 m$\mathring{\rm A}$ and
90 m$\mathring{\rm A}$ at $\lambda\sim 4500~\mathring{\rm A}$, and EW
= 17 m$\rm\mathring{A}$ and 135 m$\rm\mathring{A}$ at $\lambda \sim
6800~\mathring{\rm A}$, respectively).  Moreover, we discarded from
the analysis also the lines with EW uncertainties larger than $20\%$,
where the uncertainty of each individual line is provided by DAOSPEC
on the basis of the fit residuals. With these limitations, the iron
abundance has been derived, on average, from $\sim 150$ FeI lines and
$\sim 13$ FeII lines.
In the computation of the final iron abundances we adopted
as reference solar value A(Fe)$_{\odot}$ = 7.50 dex \citep{grevesse98}.

Uncertainties on the derived abundances have been computed for each
target by adding in quadrature the two main error sources: {\sl (a)}
those arising from the EW measurements, which have been estimated as
the line-to-line abundance scatter divided by the square root of the
number of lines used, and {\sl (b)} the uncertainties arising from the
atmospheric parameters, computed varying by the corresponding
uncertainty only one parameter at a time, while keeping the others
fixed. The abundance variations thus obtained have been added in
quadrature.
Term (a) is of the order of less than 0.01 dex for FeI and 0.03 dex for FeII.
Since the atmospheric parameters have been estimated from photometry,
by projecting the position of each target in the CMD onto the isochrone,
we estimated term (b) from the photometric uncertainties.
By assuming a conservative uncertainty of 0.1 mag for the magnitudes of our targets
we obtained an uncertainty of about $\pm$50 K and $\pm$0.05 dex
on the final T$_{eff}$ and log~$g$, respectively.
The total uncertainties in [FeI/H] are of the order of
0.04-0.05 dex, while in [FeII/H] are of about 0.08-0.10 dex (due to
the higher sensitivity of FeII lines to $T_{\rm eff}$ and log~$g$).

\section{Iron abundance}
\label{c4_results}

\begin{figure}[h]
\centering
\includegraphics[trim=0cm 0cm 0cm 0cm,clip=true,scale=.50,angle=0]{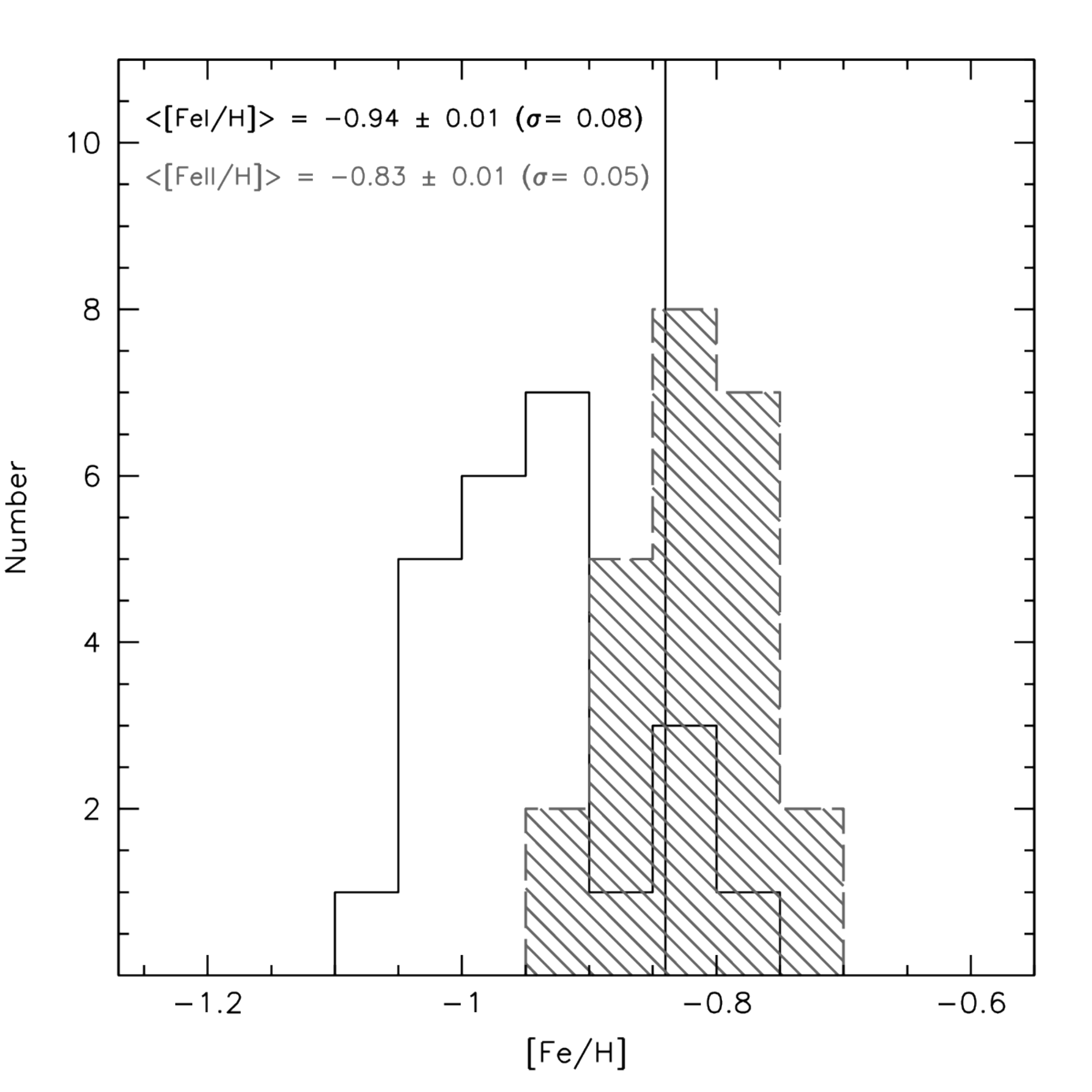}
\caption{Distribution of the iron abundance ratios measured for the 24
AGB stars in our sample, from FeI lines (empty histogram) and from
FeII lines (shaded histogram). The vertical line indicates the
average iron abundance derived from 11 RGB stars.}
\label{c4_iron}
\end{figure}

The [FeI/H] and [FeII/H] abundance ratios measured for each target
are listed in Table~\ref{c4_tab2}, together with the total uncertainties
and the number of lines used.  Their distributions are shown in Figure~\ref{c4_iron}.
A systematic difference between [FeI/H] and [FeII/H] is
evident, with the abundances derived from FeI lines being, on
average, $0.1$ dex smaller than those obtained from FeII: the mean
values of the distributions are [FeI/H] $= -0.94 \pm 0.01$ ($\sigma =
0.08$ dex) and [FeII/H] $= -0.83 \pm 0.01$ ($\sigma = 0.05$ dex).
These values are clearly incompatible each other. Moreover,
only [FeII/H] is in agreement with the metallicity quoted in the literature
and based on sub-giant or RGB stars
\citep{carretta04,alvesbrito05,koch08,carretta09b}, while the iron
abundance obtained from FeI lines is significantly smaller. The
distribution of the difference [FeI/H]$-$[FeII/H] is shown as a
function of [FeII/H] in Figure~\ref{c4_fediff} (black circles). It is quite broad,
ranging from $-0.25$ to $+0.01$ dex.

% FIGURE 2
%

\subsection{Sanity checks}
\label{c4_check}

The difference in the derived [FeI/H] and [FeII/H] abundances cannot be
easily explained (especially if considering the high quality of the acquired spectra and the very
large number of used iron lines) and it clearly needs to be understood.
In order to test the correctness of our analysis and to exclude
possible bias or systematic effects, we therefore performed a number
of sanity checks.

\subsubsection{Checks on the chemical analysis procedure}
To test the reliability of our chemical analysis, we
studied a sample of RGB stars in 47 Tuc, the standard star 104 Tau,
Arcturus and the Sun by following the same procedure adopted for the
AGB targets discussed above (i.e., by using the same linelist, model
atmospheres, and method to infer the atmospheric parameters).

\emph{RGB stars in 47 Tuc --} We measured the iron abundance for a
sample of 11 RGB stars in 47 Tuc, for which high-resolution
(R$\sim$45000) FLAMES-UVES spectra are available in the ESO Archive
(Program ID: 073.D-0211).
The location of these stars in the ($V,V-I$) CMD from \cite{ferraro04b}
is shown in Figure~\ref{c4_cmd} (bottom panel).
Their atmospheric parameters and [FeI/H] and [FeII/H] abundance ratios
are listed in Table~\ref{c4_tab3}.
The distribution of the [FeI/H]-[FeII/H] differences is shown
in Figure~\ref{c4_results} (large gray squares).
We found an average [FeI/H]$_{\rm RGB} = -0.83 \pm
0.01$ dex ($\sigma= 0.02$ dex) and [FeII/H]$_{\rm RGB} = -0.84 \pm
0.01$ dex ($\sigma= 0.03$ dex).
These values are fully consistent with previous determinations.
In fact, the careful comparison with the results of \cite{carretta09b}, who
analyzed the same 11 RGB spectra, shows that the average differences
in the adopted parameters are: $\Delta$ T$_{\rm eff}$ = 36 $\pm$ 8 K
($\sigma$ = 26 K), $\Delta$log~$g$ = 0.07 $\pm$ 0.01 ($\sigma$ = 0.04),
$\Delta$ v$_{\rm turb}$ = 0.01 $\pm$ 0.04 km s$^{-1}$ ($\sigma$ = 0.14 km s$^{-1}$).
By taking into account also the differences among the
adopted atomic data, model atmospheres and procedure to measure the
EWs, the derived abundances turn out to be in very good agreement
within the uncertainties, the mean difference between our values and
those of \cite{carretta09b} being $\Delta$[FeI/H] = --0.06 $\pm$ 0.03
dex and $\Delta$[FeII/H] = --0.02 $\pm$ 0.03 dex.

% FIGURE 3
%
\begin{figure}[h]
\centering
\includegraphics[trim=0cm 3cm 0cm 3cm,clip=true,scale=.55,angle=270]{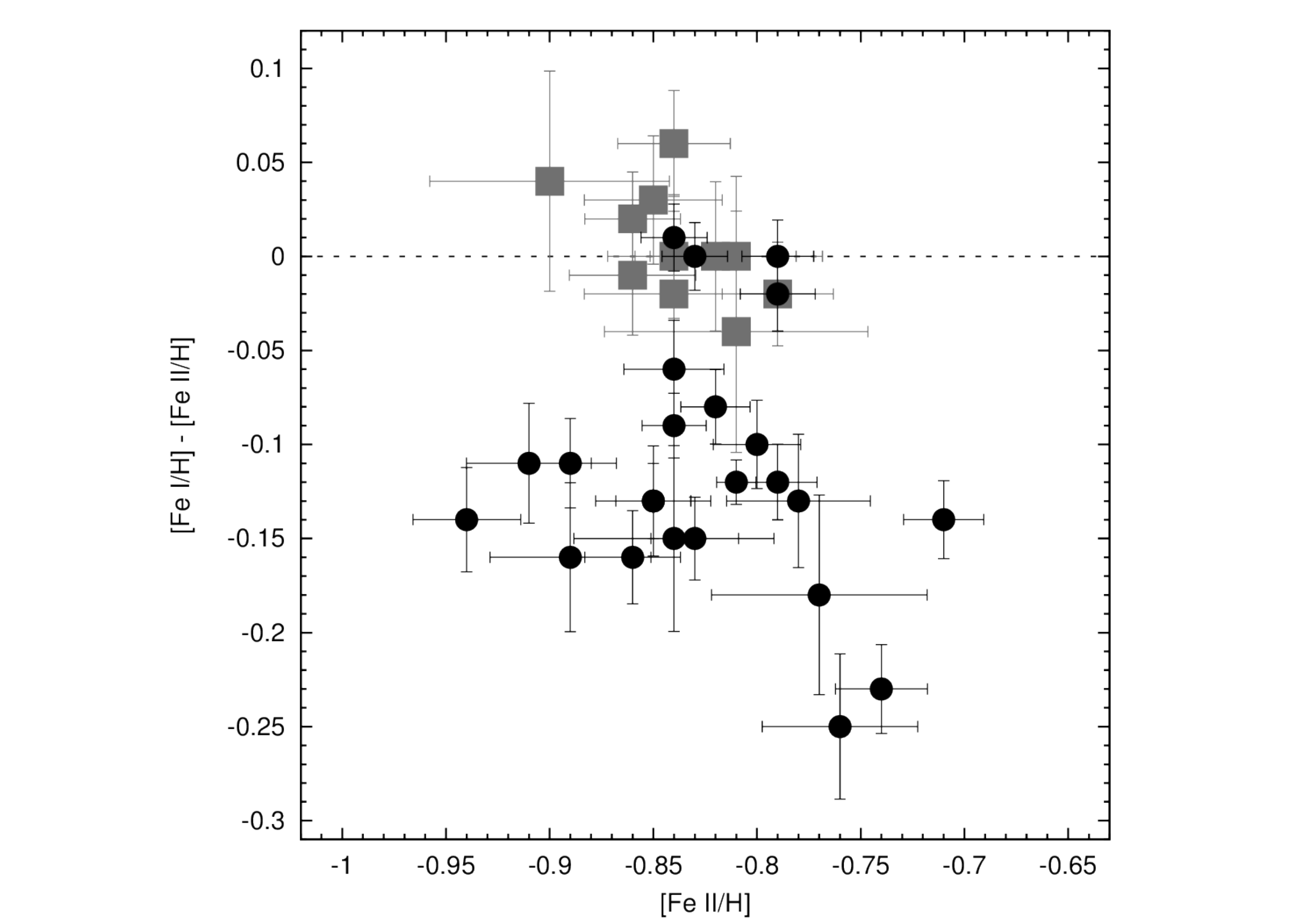}
\caption{Difference between [FeI/H] and [FeII/H] as a function of
[FeII/H] for the 24 AGB stars (black circles) and the 11 RGB stars
(gray squares) in our 47 Tuc samples.}
\label{c4_fediff}
\end{figure}
%
%

%%%%%%%%%%%%%%%%%%%%%%%%%%%%%%%%%%%%%%%%%%%%%%%%%%%%%%%%%%%

\emph{104 Tau --} We applied the same procedure of data reduction and
spectroscopic analysis to the star 104 Tau (HD 32923), which was
observed during the first observing night as a radial velocity
standard star. The spectrum has been reduced with the same set of
calibrations (i.e., bias, flat-fields, Th-Ar-Ne lamp) used for the
main targets.  We obtained a radial velocity of $20.96 \pm 0.02$ km
s$^{-1}$, which is consistent with the value ($20.62 \pm 0.09$ km
s$^{-1}$) quoted by \cite{nidever02}.  We adopted the average values
of $T_{\rm eff}$ and log~$g$ provided by \cite{takeda05} and
\cite{ramirez09} in previous analyses of the star ($T_{\rm eff}$=
5695 K and log~$g= 4.05$), while $v_{\rm turb}$ has been constrained
spectroscopically.  We derived [FeI/H] $= -0.17 \pm 0.01$ ($\sigma$ =
0.09 dex) and [FeII/H] $= -0.20 \pm 0.02$ ($\sigma$ = 0.07 dex), well
matching, within a few hundredth of dex, the values quoted by
\cite{takeda05} and \cite{ramirez09}.

%---------------------------------------------------------------------- LINELIST

\emph{Arcturus and the Sun --} In order to test the robustness of the
used linelist (and in particular to check for possible systematic
offsets due to the adopted oscillator strengths of FeI and FeII
lines), we adopted the same procedure to measure the iron abundance of
Arcturus (HD 124897) and the Sun, both having well established
atmospheric parameters.  In the case of Arcturus we retrieved a FEROS
spectrum from the ESO archive (Program ID: 074.D-0016), adopting 
$T_{\rm eff}$ = 4300 K, log~$g$ = 1.5, $v_{\rm turb}$ = 1.5 km s$^{-1}$
and [M/H]$ = -0.5$ as derived by \cite{lecureur07}.
We obtained [FeI/H]$ = -0.56 \pm 0.01$
dex and [FeII/H]$ = -0.57 \pm 0.01$ dex, in very good agreement with
previous determinations \citep{fulbright06,lecureur07,ramirez11}.
We repeated the same test on a FLAMES-UVES ($R\sim 45000$) twilight
spectrum of the Sun\footnote{http://www.eso.org/observing/dfo/quality/GIRAFFE/pipeline/solar.html},
adopting $T_{\rm eff}$ = 5777 K, log~$g$ = 4.44 and $v_{\rm turb}$ =
1.0 km s$^{-1}$ and finding absolute Fe abundances of 7.49 $\pm$ 0.01
dex and 7.50 $\pm$ 0.02 dex from neutral and single ionized iron lines,
respectively.

%---------------------------------------------------------------------- PARAMETRI
\subsubsection{Checks on the atmospheric parameters}

\emph{Effective temperature and gravity from spectroscopy --} We
checked whether atmospheric parameters derived spectroscopically could
help to reconcile the [FeI/H] and [FeII/H] abundance ratios.  The
adopted photometric estimates of $T_{\rm eff}$ well satisfy the
{\it excitation balance} (i.e. there is no slope between abundances and
excitation potential). Hence very small (if any) adjustments, with a
negligible impact on the derived abundances, can be admitted.
Instead, the values adopted for the surface gravity can have a
significant impact on the difference between [FeI/H] and [FeII/H].
In fact, the abundances derived from FeII lines are a factor of $\sim
5$ more sensitive to variations of log~$g$ with respect to those
obtained from FeI lines: for instance, a variation of $-0.1$ dex in
log~$g$ leads to negligible variation ($\sim -0.01$ dex) in the FeI
abundance, while [FeII/H] decreases by 0.05 dex.  Hence, a lower
value of the surface gravity could, in principle, erase the difference
between [FeI/H] and [FeII/H].

We found that the derived spectroscopic gravities are on average lower
than the photometric ones by 0.25 dex, with a maximum difference of
$\sim$0.5 dex.  As an example, the photometric analysis of star \#100103
provides [FeI/H]$ = -1.01 \pm 0.01$ ($\sigma$ = 0.11 dex) and
[FeII/H] $= -0.76 \pm 0.04$ ($\sigma$ = 0.13 dex), with $T_{\rm eff}$ =
4450 K, log~$g$ = 1.60 and $v_{\rm turb}$ = 1.15 km s$^{-1}$. When a
fully spectroscopic analysis is performed (thus optimizing all the
atmospheric parameters simultaneously), we obtain $T_{\rm eff}$ = 4475
K, log~$g$ = 1.10 and $v_{\rm turb}$ = 1.25 km s$^{-1}$, and the
derived abundances are [FeI/H] $= -1.09 \pm 0.01$ and [FeII/H] =
$-1.10 \pm 0.02$.  As expected, the spectroscopic values of $T_{\rm
eff}$ and $v_{\rm turb}$ are very similar to those adopted in the
photometric analysis.  A large difference is found for log~$g$, but
the final iron abundances are both too low with respect to the
literature and the 11 RGB star values to be considered acceptable.
Similar results are obtained for all the other AGB targets in our
sample.

Small surface gravity values as those found from the fully
spectroscopic analysis would require that stars reach the AGB phase
with an average mass of 0.4~$M_\odot$ (keeping $T_{\rm eff}$ and
luminosity fixed). This is lower than the value expected by
considering a main sequence turnoff mass of 0.9~$M_\odot$ and a $\sim
0.25~M_\odot$ mass loss during the RGB phase \citep{origlia07,origlia14}. 
Note that for some stars where [FeI/H]$-$[FeII/H]$\le
-0.20$ dex, the derived spectroscopic gravity would require a stellar
mass of $\sim 0.2~M_\odot$, which is even more unlikely for GC stars
in this evolutionary stage, also by taking into account the mass loss
rate uncertainties \citep{origlia14}.  Alternatively, these values of
log~$g$ can be obtained by assuming a significantly larger (by
about 0.5 mag) distance modulus.  However, this would be incompatible
with all the previous distance determinations for 47Tuc
\citep[see e.g.][]{ferraro99a,mclaughlin06,bergbuch09}.

\emph{Effective temperature and gravity from a different photometric
approach --} We repeated the analysis by adopting effective
temperatures estimated from the de-reddened color of each
target and the $(V-I)_0$-$T_{\rm eff}$ relation provided by
\cite{alonso99}, based on the Infrared Flux Method \citep[see][and
references therein]{blackwell90}.  Because this color-$T_{\rm eff}$
relation is defined in the Johnson photometric system, we converted
the target magnitudes in that system, following the prescriptions of
\cite{sirianni05}.  Moreover, gravities have been computed from the
Stefan-Boltzmann relation, by using the derived values of $T_{\rm eff}$,
the luminosities obtained from the observed $V$-band
magnitudes, assuming a mass of 0.8~$M_\odot$ for all the
stars (according to the best-fit isochrone discussed above) and
adopting the bolometric corrections computed according to
\cite{buzzoni10}. The average difference between the $T_{\rm eff}$
values obtained from the isochrone and those derived from the
\cite{alonso99} relation is of 3 K ($\sigma$ = 50 K). For gravities we
obtained an average difference of 0.05 dex ($\sigma$ = 0.03 dex) and
for the microturbulent velocities we found 0.01 km s$^{-1}$ ($\sigma$
= 0.04 km s$^{-1}$).  We repeated the chemical analysis with the new
parameters, finding that they do not alleviate the difference between
the average [FeI/H] and [FeII/H] abundance ratios: we obtained [FeI/H] $= -0.94
\pm 0.01$ dex ($\sigma$ = 0.06 dex) and [FeI/H] $= -0.84 \pm 0.01$
dex ($\sigma$ = 0.07 dex).  Thus, the iron abundances estimated from
FeI lines remain systematically lower than those obtained from FeII
lines and those found in the RGB stars.

\emph{Microturbulent velocity --} We note that the (spectroscopically)
derived values of $v_{\rm turb}$ span a large range (between 1 and 2
km s$^{-1}$ for most of the targets). Also, a small trend between the
average abundances and $v_{\rm turb}$ is detected, [FeI/H] increasing
by 0.15 dex/km s$^{-1}$ and [FeII/H] varying by 0.08 dex/km s$^{-1}$.
The very large number of lines ($\sim$150) used to constrain $v_{\rm turb}$,
as well as the wide range of line strengths covered by the
selected transitions, ensure that no bias due to small number
statistics or small range of line strengths occurs in the
determination of $v_{\rm turb}$ (note that no specific trend between
[Fe/H] and $v_{\rm turb}$ is found among the RGB stars). Also, the
values of $v_{\rm turb}$ do not change significantly changing the
range of used reduced EWs (see Section~\ref{c4_chem}).

We checked the impact of a different $v_{\rm turb}$ scale, adopting
the $v_{\rm turb}$--log~$g$ relation provided by \cite{kirby09}.
Because our targets have very similar gravities, they have ultimately
the same value of $v_{\rm turb}$ ($\sim$ 1.7 km s$^{-1}$), and the
situation worsens: in several stars the dispersion around the mean
abundance significantly increases (up to $\sim$0.3 dex, in comparison
with $\sigma$ = 0.15 dex found with the spectroscopic estimate of
$v_{\rm turb}$).  This is a consequence of the trends found between
abundances and line strengths introduced by not optimized $v_{\rm
turb}$.  The new average abundances of the entire sample are
[FeI/H] $= -1.03 \pm 0.04$ dex ($\sigma$ = 0.18 dex) and [FeII/H] $=
-0.89 \pm 0.02$ dex ($\sigma$ = 0.12 dex). Hence, with a different
assumption about $v_{\rm turb}$ not only the star-to-star dispersion
increases by a factor of 2 for both the abundance ratios, but, also,
the systematic difference between [FeI/H] and [FeII/H] remains in
place.

\emph{Model atmospheres --} The plane-parallel geometry
is adopted both in the ATLAS9 model atmospheres and 
in the line-formation calculation performed by GALA.
As pointed out by \cite{heiter06}, that investigated the impact 
of the geometry on the abundance analysis of giant stars, 
the geometry has a small effect on line formation. 
In order to quantify these effects, we reanalyzed the target stars by using 
the last version of the MARCS model atmospheres \citep{gustafsson08},
which adopt spherical geometry. 
The average abundance differences between the analysis performed with 
MARCS and that performed with ATLAS9 are of --0.005 dex ($\sigma$ = 0.01 dex)
and +0.02 dex ($\sigma$ = 0.04 dex) for FeI and FeII, respectively.
Hence, the use of MARCS model atmospheres does not change our finding
about FeI and FeII abundances (both in AGB and in RGB stars). 
Note that \cite{heiter06} conclude that abundances derived 
with spherical models and plane-parallel transfer are in excellent agreement 
with those obtained with a fully spherical treatment.

\section{Discussion}
\label{c4_discussion}

\subsection{A possible signature of NLTE effects?}

For the 24 AGB stars studied in 47 Tuc, the iron abundance obtained
from single ionized lines well matches that measured in RGB stars
(from both FeI and FeII lines).  Instead, systematically lower iron
abundances are found for the AGB sample from the analysis of FeI.
All the checks discussed in Section~\ref{c4_check} confirm that such a
discrepancy is not due to some bias in the analysis or to the adopted
atmospheric parameters, and there are no ways to reconcile the
abundances from Fe lines with those observed in the RGB
stars.

The only chemical analyses performed so far on AGB stars in 47 Tuc
have been presented by \cite{wylie06} and \cite{worley09}. In both
cases all the parameters have been constrained spectroscopically (in
particular, log~$g$ is obtained by forcing [FeI/H] and [FeII/H] to
be equal within the uncertainties).  \cite{wylie06} analysed 5 AGB
stars (brighter than those discussed in this work), finding [FeI/H]
$= -0.60 \pm 0.06$ dex and [FeII/H] =$ -0.64 \pm 0.10$ dex.  The same
methodology to derive the parameters has been used by \cite{worley09}
to analyse a bright AGB star, finding [FeI/H]$ = -0.72 \pm 0.16$ dex
and [FeII/H]$ = -0.74 \pm 0.08$ dex.  Unfortunately, the spectroscopic
determination of the gravity does not allow to understand whether also
for these AGB stars a real discrepancy of [FeI/H] and [FeII/H] does
exist.

A natural explanation for the negative values of [FeI/H]$-$[FeII/H]
measured for our AGB sample would be that these stars suffer for
departures from the LTE condition, which mainly affects the less
abundant species (in this case FeI), while leaving virtually
unaltered the dominant species \citep[i.e. FeII;][]{mashonkina11}.
In late-type stars, NLTE effects are mainly driven by overionization
mechanisms, occurring when the intensity of the radiation field
overcomes the Planck function \citep[see][for a complete review of
these effects]{asplund05}.  These effects are predicted to increase
for decreasing metallicity and for decreasing atmospheric densities
(i.e., lower surface gravities at a given $T_{\rm eff}$), as pointed
out by a vast literature
\citep[see e.g.][]{thevenin99,asplund05,mashonkina11,lind12,bergemann14}.
At the metallicity of 47 Tuc, significant deviations are expected only
for stars approaching the RGB-Tip. \cite{bergemann12} and
\cite{lind12} computed a grid of NLTE corrections for a sample of Fe
I and FeII lines in late-type stars over a large range of metallicity.
Assuming the atmospheric parameters of the 11 RGB stars in our sample
and the measured EWs of the iron lines in common with their grid (25
FeI and 9 FeII lines), the predicted NLTE corrections are 
[Fe/H]$_{\rm NLTE}-$[Fe/H]$_{\rm LTE}\simeq +0.04$ dex.  This is
consistent with no significant differences between [FeI/H] and [FeII/H]
found in our analysis (Section~\ref{c4_check}) and in previous
studies \citep[see e.g.][]{carretta04b,koch08,carretta09b}.  Instead, a
larger difference ([FeI/H]$-$[FeII/H]$ = -0.08$ dex) has been found
for the brightest RGB stars in 47 Tuc \citep{koch08}, as expected.
\footnote{For sake of comparison, \cite{mucciarelli13c} analysed RGB
stars close to the RGB Tip in the metal-poor GC NGC 5694 ([Fe/H]
$\sim -2.0$ dex), finding an average difference between [FeI/H] and
[FeII/H] of $-0.14$ dex, consistent with the expected
overionization effects.}

However, if we use the same grid to estimate the NLTE corrections for
our sample of AGB stars, we find [Fe/H]$_{\rm NLTE}-$[Fe/H]$_{\rm LTE}= +0.06$ dex.
This value is consistent with the NLTE corrections
predicted for RGB stars and smaller than the difference we observe between
FeI and FeII in our sample of AGB stars.
Interestingly, a situation similar to that encountered in
the present work has been met by \cite{ivans01} in the spectroscopic
analysis of giant stars in the GC M5.  Their sample includes 6 AGB and
19 RGB stars, ranging from the luminosity level of the AGB clump, up
to the RGB-Tip. Also in their analysis, [FeI/H] in AGB stars is
systematically lower (by about 0.15 dex) than [FeII/H], while no
differences are found for the RGB stars.  The authors performed
different kinds of analysis, finding that the only way to reconcile
the iron abundance in AGB stars with the values obtained in RGB
stars is to adopt the photometric gravities and rely on the FeII
lines only, which are essentially insensitive to LTE departures.  Our
findings, coupled with the results of \cite{ivans01} in M5, suggest
that the NLTE effects could depend on the evolutionary stage (being
more evident in AGB stars with respect to RGB stars), also at
metallicities where these effects should be negligible (like in the
case of 47 Tuc that is more metal-rich than M5).
This result is somewhat surprising, because a
dependence of NLTE effects on the evolutionary stage is not
expected by the theoretical models.

%%%%% 4 AGB normali %%%%%%%%%%%%%%%%%%%%%%%%%%%%%%%%%%%%%%%%%%%%%%%%%%

\begin{figure}[h]
\centering
\includegraphics[trim=0cm 0cm 0cm 0cm,clip=true,scale=.55,angle=270]{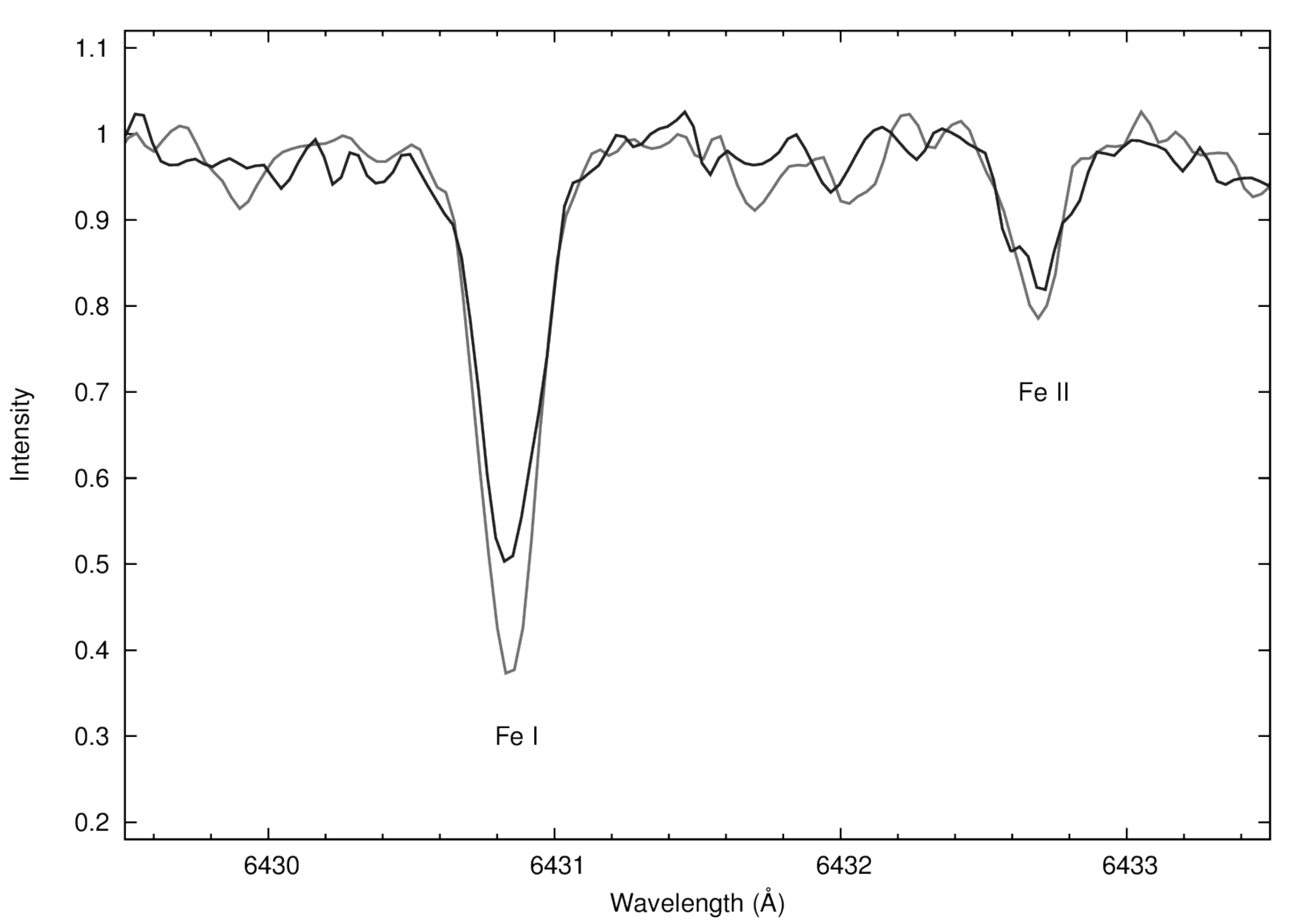}
\caption{Comparison between the normalized spectra of target \#100171
(grey line) and \#100174 (black line). The position of a FeI line
and a FeII line is marked.}
\label{c4_spec}
\end{figure}

In our sample, we identify 4 AGB stars (namely \#100169, \#100171, \#200021 and \#200023)
where the absolute difference between [FeI/H] and [FeII/H] is quite small
(less than $0.05$ dex; see Figure~\ref{c4_fediff}). According to the different behaviour observed
between the AGB and the RGB samples, one could suspect that these
objects are RGB stars. However, we checked their position on the CMD
also by using an independent photometry \citep{sarajedini07},
confirming that these 4 targets are indeed genuine AGB stars.
Figure~\ref{c4_spec} compares two iron lines of the spectra of targets \#100171 and \#100174,
which are located in the same position of the CMD (thus being
characterized by the same atmospheric parameters), 
but have [FeI/H]$-$[FeII/H]$ = 0$ and $-0.16$ dex, respectively. Clearly, the Fe
II lines of the two stars have very similar depths, suggesting the
same iron abundance, while the FeI line of \#100174 (black spectrum) is significantly
shallower than that in the other star.  This likely suggests that NLTE
effects among the AGB stars have different magnitudes, the
overionization being more or less pronounced depending on the star.

The origin of this behaviour, as well as the unexpected occurrence of
NLTE effects in AGB stars of these metallicity and atmospheric
parameters, are not easy to interpret and their detailed investigation
is beyond the scope of this paper.  Suitable theoretical models of the
line formation under NLTE conditions in AGB stars should be computed
in order to explain the observed difference in the FeI and FeII
abundances.  We cannot exclude that some
inadequacies of the 1-dimensional model atmospheres can play a role in
the derived results. Up to now, 3-dimensional hydrodynamics
simulations of the convective effects in AGB stars have been performed
only for a typical AGB star during the thermal pulses phase and with
very low $T_{\rm eff}$, $\sim$2800 K \citep{freytag08}.
Similar sophisticated models for earlier and warmer phases of the AGB are urged.

% FIGURE 4
%

\subsection{Impact on traditional chemical analyses}
\label{c4_tradit}

This result has a significant impact on the approach traditionally
used for the chemical analysis of AGB stars.  In particular, two main
aspects deserve specific care:

(1) the FeI lines should not be used to determine the iron abundance
of AGB stars. In fact, when photometric parameters are adopted,
the [FeI/H] abundance ratio can be systematically lower than that
obtained from FeII lines.  Indeed, the most reliable route to derive
the iron abundance in AGB stars is to use the FeII lines, that are
essentially unaffected by NLTE effects and provide the same abundances
for both RGB and AGB objects. This strategy requires high-resolution,
high-quality spectra, because of the low number of FeII lines
available in the optical range (smaller by a factor of $\sim 10$ with
respect to the number of FeI lines).  This is especially true 
at low metallicity since the lines are shallower and the NLTE effects 
are expected to be stronger;

(2) another point of caution is that AGB stars must be analysed by
adopting the photometric gravities (derived from a theoretical
isochrone or through the Stefan-Boltzmann equation) and not by using
the spectroscopic method of the ionization balance (at variance with
the case of RGB stars, where this approach is still valid).  This
method, which is widely adopted in the chemical analysis of optical
stellar spectra, constrains log~$g$ by imposing 
the same abundance for given species as obtained from
lines of two different ionization stages.  Variations of log~$g$ lead
to variations in the abundances measured from ionized lines (which are
very sensitive to the electronic pressure), while the neutral lines
are quite insensitive to these variations.  Because of the
systematically lower FeI abundance, this procedure leads to
improbably low surface gravities in AGB stars (as demonstrated in
Section~\ref{c4_check}).

If the FeI line are used as main diagnostic of the iron abundance, a
{\sl blind} analysis, where AGB and RGB stars are not analysed
separately, can lead to a spurious detection of large iron spreads in
GCs.  In light of these considerations, the use of FeII lines is
recommended to determine the metallicity of AGB objects, regardless of
their metallicity and luminosity.  On the other hand, the systematic
difference between FeI abundances in AGB and RGB stars could, in
principle, be used to recognize AGB stars when reliable photometric
selections cannot be performed.

\subsection{Searching for evolved BSSs among AGB stars: a new diagnostic?}
\label{c4_ebss}

The discovery of such an unexpected NLTE effect in AGB stars might
help to identify possible e-BSSs among AGB stars.  In fact, BSSs spend
their RGB phase in a region of the CMD which is superimposed to that
of the cluster AGB \citep[e.g.,][]{beccari06,dalessandro09}.
Thus, e-BSSs are indistinguishable from genuine AGB stars in terms of colors and
magnitudes, but they have larger masses at comparable radii.  Hence,
their surface gravity is also expected to be larger (by about 0.2-0.3
dex) than that of ``canonical'' AGB stars of similar temperature and
luminosity.  Unfortunately, the spectroscopic measurement of log~$g$
cannot be used to distinguish between genuine AGB stars and e-BSSs,
since the ionization balance method cannot be applied in the presence
of the NLTE effects affecting the AGB (see Sect. \ref{c4_tradit}).
However, because of the seemingly dependence of the NLTE effects on
the evolutionary phase, the measurement of different 
values of [FeI/H] and [FeII/H] should allow to recognize genuine AGB
stars from e-BSSs evolving along their RGB within a sample of objects
observed in the AGB clump of a GC.

Of course, this expectation holds only if the lack of NLTE
effects found for low-mass ($\sim 0.8~M_\odot$) RGB stars also holds
for larger masses ($\sim 1.2~M_\odot$), typical of e-BSSs in
GCs.  In order to check this hypothesis, we retrieved from the ESO
archive FLAMES-UVES spectra for 7 giant stars in the open cluster
Berkeley 32 (ID Program: 074.D-0571). This cluster has an age of $\sim$6-7 Gyr
\citep{dorazi06}, corresponding to a turnoff mass of $\sim
1~M_{\odot}$, comparable with the typical masses of the BSSs observed
in Galactic GCs \citep[e.g.,][]{shara97,demarco05,ferraro06,lanzoni07,fiorentino14}.
We analyzed the spectra following the same procedure used for the AGB
stars in 47 Tuc.  The derived abundances are [FeI/H]$ = -0.34 \pm
0.01$ dex ($\sigma$ = 0.04 dex) and [FeII/H]$ = -0.38 \pm 0.02$ dex
($\sigma$ = 0.06 dex), in nice agreement with the results of
\cite{sestito06} based on the same dataset. The small difference
between the abundances from FeI and FeII lines confirms the evidence
arising from the RGB sample of 47 Tuc: in metal-rich RGB stars no sign
of overionization is found, also for stellar masses larger than those
typical of Galactic GC stars.

Within this framework, we checked whether some e-BSSs could be hidden
in the analysed sample of putative AGB stars, especially among the
four objects with negligible difference between FeI and FeII
abundances. To this end we derived new atmospheric parameters for all
targets projecting their position in the CMD on a grid of evolutionary
BaSTI tracks crossing the mean locus of the studied stars: these are
the RGB tracks for stellar masses of 1.2, 1.3 and 1.4~$M_\odot$
\citep[see][]{beccari06}. On average, the new values of $T_{\rm eff}$
are slightly larger than those obtained in Section~\ref{c4_chem}, while
gravities are systematically larger, by up to $\sim$0.3 dex.  For
increasing mass of the adopted track, the general behaviour is that
[FeI/H] slightly increases (mainly because of a small increase of
$T_{\rm eff}$), while [FeII/H] decreases (because of the combined
growth of both $T_{\rm eff}$ and log~$g$).  However, none of the measured 
AGB stars show [FeI/H]$\sim$[FeII/H]$\sim$[Fe/H]$_{\rm RGB}$ with the new
sets of parameters. This suggests that no e-BSSs are hidden within our
observed AGB sample.  In particular, we found that for the four AGB
stars with no evidence of overionization, the [FeII/H] abundance
ratios derived with the new parameters are larger than those obtained
from [FeI/H] (due to the increase in log~$g$) and for RGB stars.
This indicates that the atmospheric parameters estimated from the AGB
portion of the isochrone are the most appropriate and these four
objects are indeed genuine AGB stars.  As an additional check, we
measured the iron abundances assuming atmospheric parameters from
theoretical tracks of massive (1-2~$M_\odot$) AGB stars. In this case,
the situation improves.  In particular, by using an AGB track of 1.2~$M_\odot$, 
because of the combined effect of larger $T_{\rm eff}$
($\sim$+100 K) and log~$g$ ($\sim$+0.2), we find 
average values of [FeI/H]$ = -0.85$ dex and [FeII/H]$ = -0.81$ dex.
However, this scenario is unlikely because the probability
to detect an e-BSS during its AGB phase is a factor of $\sim 10$ lower
than the probability to observe it during the RGB phase (consistently
with the time duration of these evolutionary phases).

As a general rule, however, we stress that, the dependence of NLTE
effects on the evolutionary stage (if confirmed) can be used as a
diagnostic of the real nature of the observed AGB stars and to
identify e-BSSs hidden in a putatively genuine AGB sample.

%------------------------------------------------------------------------------

\section{Summary}
\label{c4_summary}

We have measured the iron abundance of 24 AGB stars members of the GC
47 Tuc, by using high-resolution FEROS spectra.  By adopting photometric
estimates of $T_{\rm eff}$ and log~$g$, we derived average iron
abundances [FeI/H]$ = -0.94 \pm 0.01$ dex ($\sigma$ = 0.08 dex) and
[FeII/H]$ = -0.83 \pm 0.01$ dex ($\sigma$ = 0.05 dex).  Thus, while
the abundance estimated from ionized lines ($-0.83$ dex) well matches
the one obtained for RGB stars, the values measured from neutral lines
appear to be systematically lower.
We carefully checked all the steps of our
chemical analysis procedure and the adopted atmospheric parameters,
finding no ways to alleviate this discrepancy.

Such a difference is compatible with the occurrence of NLTE effects
driven by iron overionization, confirming the previous claim by
\cite{ivans01} for a sample of AGB stars in M5.  Our findings suggest that
$(i)$ the departures from the LTE approximation can be more
significant than previously thought, even at relatively high
metallicities ([Fe/H]$\sim -0.83$) and for stars much fainter
than the RGB-Tip, $(ii)$ iron overionization can be more or less
pronounced depending on the star (in fact, four stars in our sample
turn out to be unaffected), and $(iii)$ these effects depend on the
evolutionary stage (they are not observed among RGB stars).  We
discussed the impacts of this effect on the traditional chemical
analysis of AGB stars: if FeI lines are used and/or surface gravities
are derived from the ionization balance, artificial under-estimates
and/or spreads of the iron abundances can be obtained.  If the
dependence of these NLTE effects on the evolutionary stage is
confirmed, the systematic difference between [FeI/H] and [FeII/H]
abundance ratios can, in principle, be used to identify e-BSSs within
a sample of genuine AGB stars (these two populations sharing the same
locus in the CMD).

From the theoretical point of view, new and accurate models are needed
to account for these findings, in particular to explain the dependence
on the evolutionary stage.  Observationally, further analyses of
high-resolution spectra of AGB stars are crucial and urged to firmly
establish the occurrence of these effects and to investigate their
behaviour as a function of other parameters, like the cluster
metallicity, the stellar mass and the stellar luminosity.

%%%%%%%%%%%%%%%%%%%%%%%%%%%%%%%%%%%%%%%%%%%%%%%%%%%%%%%%%%%%%%%%%%%% TABLES

%\begin{landscape}
\begin{deluxetable}{rrrccr}
\tiny
\tablecolumns{6}
\tablewidth{0pt}
\tablecaption{Photometric properties and radial velocities of the AGB sample}
\tablehead{\colhead{ID} & \colhead{RA} & \colhead{Dec} & \colhead{$m_{\rm F606W}$}
& \colhead{$m_{\rm F814W}$} & \colhead{RV} \\
& (J2000) & (J2000) & \colhead{} & \colhead{} & \colhead{(km s$^{-1}$)}  }
\startdata
%\hline
 & & & & & \\
%\hline
 100094 & 6.1013041 & --72.0745785 & 12.50 & 11.61 & --28.20 $\pm$ 0.02 \\
 100103 & 6.0415972 & --72.0787949 & 12.58 & 11.75 & --19.03 $\pm$ 0.03 \\
 100110 & 6.0212020 & --72.0791089 & 12.61 & 11.76 & --24.71 $\pm$ 0.02 \\
 100115 & 6.0253911 & --72.0763508 & 12.66 & 11.81 &   +9.42 $\pm$ 0.03 \\
 100118 & 6.0021239 & --72.0797989 & 12.66 & 11.83 & --22.55 $\pm$ 0.03 \\
 100119 & 5.9937010 & --72.1048573 & 12.66 & 11.82 &  --0.83 $\pm$ 0.02 \\
 100120 & 6.0686535 & --72.0977710 & 12.67 & 11.83 & --36.76 $\pm$ 0.02 \\
 100125 & 6.0229324 & --72.0843986 & 12.68 & 11.85 & --28.72 $\pm$ 0.03 \\
 100133 & 6.0062382 & --72.0914060 & 12.74 & 11.92 & --12.33 $\pm$ 0.02 \\
 100136 & 6.0116940 & --72.0848407 & 12.75 & 11.94 & --13.21 $\pm$ 0.02 \\
 100141 & 6.0474465 & --72.0917594 & 12.77 & 11.93 & --13.46 $\pm$ 0.02 \\
 100142 & 6.0474090 & --72.1034811 & 12.77 & 11.93 & --20.41 $\pm$ 0.03 \\
 100148 & 6.0283392 & --72.0802692 & 12.80 & 11.99 & --26.31 $\pm$ 0.03 \\
 100151 & 5.9726093 & --72.1059031 & 12.81 & 12.00 &  --8.50 $\pm$ 0.03 \\
 100152 & 6.0354605 & --72.0975184 & 12.81 & 11.97 &  --2.45 $\pm$ 0.02 \\
 100154 & 6.0171898 & --72.0853225 & 12.82 & 12.00 &  --4.70 $\pm$ 0.03 \\
 100161 & 6.0075116 & --72.0971679 & 12.84 & 12.00 & --22.37 $\pm$ 0.02 \\
 100162 & 6.0212298 & --72.0768974 & 12.85 & 12.01 & --41.45 $\pm$ 0.03 \\
 100167 & 6.0406724 & --72.0919126 & 12.87 & 12.03 & --13.98 $\pm$ 0.04 \\
 100169 & 6.0416117 & --72.1082069 & 12.87 & 12.04 & --21.38 $\pm$ 0.02 \\
 100171 & 6.0479798 & --72.0906284 & 12.88 & 12.05 & --12.69 $\pm$ 0.02 \\
 100174 & 6.0424039 & --72.0857888 & 12.90 & 12.07 & --13.81 $\pm$ 0.03 \\
 200021 & 6.1149444 & --72.0892243 & 12.74 & 11.89 & --21.07 $\pm$ 0.02 \\
 200023 & 6.1184182 & --72.0838295 & 12.85 & 12.02 & --22.19 $\pm$ 0.02 \\
\enddata
\tablecomments{Identification number, coordinates, $m_{\rm F606W}$ and $m_{\rm F814W}$
magnitudes \citep{beccari06}, and radial velocities for the 24
AGB stars analyzed.}
\label{c4_tab1}
\end{deluxetable}
%\end{landscape}

%\begin{landscape}
\begin{deluxetable}{cccccccc}
\tiny
\tablecolumns{8}
\tablewidth{0pt}
\tablecaption{Atmospheric parameters and iron abundances of the AGB sample}
\tablehead{\colhead{ID} & \colhead{$T_{\rm eff}^{\rm phot}$} & \colhead{log~$g^{\rm phot}$}
&  \colhead{$v_{\rm turb}$}
& \colhead{[FeI/H]} & \colhead{n(FeI)} & \colhead{[FeII/H]} & \colhead{n(FeII)} \\
& \colhead{(K)} & \colhead{(dex)} &  \colhead{(km s$^{-1}$)} & \colhead{(dex)} & & \colhead{(dex)} & }
\startdata
%\hline
 & & & & & & &  \\
%\hline
 100094  &  4425  &   1.55  &    2.00  &  --0.91$\pm$0.04  &  134  &  --0.79$\pm$0.08  &  13  \\
 100103  &  4450  &   1.60  &    1.15  &  --1.01$\pm$0.05  &  170  &  --0.76$\pm$0.10  &  14  \\
 100110  &  4475  &   1.60  &    1.10  &  --0.98$\pm$0.04  &  165  &  --0.85$\pm$0.07  &  12  \\
 100115  &  4500  &   1.65  &    0.55  &  --0.99$\pm$0.05  &  171  &  --0.84$\pm$0.09  &  13  \\
 100118  &  4500  &   1.65  &    1.30  &  --0.98$\pm$0.05  &  161  &  --0.85$\pm$0.08  &  15  \\
 100119  &  4500  &   1.65  &    1.80  &  --0.85$\pm$0.04  &  138  &  --0.71$\pm$0.07  &  15  \\
 100120  &  4500  &   1.65  &    1.70  &  --0.97$\pm$0.05  &  138  &  --0.74$\pm$0.07  &  13  \\
 100125  &  4500  &   1.65  &    0.95  &  --1.08$\pm$0.04  &  171  &  --0.94$\pm$0.08  &  14  \\
 100133  &  4550  &   1.70  &    1.50  &  --0.98$\pm$0.04  &  147  &  --0.83$\pm$0.08  &  14  \\
 100136  &  4550  &   1.70  &    1.30  &  --0.93$\pm$0.04  &  166  &  --0.84$\pm$0.07  &  15  \\
 100141  &  4550  &   1.70  &    1.65  &  --0.93$\pm$0.04  &  155  &  --0.81$\pm$0.07  &  11  \\
 100142  &  4550  &   1.70  &    1.60  &  --0.90$\pm$0.05  &  140  &  --0.80$\pm$0.07  &  12  \\
 100148  &  4575  &   1.75  &    0.95  &  --1.05$\pm$0.04  &  173  &  --0.89$\pm$0.08  &  13  \\
 100151  &  4575  &   1.75  &    1.85  &  --0.90$\pm$0.05  &  141  &  --0.82$\pm$0.07  &  12  \\
 100152  &  4575  &   1.75  &    1.80  &  --0.91$\pm$0.04  &  156  &  --0.78$\pm$0.07  &  14  \\
 100154  &  4575  &   1.75  &    1.20  &  --1.00$\pm$0.04  &  159  &  --0.89$\pm$0.07  &  15  \\
 100161  &  4575  &   1.75  &    1.40  &  --0.90$\pm$0.05  &  158  &  --0.84$\pm$0.08  &  13  \\
 100162  &  4600  &   1.75  &    0.75  &  --1.02$\pm$0.05  &  174  &  --0.91$\pm$0.07  &  13  \\
 100167  &  4600  &   1.80  &    1.20  &  --0.95$\pm$0.05  &  158  &  --0.77$\pm$0.09  &  13  \\
 100169  &  4600  &   1.80  &    1.80  &  --0.79$\pm$0.04  &  154  &  --0.79$\pm$0.07  &  14  \\
 100171  &  4600  &   1.80  &    1.60  &  --0.83$\pm$0.04  &  155  &  --0.83$\pm$0.07  &  11  \\
 100174  &  4600  &   1.80  &    1.10  &  --1.02$\pm$0.06  &  144  &  --0.86$\pm$0.08  &  13  \\
 200021  &  4550  &   1.70  &    1.90  &  --0.83$\pm$0.04  &  145  &  --0.84$\pm$0.07  &  14  \\
 200023  &  4575  &   1.75  &    1.80  &  --0.81$\pm$0.04  &  147  &  --0.79$\pm$0.07  &  15  \\
\hline
 &  &  &  & $\langle$[FeI/H]$\rangle$&  & $\langle$[FeII/H]$\rangle$ &   \\
 &  &  &  & --0.94$\pm$0.01  &  & --0.83$\pm$0.01  &  \\
\enddata
\tablecomments{ Identification number, photometric temperature and
  gravities, microturbulent velocities, [Fe/H] abundance ratios with
  total uncertainty and number of used lines, as measured from 
  neutral and single ionized lines.  For all the stars a global
  metallicity of [M/H]$ = -1.0$ dex has been assumed for the model
  atmosphere. The adopted solar value is 7.50 \citep{grevesse98}.}
\label{c4_tab2}
\end{deluxetable}
%\end{landscape}

\begin{deluxetable}{cccccccc}
\tiny
\tablecolumns{8}
\tablewidth{0pt}
\tablecaption{Atmospheric parameters and iron abundances of the RGB sample}
\tablehead{\colhead{ID} & \colhead{$T_{\rm eff}^{\rm phot}$} & \colhead{log~$g^{\rm phot}$}
&  \colhead{$v_{\rm turb}$}
& \colhead{[FeI/H]} & \colhead{n(FeI)} & \colhead{[FeII/H]} & \colhead{n(FeII)} \\
& \colhead{(K)} & \colhead{(dex)} &  \colhead{(km s$^{-1}$)} & \colhead{(dex)} & & \colhead{(dex)} & }
\startdata
 & & & & & & &  \\
 5270  &  4035  &  1.10  &  1.50  &  --0.85$\pm$0.05  &  140  &  --0.81$\pm$0.13  &  12 \\
12272  &  4130  &  1.25  &  1.50  &  --0.87$\pm$0.05  &  147  &  --0.86$\pm$0.11  &  13 \\
13795  &  4170  &  1.35  &  1.60  &  --0.81$\pm$0.05  &  141  &  --0.79$\pm$0.11  &  14 \\
14583  &  4305  &  1.60  &  1.50  &  --0.81$\pm$0.05  &  150  &  --0.81$\pm$0.10  &  13 \\
17657  &  4005  &  1.05  &  1.50  &  --0.86$\pm$0.04  &  133  &  --0.90$\pm$0.10  &  12 \\
18623  &  4250  &  1.50  &  1.50  &  --0.84$\pm$0.05  &  144  &  --0.86$\pm$0.10  &  12 \\
20002  &  4200  &  1.40  &  1.50  &  --0.86$\pm$0.05  &  147  &  --0.84$\pm$0.10  &  12 \\
23821  &  4250  &  1.50  &  1.20  &  --0.84$\pm$0.04  &  147  &  --0.84$\pm$0.09  &  14 \\
34847  &  4095  &  1.20  &  1.40  &  --0.82$\pm$0.05  &  141  &  --0.82$\pm$0.12  &  13 \\
36828  &  4215  &  1.40  &  1.40  &  --0.78$\pm$0.05  &  142  &  --0.84$\pm$0.11  &  11 \\
41654  &  4130  &  1.25  &  1.50  &  --0.82$\pm$0.05  &  142  &  --0.85$\pm$0.11  &  13 \\
\hline
 &  &  &  & $\langle$[FeI/H]$\rangle$&  & $\langle$[FeII/H]$\rangle$ &   \\
 &  &  &  & --0.83$\pm$0.01  &  & --0.84$\pm$0.01  &  \\
\enddata
\tablecomments{Columns are as in Table 2. For all the stars a global
metallicity of [M/H]$ = -1.0$ dex has been assumed for the model
atmosphere. The adopted solar value is 7.50 \citep{grevesse98}.}
\label{c4_tab3}
\end{deluxetable}

%% file: c5/ms_5.tex
% CHAPTER 5

\chapter{The Origin of the Spurious Iron Spread in the Globular Cluster NGC3201}

\label{c5}

{\bf Published in Mucciarelli et al. 2015, ApJ, 801, 69M}

{\it NGC3201 is a globular cluster suspected to have an intrinsic spread in  
the iron content.
We re-analysed a sample of 21 cluster stars observed with UVES-FLAMES at the Very Large 
Telescope and for which Simmerer et al. found a 0.4 dex wide [Fe/H] distribution with a 
metal-poor tail.
We confirmed that when spectroscopic gravities are adopted, the derived [Fe/H] distribution spans $\sim$0.4 dex.
On the other hand, when photometric gravities are used, the metallicity distribution 
from FeI lines remains large, while that derived from FeII lines is narrow and compatible 
with no iron spread.
We demonstrate that the metal-poor component claimed by Simmerer et al.
is composed by asymptotic giant branch stars that could be affected by non local thermodynamical
equilibrium effects driven by iron overionization. 
This leads to a decrease of the FeI abundance, while leaving the FeII abundance unaltered.
A similar finding has been already found in asymptotic giant branch 
stars of the globular clusters M5 and 47Tucanae.
We conclude that NGC3201 is a normal cluster, with no evidence of intrinsic iron spread.}

\section{Observations}

High-resolution spectra taken with UVES-FLAMES@VLT \citep{pasquini00} for 21 giant stars members of NGC3201
have been retrieved from the ESO archive. The spectra have been acquired with the UVES grating 580 
Red Arm CD\#3, that provides a high spectral resolution (R$\sim$45000) and a large spectral coverage 
($\sim$4800-6800 \AA). The spectra have been reduced using the dedicated ESO
pipeline\footnote{http://www.eso.org/sci/software/pipelines/}, performing bias subtraction, flat-fielding, 
wavelength calibration, spectral extraction and order merging. In each exposure one fiber is dedicated 
to sample the sky background and used to subtract this contribution from each individual spectrum.

Spectroscopic targets have been identified in our photometric catalog, obtained by combining 
high resolution images acquired with the HST-ACS camera and wide-field images 
acquired with the ESO-WFI imager. Both the photometric datasets have been obtained through the V and 
I filters. A total of 13 targets lie in the innermost cluster region, covered by ACS, while 8 stars are 
in the external region, covered by WFI.
The membership of all the targets is confirmed by their very high radial velocity 
($<RV_{helio}>=+494.6\pm0.8$ km s$^{-1}$, $\sigma=3.6$ km s$^{-1}$) that allows to easily distinguish 
the cluster members from the surrounding field stars.

\begin{figure}[h]
\centering
\plottwo{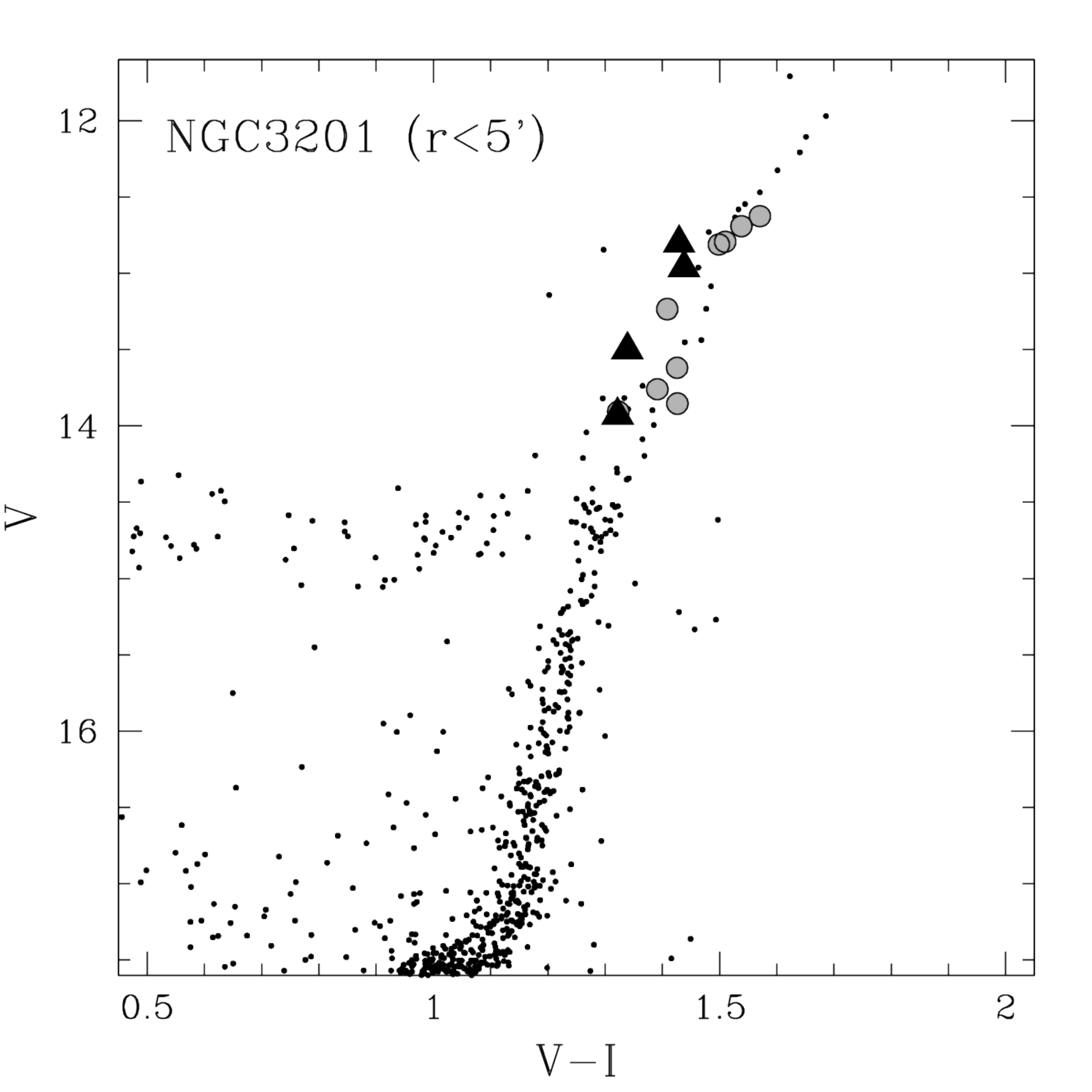}{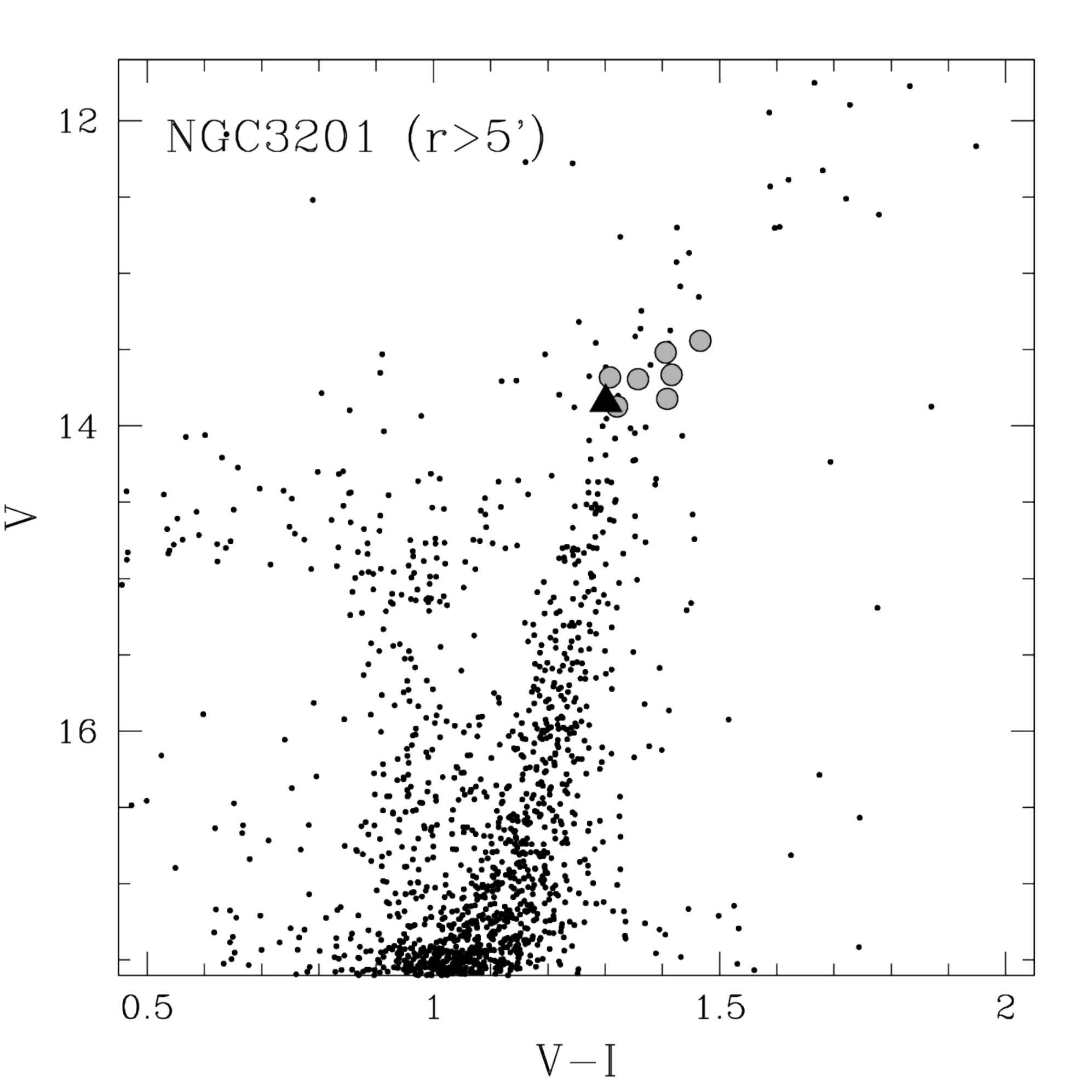}
\caption{CMDs for central and external regions of NGC3201 (left and right panels, respectively), 
corrected for differential reddening. Large circles are the spectroscopic 
targets flagged as {\sl metal-rich} ([Fe/H]$>$--1.58 dex) in \citet{simmerer13}, while 
large triangles those identified as {\sl metal-poor} ([Fe/H]$\le$--1.58 dex).}
\label{cmd}
\end{figure}

The position of the targets in the color-magnitude diagrams (CMDs) is shown in Figure~\ref{cmd}.
These CMDs have been corrected for differential reddening using the method described 
in \citet{massari12} and adopting the extinction law by \citet{cardelli89}. 
In order to calculate guess values for the atmospheric parameters of the target stars, 
we fitted the CMDs with an appropriate theoretical isochrone from the BaSTI dataset 
\citep{pietrinferni06}, computed with an age of 11 Gyr \citep{marin09}, 
Z = 0.001 and $\alpha$-enhanced chemical mixture, finding a color excess $E(B-V)$ = 0.31 mag and 
a true distance modulus $(m-M)_0$ = 13.35 mag.

Table~\ref{tab1_c5} lists the main information about the targets, by adopting the same 
identification numbers used by \citet{simmerer13} who adopted the original names by \citet{cote94}.

% FIGURE 1
%

\section{Analysis}

Iron abundances have been derived with the package 
GALA\footnote{http://www.cosmic-lab.eu/gala/gala.php}\citep{mucciarelli13}
by matching the measured and theoretical equivalent widths (EWs).
Model atmospheres have been calculated with the code 
ATLAS9\footnote{http://wwwuser.oats.inaf.it/castelli/sources/atlas9codes.html}.
We selected FeI and FeII lines predicted to be unblended at the UVES resolution and at the 
typical atmospheric parameters and metallicity of the observed stars, through the careful inspection 
of synthetic spectra calculated with the SYNTHE package \citep{sbordone05}. Atomic data 
of the transitions of interest are from the last release of the Kurucz/Castelli linelist
\footnote{http://wwwuser.oats.inaf.it/castelli/linelists.html}. The final iron abundances are based 
on $\sim$130-150 FeI and $\sim$15-20 FeII lines.
EWs have been measured with DAOSPEC \citep{stetson08}, run iteratively  
by means of the package 4DAO\footnote{http://www.cosmic-lab.eu/4dao/4dao.php}\citep{mucciarelli13b}.
EW, oscillator strength and excitation potential for all the measured transitions are listed 
in Table~\ref{tab2_c5} (available in its entirety in the online version).

\subsection{Analysis with spectroscopic gravities}

First, we performed a fully spectroscopic analysis, as done by 
\citet{simmerer13}, in order to verify whether we obtain the same evidence of 
a metallicity dispersion. In this analysis the atmospheric parameters have been constrained 
as follows:
{\sl (a)}~for the effective temperatures ($T_{\rm eff}$) we requested that no trend exists between abundances 
and excitation potential, 
{\sl (b)}~for the surface gravities ({\sl log~g}) we imposed that the same abundance is obtained 
(within the uncertainties) from FeI and FeII lines, 
{\sl (c)}~for the microturbulent velocity ($v_{\rm turb}$) we requested that no trend exists between abundances 
from FeI lines and the reduced line strength. The derived values of $v_{\rm turb}$ are based on $\sim$130-150 FeI lines 
distributed over a large interval of reduced EWs, with log(EW/$\lambda$) ranging between --5.6 and --4.7.

We derived an average [FeI/H]=--1.46$\pm$0.02 dex ($\sigma$=~0.10 dex), with a 
distribution ranging from --1.62 dex to --1.27 dex. 
The [FeI/H] and [FeII/H] abundance distributions are shown in Figure~\ref{histo} 
as generalized histograms.
This result well matches that obtained by \citet{simmerer13} that find an average abundance 
[Fe/H]=--1.48$\pm$0.02 dex ($\sigma$=~0.11 dex)\footnote{Note that \citet{simmerer13} adopted as solar reference value 
7.56 obtained from their own solar analysis, while we used 7.50 by \citet{grevesse98}. 
Througout the paper we refer to the abundances by \citet{simmerer13} corrected 
for the different solar zero-point.} with a comparable iron range ($\Delta[Fe/H]\sim$0.4 dex). 
This spectroscopic analysis fully confirms the claim by \citet{simmerer13}: when 
analysed with atmospheric parameters derived following the constraints listed above, 
the stars of NGC3201 reveal a clear star-to-star scatter in the iron content.
The 5 stars labelled as {\sl metal-poor} 
by \citet{simmerer13}, with [Fe/H]$<$--1.58 dex, are the most metal-poor also in our 
metallicity distribution.

\subsection{Analysis with photometric gravities}

As pointed out by \citet{lapenna14}, possible NLTE effects in AGB stars 
can be easily detected by assuming photometric values for {\sl log~g} and measuring
FeI and FeII independently\footnote{Recently, \citet{johnson15} analysed a sample 
of 35 AGB stars in 47 Tucanae, finding no clear evidence of NLTE effects.
Even if a detailed comparison between the two analyses is not the scope of this paper, 
we highlight some main differences between the two works: the spectral resolution 
\citep[48000 in \citet{lapenna14} and 22000 in][]{johnson15}, the number of FeII lines 
\citep[13 in \citet{lapenna14} and 4 in][, on average]{johnson15}
and the adopted linelists for AGB and RGB stars 
(\citet{lapenna14} adopted the same linelist for both the groups of stars, at variance 
with \citet{johnson15} that used a linelist consistent, but not exactly the same, 
with that adopted by \citet{cordero14} where the reference RGB stars are discussed).}.
In order to see whether this effect is present in the NGC3201 data, we adopted the 
following procedure.
$T_{\rm eff}$ has been derived spectroscopically, by imposing the 
excitation equilibrium as described above. Thanks to the high quality 
of the spectra (with S/N ratio per pixel higher than 100) and the 
large number of FeI lines distributed over a large range of excitation potentials, 
very accurate spectroscopic $T_{\rm eff}$ can be estimated, with internal uncertainties 
of about 20-30 K.
As a guess value we adopted $T_{\rm eff}$ calculated from the 
$(V-I)_0 - T_{\rm eff}$ calibration by \citet{alonso99} and 
assuming a color excess $E(B-V)$=0.31 mag.
Gravities have been derived from the Stefan-Boltzmann equation,
assuming $E(B-V)$=0.31 mag, $(m-M)_0$=13.35 mag, 
bolometric corrections from \citet{alonso99} and a mass of 0.82 $M_{\odot}$ 
(according to the best-fit isochrone, the latter is 
a suitable value for RGB stars brighter than the RGB Bump magnitude level).
Because $T_{\rm eff}$ is derived spectroscopically, 
the gravity is recomputed through the Stefan-Boltzmann equation in each iteration 
according to the new value of $T_{\rm eff}$.
The mass value of 0.82 $M_{\odot}$ is appropriate for RGB stars 
but probably too high for AGB stars, because of mass loss phenomena during 
the RGB phase \citep{rood73,origlia02,origlia07,origlia14}. 
In fact, \citet{gratton10b} provide the masses for a sample of HB stars in NGC3201, 
finding values between 0.62 and 0.71 $M_{\odot}$. We initially 
analysed all the targets assuming the mass of a RGB star. Then, the AGB candidates, 
selected according to their position in $T_{\rm eff}$--log~g plane (as discussed in Section~\ref{discuss})
have been re-analysed by assuming the median value 
(0.68 $M_{\odot}$) of the HB stars estimated by \citet{gratton10b}.

This method allows us to take advantage of the high-quality of the spectra, 
deriving accurate $T_{\rm eff}$ thanks to the large number of transitions 
spanning a large range of excitation potentials. On the other hand, this approach 
does not require a fully spectroscopic determination of {\sl log~g}
that is instead calculated using both photometric information and 
spectroscopic $T_{\rm eff}$, avoiding any possible bias 
related to NLTE effects.

\begin{figure}[h]
\centering
\plottwo{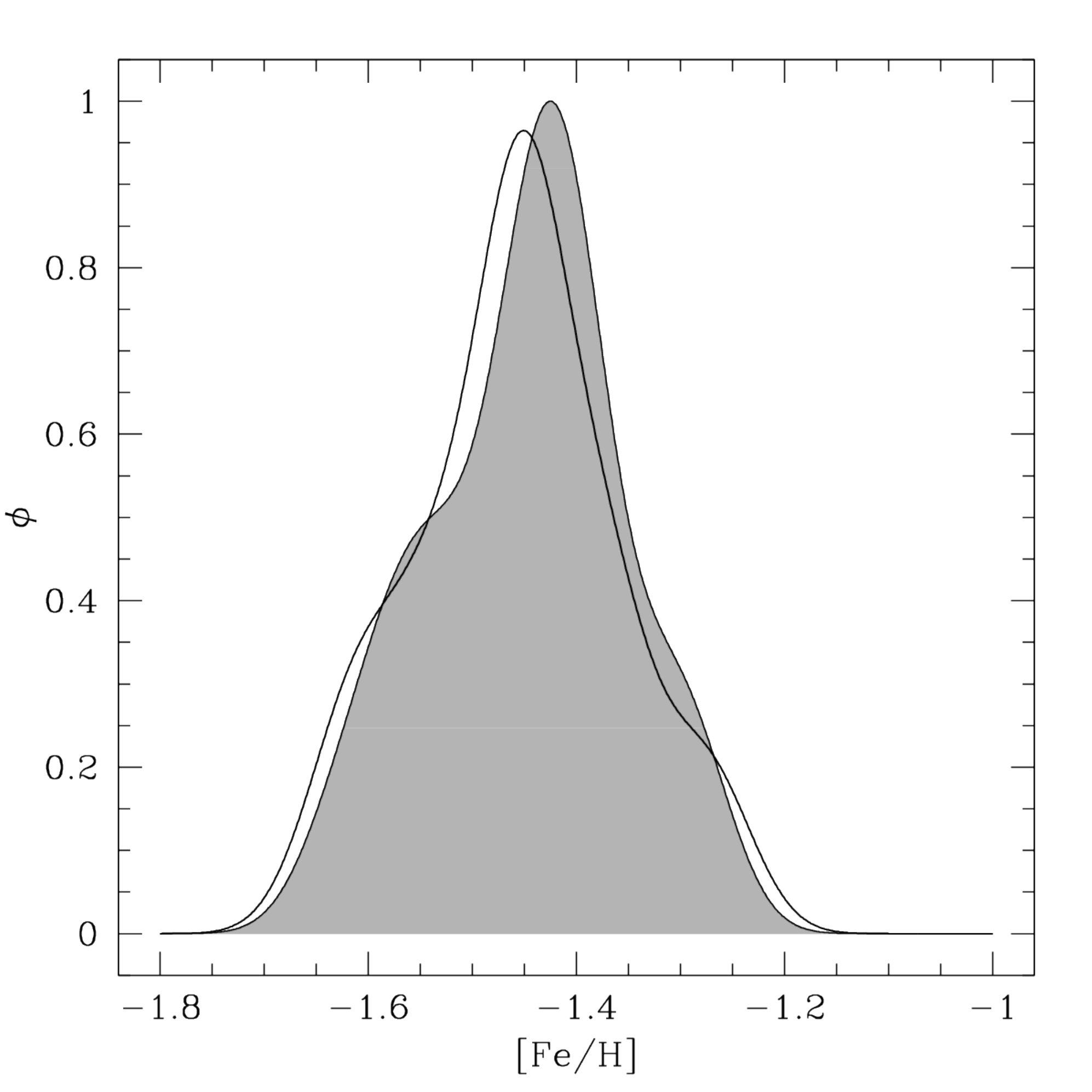}{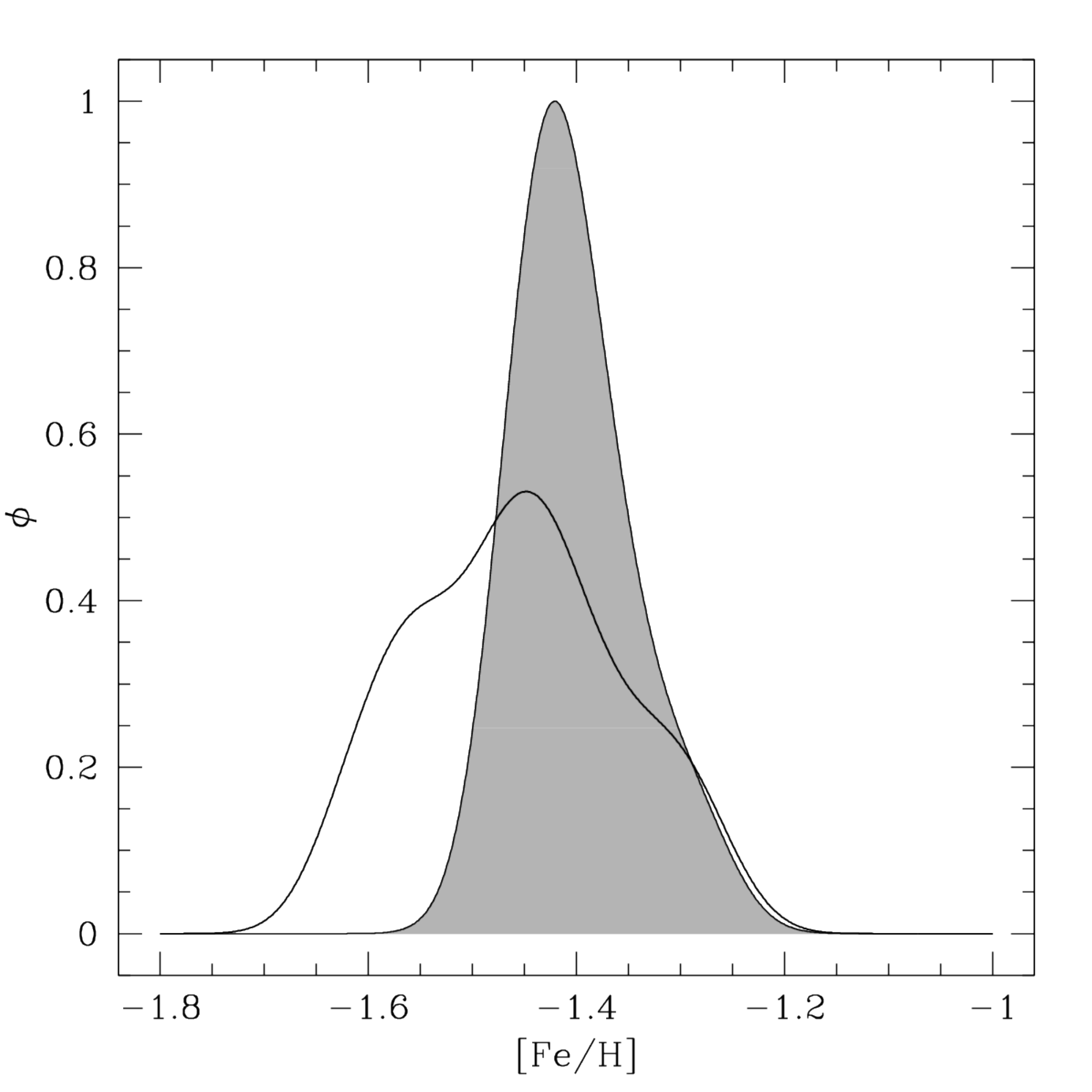}
\caption{Generalized histograms for [FeI/H] (empty histogram) and [FeII/H] (grey histogram) 
obtained from the analysis performed with spectroscopic gravities (left panel) and with
photometric gravities (right panel).}
\label{histo}
\end{figure}

The [FeI/H] and [FeII/H] abundances obtained with this method are listed in Table~\ref{tab1_c5}.
The right panel of Figure~\ref{histo} shows the [FeI/H] and [FeII/H] distributions represented as generalized 
histograms obtained from this analysis. The two distributions turn out to be quite 
different. The iron distribution obtained from FeI lines resembles that obtained with the spectroscopic 
parameters (left panel of Figure~\ref{histo}), with an average value of  [FeI/H]=--1.46$\pm$0.02 dex
($\sigma$=~0.10 dex),
while the distribution obtained from FeII lines has a narrow gaussian-shape 
([FeII/H]$>$=--1.40$\pm$0.01 dex, $\sigma$=~0.05 dex) pointing to a quite 
homogeneous iron content.

% FIGURE 2
%

\subsection{Uncertanties}
 
Internal uncertainties in the derived Fe abundances have been calculated by adding in quadrature 
two sources of uncertainties:\\
(1)~those arising from the EW measurement. For each target, we estimated this term as the line-to-line dispersion normalized 
to the root mean square of the number of lines. Because of the high quality of the used spectra, the 
line-to-line scatters are smaller than 0.1 dex, leading to internal uncertainties of about 0.005-0.008 dex 
for [FeI/H] and of about 0.010-0.025 dex for [FeII/H].\\ 
(2)~those arising from the atmospheric parameters. To estimate this term, we follow the approach described by 
\citet{cayrel04} to take into account the covariance terms due to the correlations among the atmospheric parameters.
For each target, the temperature has been varied by $\pm1\sigma_{T_{\rm eff}}$, the gravity has been 
re-calculated through the Stefan-Boltzmann equation 
adopting the new values of $T_{\rm eff}$ and the microturbulent velocity re-optmized spectroscopically.

Table~\ref{tab1_c5} lists the total uncertainty including both the terms (1) and (2). 
Also, Table~\ref{tab3_c5} shows for two representative targets (one RGB and one AGB star) the 
abundance uncertainty obtained following the prescriptions by \citet{cayrel04} (second column) and 
those obtained with the usual method of independently varying each parameter 
(an approach that obviously does not take into account the correlation among the parameters and 
can over-estimate the total uncertainty).

\section{Discussion}
\label{discuss}

In this paper we present a new analysis of the UVES-FLAMES spectra of 21 member stars of NGC3201 
already discussed in \citet{simmerer13}. 
The Fe abundances have been calculated both using spectroscopic gravities 
(obtained by imposing the ionization balance between FeI and FeII abundances) and 
photometric ones (obtained through the Stefan-Boltzmann equation). 
The two methods provide different results concerning [FeI/H] and [FeII/H]. 
In particular, the use of spectroscopic {\sl log~g} provides a wide [Fe/H] distribution 
(al large as $\sim$0.4 dex), in agreement with the finding of \citet{simmerer13} 
who adopted the same method. On the other hand, when photometric gravities are used,
the [FeI/H] distribution remains quite large, while that of [FeII/H] is narrow. 
We compute the intrinsic spread of the two Fe distributions adopting the Maximum Likelihood 
algorithm described in \citet{mucciarelli12}. Concerning [FeI/H] we derive $\sigma_{int}=0.09\pm0.01$ dex, 
while for [FeII/H] $\sigma_{int}=0.00\pm0.02$ dex. Hence, the [FeII/H] distribution is compatible with 
no iron spread.

\begin{figure}[h]
\centering
\plottwo{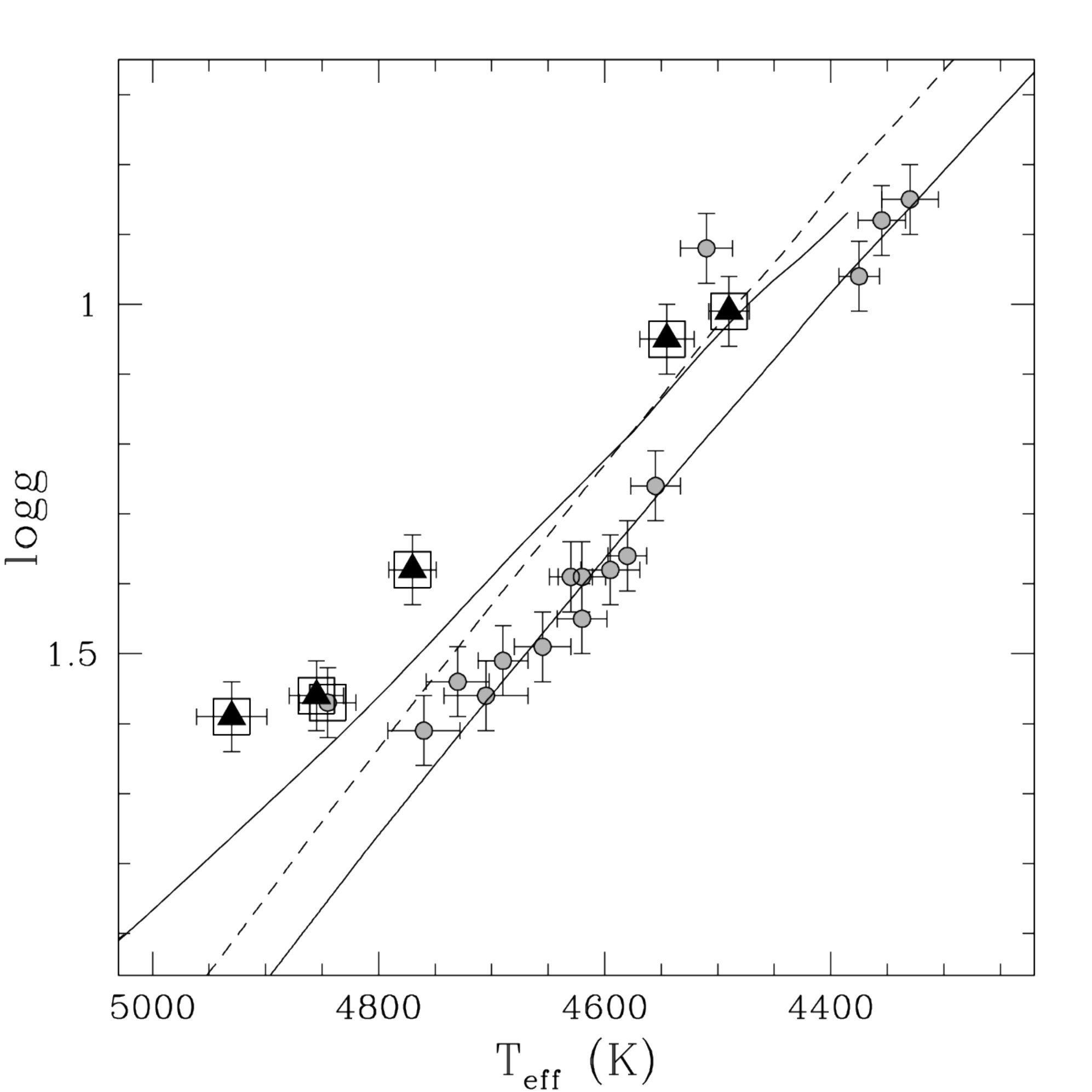}{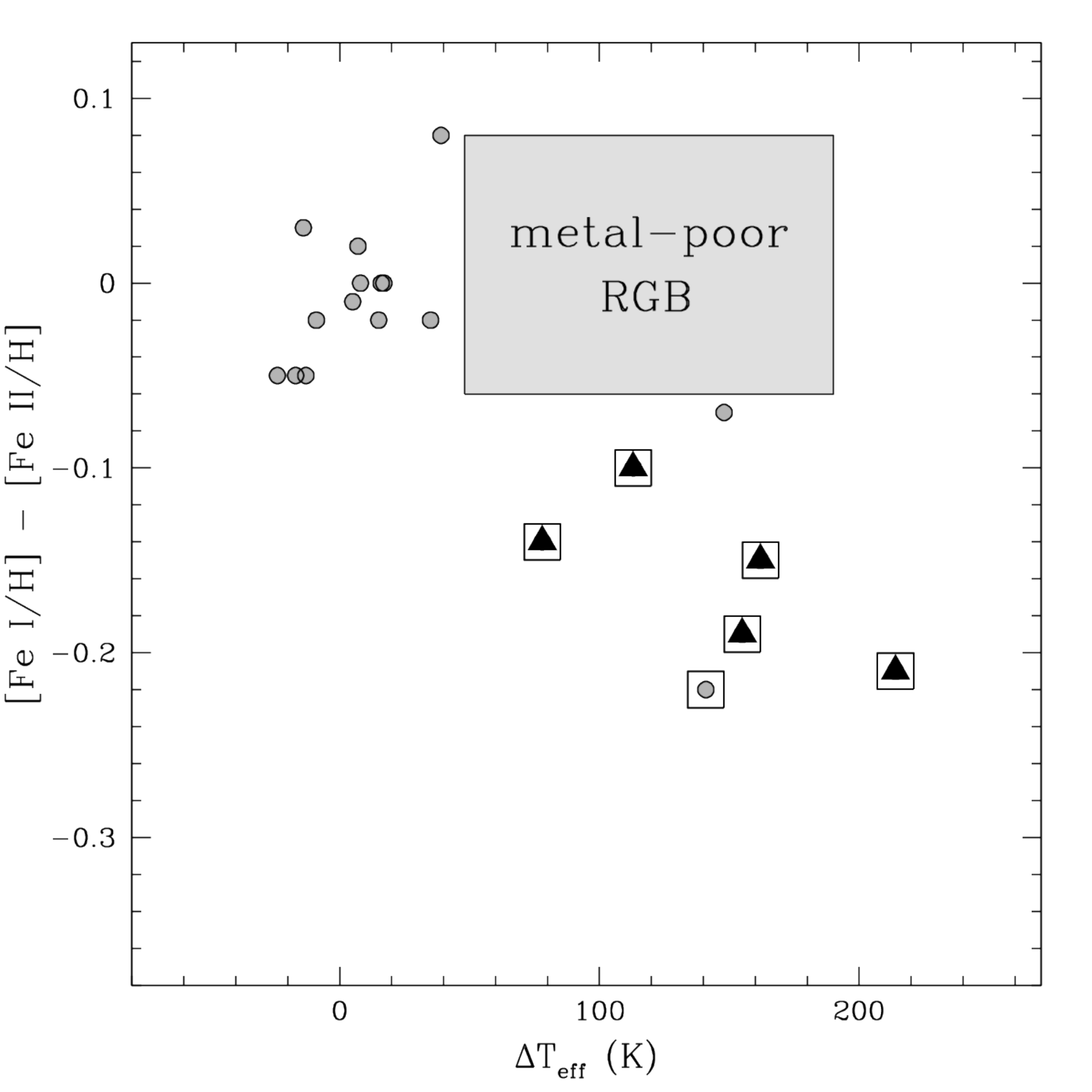}
\caption{{\sl Left panel}: position of the targets in the $T_{\rm eff}$--{\sl log~g} plane. 
A BaSTI isochrone with age of 11 Gyr, Z=0.001 and $\alpha$-enhancement chemical 
mixture (solid line) is superimposed for sake of comparison.
The dashed line indicates the position of the RGB for a BaSTI isochrone with age of 11 Gyr and Z=0.0006.
Grey triangles are the metal-poor stars of \citet{simmerer13}. 
Empty grey squares are the stars in our analysis with [FeI/H] - [FeII/H]$<$--0.1 dex. 
{\sl Right panel}: Behaviour of [FeI/H] - [FeII/H] as a function of the 
difference between the spectroscopic temperature of each target and the photometric 
value estimate from the isochrone with Z=0.001 shown in the left panel. 
Same symbols as in the left panel. The grey light area indicates the expected mean locus 
for metal-poor ([Fe/H]$\sim$--1.6 dex) RGB stars.}
\label{iso}
\end{figure}

\citet{simmerer13} highlight that the 5 most metal-poor stars of their sample are bluer
than the other stars, as expected in cases of a lower metallicity. 
The left panel of Figure~\ref{iso} shows the position of the targets in the $T_{\rm eff}$--{\sl log~g} plane, 
with superimposed, as reference, two isochrones with the same age but different 
metallicity: Z=~0.001 (solid line) and Z=~0.0006 (dashed line). 
The RGB of the isochrone with Z=~0.0006 overlaps the position of the AGB 
of the Z=~0.001 isochrone. 
Seven targets (including the 5 candidate metal-poor stars) are located 
in a position compatible with both the scenarios: metal-poorer RGB or AGB at the 
cluster metallicity. 
These seven targets have average abundances of [FeI/H]=--1.57$\pm$0.01 dex and 
[FeII/H]=--1.41$\pm$0.01 dex, while the RGB stars have [FeI/H]=--1.42$\pm$0.02 dex and 
[FeII/H]=--1.40$\pm$0.01 dex.

% FIGURE 3
%

However, if these stars were metal-poor RGB stars,
the iron abundance derived from FeI lines should be in agreement 
with that obtained from FeII, since NLTE effects are not observed 
in RGB stars of comparable luminosity and metallicity. 
The right panel of Figure~\ref{iso} shows the behaviour of ([FeI/H] - [FeII/H]) as a function 
of the difference between the spectroscopic values of $T_{\rm eff}$ and those obtained from 
the projection along the RGB of the best-fit isochrone.
The stars located along the RGB ($\Delta T_{\rm eff}\sim$0) have similar [FeI/H] and [FeII/H] 
abundances, compatible with no NLTE effects. The stars hotter than the 
reference RGB have differences between [FeI/H] and [FeII/H] 
ranging from --0.07 to --0.22 dex, with a mean value of --0.15 dex.
The grey region marks the expected position for metal-poorer RGB stars: they should be 
hotter than the reference RGB, but with [FeI/H] - [FeII/H]$\sim$0, as commonly measured 
in the RGB stars.
This reveals the true nature of these stars: they are genuine AGB stars, with 
the same metallicity of the cluster (as measured from their FeII lines) 
but affected by NLTE effects leading to a systematic decrease of [FeI/H]. 
This is the same effect observed by \citet{ivans01} and \citet{lapenna14} in the AGB stars 
of M5 and 47~Tucanae, respectively.

\begin{figure}[h]
\centering
\includegraphics[angle=-90,scale=0.55]{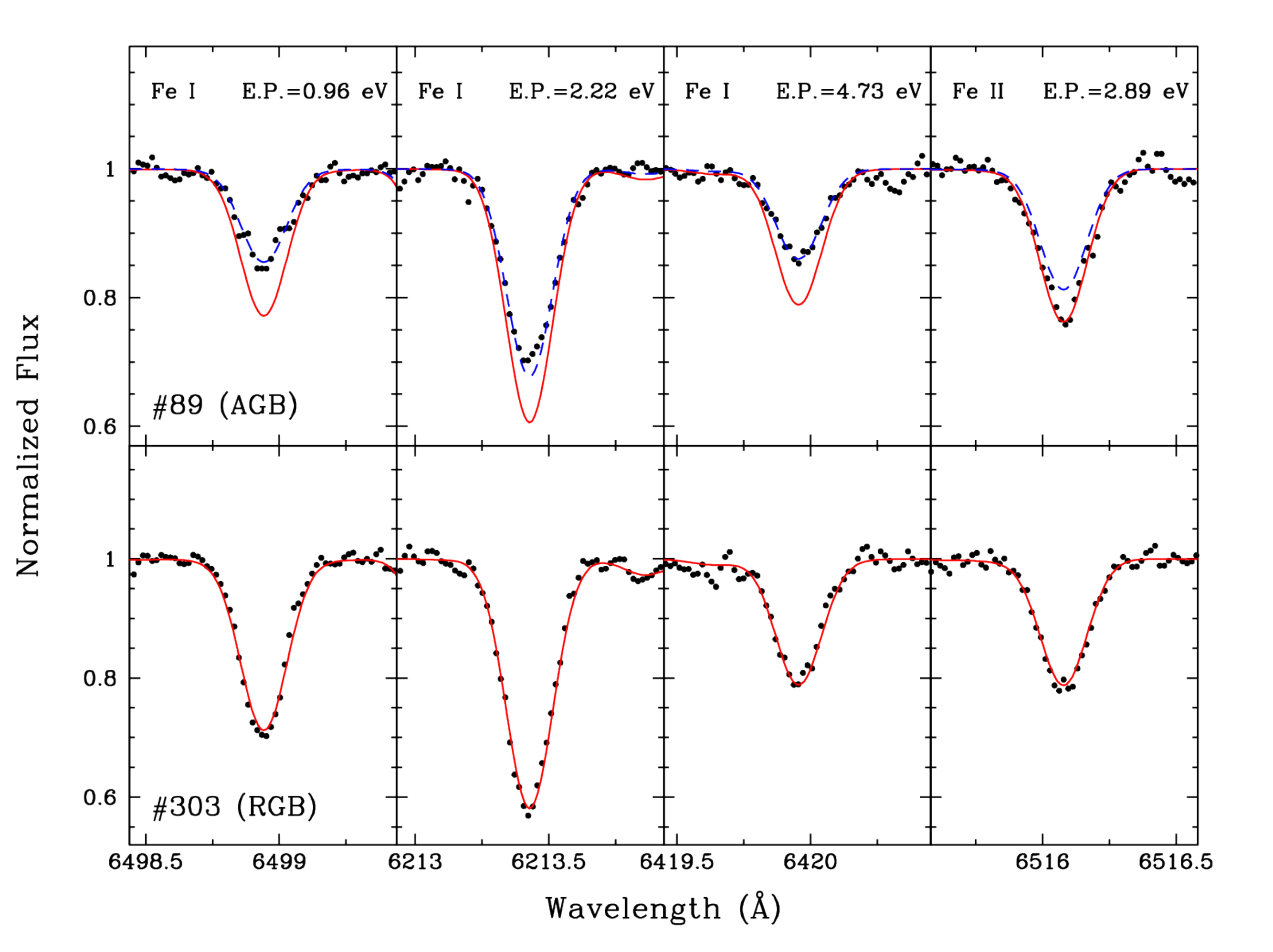}
\caption{Spectral regions around three FeI lines with different excitation potential and 
one FeII line, for the AGB star \#89 (upper panels) and the RGB star \#303 (lower panels). 
Synthetic spectra calculated with the corresponding atmospheric parameters (see Table~\ref{tab1_c5}) 
and adopting the average iron abundance derived from FeII lines are superimposed as red curves. 
The blue dashed curve shown in the upper panels is the synthetic spectrum calculated
with the iron abundance derived from FeI lines.}
\label{spec}
\end{figure}

A direct inspection of the spectra reveals the different behaviour of FeI and FeII lines 
in AGB and RGB stars. Figure~\ref{spec} shows three FeI lines (chosen with different excitation 
potential) and one FeII line in the spectra of the AGB star \#89 (upper panels) and of the 
RGB star \#303 (lower panels). Synthetic spectra calculated with the appropriate atmospheric parameters 
and the metallicity derived from FeII lines (red lines).
In the upper panels we also show the synthetic spectrum computed with 
the average [FeI/H] (blue dashed line).
Clearly, the synthetic spectrum assuming the [FeII/H] abundance well reproduces all the 
observed lines in the case of the RGB star, while it fails to fit the FeI lines observed in the 
AGB star, regardless of the excitation potential (pointing out that 
this effect cannot be attributed to inadequacies in the adopted $T_{\rm eff}$). On the other hand, the abundance derived 
from FeI lines is too low to well reproduce the depth of the AGB FeII line plotted in 
Figure~\ref{spec}. This clearly demonstrates a different behaviour of iron lines in AGB 
and RGB stars.

It is worth noting that this behaviour is somewhat puzzling, because theoretical models do not 
predict significant differences in the NLTE corrections for stars in the parameter space 
covered by our targets. For instance, the grid of NLTE corrections computed by \citet{bergemann12} and 
\citet{lind12} predicts that the FeI lines in AGB and RGB stars should be affected 
in a very similar way at the metallicity of NGC3201. However, some additional effects/mechanisms could play a role 
in the AGB photospheres, leading to the departure from the LTE condition, which are not yet 
accounted for in the available theoretical calculations.

% FIGURE 4
%

We checked whether the mass assumed for AGB stars could change our conclusions.
The mass distribution of HB stars in NGC3201 provided by \citet{gratton10b} ranges from 0.62 to 
0.71 $M_{\odot}$, with a median value of 0.68 $M_{\odot}$. 
The minimum and maximum mass values correspond to a difference in  
log~g of 0.06, leading to a variation in [FeII/H] of only 0.02-0.025 dex 
(and a variation in [FeI/H] of 0.002-0.005 dex).
On the other hand, if we adopt the RGB mass (0.82 $M_{\odot}$) for all targets, 
as it is often done because of the difficulty to observationally distinguish 
between RGB and AGB stars, the values of [FeII/H] increase by only 0.03 dex 
with respect to estimate obtained assuming 0.68 $M_{\odot}$.
Hence, the precise value of the adopted mass (within a reasonable mass range) cannot reconcile the difference 
between [FeI/H] and [FeII/H].

As additional check, for each AGB stars, the stellar mass has been varied until 
the ionization equilibrium was satisfied. 
The derived values range from $\sim$0.2 and $\sim$0.5 $M_{\odot}$: such masses are too low with 
respect to the mass distribution of the HB stars derived by \citet{gratton10b}. 
In particular, for the stars \#89, \#181 and \#240, that exhibit the largest difference between 
[FeI/H] and [FeII/H] ($\sim$--0.2 dex), a satisfying ionization equilibrium can be reached only
with masses smaller than 0.2-0.25 $M_{\odot}$, which are very unlikely values 
for globular cluster AGB stars.

\section{Conclusions}

We demonstrated that the observed intrinsic star-to-star Fe scatter 
in the GC NGC3201 is due to unaccounted NLTE effects in the spectroscopic analysis of some AGB stars included 
in the sample. These stars suffer from NLTE effects driven by the iron overionization, 
a mechanism that affects mainly the less abundant species like FeI, but has no a
significant effect on the dominant species (e.g. FeII).
When the gravity of these stars is obtained spectroscopically, forcing to have 
the same abundance from FeI and FeII lines, the derived [Fe/H] abundance 
turns out to be under-estimated.

Our findings confirm the conclusion by \citet{lapenna14} that the chemical analysis of samples of stars 
including both AGB and RGB stars, and based on spectroscopic gravities, can lead to 
spurious broadening of the iron distribution.

{\sl We conclude that NGC3201 is a normal GC, without evidence of intrinsic iron scatter.}
In light of this result, it is not necessary to suppose that NGC3201 was more massive in the past 
to retain the SN ejecta, as invoked by \citet{simmerer13}.

%%%%%%%%%%%%%%%%%%%%%%%%%%%%%%%%%%%%%%%%%%%%%%%%%%%%%%%%%%%%%%%%%%%% TABLES

\begin{landscape}
\begin{deluxetable}{lcccccccccc}
\tiny
\tablecolumns{11} 
\tablewidth{0pc}  
\tablecaption{}
\tablehead{ 
\colhead{Star} &   RA & Dec &  V  &  I & $RV_{helio}$    & $T_{\rm eff}$ &  {\sl log~g} & $v_{\rm turb}$  &  [FeI/H] & [FeII/H]\\
  &   (J2000)  &  (J2000) &  &   & (km/s)  & (K)  &   & (km/s) & (dex) & (dex)}
\startdata 
 63 &	154.3084680  &   -46.4125790  &  13.76  &  12.42  & 495.9$\pm$0.6        &  4730 &  1.54	&   1.45     &  -1.27$\pm$0.03   & -1.35$\pm$0.05  \\	 
 89 &	154.3346190  &   -46.3860090  &  13.87  &  12.56  & 498.8$\pm$0.5        &  4855 &  1.63	&   1.60     &  -1.57$\pm$0.02   & -1.35$\pm$0.03  \\	   
 91 &	154.3359600  &   -46.3355290  &  13.81  &  12.41  & 490.4$\pm$0.6        &  4705 &  1.56	&   1.50     &  -1.36$\pm$0.05   & -1.35$\pm$0.05  \\	 
105 &	154.3439854  &   -46.4201000  &  13.94  &  12.59  & 493.9$\pm$0.9        &  4760 &  1.61	&   1.45     &  -1.31$\pm$0.03   & -1.29$\pm$0.05  \\	
124 &	154.3601786  &   -46.4140432  &  12.85  &  11.32  & 498.5$\pm$0.2        &  4375 &  0.96	&   1.55     &  -1.37$\pm$0.02   & -1.40$\pm$0.05  \\	
129 &	154.3623900  &   -46.4312480  &  13.54  &  12.13  & 491.8$\pm$0.4        &  4580 &  1.36	&   1.50     &  -1.48$\pm$0.02   & -1.43$\pm$0.04  \\	
181 &	154.3876490  &   -46.4126379  &  13.89  &  12.58  & 497.5$\pm$0.4        &  4920 &  1.65	&   1.65     &  -1.62$\pm$0.03   & -1.37$\pm$0.04  \\  
200 &	154.3949420  &   -46.3972062  &  12.79  &  11.30  & 495.1$\pm$0.4        &  4515 &  0.99	&   1.85     &  -1.50$\pm$0.02   & -1.40$\pm$0.04  \\  
222 &	154.4031199  &   -46.4257714  &  12.69  &  11.15  & 491.0$\pm$0.5        &  4355 &  0.88	&   1.65     &  -1.47$\pm$0.02   & -1.45$\pm$0.05  \\  
231 &	154.4067690  &   -46.4012923  &  13.52  &  12.18  & 491.2$\pm$0.5        &  4785 &  1.45	&   1.60     &  -1.54$\pm$0.02   & -1.39$\pm$0.03  \\	  
240 &	154.4091582  &   -46.4277707  &  13.91  &  12.59  & 495.2$\pm$0.6        &  4855 &  1.64	&   1.55     &  -1.61$\pm$0.03   & -1.38$\pm$0.04  \\	 
244 &	154.4097946  &   -46.4024205  &  13.77  &  12.38  & 492.2$\pm$0.3        &  4690 &  1.51	&   1.50     &  -1.44$\pm$0.02   & -1.44$\pm$0.05  \\	   
249 &	154.4104867  &   -46.4270988  &  12.99  &  11.54  & 498.1$\pm$0.3        &  4545 &  1.11	&   1.60     &  -1.56$\pm$0.02   & -1.42$\pm$0.04  \\	 
277 &	154.4198397  &   -46.4124756  &  13.60  &  12.18  & 497.6$\pm$0.3        &  4620 &  1.39	&   1.50     &  -1.40$\pm$0.02   & -1.42$\pm$0.05  \\	
279 &	154.4208433  &   -46.3954964  &  13.34  &  11.89  & 499.0$\pm$0.3        &  4555 &  1.26	&   1.50     &  -1.43$\pm$0.03   & -1.43$\pm$0.04  \\	
303 &	154.4288640  &   -46.3800960  &  13.58  &  12.20  & 491.7$\pm$0.4        &  4630 &  1.39	&   1.50     &  -1.45$\pm$0.02   & -1.45$\pm$0.04  \\	
308 &	154.4304349  &   -46.4108231  &  13.76  &  12.37  & 487.2$\pm$0.4        &  4620 &  1.45	&   1.45     &  -1.50$\pm$0.02   & -1.45$\pm$0.04  \\  
312 &	154.4342122  &   -46.4240651  &  12.71  &  11.10  & 494.9$\pm$0.3        &  4330 &  0.85	&   1.65     &  -1.44$\pm$0.03   & -1.44$\pm$0.06  \\  
332 &	154.4478331  &   -46.3982898  &  12.96  &  11.47  & 499.7$\pm$0.5        &  4500 &  1.08	&   1.65     &  -1.54$\pm$0.02   & -1.38$\pm$0.04  \\  
344 &	154.4544130  &   -46.4162830  &  13.82  &  12.41  & 490.7$\pm$0.7        &  4655 &  1.49	&   1.40     &  -1.31$\pm$0.03   & -1.29$\pm$0.05  \\  
374 &	154.5004800  &   -46.5194470  &  13.64  &  12.09  & 498.7$\pm$0.6        &  4595 &  1.38	&   1.50     &  -1.41$\pm$0.03   & -1.36$\pm$0.05  \\	  
\hline
    &	             &                &      &  &             &       &         &            &  $<$[FeI/H]$>$        &    $<$[FeII/H]$>$     \\	 
    &	             &                &      &  &             &       &         &            &   --1.46$\pm$0.02     &   --1.40$\pm$0.01     \\
\enddata 
\tablecomments{Main information of the target stars. 
Identification numbers are the same adopted by \citet{simmerer13}.
[FeI/H] and [FeII/H] have been obtained adopting photometric gravities.}
\label{tab1_c5}
\end{deluxetable}
\end{landscape}

\begin{deluxetable}{cccccc}
\tiny
\tablecolumns{6} 
\tablewidth{0pc}  
\tablecaption{Star identification number, wavelength, oscillator strength, excitation potential and 
measured EWs for all the used transitions.}
\tablehead{ 
\colhead{Star} &   $\lambda$ & Ion &  log(gf)  & E.P. & EW   \\
  &   (\AA) &  &  & (eV)  & (m\AA)}
\startdata 
 63  &   4791.246   &	FeI       &  -2.435  &  3.270  &  26.40    \\    
 63  &   4834.507   &	FeI  	  &  -3.330  &  2.420  &  27.90    \\      
 63  &   4842.788   &	FeI  	  &  -1.530  &  4.100  &  16.30    \\    
 63  &   4892.859   &	FeI  	  &  -1.290  &  4.220  &  29.50    \\   
 63  &   4911.779   &	FeI  	  &  -1.760  &  3.930  &  19.50    \\   
 63  &   4917.230   &	FeI  	  &  -1.160  &  4.190  &  35.90    \\   
 63  &   4918.013   &	FeI  	  &  -1.340  &  4.230  &  28.30    \\  
 63  &	 4950.106   &	FeI  	  &  -1.670  &  3.420  &  54.10    \\  
 63  &	 4962.572   &	FeI  	  &  -1.182  &  4.180  &  32.30    \\  
 63  &	 4969.917   &	FeI  	  &  -0.710  &  4.220  &  46.00    \\	  
 63  &	 4985.253   &	FeI  	  &  -0.560  &  3.930  &  75.30    \\	 
 63  &	 5002.793   &	FeI  	  &  -1.530  &  3.400  &  67.20    \\	   
 63  &	 5014.942   &	FeI  	  &  -0.303  &  3.940  &  86.70    \\	 
 63  &	 5022.236   &	FeI  	  &  -0.560  &  3.980  &  76.40    \\	
 63  &   5028.126   &	FeI  	  &  -1.123  &  3.570  &  68.30    \\  
\enddata 
\tablecomments{This table is available in its entirety in a machine-readable form 
in the online journal. A portion is shown here for guidance regarding its form and content.}
\label{tab2_c5}
\end{deluxetable}

\begin{deluxetable}{ccccc}
\tiny
\tablecolumns{5} 
\tablewidth{0pc}  
\tablecaption{Abundance uncertanties due to the atmospheric parameters 
for the stars \#63 and \#89.}
\tablehead{ 
\colhead{Ion} &   Parameters    & $\delta T_{\rm eff}$ &  $\delta$log~g  & $\delta v_{\rm turb}$    \\
              &   Uncertainty   &   $\pm$50 K     &    $\pm$0.1     &  $\pm$0.1 km/s \\
	      &     (dex)       &     (dex)       &    (dex)        &   (dex)  }
\startdata
  &  &  \#63 (RGB) &  & \\
\hline 
 F~I   & $\pm$0.03 & $\pm$0.05 & $\pm$0.00 & $\mp$0.03 \\
 FeII  & $\pm$0.04 & $\mp$0.02 & $\pm$0.04 & $\pm$0.03 \\
\hline
  &  &  \#89 (AGB) &  & \\
\hline 
 FeI   & $\pm$0.02 & $\pm$0.04 & $\pm$0.00 & $\mp$0.03 \\
 FeII  & $\pm$0.03 & $\mp$0.02 & $\pm$0.04 & $\pm$0.02 
\enddata 
\tablecomments{The second column is the total uncertainty 
calculated according to \citet{cayrel04}. The other columns list the 
abundance variations related to the variation of only one parameter.}
\label{tab3_c5}
\end{deluxetable}

%% file: c6/ms_6.tex
% CHAPTER 6

\chapter{A Chemical {\sl Trompe-l'\oe{}il}: No Iron Spread in the Globular Cluster M22}

\label{c6}

{\bf Published in Mucciarelli et al. 2015, ApJ, 809, 128M}

{\it We present the analysis of high-resolution spectra obtained with UVES and UVES-FLAMES 
at the Very Large Telescope of 17 giants in the globular cluster M22, 
a stellar system suspected to have an intrinsic 
spread in the iron abundance. We find that when surface gravities are derived 
spectroscopically (by imposing to obtain the same iron abundance from FeI and FeII lines) 
the [Fe/H] distribution spans $\sim$0.5 dex, according to previous analyses. 
However, the gravities obtained in this way correspond to
unrealistic low stellar masses (0.1-0.5 $M_{\odot}$) for most of the surveyed giants.
Instead, when photometric gravities are adopted, the [FeII/H] distribution
shows no evidence of spread at variance with the [FeI/H] distribution. 
This difference has been recently observed in other clusters and
could be due to non-local thermodynamical equilibrium effects driven by 
over-ionization mechanisms, that mainly affect the neutral species (thus providing 
lower [FeI/H]) but leave [FeII/H] unaltered. We confirm that 
the s-process elements show significant star-to-star variations and their abundances 
appear to be correlated with the difference between [FeI/H] and [FeII/H].
This puzzling finding suggests that the peculiar chemical composition of some cluster 
stars may be related to effects able to spuriously decrease [FeI/H].
We conclude that M22 is a globular cluster with no evidence of intrinsic iron spread,
ruling out that it has retained the supernovae ejecta in its gravitational potential well.}

\section{Observations}

The spectroscopic dataset analysed here is the same used by \cite[][, hereafter M09]{marino09} and
includes six giant stars observed with UVES@VLT \citep{dekker00} on 18-21 March 2002, 
and 11 giant stars observed with UVES-FLAMES@VLT \citep{pasquini00} on 24-26 May 2003,
adopting the Red Arm 580 grating that ranges from $\sim$4800 to 
$\sim$6800 \AA\  with a typical spectral resolution R=~47000. 
All the spectra have been reduced with the dedicated ESO pipelines, including 
bias subtraction, flat-fielding, wavelength calibration, spectral extraction and 
order merging. The typical signal-to-noise ratio per pixel of the acquired spectra 
is $\sim$150 at $\sim$6000 \AA .

\begin{figure}[h]
\centering
\includegraphics[width=0.70\textwidth]{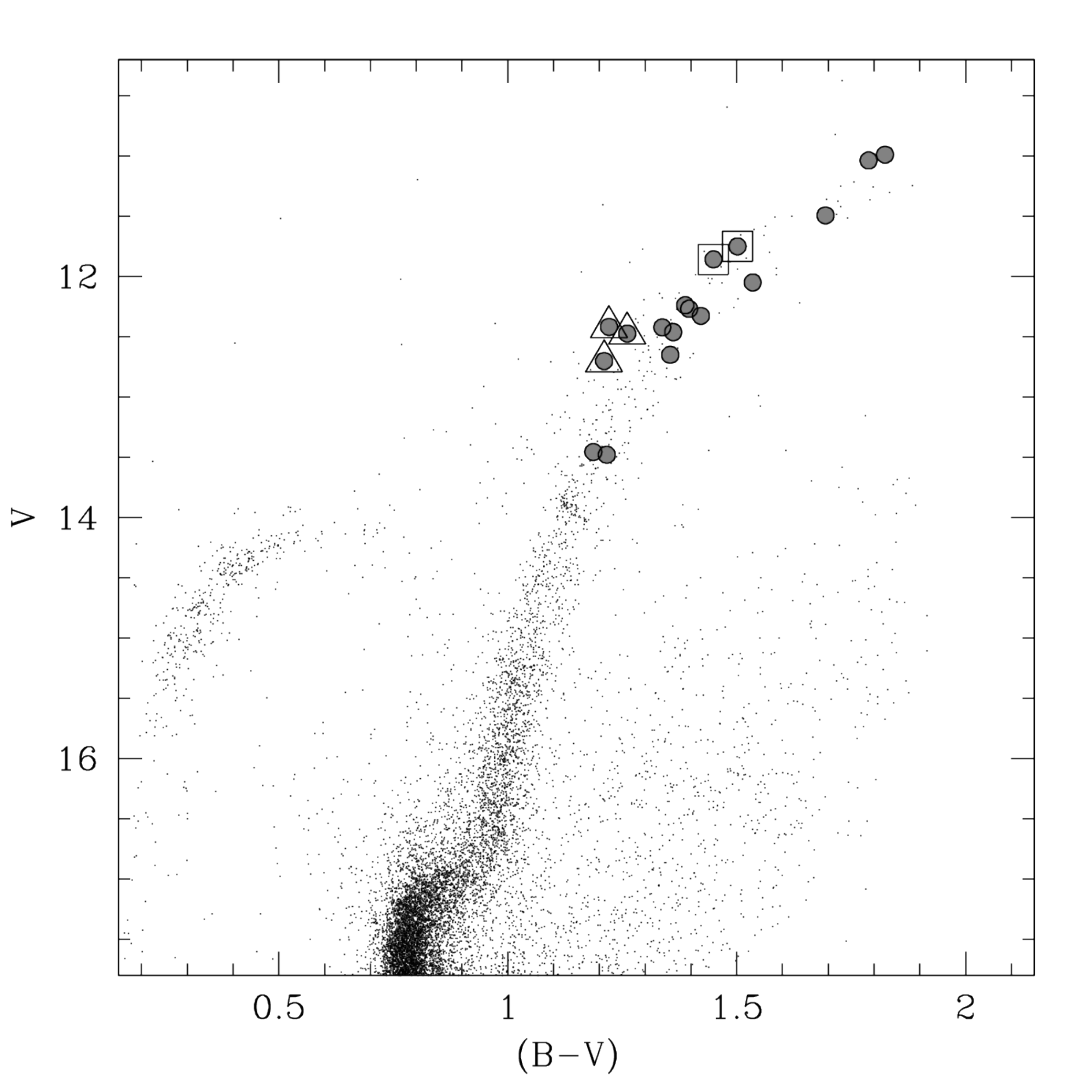}
\caption{(B-V,V) color-magnitude diagram of M22 \citep{kunder13} with marked as grey circles the spectroscopic targets. 
Empty triangles are the likely candidate AGB stars, while empty squares are possible (but not sure) AGB stars.}
\label{cmd}
\end{figure}

The target stars, originally selected from the photometric catalog by \citet{monaco04}, have been 
cross-identified in the UBVI ground-based catalog described in \citet{kunder13} and 
in the $\rm {JHK_s}$ 2MASS catalog \citep{skrutskie06}. 
Their position in the (V,B-V) CMD is shown in Figure~\ref{cmd}.
Main information about the targets is available in M09.

For each target the correction for differential reddening has been derived as in 
\citet{milone12}, adopting the extinction law by \citet{cardelli89}. 
We found that the maximum variation 
of E(B-V) across the area covered by the observed targets is of $\sim$0.07 mag, in nice agreement 
with \citet{monaco04} who quoted a maximum variation of $\sim$0.06 mag. 
The true distance modulus, ${\rm (m-M)}_0$=~12.65 mag, and the color excess, E(B-V)=~0.34 mag, of the cluster have 
been estimated by fitting the CMD with an isochrone from the BaSTI dataset \citep{pietrinferni06}, 
computed with an age of 12 Gyr, a metallicity Z=~0.0006 and $\alpha$-enhanced chemical mixture 
(corresponding to an iron content of [Fe/H]=--1.84 dex, in agreement with M09). 
The color excess is the same quoted by \citet{harris96}, based on the photometry by \citet{cudworth86}, 
while we derived a slightly fainter ($\sim$0.11 mag) distance modulus.

% FIGURE 1
%

\section{Iron abundance}
\label{analysis}

The iron abundances have been derived by comparing observed and theoretical equivalent widths (EWs) 
by means of the code GALA \citep{mucciarelli13}. 
EWs have been measured with the code DAOSPEC \citep{stetson08} run through the wrapper 4DAO \citep{mucciarelli13b} 
that allows a visual inspection of the best-fit Gaussian profile for each individual line.
Model atmospheres have been computed with the code 
ATLAS9\footnote{http://wwwuser.oats.inaf.it/castelli/sources/atlas9codes.html} 
assuming 1-dimensional, plane-parallel geometry, no overshooting in the computation 
of the convective flux and adopting 
the new opacity distribution functions by \citet{castelli04} computed with an enhanced 
chemical composition for the $\alpha$-elements 
(while for all the other elements a solar [X/Fe] abundance ratio is assumed). 
The metallicity [M/H] of the model atmosphere for each star 
has been chosen according to the average [Fe/H] derived from FeII lines, 
being most of the iron in the ionized stage in the atmosphere of late-type stars.

%GUESS PARAMETERS
First guess parameters for effective temperature (T$_{eff}$) and surface gravities (log~$g$)
have been calculated from the photometry.
T$_{eff}$ has been derived from the color-T$_{eff}$ transformations by \citet{alonso99}, 
by averaging the values obtained from different de-reddened broad-band colors, namely
$\rm (U-B)_0$, $\rm (B-V)_0$, $\rm (V-I)_0$, $\rm (V-K_{s})_0$ and $\rm (J-K_{s})_0$.
Surface gravities have been derived through the Stefan-Boltzmann relation, 
assuming the average T$_{eff}$ , the bolometric corrections by \citet{alonso99} computed 
with the average T$_{eff}$ and the stellar masses obtained from the best-fit isochrone.
For most of the stars we adopted a mass of 0.78 $M_{\odot}$ 
(appropriate for RGB stars, according to the best-fit theoretical isochrone). 
Three targets are identified as likely AGB stars, 
according to their positions in the optical CMDs 
(they are marked as empty triangles in Figure~\ref{cmd}).
We assumed for these stars a mass of 0.65 $M_{\odot}$, corresponding to the median value of the
mass distribution of the horizontal branch stars of M22 
(obtained by using the zero age horizontal branch models of the BaSTI database). 
The position of two other stars (marked as empty squares in Figure~\ref{cmd}) could also be compatible 
with the AGB but the small color separation from the RGB makes it difficult to unambiguously assign 
these targets to a given evolutionary sequence.
For these two stars we assume conservatively a mass of 0.78 $M_{\odot}$ and checking that the impact 
of a different mass on the iron abundances is very small: assuming the AGB mass, [FeII/H] 
changes by $\sim$0.03 dex, while [FeI/H] does not change.

%LINELIST
Because the targets span relatively large ranges in the parameter space 
($\delta$T$_{eff}$$\sim$700 K and $\delta$log~$g$$\sim$1.5 dex, according to the photometric estimates), 
the use of an unique linelist is inadvisable, because the line blending conditions 
vary with the evolutionary stage of the stars.
Hence, a suitable linelist has been defined for each individual target, by using 
a specific synthetic spectrum calculated with the code SYNTHE \citep[see][for details]{sbordone05}, 
adopting the photometric parameters and including 
only transitions predicted to be unblended and detectable in the observed spectrum.
Each linelist has been refined iteratively: after a first analysis, the selected 
transitions have been checked with synthetic spectra calculated with the new parameters 
and including the precise chemical composition obtained from the analysis. 
The oscillator strengths for FeI lines are from the compilation by \citet{fuhr88} and \citet{fuhr06}, 
while for FeII lines we adopted the recent atomic data by \citet{melendez09}. 
Concerning the van der Waals damping constants, the values calculated by \citet{barklem00} are adopted 
whenever possible, while for other transitions they were computed 
according to the prescriptions of \citet{castelli05}. The reference solar value is 7.50 \citep{grevesse98}.
EW, excitation potential and oscillator strength are listed in Table~\ref{tab1_c6}.

The iron abundances have been derived from 130-200 FeI lines and 15-20 FeII lines, 
leading to internal uncertainties arising from the EW measurements 
(estimated as the line-to-line scatter divided to the square root of the number of used lines)
of the order of 0.01 dex (or less) for FeI and 0.01-0.02 dex for FeII.
The chemical analysis has been performed with three different approaches to constrain T$_{eff}$ and log~$g$,
while the microturbulent velocities (v$_{turb}$) have been constrained by imposing 
no trend between the iron abundance and the line strength, expressed as $\log(EW/\lambda)$. 
The total uncertainty in the chemical abundance has been computed by summing in quadrature 
the internal uncertainty and that arising from the atmospheric parameters, the latter being estimated 
according to the different method adopted (as discussed below).
Table~\ref{tab2_c6} summarises the average [FeI/H] and [FeII/H] abundances obtained with the different methods.

\subsection{Method (1): spectroscopic T$_{eff}$ and log~$g$}

The values of T$_{eff}$ have been derived by erasing any trend between the iron abundance obtained from FeI lines 
 and the excitation potential ($\chi$), while
log~$g$ have been derived by requiring the same abundance from FeI and FeII lines.
Because of the large number of FeI lines, well distributed over a wide range of $\chi$ values, 
the spectroscopic T$_{eff}$ are constrained with internal uncertainties of about 30-50 K, while 
the internal uncertainties on log~$g$ are $\sim$0.03-0.05. 
Uncertainties in v$_{turb}$ are of about 0.1 km/s (this value is valid also for the 
other methods where the same approach is used to derive v$_{turb}$). 
We assumed a typical uncertainty of $\pm$0.05 dex in the metallicity [M/H] of the model atmosphere; this 
has a negligible impact on [FeI/H] but leads to variations of $\pm$0.02-0.04 dex in [FeII/H].

% FIGURE 2
%
\begin{figure}[h]
\centering
\includegraphics[width=0.90\textwidth]{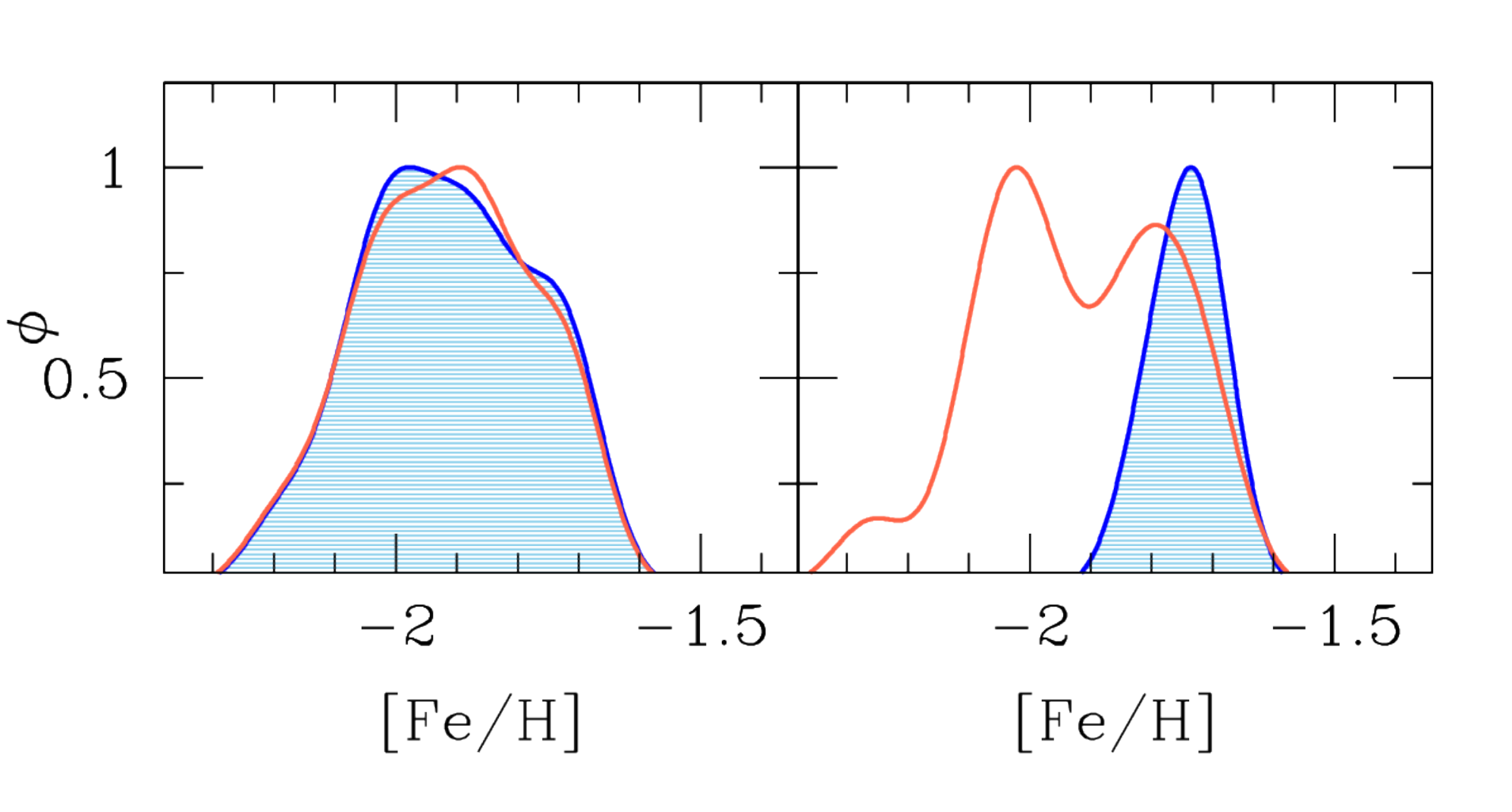}
\caption{Generalized histograms for [FeI/H] (empty red histogram) and [FeII/H] (blue histogram) obtained 
from the analysis performed with spectroscopic gravities (method (1), left panel) and with photometric 
gravities (method (2), right panel).}
\label{histo1}
\end{figure}

The [Fe/H] distributions thus derived from neutral and single ionized lines are 
shown in the left panel of Figure~\ref{histo1}, as generalized histograms. 
The two distributions are, by construction, very similar to each other (because of the adopted 
constraint to derive log~$g$) and $\sim$0.5 dex wide, 
with an average value of [Fe/H]=--1.92$\pm$0.03 ($\sigma$=0.13 dex) for both [FeI/H] and [FeII/H].
In order to evaluate whether the observed scatter is compatible with an intrinsic spread, 
we adopted the maximum likelihood (ML) algorithm described in \citet{mucciarelli12},
which provides the intrinsic scatter ($\sigma_{int}$) of the metallicity distributions 
by taking into account the uncertainties of each individual star. 
Both the iron distributions have a non-zero scatter, with $\sigma_{int}$=0.13$\pm$0.02 dex. 
This result is qualitatively similar to that of M09, who obtained
a broad Fe distribution adopting the same approach to derive the atmospheric parameters.

\subsection{Method (2): spectroscopic T$_{eff}$ and photometric log~$g$}

The values of T$_{eff}$ have been constrained spectroscopically, as done in the method (1), 
while those of log~$g$ have been derived through the Stefan-Boltzmann relation. 
In the computation of log~$g$, we adopted the distance modulus, stellar masses, color excess and 
bolometric corrections used for the guess parameters, together with the spectroscopic T$_{eff}$.
The internal uncertainty of the photometric log~$g$ has been computed including 
the uncertainties in the adopted T$_{eff}$, stellar mass, magnitudes and differential reddening corrections, 
leading to a total uncertainty of about 0.05 dex. 
Errors in distance modulus and color excess have been neglected because they impact 
systematically all the stars, while we are interested in the star-to-star uncertainties only.
This approach allows to benefit at best from all the spectroscopic and photometric pieces of information in hand, 
minimizing the impact (mainly on T$_{eff}$) of the uncertainties in the differential and absolute reddening. 
The atmospheric parameters and the [FeI/H] and [FeII/H] abundance ratios 
derived with this method are listed in Table~\ref{tab3_c6}.

By adopting this method, which (at odds with the previous one) does not impose ionization balance,
we find that, for most of the targets, a large difference between [FeI/H] and [FeII/H].
The [FeI/H] and [FeII/H] distributions are shown in the right panel of Figure~\ref{histo1}. 
At variance with the previous case, the two distributions look very different: 
the distribution of [FeI/H] spans a range of $\sim$0.5 dex,
with an average value of --1.92$\pm$0.04 ($\sigma$=0.16 dex), while the [FeII/H] distribution 
is narrow and symmetric, with an average value of --1.75$\pm$0.01 dex ($\sigma$=~0.04 dex). 
The ML algorithm provides an intrinsic spread $\sigma_{int}$=0.15$\pm$0.02 dex for the [FeI/H] distribution, 
while the [FeII/H] distribution is compatible with a negligible intrinsic scatter ($\sigma_{int}$=~0.00$\pm$0.02 dex).

% FIGURE 3
%
\begin{figure}[h]
\centering
\includegraphics[angle=-90,scale=0.55]{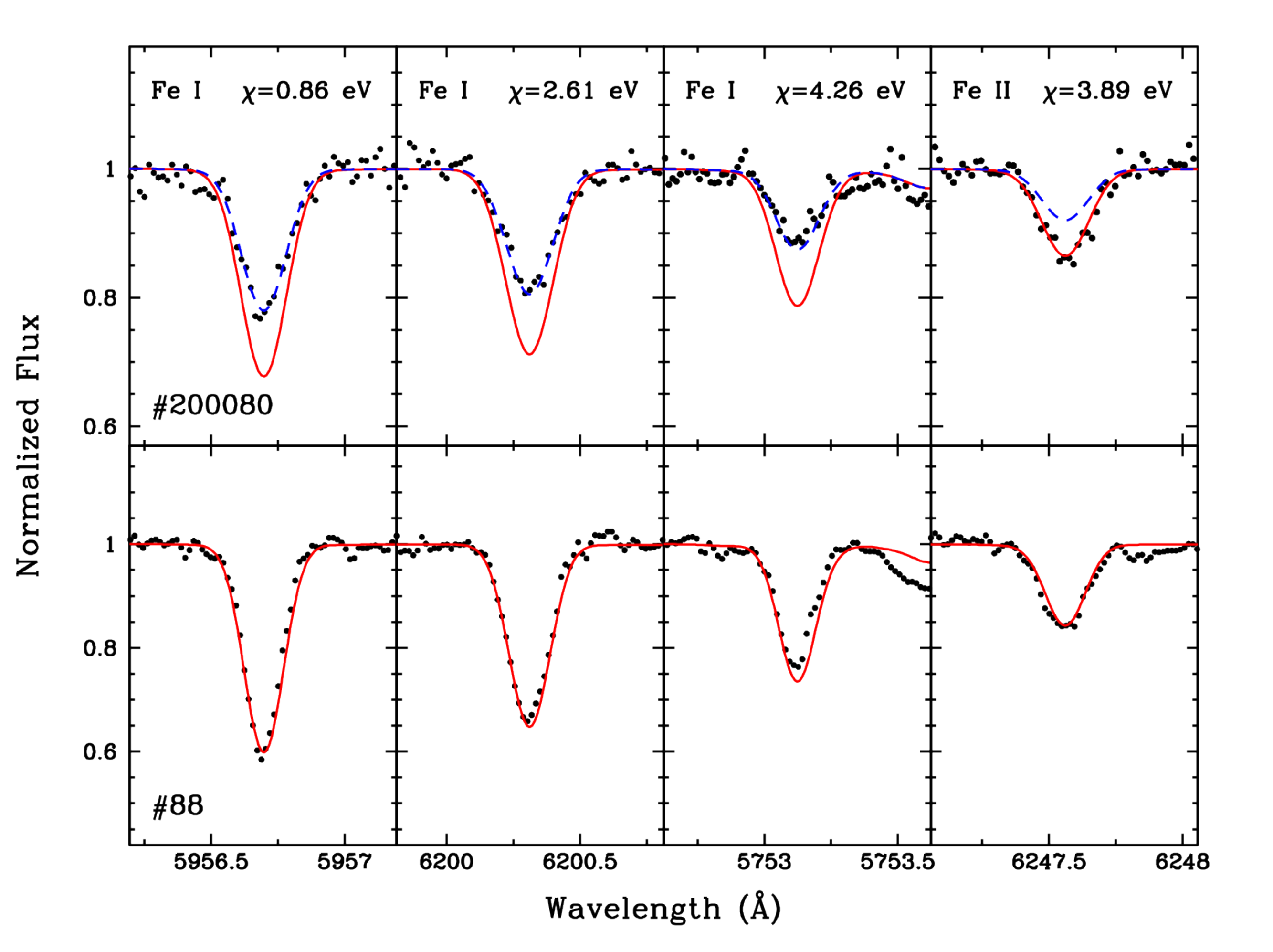}
\caption{Spectral regions around three FeI lines with different excitation potential and 
one FeII line, for the target stars \#200080 (upper panels) and \#88 (lower panels). 
Synthetic spectra calculated with the corresponding atmospheric parameters (see Table~\ref{tab4_c6}) 
and adopting the average iron abundance derived from FeII lines are superimposed as red curves. 
The blue dashed curve shown in the upper panels 
is the synthetic spectrum calculated with the iron abundance derived from FeI lines.}
\label{spec2}
\end{figure}

To illustrate this difference between [FeI/H] and [FeII/H], 
Figure~\ref{spec2} shows some FeI and FeII lines in the spectra of stars \#200080
(where [FeI/H] is 0.29 dex lower than [FeII/H]) and \#88 (where 
[FeI/H] and [FeII/H] differ by 0.05 dex only). In the first case, 
the synthetic spectrum calculated with the average abundance derived from FeII lines 
(red solid line) is not able to reproduce the FeI lines.
The latter are always weaker than those of the synthetic spectrum, 
regardless of their $\chi$ and line strength, thus suggesting that the discrepancy 
is not due to inaccuracies in T$_{eff}$ and/or v$_{turb}$ (otherwise a 
better agreement would have been found for high-$\chi$ lines, less sensitive to T$_{eff}$, 
and/or for weak lines, less sensitive to v$_{turb}$).
On the other hand, the synthetic spectrum computed with the [FeI/H] abundance 
(blue dashed line) does not fit the FeII line, 
that is stronger than that predicted by the synthetic spectrum.
In the case of star \#88 the situation is different and 
an unique Fe abundance is able to well reproduce both FeI and FeII lines.

% FIGURE 4
%
\begin{figure}[h]
\centering
\includegraphics[angle=0,scale=0.55]{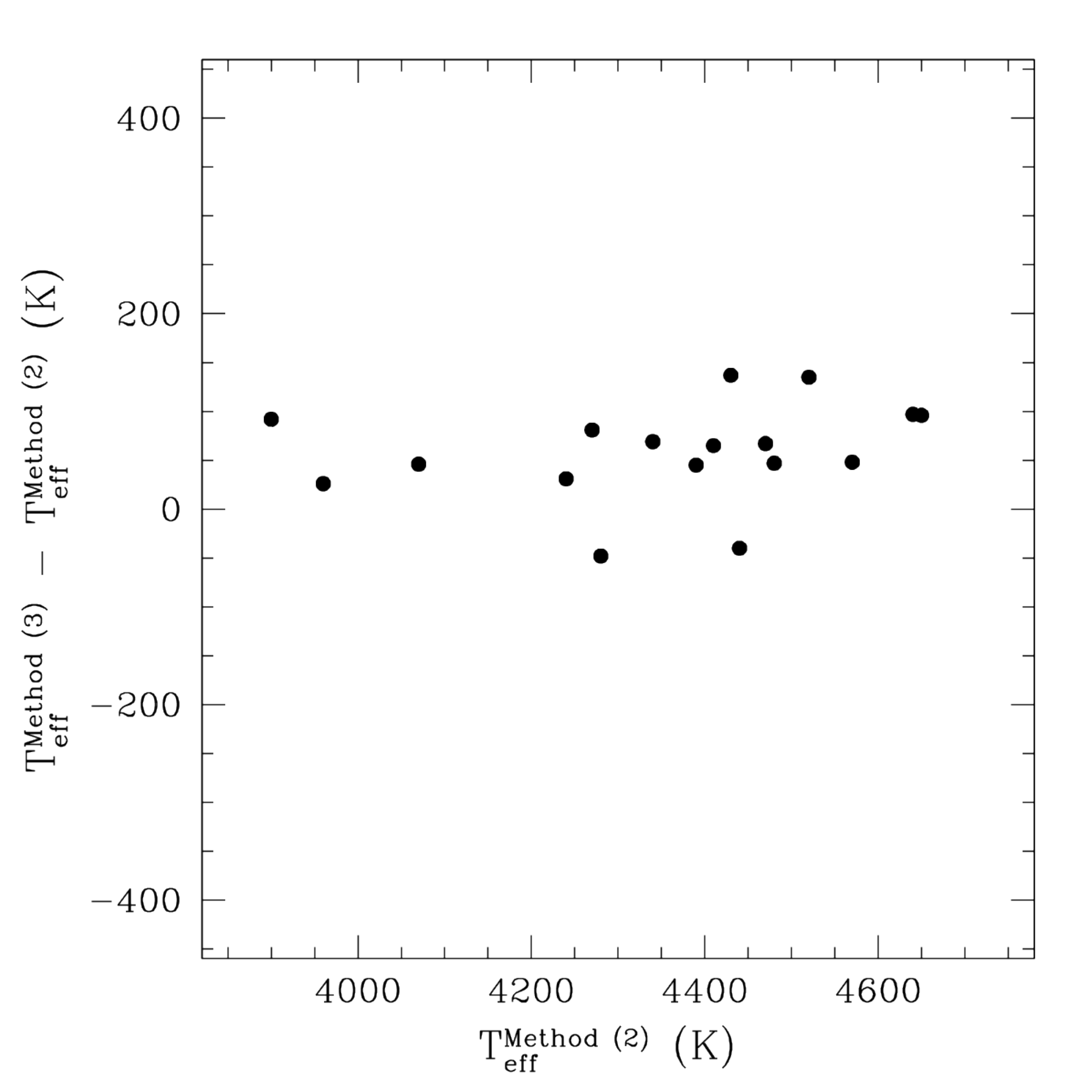}
\caption{ Behaviour of the difference between T$_{eff}$ as derived with
method (3) and (2) as a function of those derived with method (2).}
\label{difft}
\end{figure}

\subsection{Method (3): photometric T$_{eff}$ and log~$g$}

As an additional check, the analysis has been performed keeping T$_{eff}$ and log~$g$ fixed
at the guess values derived from the photometry (see Section~\ref{analysis}),
and optimizing spectroscopically only v$_{turb}$. 
This set of parameters is very similar to that obtained with 
method (2), with the average differences, in the sense of method (3) - method (2), 
of +58$\pm$12 K ($\sigma$=~50 K) in T$_{eff}$, +0.02$\pm$0.005 ($\sigma$=0.02) 
in log~$g$ and +0.04$\pm$0.02 km$s^{-1}$ ($\sigma$=~0.08 km$s^{-1}$) in v$_{turb}$ . 
In particular, we note that spectroscopic and photometric T$_{eff}$ agree very well, and their 
differences do not show trends with the photometric T$_{eff}$, as visible in Figure~\ref{difft} where 
the difference between T$_{eff}$ from method (3) and (2) are shown as a function of the spectroscopic T$_{eff}$.

% FIGURE 5
%
\begin{figure}[h]
\centering
\includegraphics[angle=0,scale=0.60]{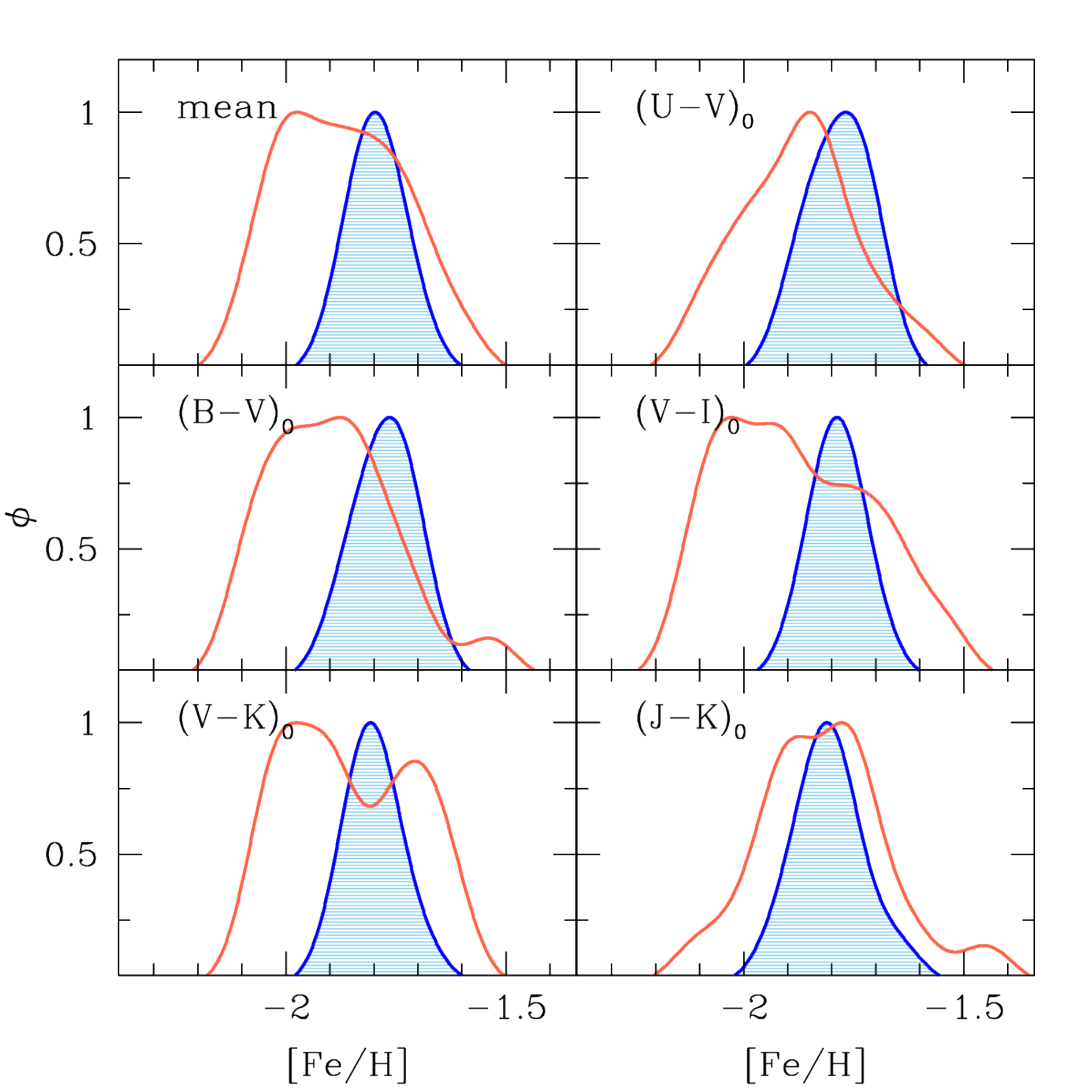}
\caption{Generalized histograms for [FeI/H] and [FeII/H] (same colors of Figure~\ref{histo1}) obtained 
with the method (3) (photometric T$_{eff}$ and log~$g$), adopting the mean parameters (left-upper panel) and 
those derived from individual broad-band colors. }
\label{histo2}
\end{figure}

Figure~\ref{histo2} shows the Fe abundance distributions obtained with 
the average photometric parameters (upper-left panel) and using
the individual broad-band colors $\rm (U-B)_0$, $\rm (B-V)_0$, $\rm (V-I)_0$, $\rm (V-K_{s})_0$, $\rm (J-K_{s})_0$. 
In all cases, the [FeII/H] distribution is single-peaked and narrow,
well consistent with that obtained from method (2).
Instead, whatever color is adopted, the [FeI/H] distribution always has a much larger (by a factor of 2-3) 
dispersion than that obtained for [FeII/H], similar to the 
finding of method (2). 
In particular, when we consider the average photometric parameters, 
the ML algorithm provides intrinsic scatter of 0.12$\pm$0.02 for [FeI/H] and 
0.00$\pm$0.02 for [FeII/H].
Because the results obtained with this method agree with those obtained with method (2), 
and the star-to-star uncertainties in spectroscopic T$_{eff}$ are smaller than 
the photometric ones (which are also affected by the uncertainties on the differential reddening corrections),
in the following we refer only to method (2) as alternative approach to method (1).

%------------------------------------------------------------------ COMPARISON

\section{A sanity check: NGC6752}
\label{refe}

\begin{figure}[h]
\centering
\includegraphics[angle=0,scale=0.50]{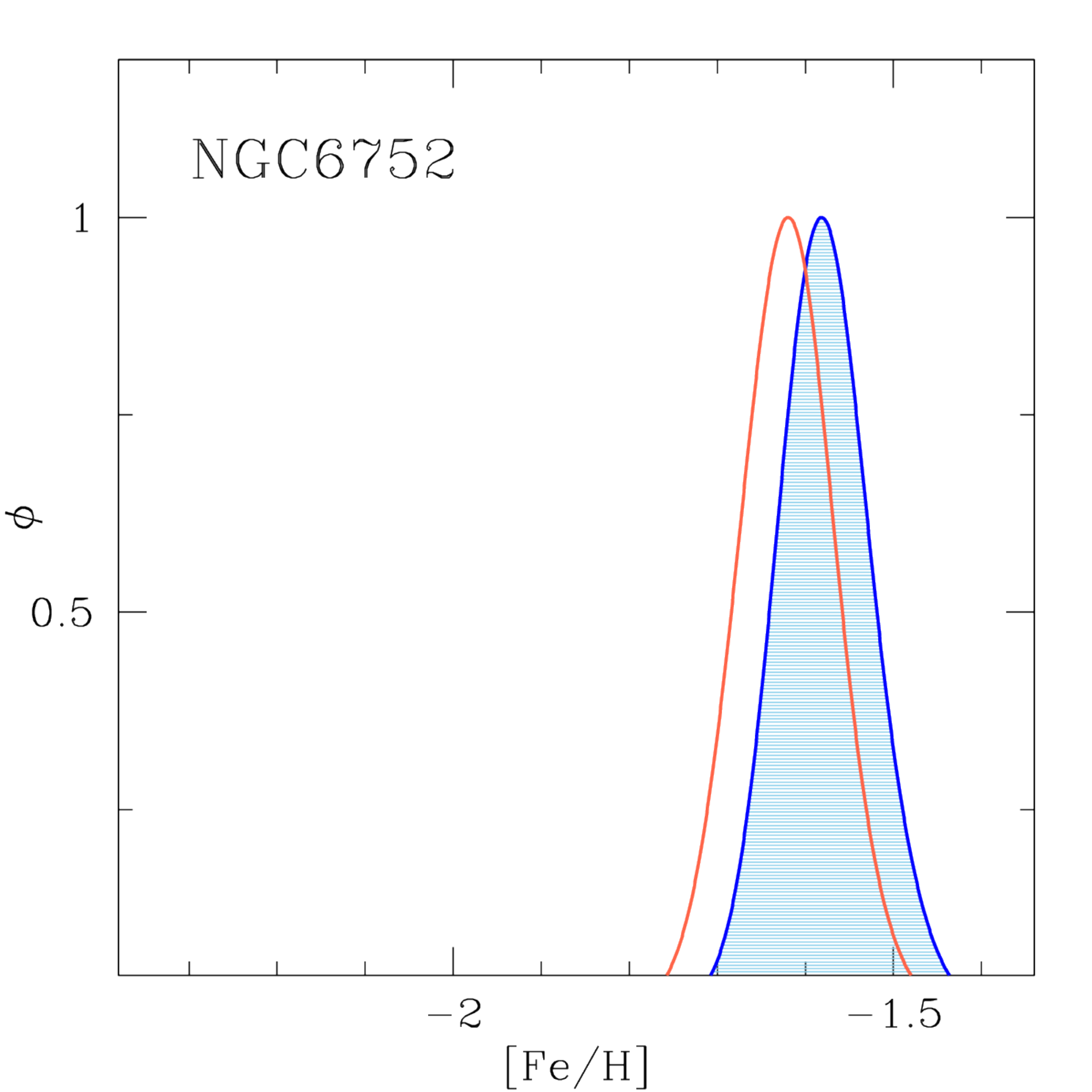}
\caption{Generalized histograms for [FeI/H] and [FeII/H] (same colors of Figure~\ref{histo1}) 
for a sample of 14 RGB stars in the GC NGC6752. The analysis has been performed adopting 
spectroscopic T$_{eff}$ and photometric log~$g$, the same method used for the right panel of Figure~\ref{histo1}.}
\label{6752md}
\end{figure}

As a sanity check, UVES-FLAMES archival spectra of 14 RGB stars in the GC NGC6752 
observed with the Red Arm 580 grating have been analysed following the same procedure used for M22.
NGC6752 is a well-studied GC that can be considered as a standard example of {\sl genuine} 
GC, with no intrinsic iron spread \citep[see e.g.][]{yong05,carretta09b}\footnote{ \citet{yong13} performed 
a strictly differential line-by-line analysis on 37 RGB stars of NGC6752 by using high-quality UVES spectra, 
finding an observed spread in [Fe/H], 0.02 dex, larger of a factor of 2 than the internal uncertainties.
This small intrinsic spread could reflect He variations and/or real inhomogeneities in the cluster iron content. 
Because such chemical inhomogeneities can be revealed only when the internal uncertainties are smaller than 
$\sim$0.02 dex, for our purposes we can consider NGC6752 as a {\sl genuine} GC.}
and with a metallicity comparable with that of M22.
This approach allows to remove any systematics due to the adopted atomic data, solar 
reference values, model atmospheres, method to measure EWs and to derive the atmospheric 
parameters.
When the parameters are derived following method (2),
we derive average abundances [FeI/H]=--1.62$\pm$0.01 dex ($\sigma$=~0.04 dex) and 
[FeII/H]=--1.58$\pm$0.01 dex ($\sigma$=~0.04 dex), in good agreement with the previous 
estimates available in the literature.
In this case, the two iron distributions (shown in Figure~\ref{6752md}) have small observed dispersions, 
both compatible with a negligible scatter within the uncertainties, as demonstrated by the ML algorithm. 
The two distributions are compatible with each other also in terms of their shape, 
at variance with those of M22. 
The same results are obtained when the parameters are all derived spectroscopically.
This test demonstrates that: (i)~the different [FeI/H] and [FeII/H] distributions obtained 
for M22 with methods (2) and (3) are not due to the adopted procedure; 
(ii)~in a {\rm normal} GC the shape of [FeI/H] and [FeII/H] distributions are not significantly different.

% FIGURE 6
%
%
%

%------------------------------------------------------------------ RESULTS

\section{No iron spread in M22}

The new analysis of the sample of giant stars already discussed in M09 leads to an unexpected result:
an iron abundance spread in M22 is found when FeI lines are 
used, independently of the adopted spectroscopic or photometric gravities.
{\sl This scatter totally vanishes when the iron abundance is derived from FeII lines 
and photometric gravities are used.} In the case of spectroscopic gravities, 
the abundances from FeII lines are forced to match those from FeI lines, thus 
producing a broad [FeII/H] distribution. Given that the adoption of 
photometric gravities leads to a broad [FeI/H] distribution and a narrow, mono-metallic [FeII/H] distribution,
which one should we trust?
In principle, FeII lines are most trustworthy than FeI lines to determine the iron abundance, 
because FeII is a dominant species in the atmospheres of late-type stars 
(where iron is almost completely ionized) and its lines 
are unaffected by NLTE effects, at variance with the FeI lines 
\citep[see e.g.][]{kraft03,mashonkina11}.

The analysis of the results shown in Figure~\ref{histo1} and \ref{histo2} suggests that the adoption of method (1) 
tends to produce an artificial spread of [FeII/H] toward low metallicities. 
Since [FeII/H] strongly depends on the adopted values of logg, this implies that gravities are severly 
underestimated in method (1). 
This bias is clearly revealed when the stellar masses corresponding to the spectroscopic values of
log~$g$ values are computed.
We estimated the stellar masses by inverting the Stefan-Boltzmann equation and assuming the spectroscopic 
log~$g$ derived with method (1). 
The derived masses range from 0.12 to 0.79 $M_{\odot}$, with a 
mean value of 0.46 $M_{\odot}$ and a dispersion of 0.2 $M_{\odot}$. 
Note that $\sim$70\% of the stars have masses below 0.6 $M_{\odot}$. Such low values, 
as well as the large dispersion of the mass distribution, are unlikely for a sample 
dominated by RGB stars, with expected masses close to 0.75-0.80 $M_{\odot}$.
In particular, 10 target stars have log~$g$ that would require masses below 0.5 $M_{\odot}$, 
thus smaller than the typical mass of the He-core of GC giant stars at the luminosity level 
of our targets. Such very low masses cannot be justified even in light of the uncertainties 
in the mass loss rate \citep{origlia14}. A similarly wide mass distribution is obtained 
by adopting the spectroscopic parameters by M09, leading to a mass range between 0.34 and 1.19 $M_{\odot}$.
In that case, three stars have masses larger than 0.8 $M_{\odot}$, corresponding to 
the typical mass of a turnoff star of M22. 
For comparison, the masses derived from the spectroscopic log~$g$ of the spectral sample 
of NGC6752 (see Section~\ref{refe}) cover a small and well reasonable range, from 0.65 to 0.85 $M_{\odot}$, 
with an average value of 0.75 $M_{\odot}$ ($\sigma$=0.06 $M_{\odot}$).

% FIGURE 7
%
\begin{figure}[h]
\centering
\includegraphics[angle=0,scale=0.50]{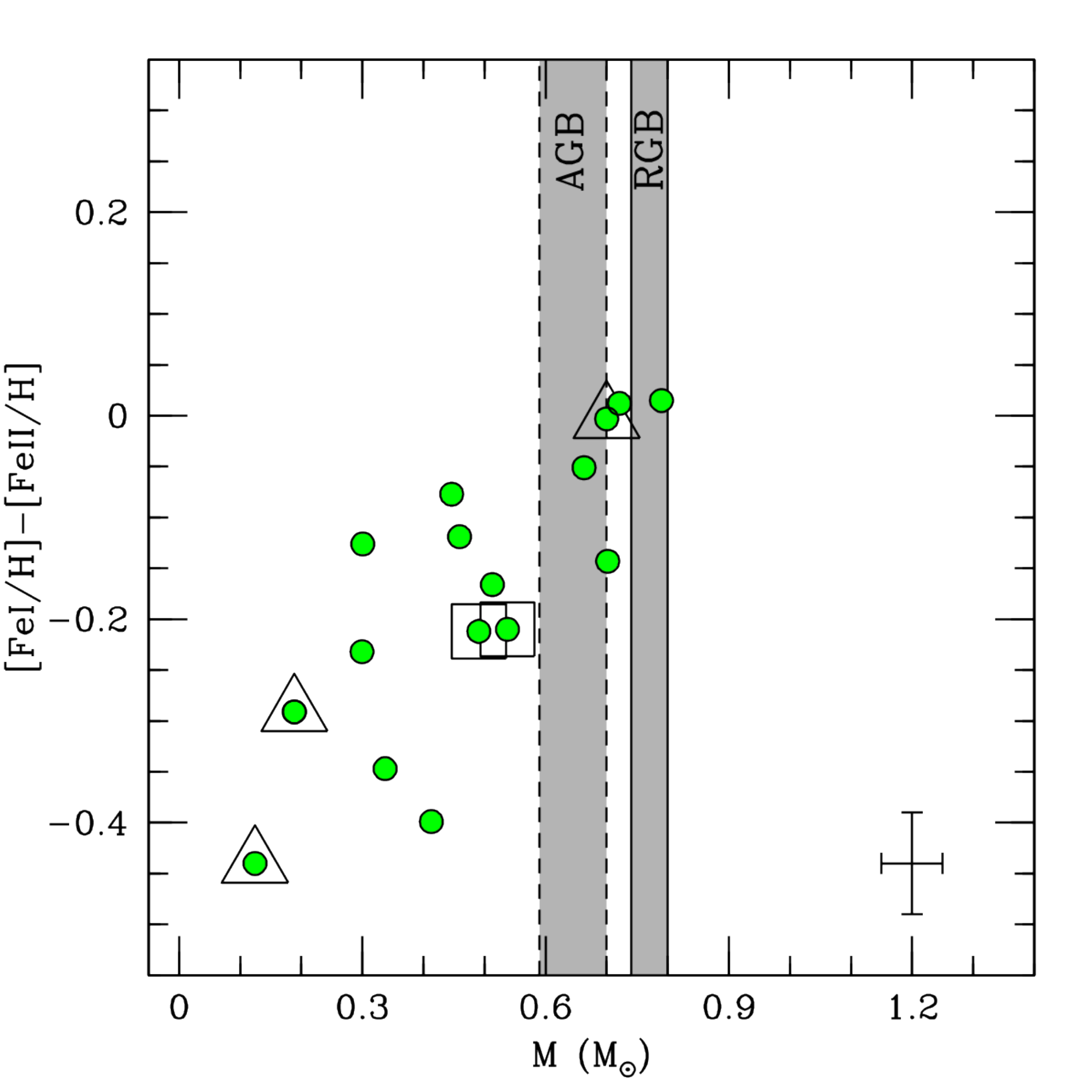}
\caption{Behavior of the difference [FeI/H]-[FeII/H], as derived with method (2), of the spectroscopic 
targets as a function of the stellar masses inferred from the spectroscopic log~$g$ in method (1). 
The two shaded grey regions mark the mass range expected for AGB and RGB stars. 
Same symbols of Figure~\ref{cmd}.}
\label{mass}
\end{figure}

Figure~\ref{mass} shows the behavior of the difference [FeI/H]-[FeII/H], as derived with method (2),
as a function of the stellar masses, as derived from the spectroscopic gravities in method (1). 
The mass intervals expected for RGB and AGB stars 
in the luminosity range of our spectroscopic targets are shown as grey shaded regions. 
A clear trend between the [FeI/H]-[FeII/H] difference and the stellar mass is found.
The stars with the largest difference between FeI and FeII abundances are also those 
where the spectroscopic log~$g$ requires an unrealistically low mass, while for the stars where 
[FeI/H] is consistent with [FeII/H] the spectroscopic log~$g$ provide masses in reasonable agreement 
with the theoretical expectations.
This demonstrates that the spectroscopic gravities needed to 
force [FeII/H] matching the low-abundance tail of the [FeI/H] distribution
lead to unreliable stellar masses.
Since this is the only case in which [Fe II/H] shows significant
spread, we have to conclude that the observed large iron distribution is 
not real. The correct diagnostic of iron content therefore are the
Fe II lines analyzed under the assumption of photometric
gravities. These always lead to a narrow iron distribution (see
Figures~\ref{histo1} and \ref{histo2}), thus implying that no iron spread is observed in M22.

% FIGURE 8
%
\begin{figure}[h]
\centering
\includegraphics[width=0.90\textwidth]{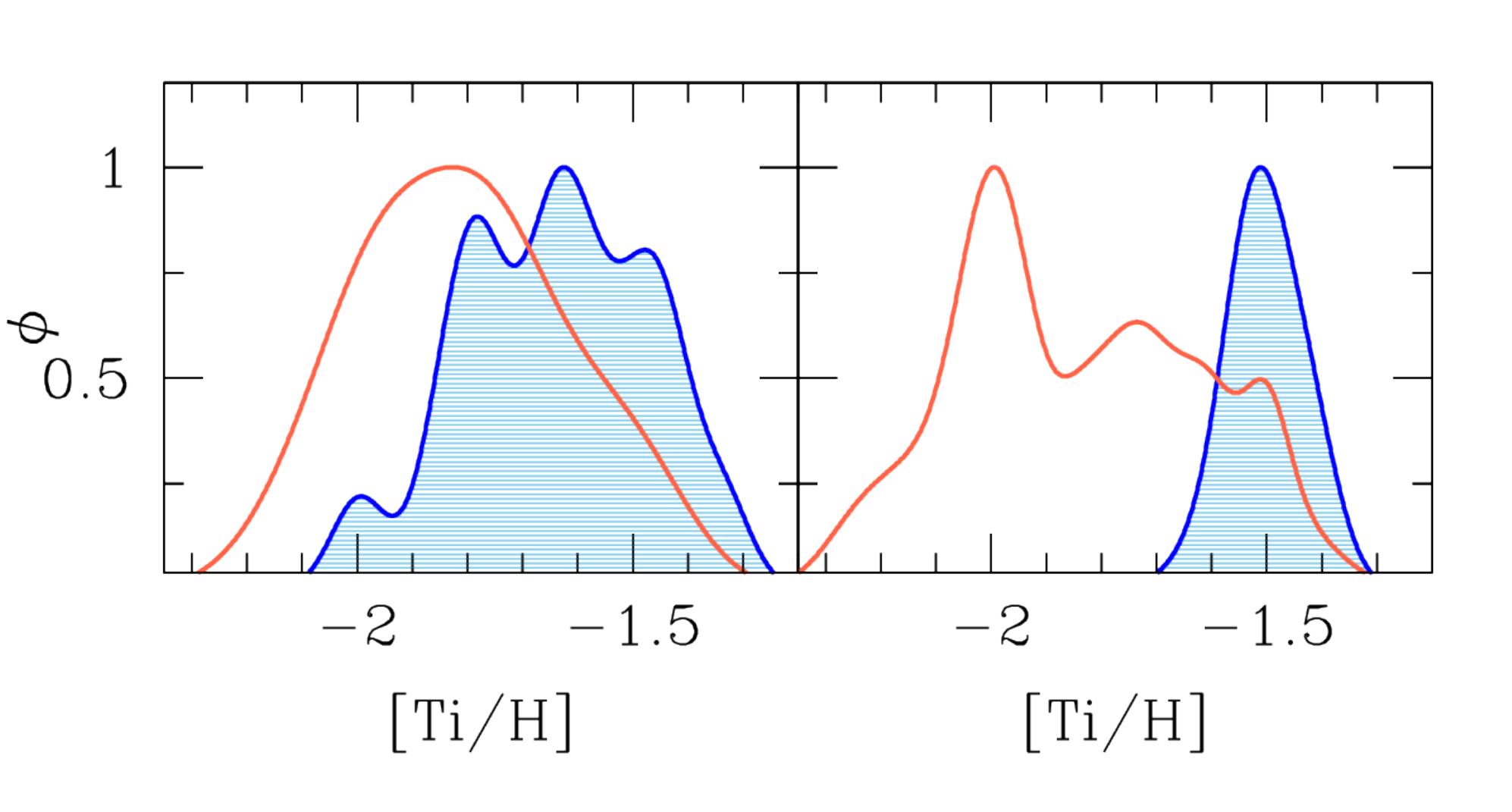}
\caption{Generalized histograms for [TiI/H] (empty red histogram) and [TiII/H] (blue histogram) obtained adopting the
spectroscopic (left panel) and photometric log~$g$ (right panel).}
\label{tid}
\end{figure}
%
%

%%%% TITANIUM
An additional confirmation of the different behavior of neutral and ionized lines in our sample is provided by 
the analysis of the titanium transitions, because this element is one of the few species that provides 
a large number of both neutral and single ionized lines. 
The oscillator strengths are from \citet{martin88} and \citet{lawler13} for TiI lines and 
from \citet{wood13} for TiII lines. The [TiI/H] and [TiII/H] abundances exhibit 
the same behavior discussed above for the Fe abundances. When the spectroscopic gravities 
are used, both the distributions are broad, with an observed scatter of $\sim$0.2 dex (see left panel of Figure~\ref{tid}). 
On the other hand, when the photometric gravities are adopted (see Table~\ref{tab3_c6}),
the [TiII/H] distribution is consistent with null intrinsic scatters,
while that of [TiI/H] remains broad and skewed toward low abundances (right panel of Figure~\ref{tid}).
We note that the difference [TiI/H]-[TiII/H] strongly correlates with the difference 
[FeI/H]-[FeII/H], with a Spearman rank correlation coefficient $C_S$=+0.956 that provides 
a probability of $\sim10^{-8}$ that the two quantities are not correlated.
Hence, the analysis of [TiI/H] and [TiII/H] reinforces the scenario where the abundances from 
neutral lines in most of the M22 stars are biased,
providing distributions (artificially) larger than those from single ionized lines.

\section{The s-process elements abundance}

M09 and \citet{marino11b} found that M22 has, together with a dispersion in the iron content, 
an intrinsic spread in the abundances of s-process elements. In light of the results described above, 
we derived abundances also for these 
elements, by adopting the parameters obtained with method (2) and measuring 
YII, BaII, LaII and NdII lines.
For Y and Nd the abundances have been obtained with GALA from the EW measurement, 
as done for the Fe and Ti lines, and adopting the oscillator strengths available in the 
Kurucz/Castelli linelist. BaII and LaII lines are affected by hyperfine and isotopic splittings.
The linelists for the LaII lines are from \citet{lawler01}, while those for 
the BaII lines from the NIST database \footnote{http://physics.nist.gov/PhysRefData/ASD/lines\_form.html}.
Only for these two elements, the abundances have been derived 
with our own code SALVADOR (A. Mucciarelli et al. in preparation) that performs a
$\chi^{2}$-minimization between observed and synthetic spectra calculated with 
the code SYNTHE.

For all these elements, we found that the absolute abundances show large star-to-star variations, 
with observed scatters between $\sim$0.2 and $\sim$0.3 dex, depending on the element.
These spreads are not compatible within the uncertainties.
Because of the possible occurrence of the NLTE effects, 
the abundance ratios [X/Fe] (see Table~\ref{tab4_c6})
have been estimated by using the FeII abundances as reference; in fact, for these elements 
the chemical abundances have been derived only from single ionized transitions,
which are less sensitive to the overionization 
\citep[or sensitive to it in a comparable way to the FeII lines; see e.g. the discussion in][]{ivans01}. 
The [X/FeII] abundance ratios show significant intrinsic spreads, 
as confirmed by the ML algorithm.
Note that, if we adopt FeI abundances as reference, the [X/FeI] abundance ratios 
still display an intrinsic scatter, because the observed spread in the absolute 
abundances for these s-process elements is larger than that measured from the FeI lines.

% FIGURE 9
%
\begin{figure}[h]
\centering
\includegraphics[width=0.70\textwidth]{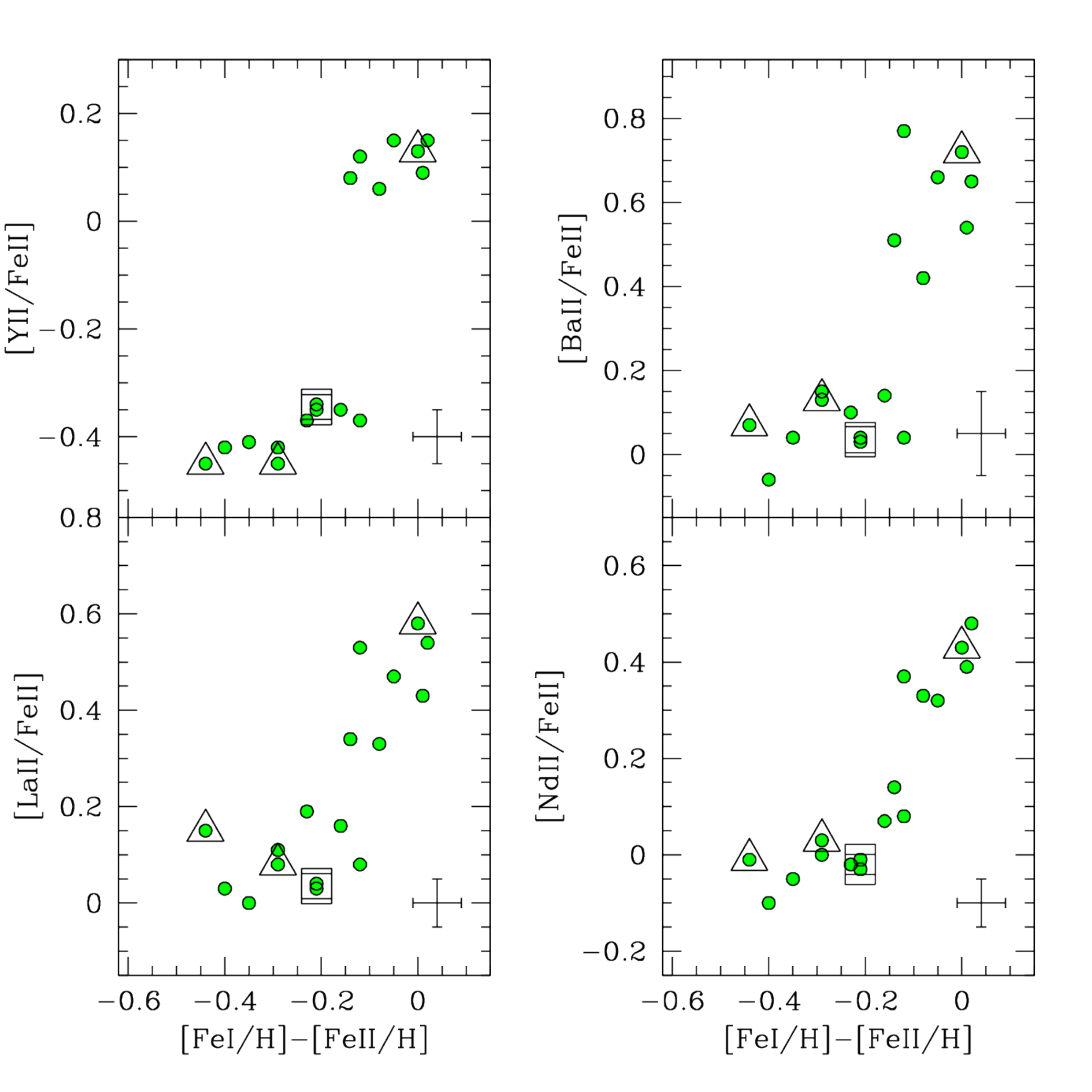}
\caption{Behavior of the abundance of the s-process elements Y, La, Ba and Nd as
a function of the difference between [FeI/H] and [FeII/H]. Same symbols of Figure~\ref{cmd}.}
\label{spro1}
\end{figure}

\begin{figure}[h]
\centering
\includegraphics[width=0.70\textwidth]{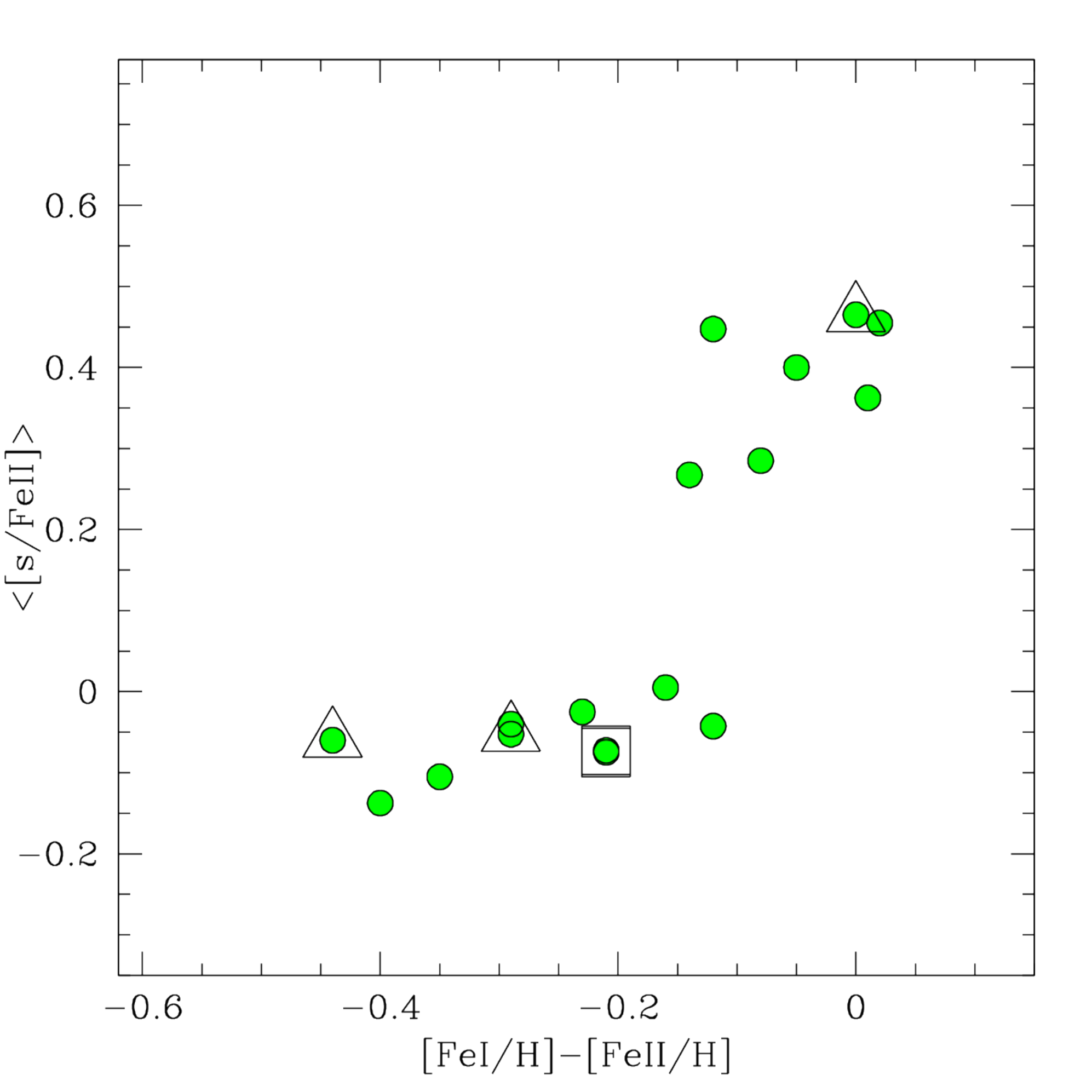}
\caption{Behavior of the average abundance of s-process elements 
(derived by averaging together the abundances of Y, La, Ba and Nd)
as a function of the difference between [FeI/H] and [FeII/H].
Same symbols of Figure~\ref{cmd}.}
\label{spro2}
\end{figure}

Figure~\ref{spro1} shows the behavior of each s-process element abundance ratio 
as a function of the difference between [FeI/H] and [FeII/H]. 
In all the cases, a clear trend between [X/FeII] and [FeI/H]-[FeII/H] 
is detected, in the sense that the stars characterized by higher s-process abundances 
display a better agreement between FeI and FeII. In the case of Y, Ba and Nd, we find two distinct and 
well separated groups of stars, while for La the behavior is continuous, 
with no clear gap.
Finally, Figure~\ref{spro2} plots the behavior of $<$[s/FeII]$>$, obtained by averaging 
together the four abundance ratios, as a function of [FeI/H]-[FeII/H], confirming the existence 
of two groups of stars, with different [s/Fe] and [FeI/H] (but the same [FeII/H]). 
This finding resembles the results by M09 who identify two groups
of stars, named {\sl s-poor} and {\sl s-rich}.

% FIGURE 10
%
%
%

%------------------------------------------------------------------

\section{Discussion: re-thinking M22}

The main results and conclusions of this work are summarized as follows:
\begin{itemize}

%%%%%% GENERAL
\item 
The new analysis of M22 presented here demonstrates that this GC is mono-metallic and that 
the previous claim of a metallicity scatter was due to a systematic under-estimate of the FeI abundance 
combined with the use of spectroscopic gravities.
When photometric log~$g$ are adopted, the FeII lines provide the same abundance for all the stars,
regardless of the adopted method to estimate T$_{eff}$.

%%%%% ORIGIN
\item 
In light of this result, the formation/evolution scenario for M22 must be 
deeply re-thought. The homogeneity in its iron content 
suggests that M22 was not able to retain the SN ejecta in its gravitational well. 
Hence, it is not necessary to invoke that the cluster was significantly more massive at its birth
and that it subsequently lost a large amount of its mass.
The observed unimodal [FeII/H] distribution rules out 
the possibility that M22 is the remnant of a now disrupted dwarf galaxy, 
because these systems are characterized by a wide range of metallicity, due to
the prolonged star-formation activity \citep[see][and references therein]{tolstoy09}. 
Also, comparisons between M22 and $\omega$ Centauri \citep{dacosta11} are 
undermined by the homogeneity in the [FeII/H] abundance of M22. 
On the other hand, M22 cannot be considered as a {\sl genuine} GC, because of
the intrinsic spread in heavy s-process elements abundances, pointing out the occurence of 
a peculiar chemical enrichment (probably from AGB stars) in this cluster, at variance with 
most of the GCs where s-process elements do not show intrinsic scatters \citep{dorazi10}.

%%%%%% MERGING
\item
M09 and \citet{marino11b} discussed the possibility that M22 is the product of a merging between two 
GCs with different chemical composition. In light of our new analysis, 
this scenario appears unlikely, even if it cannot be totally ruled out.
In this framework, M22 should form from the merging between two clusters with 
the same Fe content, but characterized by different s-process element abundances. 
While clusters with comparable metallicity and different s-process abundance are indeed observed 
\citep[for instance M4 and M5;][]{ivans99,ivans01}, in this scenario
the cluster with normal s-process abundances should be composed mainly by stars with a large difference 
between [FeI/H] and [FeII/H], while the second cluster should have stars with enhanced s-process abundances 
and similar [FeI/H] and [FeII/H] (see Figure~\ref{spro2}).

%%%% NLTE
\item
As a possible working hypothesis to explain the observed behavior of [FeI/H] and [FeII/H], 
we note that the difference between [FeI/H] and [FeII/H]
is qualitatively compatible with the occurrence of NLTE
effects driven by overionization. These effects are known to affect mainly the less abundant species,
like FeI, and to have a negligible/null impact on the dominant species, like FeII 
\citep[see e.g.][]{thevenin99,mashonkina11,fabrizio12}. 
Under NLTE conditions, the spectral lines of neutral ions are weaker than in LTE. 
Hence, when the line formation is calculated in LTE conditions (as done in standard analyses), 
the resulting abundance of neutral lines will be correspondingly lower.\\
The same interpretative scheme can be applied to M22. A large and intrinsically broad 
Fe distribution is obtained only from FeI lines, according to the systematic 
underestimate of the Fe abundance obtained when lines affected by overionization are 
analysed in LTE. On the other hand, FeII lines are not affected by NLTE and they 
provide (when photometric log~$g$ are used) the correct abundance, leading to a 
narrow abundance distribution.

%%%% AGB 
\item
The mismatch between [FeI/H] and [FeII/H] observed in M22 resembles those found in the GCs
M5 \citep{ivans01}, 47Tucanae \citep{lapenna14} and NGC3201 \citep{mucciarelli15}.
In these cases, the different behavior observed for FeI and FeII lines 
is restricted to AGB stars only, where FeI lines provide abundances systematically 
lower than those from FeII lines,
while RGB stars have similar [FeI/H] and [FeII/H]. 
However, the situation is more complex in M22, because a large difference 
between [FeI/H] and [FeII/H] is observed in most of the stars and not only in AGB stars.
Among the target stars of M22, three are identified as AGB stars, 
according to their positions in the CMDs. 
Two of them have a large Fe difference, [FeI/H]-[FeII/H]=--0.29 and --0.44 dex, while 
for the third star FeI and FeII lines provide almost the same abundance. 
The other two possible AGB stars (empty squares in Figure~\ref{cmd}) have Fe differences of 
--0.21 dex. On the other hand, comparable differences are observed among some RGB stars. 
For instance, the two faintest stars of the sample (\#221 and \#224) are clearly RGB stars 
(see Figure~\ref{cmd} and \ref{mass}), because they are located at the luminosity level where the color separation between 
RGB and AGB is the largest. On the other hand, these two stars (with very similar atmospheric 
parameters and [FeII/H]) have different [FeI/H] abundances: 
star \#224 has a difference of [FeI/H]-[FeII/H]=--0.14 dex, while star \#221 has [FeI/H]-[FeII/H]=--0.29 dex. 
If departures from LTE are the reasons for the observed discrepancy between [FeI/H] and [FeII/H] (and between [TiI/H] and [TiII/H]), 
this finding challenges the available NLTE calculations \citep[see e.g.][]{lind12,bergemann12}, in which
stars with very similar parameters are expected to have the same NLTE corrections. 
New NLTE calculations should be performed to investigate this hypothesis, with the constraint 
to reproduce simultaneously the discrepancies in Fe and Ti.

%%%%% s-/CN
\item
We found that the difference between [FeI/H] and [FeII/H] is correlated with the s-process element abundances.
The behavior is quite puzzling, because the stars with an anomalous difference between [FeI/H] and 
[FeII/H] are those with {\sl normal} s-process abundances, compatible with the abundances observed 
in most of the GCs and in the Galactic field stars of similar metallicity \citep[see e.g. Figure 3 in ][]{venn04}. 
On the other hand, the stars enriched in s-process elements show a good agreement between [FeI/H] and [FeII/H].
{\sl Whatever the mechanism responsible to spread [FeI/H] is, it must be also responsible for the 
peculiar behavior of the s-process element abundances.}

%%%%% COME STUDIARE GLI SPREADS
\item
We confirm the claim already suggested by \citet{lapenna14} and \citet{mucciarelli15}: chemical analyses based on 
FeI lines and spectroscopic gravities can lead to spurious abundance spreads. 
In light of these results, any claim of intrinsic iron spread in GCs should be always confirmed with an analysis 
based on FeII lines and photometric gravities. If the abundance spread is real,
it should be detected also when FeII lines and photometric log~$g$ are adopted, 
since FeII lines are the most reliable indicators of the iron abundance. 
All the GCs with anomalous intrinsic Fe spreads observed so far \citep[see][for an updated list]{marino15} 
deserve new analyses in light of this effect, in order to firmly establish whether these spreads are
real or spurious.

\end{itemize}

%%%%%%%%%%%%%%%%%%%%%%%%%%%%%%%%%%%%%%%%%%%%%%%%%%%%%%%%%%%%%%%%%%%%%%% TABLES

\begin{deluxetable}{lccccc}
\tiny
\tablecolumns{6} 
\tablewidth{0pc}  
\tablecaption{Star identification number, wavelength, ion, excitation potential, oscillator strength 
and measured EWs for all the used transitions.}
\tablehead{ 
\colhead{Star} &  $\lambda$ & Ion & $\chi$ & log~$g$f & EW \\
               &     (\AA)  &         & (eV)   &        & (m\AA)}
\startdata 
  51 & 4805.415   &    TiI	& 2.340   &   0.070   &   30.4   \\  
  51 & 4870.126   &    TiI	& 2.250   &   0.440   &   46.7   \\ 
  51 & 4885.079   &    TiI	& 1.890   &   0.410   &   76.1   \\ 
  51 & 4909.098   &    TiI	& 0.830   &  -2.370   &   15.8   \\ 
  51 & 4913.614   &    TiI	& 1.870   &   0.220   &   64.0   \\	    
  51 & 4915.229   &    TiI	& 1.890   &  -0.910   &   11.8   \\ 
  51 & 4919.860   &    TiI	& 2.160   &  -0.120   &   30.1   \\ 
  51 & 4926.148   &    TiI	& 0.820   &  -2.090   &   20.9   \\ 
  51 & 4937.726   &    TiI	& 0.810   &  -2.080   &   27.2   \\  
  51 & 4997.096   &    TiI	& 0.000   &  -2.070   &   90.7   \\  
  51 & 5009.645   &    TiI	& 0.020   &  -2.200   &   83.1   \\  
  51 & 5016.161   &    TiI	& 0.850   &  -0.480   &  107.5   \\  
  51 & 5020.026   &    TiI	& 0.840   &  -0.330   &  117.0   \\	
  51 & 5036.464   &    TiI	& 1.440   &   0.140   &   98.9   \\	
  51 & 5038.397   &    TiI	& 1.430   &   0.020   &   95.9   \\  
  51 & 5043.584   &    TiI	& 0.840   &  -1.590   &   46.6   \\  
  51 & 5045.415   &    TiI	& 0.850   &  -1.840   &   29.7   \\	
  51 & 5052.870   &    TiI	& 2.170   &  -0.270   &   23.0   \\	
  51 & 5062.103   &    TiI	& 2.160   &  -0.390   &   15.6   \\  
  51 & 5065.985   &    TiI	& 1.440   &  -0.970   &   38.1   \\  
\enddata 
\tablecomments{This table is available in its entirety in machine-readable form.}
\label{tab1_c6}
\end{deluxetable}

\begin{deluxetable}{lcccccc}
\tiny
\tablecolumns{7} 
\tablewidth{0pc}  
\tablecaption{Observed and intrinsic scatters for [FeI/H] and [FeII/H] as derived from the ML algorithm and from the
three methods described in the paper.}
\tablehead{ 
 & [FeI/H] & $\sigma_{obs}$ & $\sigma_{int}$  & [FeII/H] & $\sigma_{obs}$ & $\sigma_{int}$  \\
 &  &  &  & & & }
\startdata
 {\rm Method 1}      &  --1.92$\pm$0.03   &  0.14 & 0.13$\pm$0.02  &  --1.90$\pm$0.03 & 0.14  & 0.13$\pm$0.02 \\
 {\rm Method 2}      &  --1.92$\pm$0.04   &  0.16 & 0.15$\pm$0.02  &  --1.75$\pm$0.01 & 0.04  & 0.00$\pm$0.02 \\
 {\rm Method 3}      &  --1.86$\pm$0.03   &  0.13 & 0.12$\pm$0.02  &  --1.81$\pm$0.01 & 0.05  & 0.00$\pm$0.02 \\
\enddata 
%\tablecomments{$~~~~~$}
\label{tab2_c6}
\end{deluxetable}

%%%%%%%%%%%%%%%%%%% METHOD 1 (mancano ABU-err)

\begin{landscape}
\begin{deluxetable}{lcccccccc}
\tiny
\tablecolumns{9} 
\tablewidth{0pc}  
\tablecaption{Atmospheric parameters, [FeI/H], [FeII/H], [TiI/H] and [TiII/H] abundances for the 
spectroscopic targets of M22, as derived with method (1). The last line lists the average abundances with the statistical error.}
\tablehead{ 
\colhead{Star} &   T$_{eff}$ & log~$g$ & v$_{turb}$ & [FeI/H] & [FeII/H] & [TiI/H] &  [TiII/H] & Notes\\
  &   (K)  &    &(km/s) & &  &  &  &}
\startdata
      51  &  4280  &	1.00	   &  1.70   &  --1.70$\pm$0.02   &   --1.71$\pm$0.04	&    --1.50$\pm$0.04  &  --1.34$\pm$0.04  &   \\ 
      61  &  4430  &	0.95	   &  1.70   &  --1.85$\pm$0.05   &   --1.84$\pm$0.04	&    --1.74$\pm$0.06  &  --1.61$\pm$0.04  &   \\ 
      71  &  4405  &	0.97	   &  1.50   &  --1.90$\pm$0.04   &   --1.89$\pm$0.04	&    --1.77$\pm$0.04  &  --1.59$\pm$0.04  &  \\  
      88  &  4450  &	1.20	   &  1.50   &  --1.78$\pm$0.05   &   --1.74$\pm$0.05	&    --1.67$\pm$0.08  &  --1.46$\pm$0.04  &   \\ 
     221  &  4570  &	1.13	   &  1.40   &  --2.04$\pm$0.04   &   --2.04$\pm$0.04	&    --2.00$\pm$0.04  &  --1.86$\pm$0.04  &   \\ 
     224  &  4670  &	1.75	   &  1.40   &  --1.87$\pm$0.04   &   --1.78$\pm$0.04	&    --1.76$\pm$0.05  &  --1.47$\pm$0.04  &    \\	
  200005  &  3920  &	0.00	   &  2.20   &  --2.10$\pm$0.02   &   --1.92$\pm$0.06	&    --1.97$\pm$0.05  &  --1.74$\pm$0.05  &  \\  
  200006  &  3910  &	0.04	   &  2.10   &  --1.84$\pm$0.03   &   --1.78$\pm$0.05	&    --1.69$\pm$0.05  &  --1.55$\pm$0.04  &  \\  
  200025  &  4060  &	0.57	   &  1.90   &  --1.72$\pm$0.02   &   --1.75$\pm$0.05	&    --1.48$\pm$0.05  &  --1.45$\pm$0.04  &  \\ 
  200031  &  4290  &	0.72	   &  1.80   &  --1.96$\pm$0.03   &   --1.97$\pm$0.04	&    --1.85$\pm$0.05  &  --1.64$\pm$0.04  &  AGB? \\  
  200043  &  4300  &	0.73	   &  1.70   &  --1.94$\pm$0.04   &   --1.95$\pm$0.04	&    --1.86$\pm$0.04  &  --1.69$\pm$0.05  &  AGB? \\ 
  200068  &  4400  &	0.82	   &  1.60   &  --2.00$\pm$0.04   &   --2.00$\pm$0.04	&    --1.93$\pm$0.05  &  --1.74$\pm$0.04  &  \\  
  200076  &  4390  &	0.80	   &  1.60   &  --2.05$\pm$0.04   &   --2.06$\pm$0.04	&    --2.01$\pm$0.05  &  --1.82$\pm$0.04  &  \\  
  200080  &  4520  &	0.70	   &  1.70   &  --2.01$\pm$0.04   &   --2.00$\pm$0.03	&    --1.96$\pm$0.05  &  --1.78$\pm$0.04  &  AGB \\	
  200083  &  4440  &	1.20	   &  1.50   &  --1.73$\pm$0.04   &   --1.75$\pm$0.04	&    --1.60$\pm$0.05  &  --1.46$\pm$0.05  &  AGB\\  
  200101  &  4400  &	0.90	   &  1.50   &  --1.89$\pm$0.05   &   --1.91$\pm$0.05	&    --1.77$\pm$0.05  &  --1.64$\pm$0.05  &  \\  
  200104  &  4490  &	0.59	   &  1.70   &  --2.19$\pm$0.06   &   --2.19$\pm$0.03	&    --2.15$\pm$0.08  &  --2.00$\pm$0.03  &  AGB \\
\hline
          &        &	 	   &         &  --1.92$\pm$0.03   &   --1.91$\pm$0.03	&    --1.81$\pm$0.04  &  --1.64$\pm$0.04  &   \\
\enddata 
%\tablecomments{$~~~~~$}
\label{tab3_c6}
\end{deluxetable}
\end{landscape}

%%%%%%%%%%%%%%%%%%% METHOD 2

\begin{landscape}
\begin{deluxetable}{lcccccccc}
\tiny
\tablecolumns{9} 
\tablewidth{0pc}  
\tablecaption{Atmospheric parameters, [FeI/H], [FeII/H], [TiI/H] and [TiII/H] abundances for the 
spectroscopic targets of M22, as derived with method (2). The last line lists the average abundances with the statistical error.}
\tablehead{ 
\colhead{Star} &   T$_{eff}$ & log~$g$ & v$_{turb}$ & [FeI/H] & [FeII/H] & [TiI/H] &  [TiII/H] & Notes\\
  &   (K)  &    &(km/s) & &  &  &  &}
\startdata
      51  &  4280  &	0.99	   &  1.70   &  --1.70$\pm$0.02   &   --1.72$\pm$0.04	&    --1.50$\pm$0.04  &  --1.41$\pm$0.05  &   \\ 
      61  &  4440  &	1.17	   &  1.60   &  --1.84$\pm$0.05   &   --1.72$\pm$0.04	&    --1.76$\pm$0.06  &  --1.49$\pm$0.05  &   \\ 
      71  &  4390  &	1.15	   &  1.60   &  --1.94$\pm$0.03   &   --1.78$\pm$0.04	&    --1.85$\pm$0.05  &  --1.51$\pm$0.05  &  \\  
      88  &  4470  &	1.30	   &  1.50   &  --1.77$\pm$0.05   &   --1.72$\pm$0.05	&    --1.66$\pm$0.08  &  --1.44$\pm$0.06  &   \\ 
     221  &  4640  &	1.81	   &  1.30   &  --2.00$\pm$0.05   &   --1.71$\pm$0.04	&    --1.98$\pm$0.06  &  --1.53$\pm$0.05  &   \\ 
     224  &  4650  &	1.80	   &  1.30   &  --1.88$\pm$0.05   &   --1.74$\pm$0.04	&    --1.79$\pm$0.07  &  --1.42$\pm$0.05  &    \\	
  200005  &  3900  &	0.30	   &  2.20   &  --2.07$\pm$0.03   &   --1.67$\pm$0.09	&    --2.07$\pm$0.10  &  --1.62$\pm$0.06  &  \\  
  200006  &  3960  &	0.34	   &  2.10   &  --1.80$\pm$0.04   &   --1.72$\pm$0.09	&    --1.63$\pm$0.10  &  --1.48$\pm$0.06  &  \\  
  200025  &  4070  &	0.63	   &  1.80   &  --1.70$\pm$0.04   &   --1.71$\pm$0.07	&    --1.45$\pm$0.08  &  --1.46$\pm$0.06  &  \\ 
  200031  &  4240  &	0.86	   &  1.80   &  --2.02$\pm$0.04   &   --1.81$\pm$0.05	&    --1.98$\pm$0.07  &  --1.53$\pm$0.05  &  AGB? \\  
  200043  &  4270  &	0.93	   &  1.70   &  --1.98$\pm$0.04   &   --1.77$\pm$0.05	&    --1.96$\pm$0.05  &  --1.56$\pm$0.06  &  AGB? \\ 
  200068  &  4340  &	1.16	   &  1.60   &  --2.10$\pm$0.05   &   --1.75$\pm$0.05	&    --2.11$\pm$0.07  &  --1.54$\pm$0.05  &  \\  
  200076  &  4410  &	1.24	   &  1.60   &  --2.05$\pm$0.04   &   --1.82$\pm$0.04	&    --2.03$\pm$0.05  &  --1.54$\pm$0.05  &  \\  
  200080  &  4570  &	1.28	   &  1.70   &  --2.03$\pm$0.05   &   --1.74$\pm$0.04	&    --1.99$\pm$0.06  &  --1.52$\pm$0.05  &  AGB \\	
  200083  &  4430  &	1.18	   &  1.60   &  --1.76$\pm$0.04   &   --1.76$\pm$0.05	&    --1.60$\pm$0.05  &  --1.48$\pm$0.05  &  AGB\\  
  200101  &  4480  &	1.37	   &  1.50   &  --1.83$\pm$0.05   &   --1.71$\pm$0.05	&    --1.71$\pm$0.06  &  --1.44$\pm$0.06  &  \\  
  200104  &  4520  &	1.36	   &  1.70   &  --2.26$\pm$0.05   &   --1.82$\pm$0.04	&    --2.23$\pm$0.07  &  --1.55$\pm$0.05  &  AGB \\	
\hline
          &        &	 	   &         &  --1.92$\pm$0.04   &   --1.75$\pm$0.01	&    --1.84$\pm$0.05  &  --1.50$\pm$0.01  &   \\
\enddata 
%\tablecomments{$~~~~~$}
\label{tab4_c6}
\end{deluxetable}
\end{landscape}

%%%%%%%%%%%%%%%%%%% METHOD 3 (mancano err e Ti)

\begin{landscape}
\begin{deluxetable}{lcccccccc}
\tiny
\tablecolumns{9} 
\tablewidth{0pc}  
\tablecaption{Atmospheric parameters, [FeI/H], [FeII/H], [TiI/H] and [TiII/H] abundances for the 
spectroscopic targets of M22, as derived with method (3). The last line lists the average abundances with the statistical error.}
\tablehead{ 
\colhead{Star} &   T$_{eff}$ & log~$g$ & v$_{turb}$ & [FeI/H] & [FeII/H] & [TiI/H] &  [TiII/H] & Notes\\
  &   (K)  &    &(km/s) & &  &  &  &}
\startdata
      51  &  4232  &	0.96	   &  1.70   &  --1.76$\pm$0.05   &   --1.72$\pm$0.06	&    --1.57$\pm$0.06  &  --1.37$\pm$0.06  &   \\ 
      61  &  4400  &	1.16	   &  1.60   &  --1.87$\pm$0.05   &   --1.73$\pm$0.05	&    --1.81$\pm$0.06  &  --1.50$\pm$0.07  &   \\ 
      71  &  4435  &	1.17	   &  1.50   &  --1.88$\pm$0.05   &   --1.82$\pm$0.05	&    --1.76$\pm$0.08  &  --1.50$\pm$0.08  &  \\  
      88  &  4537  &	1.32	   &  1.60   &  --1.71$\pm$0.06   &   --1.80$\pm$0.06	&    --1.53$\pm$0.06  &  --1.49$\pm$0.07  &   \\ 
     221  &  4737  &	1.84	   &  1.50   &  --1.90$\pm$0.05   &   --1.81$\pm$0.05	&    --1.84$\pm$0.07  &  --1.57$\pm$0.07  &   \\ 
     224  &  4746  &	1.84	   &  1.50   &  --1.80$\pm$0.04   &   --1.82$\pm$0.04	&    --1.65$\pm$0.07  &  --1.48$\pm$0.06  &    \\	
  200005  &  3992  &	0.34	   &  2.20   &  --2.08$\pm$0.04   &   --1.78$\pm$0.04	&    --1.88$\pm$0.06  &  --1.62$\pm$0.07  &  \\  
  200006  &  3986  &	0.36	   &  2.10   &  --1.80$\pm$0.05   &   --1.77$\pm$0.05	&    --1.56$\pm$0.06  &  --1.47$\pm$0.07  &  \\  
  200025  &  4116  &	0.65	   &  1.90   &  --1.68$\pm$0.05   &   --1.81$\pm$0.05	&    --1.36$\pm$0.08  &  --1.45$\pm$0.09  &  \\ 
  200031  &  4271  &	0.87	   &  1.80   &  --1.97$\pm$0.05   &   --1.88$\pm$0.05	&    --1.93$\pm$0.06  &  --1.54$\pm$0.07  &  AGB? \\  
  200043  &  4351  &	0.97	   &  1.70   &  --1.89$\pm$0.04   &   --1.84$\pm$0.05	&    --1.81$\pm$0.08  &  --1.58$\pm$0.08  &  AGB? \\ 
  200068  &  4409  &	1.19	   &  1.60   &  --2.00$\pm$0.05   &   --1.85$\pm$0.07	&    --1.98$\pm$0.06  &  --1.56$\pm$0.07  &  \\  
  200076  &  4475  &	1.27	   &  1.60   &  --1.98$\pm$0.05   &   --1.87$\pm$0.05	&    --1.92$\pm$0.06  &  --1.60$\pm$0.06  &  \\  
  200080  &  4618  &	1.30	   &  1.80   &  --1.95$\pm$0.05   &   --1.80$\pm$0.07	&    --1.90$\pm$0.06  &  --1.55$\pm$0.07  &  AGB \\	
  200083  &  4567  &	1.24	   &  1.60   &  --1.61$\pm$0.05   &   --1.84$\pm$0.05	&    --1.36$\pm$0.08  &  --1.49$\pm$0.06  &  AGB\\  
  200101  &  4527  &	1.39	   &  1.60   &  --1.79$\pm$0.06   &   --1.80$\pm$0.04	&    --1.62$\pm$0.06  &  --1.52$\pm$0.07  &  \\  
  200104  &  4655  &	1.41	   &  1.70   &  --2.04$\pm$0.07   &   --1.89$\pm$0.07	&    --1.99$\pm$0.07  &  --1.63$\pm$0.09  &  AGB \\
\hline
          &        &	 	   &         &  --1.86$\pm$0.03   &   --1.81$\pm$0.01	&    --1.73$\pm$0.05  &  --1.52$\pm$0.02  &   \\
\enddata 
%\tablecomments{$~~~~~$}
\label{tab5_c6}
\end{deluxetable}
\end{landscape}

\begin{deluxetable}{lcccc}
\tiny
\tablecolumns{5} 
\tablewidth{0pc}  
\tablecaption{Abundance ratios for the s-process elements Y, Ba, La and Nd.}
\tablehead{ 
\colhead{Star}  & [YII/FeII] & [BaII/FeII] &  [LaII/FeII] & [NdII/FeII]\\
  }
\startdata
      51  &       +0.15$\pm$0.04   &   +0.65$\pm$0.07   &   +0.54$\pm$0.04  &	  +0.48$\pm$0.04     \\ 
      61  &      --0.37$\pm$0.05   &   +0.04$\pm$0.08   &   +0.08$\pm$0.04  &	  +0.08$\pm$0.04     \\ 
      71  &      --0.35$\pm$0.05   &   +0.14$\pm$0.08   &   +0.16$\pm$0.06  &	  +0.07$\pm$0.04    \\  
      88  &       +0.15$\pm$0.05   &   +0.66$\pm$0.08   &   +0.47$\pm$0.04  &	  +0.32$\pm$0.05     \\ 
     221  &      --0.42$\pm$0.04   &   +0.15$\pm$0.10   &   +0.11$\pm$0.04  &	  +0.00$\pm$0.05     \\ 
     224  &       +0.08$\pm$0.05   &   +0.51$\pm$0.07   &   +0.34$\pm$0.06  &	  +0.14$\pm$0.04      \\	
  200005  &      --0.42$\pm$0.06   &   +0.06$\pm$0.12   &   +0.03$\pm$0.05  &	 --0.10$\pm$0.05    \\  
  200006  &       +0.06$\pm$0.08   &   +0.42$\pm$0.11   &   +0.33$\pm$0.05  &	  +0.33$\pm$0.05    \\  
  200025  &       +0.09$\pm$0.05   &   +0.54$\pm$0.06   &   +0.43$\pm$0.06  &	  +0.39$\pm$0.05     \\ 
  200031  &      --0.35$\pm$0.04   &   +0.03$\pm$0.08   &   +0.03$\pm$0.06  &	 --0.01$\pm$0.04    \\  
  200043  &      --0.34$\pm$0.05   &   +0.04$\pm$0.07   &   +0.04$\pm$0.04  &	 --0.03$\pm$0.04     \\ 
  200068  &      --0.41$\pm$0.04   &   +0.04$\pm$0.08   &   +0.00$\pm$0.05  &	 --0.05$\pm$0.04    \\  
  200076  &      --0.37$\pm$0.04   &   +0.10$\pm$0.08   &   +0.19$\pm$0.05  &	 --0.02$\pm$0.04    \\  
  200080  &      --0.45$\pm$0.04   &   +0.13$\pm$0.09   &   +0.08$\pm$0.05  &	  +0.03$\pm$0.05     \\ 
  200083  &       +0.13$\pm$0.06   &   +0.72$\pm$0.08   &   +0.58$\pm$0.04  &	  +0.43$\pm$0.04    \\  
  200101  &       +0.12$\pm$0.06   &   +0.77$\pm$0.09   &   +0.53$\pm$0.04  &	  +0.37$\pm$0.05    \\  
  200104  &      --0.45$\pm$0.04   &   +0.07$\pm$0.09   &   +0.15$\pm$0.06  &	 --0.01$\pm$0.05     \\
\enddata 
%\tablecomments{$~~~~~$}
\label{tab6_c6}
\end{deluxetable}

%% file: c7/ms_7.tex
% CHAPTER 7

\chapter{Chemical Analysis of Asymptotic Giant Branch Stars in M62}

\label{c7}

{\bf Published in Lapenna et al. 2015, ApJ, 813, 97L}

{\it We have collected UVES-FLAMES high-resolution spectra for a
sample of 6 asymptotic giant branch (AGB) and 13 red giant branch
(RGB) stars in the Galactic globular cluster M62 (NGC6266). Here we
present the detailed abundance analysis of iron, titanium, and
light-elements (O, Na, Al and Mg). For the majority (5 out 6) of
the AGB targets we find that the abundances, of both iron and
titanium, determined from neutral lines are significantly
underestimated with respect to those obtained from ionized features,
the latter being, instead, in agreement with those measured for the
RGB targets. This is similar to recent findings in other clusters
and may suggest the presence of Non-Local Thermodynamical
Equilibrium (NLTE) effects.  In the O-Na, Al-Mg and Na-Al planes,
the RGB stars show the typical correlations observed for globular
cluster stars.  Instead, all the AGB targets are clumped in the
regions where first generation stars are expected to lie, similarly
to what recently found for the AGB population of NGC6752.  While
the sodium and aluminum abundance could be underestimated as a
consequence of the NLTE bias affecting iron and titanium, the used
oxygen line does not suffer from the same effects and the lack of
O-poor AGB stars therefore is solid.  We can thus conclude that none
of the investigated AGB stars belong to the second stellar
generation of M62.  We also find a RGB star with extremely high
sodium abundance ([Na/Fe]$=+1.08$ dex).}

%%%%%%%%%%%%%%%%%%%%%%%%%%%%%%%%%%%%%%%%%%%%%%%%%%%%%%%%%%% OBSERVATIONS

\section{Observations and spectral analysis}
\label{obs}

We have observed a sample of 19 giant stars in the GC M62 by using the
UVES-FLAMES@VLT spectrograph \citep{pasquini00} within the Large
Program 193.D-0232 (PI: Ferraro).  The spectra have been acquired by
using the grating 580 Red Arm CD\#3, which provides a high spectral
resolution (R$\sim$40000) and a spectral coverage between 4800 and
$6800\rm\mathring{A}$.
The 19 targets have been sampled by means of four different fiber configurations,
in five pointings of 30 min each (one configuration has been repeated twice),
during the nights of 2014, April 16 and June 2, 3 and 19.
In each configuration, one or two fibers have been used to sample the sky
for background subtraction purposes. After careful visual
inspection, only the (19) spectra with a signal-to-noise larger than
50 have been kept in the analysis.
The spectra have been reduced by using the dedicated ESO
pipeline\footnote{http://www.eso.org/sci/software/pipelines/}
performing bias subtraction, flat-fielding, wavelength calibration,
spectral extraction and order merging. The sky background has been
subtracted from each individual spectrum.

\begin{figure}[h]
\centering
\includegraphics[trim=0cm 0cm 0cm 0cm,clip=true,scale=0.50,angle=0]{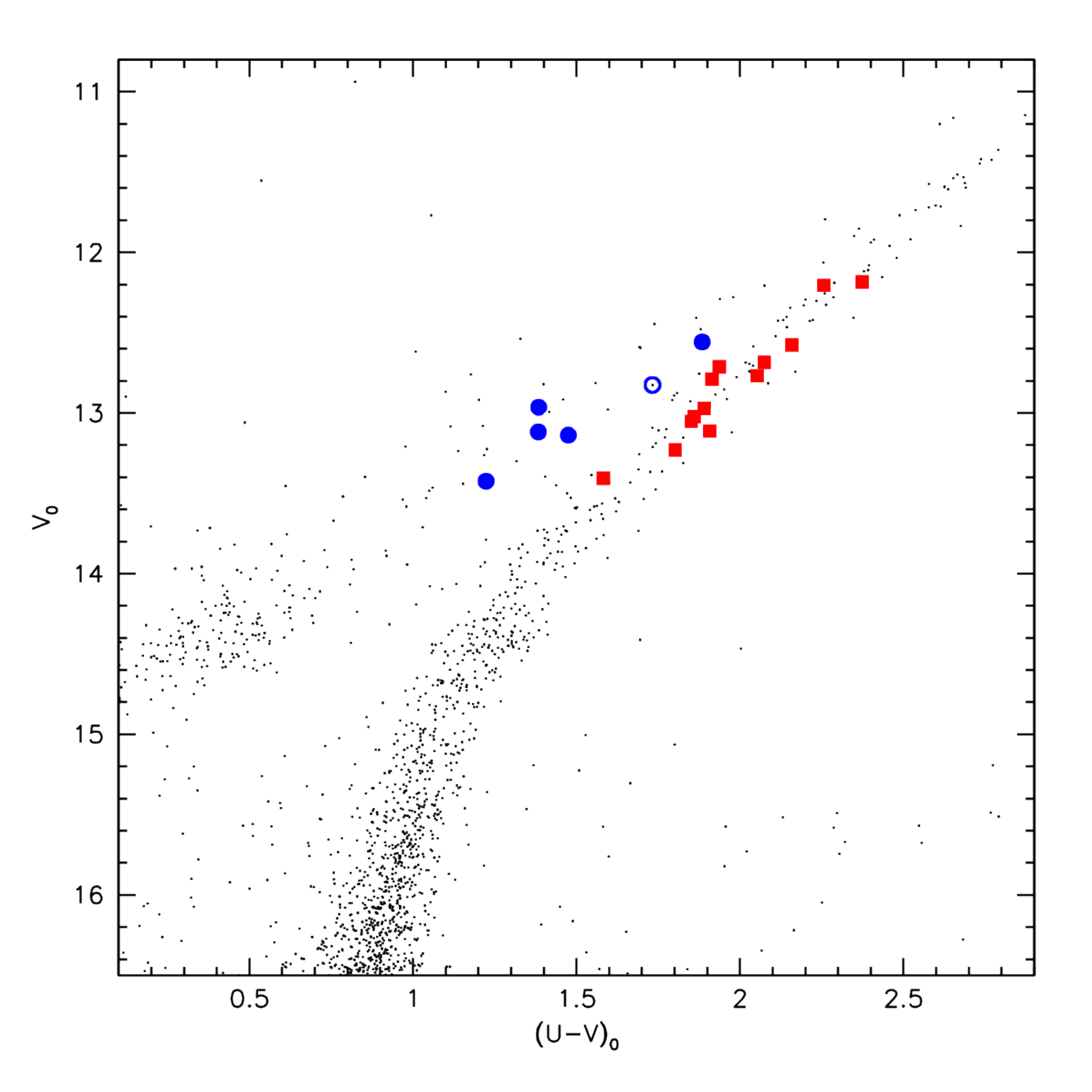}
\caption{Reddening-corrected color-magnitude diagram of M62, with the
targets of the present study highlighted: 13 RGB stars (red squares)
and 6 AGB objects (blue circles). The empty circle marks AGB star 96.}
\label{cmd}
\end{figure}

The target stars have been selected from the photometric catalog of
\citet{beccari06}, obtained from HST-WFPC2 observations. Only
stars brighter than $V=15$ and sufficiently isolated (i.e., with no
stellar sources of comparable or larger luminosity within a distance
of $2\arcsec$, and with no fainter stars within $1\arcsec$) have
been selected.  Figure~\ref{cmd} shows the $(V,U-V)$
color-magnitude diagram (CMD) corrected for differential reddening
following the procedure described in \cite{massari12} and adopting the
extinction law by \cite{mccall04}.  The final sample includes 6 AGB
and 13 RGB stars.  All the target stars are located within
$\sim$85$\arcsec$ from the cluster center.  Their identification
number, coordinates, and magnitudes are listed in Table~\ref{tab1_c7}.

% FIGURE 1
%

\subsection{Radial velocities}
\label{radial}

The radial velocities of our targets have been obtained by using the
code DAOSPEC \citep{stetson08} and by measuring the position of over
300 metallic lines distributed along the whole spectral range covered
by the 580 Red Arm of UVES-FLAMES. The uncertainties have been
computed as the dispersion of the velocities measured from each line
divided by the square root of the number of lines used, and they
turned out to be smaller than 0.05 km s$^{-1}$.  Finally, we applied
the heliocentric corrections computed with the IRAF task RVCORRECT.
For each spectrum, the zero-point of the wavelength calibration has
been accurately checked by means of a few emission lines of the
sky. The final velocities are listed in Table~\ref{tab1_c7}.  They range
from $-109.8$ km s$^{-1}$ to $-53.4$ km s$^{-1}$, with a mean value of
$-76.7 \pm 3.6$ km s$^{-1}$ and a dispersion $\sigma = 15.6$ km
s$^{-1}$. These values are in good agreement with the derivations of
\citet[][$v_r = -71.8 \pm 1.6$ km s$^{-1}$, $\sigma = 16.0$ km
s$^{-1}$]{dubath97} and \citet[][$v_r = -70.1 \pm 1.4$ km s$^{-1}$,
$\sigma = 14.3$ km s$^{-1}$]{yong14b}, the small differences
being likely due to the small statistics.

The most discrepant target (id=79), with a radial velocity of
$-109.85$ km s$^{-1}$, is still within 2$\sigma$ from the systemic
velocity of the cluster. By using the Besan\c{c}on Galactic model
\citep{robin03}, we extracted a sample of about 5300 field stars in
the direction of M62, finding a quite broad and asymmetric radial
velocity distribution, with mean $v_r \simeq -60$ km s$^{-1}$ and
dispersion $\sigma = 80$ km s$^{-1}$, which partially overlaps with
that of the cluster. On the other hand, only a few percent of the
stars studied in that region close to the Galactic bulge have
a [Fe/H] $< -$1.0 dex \citep[see e.g.][]{zoccali08,hill11,johnson13,ness13}.
Thus, taking into account the metallicity of star 79 (see below), its position in the CMD,
and its distance from the cluster center ($d \sim 38.5\arcsec$), we conclude
that it is likely a genuine cluster member and we therefore keep it
in the following analysis.

\subsection{Atmospheric parameters and stellar masses}
\label{atmos}

First guess effective temperature ($T_{\rm eff}$) and surface gravity
($\log g$) values for each target have been derived by using the
photometric information.  Temperatures have been estimated by using
the $(U-V)_{0}-T_{\rm eff}$ calibration of \cite{alonso99}.  Gravities
have been computed with the Stefan-Boltzmann equation by adopting the
color excess quoted above, a distance modulus $(m-M)_0 = 14.16$ mag
\citep{harris96} and the bolometric correction from \cite{alonso99}.
For the RGB stars we adopted a mass of 0.82 $M_{\odot}$, according to
the best fit isochrone retrieved from the PARSEC dataset
\citep{bressan12}, and computed for an age of 12 Gyr and a metallicity
Z=0.0013.  For the AGB stars we adopted a mass of 0.61 $M_{\odot}$,
according to the median value of the HB mass range estimated by \cite{gratton10b}.

Then we have performed a spectroscopic analysis as done in
\cite{lapenna14} and \cite{mucciarelli15}, constraining the
atmospheric parameters as follows: (1) spectroscopic temperatures have
been obtained by requiring that no trend exists between iron abundance
and excitation potential, (2) the gravity was derived by using the
Stefan-Boltzmann equation with the value of $T_{\rm eff}$ thus
obtained and (3) the microturbulent velocity was determined by
requiring that no trend exists between iron abundance and line
strength.  In order to evaluate the effects of a different procedure
in the derivation of the atmospheric parameters and abundances, we
have also performed a spectroscopic determination of the surface
gravities by modifying condition (2) and imposing that the same
abundance is obtained from neutral and single-ionized iron lines (ionization balance).

\subsection{Chemical abundances}
\label{chem}

The chemical abundances of Fe, Ti, Na, Al and Mg have been
derived with the package
GALA\footnote{http://www.cosmic-lab.eu/gala/gala.php}
\citep{mucciarelli13}, which adopts the classical method to
derive the abundances from the measured EWs of metallic unblended
lines. The EW and the error of each line were obtained using
DAOSPEC, iteratively launched by means of the
4DAO\footnote{http://www.cosmic-lab.eu/4dao/4dao.php} code
\citep{mucciarelli13b}.  The lines considered in the analysis have
been selected from suitable synthetic spectra at the UVES-FLAMES
resolution and computed with the SYNTHE package \citep{sbordone05} by
using the guess atmospheric parameters and the metallicity derived by
\cite{yong14b}.  The model atmospheres have been computed with the
ATLAS9\footnote{http://wwwuser.oats.inaf.it/castelli/sources/atlas9codes.html}
code. We adopted the atomic and molecular data from the last
release of the Kurucz/Castelli
compilation\footnote{http://wwwuser.oats.inaf.it/castelli/linelists.html}
and selected only the lines predicted to be unblended.
The selected lines and the atomic data adopted in the analysis are
listed in Table~\ref{tab2_c7}.

As detailed in Table~\ref{tab3_c7}, we used 100-150 FeI lines
and 7-12 FeII lines to derive the iron abundances, 25-60 lines of TiI
and 6-15 lines of TiII to derive the abundances of titanium.  For NaI,
MgI and AlI only few lines are available, namely those at
5682-5688$\rm\mathring{A}$ and 6154-6160$\rm\mathring{A}$ for NaI, the
line at 5711$\rm\mathring{A}$ and the doublet at
6318-6319$\rm\mathring{A}$ for MgI, and the doublet at
6696-6698$\rm\mathring{A}$ for AlI.  The O abundances have been
derived from spectral synthesis in order to take into account the
blending between the forbidden [OI] line at 6300.3$\rm\mathring{A}$
and a Ni transition.  For the Ni we adopted the average
  abundance obtained by \cite{yong14b}, while for stars located in the
  upper-RGB we assumed average C and N abundances according to
  \cite{gratton00}, all rescaled to the assumed solar reference values
  \citep{grevesse98}. Because in some spectra the [OI] line was
partially blended also with a telluric line, the spectra have been
cleaned by using suitable synthetic spectra obtained with the TAPAS
tool \citep{bertaux14}.  For some stars, the [OI] line is not
detectable, thus only upper limits are obtained.  As solar reference
abundances we adopted the \cite{caffau11} value for O, and those of
\cite{grevesse98} for all the other elements.

For the computation of the global uncertainties on the final
abundances we took into account two main sources of errors, which have
been added in quadrature:
\begin{itemize}
\item[1)] the error arising from the EW measurements. For each star we
  computed this term by dividing the line-to-line dispersion by the
  square root of the number of lines used.  Thanks to the high-quality
  of the spectra and to the number of lines that can be exploited,
  this term turned out to be very small, especially for FeI and TiI
  (providing up to 150 lines).  For these species the line-to-line
  scatter is smaller than 0.1 dex, leading to internal uncertainties
  lower than 0.01-0.02 dex. For FeII and TiII the number of lines
  ranges from 7 up to 15, leading to an uncertainty of about 0.02-0.03
  dex. For the other chemical species the number of measured lines is
  much smaller (1-4).  Hence, the average uncertainties are of the
  order of 0.06-0.08 dex for OI, NaI, MgI and AlI.
\item[2)] the error arising from atmospheric parameters. For the
  computation of this term we varied each parameter by the $1\sigma$
  error obtained from the previous analysis. We have found that
  representative errors for $T_{\rm eff}$, $\log g$ and $v_{turb}$ are
  $\sim50$ K, $0.1$ dex and $0.1$ km s$^{-1}$, respectively, for both
  the RGB and the AGB samples.  Thus we decided to adopt these values
  as $1\sigma$ error for all stars. We also checked the effect
  of a $\pm 0.1$ dex change in the metallicity of the model
  atmosphere, finding variations smaller than $\pm 0.01$ dex on the
  final abundances.
\end{itemize}

\section{Results}
\label{resu}

The determination of abundances and abundance ratios of the various
chemical elements is described below. The adopted atmospheric
parameters and the measured iron and titanium abundances for the
observed RGB and AGB stars are listed in Table~\ref{tab3_c7}, while the
abundances of the light-elements are listed in Table~\ref{tab4_c7}.
In Table~\ref{tab5_c7} we present the global abundance uncertainty of one
RGB and one AGB star, as well as the uncertainties obtained by varying
each atmospheric parameter independently.  Since this approach does
not take into account the correlations among different parameters, the
global error can be slightly overestimated.

Since star 96 presents an anomalous behavior with respect to the
other AGB targets, in the following analysis it is not included in
the AGB sample (thus counting five stars), and it is discussed
separately at the end of Section~\ref{sec:feti}.

%%%%%%%%%%%%%%%%%%%%%%%%%%%%%%%%% FE e TI

\subsection{Iron and titanium}
\label{sec:feti}

By using spectroscopic gravities (thus imposing that the same iron
abundance is obtained from neutral and from single-ionized lines), the
average values measured for the RGB and the AGB sub-samples are
[Fe/H]$_{\rm RGB} = -1.10 \pm 0.01$ ($\sigma = 0.04$ dex) and
[Fe/H]$_{\rm AGB} = -1.18 \pm 0.01$ ($\sigma = 0.03$ dex).  These
values are consistent (within 1-2 $\sigma$) with previous abundance
determinations of M62 giants, regardless they are on the RGB or on the
AGB: [Fe/H]$=-1.12$ dex \citep{kraft03}\footnote{We refer to the
average value computed with Kurucz models without overshooting; see
\cite{kraft03} for details.}, [Fe/H] = $-1.18 \pm 0.07$ dex
\citep{carretta09b}, and [Fe/H] = $-1.15 \pm 0.02$ dex
\citep[$\sigma = 0.05$ dex,][]{yong14b}.

% FIGURE 2
%
\begin{figure}[h]
\centering
\includegraphics[trim=0cm 0cm 0cm 0cm,clip=true,scale=0.60,angle=0]{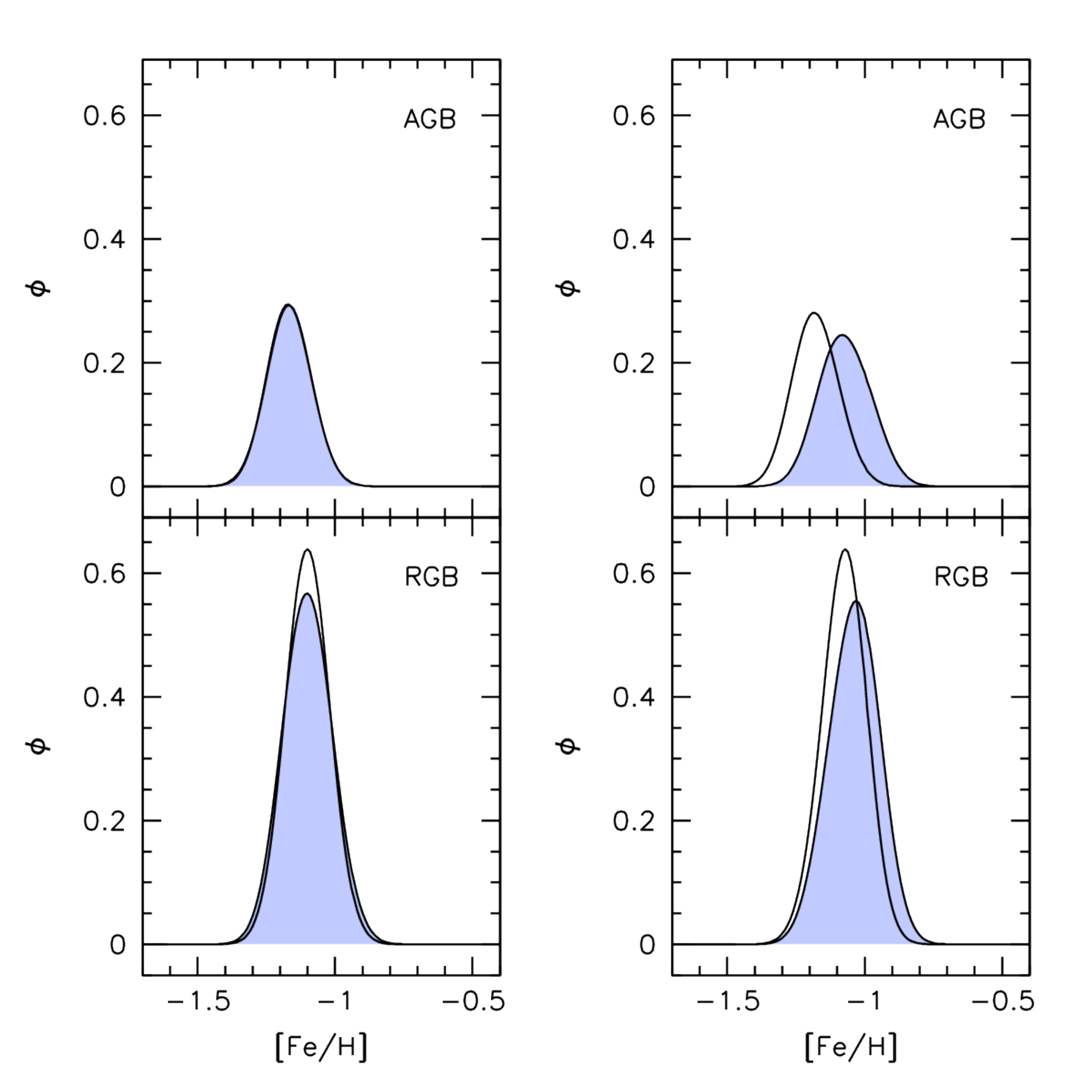}
\caption{\emph{Left panels:} generalized histograms for [FeI/H] (empty
histograms) and [FeII/H] (blue shaded histograms) obtained by
adopting spectroscopic gravities, for AGB stars (top panel) and the
RGB sample (bottom panel). \emph{Right panels:} as in the left
panels, but for the iron abundances obtained by adopting photometric
gravities.}
\label{genhist}
\end{figure}

By using photometric gravities (and not imposing ionization balance),
we determined the iron abundances separately from neutral and from
single-ionized lines. For the 13 RGB stars we obtained [FeI/H]$_{\rm
RGB} = -1.07 \pm 0.01$ dex ($\sigma = 0.04$ dex) and [FeII/H]$_{\rm
RGB} = -1.04 \pm 0.02$ dex ($\sigma = 0.06$ dex).  For the 5 AGB
stars we measured [FeI/H]$_{\rm AGB} = -1.19 \pm 0.01$ dex ($\sigma =
0.04$ dex) and [FeII/H]$_{\rm AGB} = -1.06 \pm 0.02$ dex ($\sigma =
0.06$ dex). The average difference between the values of $\log g$
derived spectroscopically and those derived photometrically are
0.09 dex ($\sigma=0.10$ dex) and 0.30 dex ($\sigma=0.20$ dex) for the RGB and the
AGB samples, respectively. Figure~\ref{genhist} shows the
generalized histograms of the iron abundances for the RGB and the AGB
samples separately, obtained by using spectroscopic (left panels) and
photometric gravities (right panels).  By construction, the
distributions of [FeI/H] and [FeII/H] essentially coincide if
spectroscopic gravities are assumed. Instead, the two distributions
significantly differ in the case of AGB stars if photometric gravities
are adopted. In particular, the average iron abundances of RGB stars
measured from neutral and single-ionized lines are consistent within
the uncertainties, while a difference of $-0.13$ dex, exceeding
5$\sigma$, is found for the AGB sample. Moreover, RGB and AGB stars
show very similar (well within 1$\sigma$) average values of [FeII/H],
while the neutral abundances of AGB stars are significantly lower (by
$0.12$ dex) than those of the RGB targets.

% FIGURE 3
%
\begin{figure}[h]
\centering
\includegraphics[trim=0cm 0cm 0cm 0cm,clip=true,scale=0.60,angle=0]{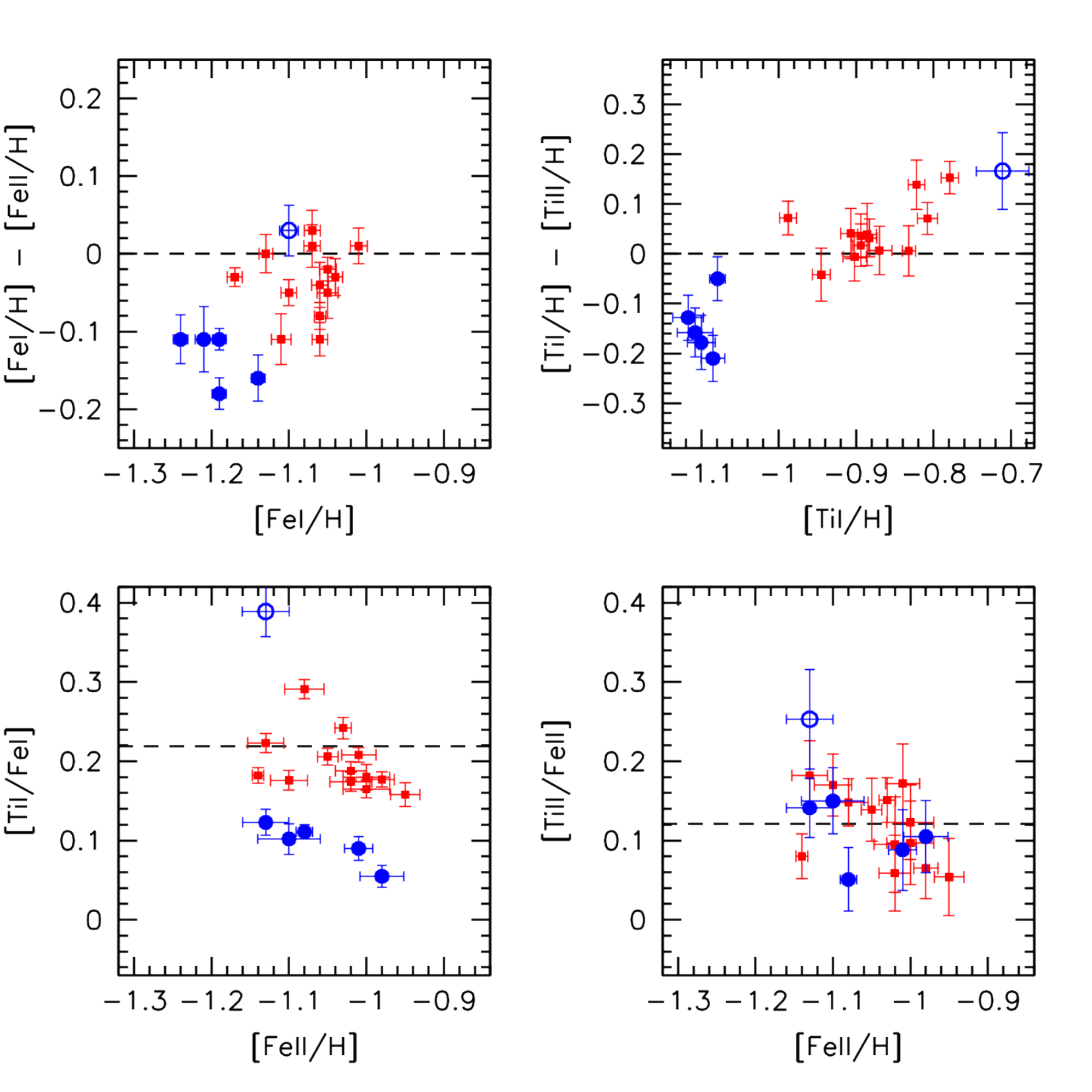}
\caption{\emph{Top panels:} difference between the chemical abundances
derived from neutral and single ionized lines, as a function of that
obtained from neutral lines, for iron (left panel) and titanium
(right panel). Symbols are as in Figure~\ref{cmd}. \emph{Bottom
panels:} [TiI/FeI] and [TiII/FeII] abundance ratios as a function
of [FeII/H] for the studied samples.}
\label{feti}
\end{figure}

When using photometric gravities similar results are obtained
for titanium, the only other chemical species presenting a large
number of neutral and single-ionized lines.
For the RGB sample we find [TiI/H]$_{\rm RGB} =
-0.88 \pm 0.01$ dex ($\sigma = 0.06$ dex) and [TiII/H]$_{\rm RGB} =
-0.92 \pm 0.01$ dex ($\sigma = 0.05$ dex). For the AGB stars we
measure [TiI/H]$_{\rm AGB} = -1.10 \pm 0.01$ dex ($\sigma = 0.02$
dex) and [TiII/H]$_{\rm AGB} = -0.95 \pm 0.02$ dex ($\sigma = 0.06$
dex). In this case, the average abundance of AGB stars from neutral
lines is lower than that of the RGB sample by 0.21 dex (while such a
difference amounts to only 0.04 dex for the RGB sample).
In Figure~\ref{feti} we report the differences between the iron (top left panel)
and the titanium (top right panel) abundances derived from neutral and
from single-ionized lines, as a function of the abundances from the
neutral species, obtained for each observed star assuming photometric
gravities.  Clearly, with the only exception of star 96 (plotted as an
empty circle in the figure), the AGB and the RGB samples occupy
distinct regions in these planes, because of systematically lower
values of the AGB abundances derived from the neutral species.

% FIGURE 4
%
\begin{figure}[h]
\centering
\includegraphics[trim=0cm 0cm 0cm 0cm,clip=true,scale=0.60,angle=0]{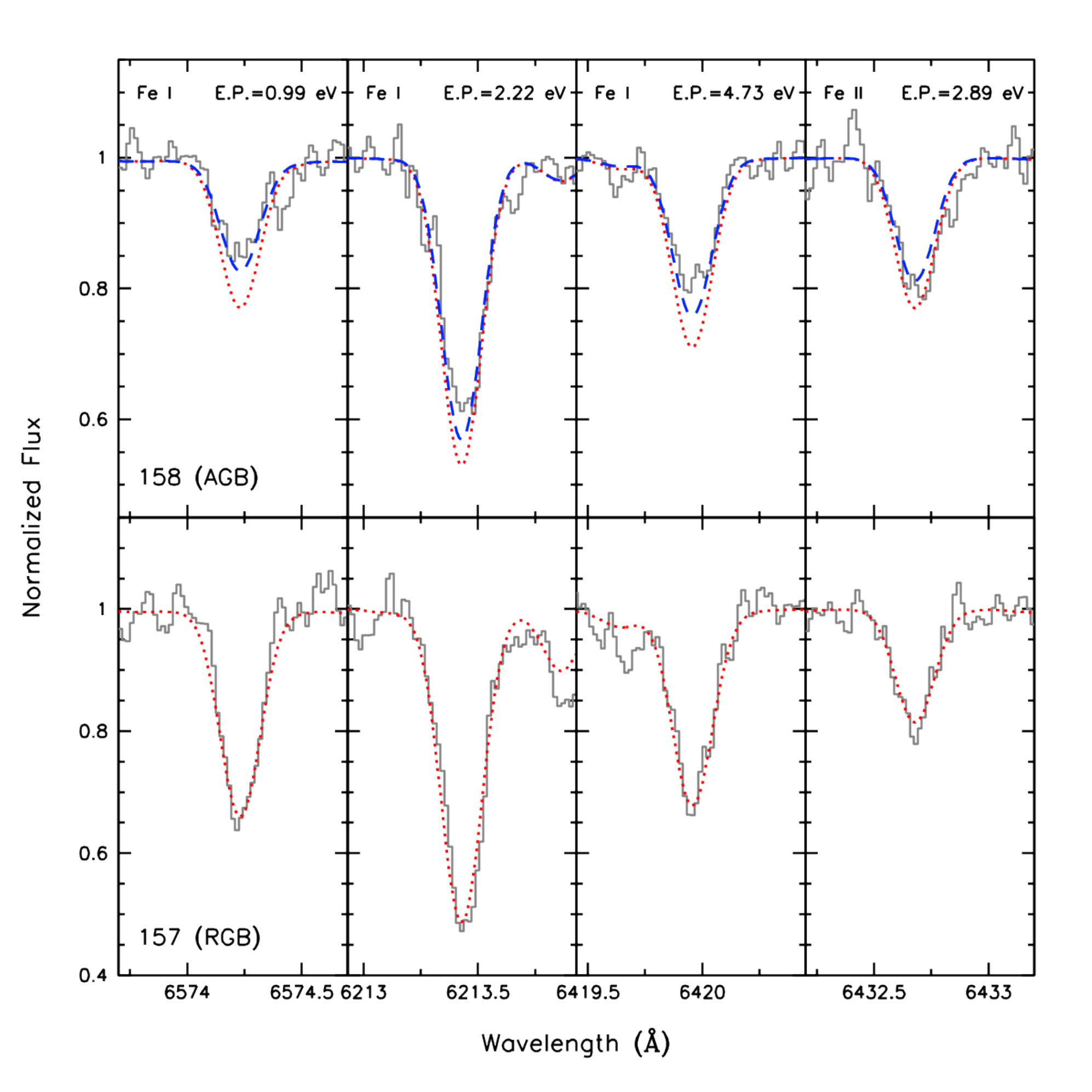}
\caption{Comparison between observed and synthetic spectra for AGB
star 158 (upper panels) and RGB star 15 (lower panels), around three
FeI lines with different excitation potentials and one FeII line
(see labels).  The observed spectra are marked with gray lines.
The synthetic spectra have been computed by using the measured [FeI/H]
(blue dashed line) and [FeII/H] (red dotted lines) abundances.
Since the two abundances are practically identical for the RGB star,
only one synthetic spectrum is shown in the lower panels.}
\label{spec}
\end{figure}

Such a difference can be also directly appreciated by visually
inspecting the line strengths in the observed spectra and their
synthetic best-fits.  As an example, in Figure~\ref{spec} we show
the observed spectra of an RGB and an AGB star around some FeI and
FeII lines, together with synthetic spectra calculated with the
appropriate atmospheric parameters and the metallicity derived from
FeII and from FeI lines. As apparent, the synthetic spectrum
computed adopting the FeII abundance well reproduces all the
observed lines in the case of the RGB star, while it fails to fit
the neutral features observed in the AGB target, independently of
the excitation potential (thus guaranteeing that the effect cannot
be due to inadequacies in the adopted temperature). On the other
hand, the abundance measured from FeI lines is too low to properly
reproduce the depth of the ionized features of the AGB star.  This
clearly demonstrates a different behaviour of iron lines in AGB and
RGB stars.

To investigate the origin of the discrepancy between FeI and FeII 
abundances obtained for the AGB sample, we checked the impact of the 
adopted stellar mass on the estimate of the photometric gravity.  
As described in Section~\ref{atmos}, for the AGB
stars we assumed a mass of 0.61 $M_\odot$, corresponding to the median
value of the distribution obtained for HB stars by \cite{gratton10b},
ranging from 0.51 to 0.67 $M_{\odot}$.  By adopting the lowest mass
(0.51 $M_{\odot}$), the average value of $\log g$ decreases by
$\sim0.08$ dex, while assuming the largest value, $\log g$ increases
by $0.04$ dex.
Such small gravity variations\footnote{Note that the
increase of $\log g$ is essentially the same ($0.05$ dex) even if
the mass provided by the best-fit isochrone (0.72 $M_\odot$) is adopted.}
have a negligible impact on the abundances derived from
the neutral iron lines, and the impact is still modest
(at a level of a few hundredths of a dex) on the abundances derived from
single-ionized lines. The only way to obtain (by construction) the
same abundance from FeI and FeII lines is to use the spectroscopic
values of $\log g$ derived from the ionization balance
(Section~\ref{atmos}). However, these gravities correspond to stellar
masses in the range 0.25-0.3 $M_\odot$, which are totally unphysical
for evolved stars in GCs.

A possible explanation of the observed discrepancy could be a
departure from LTE condition in the atmosphere of AGB stars.
In fact, lines originated by atoms in the minority ionization state
usually suffer from NLTE effects, while those originated by atoms in
the dominant ionization state are unaffected (see, e.g.,
\citealt{mashonkina11}). Thus, if this is the case, the most
reliable determination of the iron abundance is that provided by
[FeII/H], since the majority of iron atoms is in the first ionized
state in giant stars. Moreover, following \citet{ivans01}, the
degree of overionization of the neutral species should be (at least at
a first order) the same as the one affecting FeI lines.
Hence, the correct way to obtain a [X/Fe] abundance ratio is
to compute it with respect to the FeI abundance if [X/H] is derived
from minority species, and with respect to FeII if [X/H] is obtained
from majority species. In the lower panels of Figure~\ref{feti} we
present the [TiII/FeII] and the [TiI/FeI] abundance ratios as a
function of the iron abundance derived from single-ionized lines.

As expected, the abundances of AGB stars agree with those of the RGB
sample when single ionized (dominant state) titanium lines are
used. For [TiI/FeI] a systematic offset of the AGB sample toward
lower values is still observable (although reduced), thus indicating
the possible presence of residual NLTE effects. We also note
a systematic offset of +0.08 dex between [TiI/FeI] and [TiII/FeII],
especially for RGB stars.  However, taking into account that the
oscillator strength values of the TiII lines are highly uncertain and
that the offset is still reasonably small, we can conclude that the
[X/Fe] abundance ratio can be safely constrained either by neutral or
by single-ionized lines.  It is also interesting to note that the
average [TiI/Fe] and [TiII/Fe] abundance ratios (+0.16 dex and +0.25
dex, respectively) of \citet{yong14b} show a relative offset of $-0.09$
dex, which is similar to ours but in the opposite direction. This
suggests that there is an intrinsic (although small) uncertainty in
the zero point scale of the titanium abundance.

\emph{AGB star 96 --} As apparent from Figure~\ref{feti}, AGB star 96
shows a difference between neutral and ionized abundances, both for
iron and titanium, which is incompatible with those found for the
other AGB targets, and which is much more similar to the values
measured for RGB stars. Interestingly, star 96 presents
atmospheric parameters compatible with those spanned by the RGB
targets (but with a surface gravity which is 0.15-0.2 lower than
that of RGB stars at the same temperature). This case is similar
to that encountered in 47Tuc, where the FeI abundance of a small
sub-sample of AGB stars (4 out of 24) has been found to agree with the
value obtained from ionized lines, thus suggesting that one of
the possible explanations could be the lack of LTE departures for
these objects \citep{lapenna14}. Also in M22, one (out of five)
AGB star shows a perfect agreement between [FeI/H] and [FeII/H], while
the other AGB stars show systematically low [FeI/H] values
\citep{mucciarelli15_m22}.

%%%%%%%%%%%%%%%%%%%%%%%%%%%%%%%%% O, NA, MG e AL

\subsection{Oxygen, sodium, magnesium and aluminum}
In most Galactic GCs, the abundances of oxygen, sodium, magnesium and
aluminium are known to show large star-to-star variations, organized
in clear correlations (see \citealt{gratton12} for a review).  These
are usually interpreted as the signature of self-enrichment processes
occurring in the early stages of GC evolution and giving rise to at
least two stellar generations with a very small (if any) age
difference, commonly labelled as first and second generations (FG and
SG, respectively).  In particular, the variations observed in O, Na,
Mg and Al are thought to be due to the ejecta from still unclear
polluters, like massive AGB stars, fast-rotating massive stars and/or
massive binaries \citep{fenner04, ventura05, decressin07, demink09,
marcolini09, bastian13, bastian15}.

% FIGURE 5
%
\begin{figure}[h]
\centering
\includegraphics[trim=0cm 0cm 0cm 0cm,clip=true,scale=0.65,angle=0]{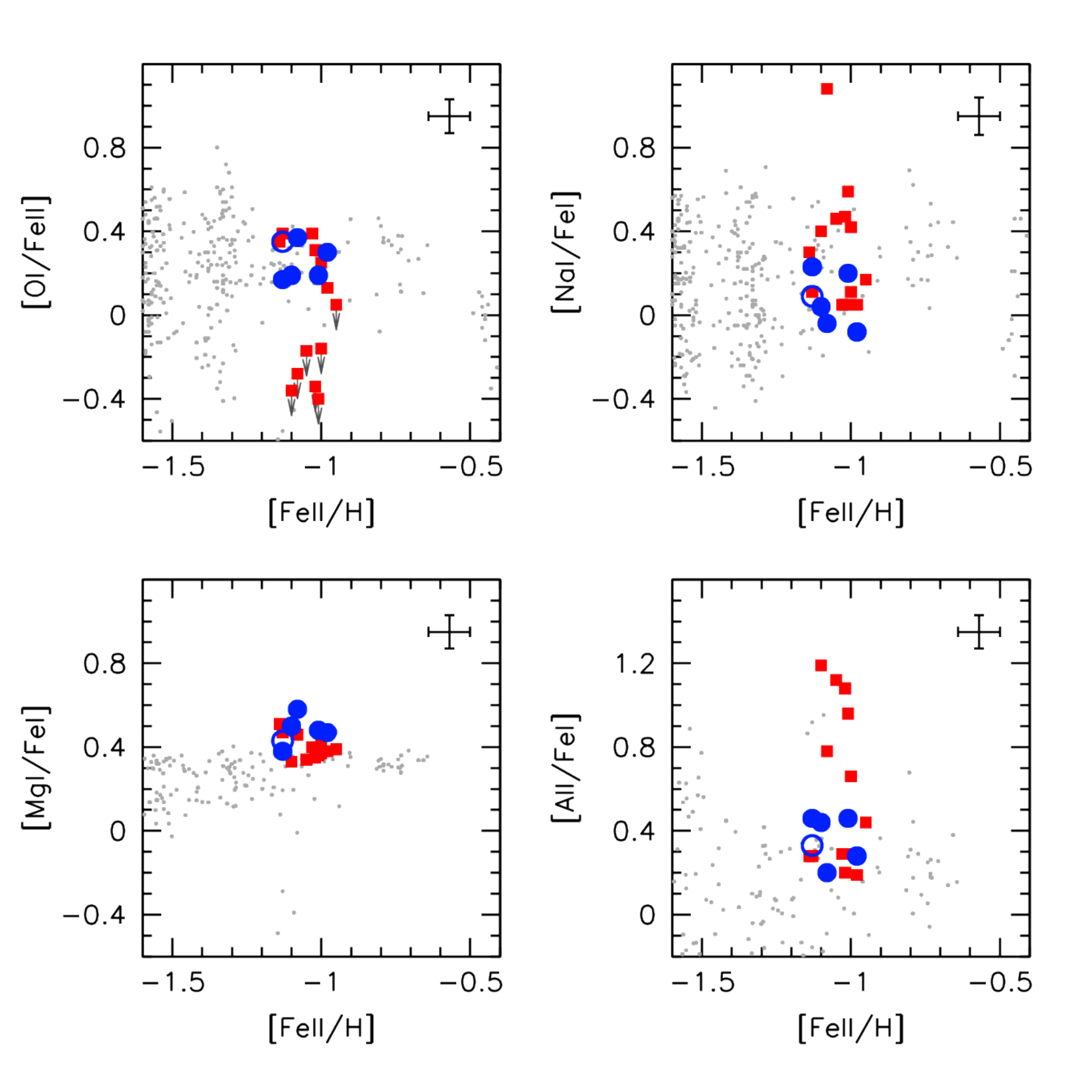}
\caption{From top-left to bottom-right, oxygen, sodium, magnesium and
aluminum abundance ratios as a function of [FeII/H] for the studied
sample of stars (same symbols as in Figure~\ref{cmd}). For a
sub-sample of (O-poor) RGB stars, only upper limits to the oxygen
abundance could be measured from the acquired spectra (see
arrows). Representative error bars are marked in the top-right
corner of each panel.  The values measured in large samples of giant
stars in 20 Galactic GCs (from GIRAFFE and UVES spectra by
\citealp{carretta09a,carretta09c,carretta14}),
rescaled to the solar values adopted in this work, are shown for
reference as grey dots.}
\label{light}
\end{figure}

\begin{figure}[h]
\centering
\includegraphics[trim=0cm 10cm 0cm 0cm,clip=true,scale=0.70,angle=0]{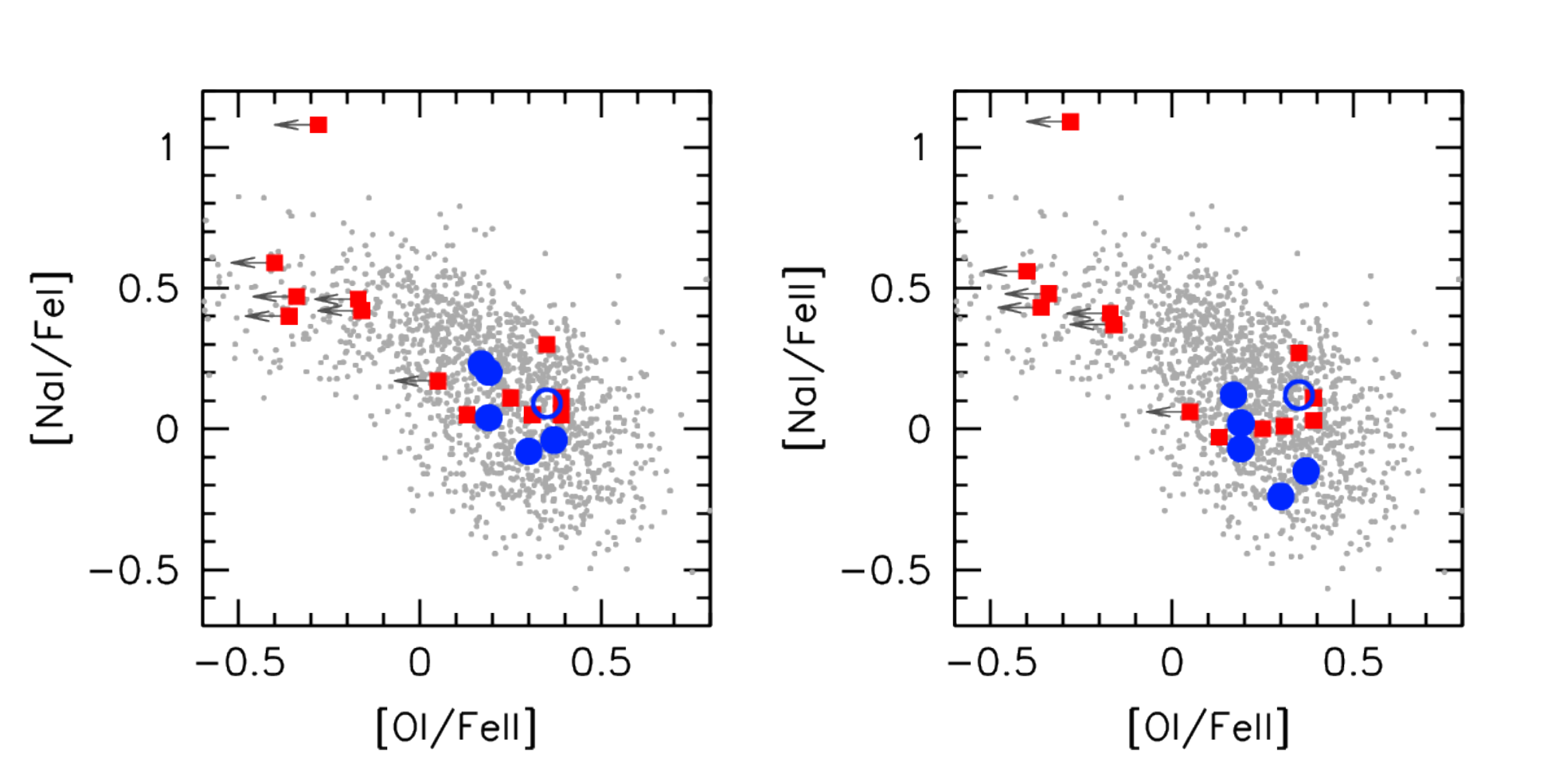}
\caption{Oxygen-sodium anti-correlation measured for the observed
stars (same symbols as in Figure~\ref{light}). The corrections for
NLTE effects provided by \citet{gratton99} have been applied to the
Na abundances. This is then expressed with respect to FeI and to
FeII in the left and right panels, respectively. Grey dots
are as in Figure~\ref{light}.}
\label{ONa}
\end{figure}

\begin{figure}[h]
\centering
\includegraphics[trim=0cm 10cm 0cm 0cm,clip=true,scale=0.70,angle=0]{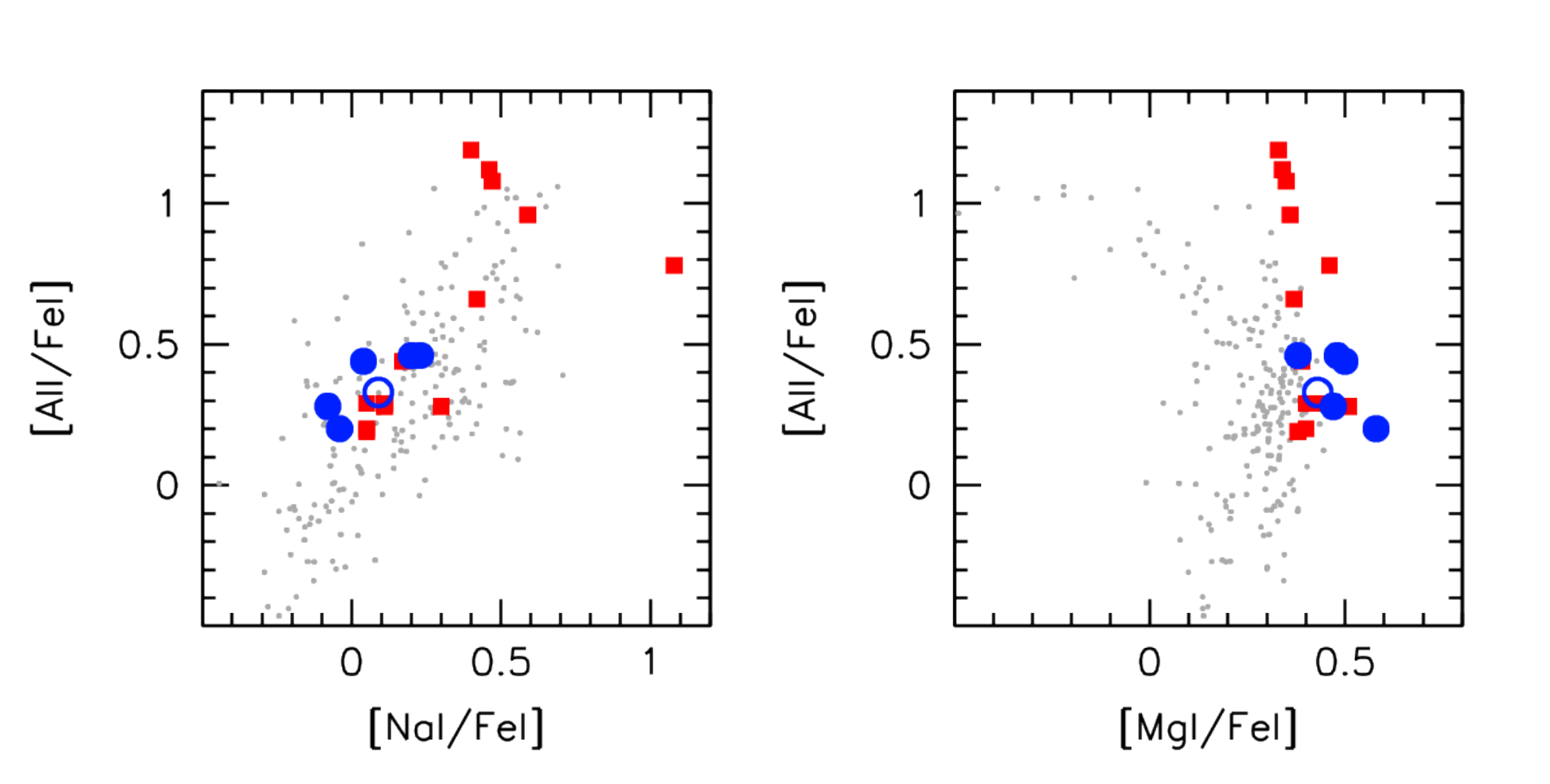}
\caption{Aluminum-sodium correlation (left panel) and
aluminum-magnesium anti-correlations (right panel) for the observed
stars (same symbols as in Figure~\ref{light}). Grey dots as in Figure~\ref{light}.}
\label{Al}
\end{figure}

To verify the presence of these key features also in our sample of
giants, we derived the abundances of O, Na, Al, and Mg from the
observed spectra. The results are shown in Figure~\ref{light}, where
all abundance ratios are plotted as a function of the iron content as
measured from the ionized lines. Since the oxygen abundance derived
from the forbidden [OI] line at 6300.3 $\rm\mathring{A}$ is not
affected by NLTE, its abundance ratio is expressed with respect to the
``true'' iron content (measured from FeII lines).  Instead, the other
species are known to suffer from NLTE effects and their abundances are
therefore plotted with respect to FeI (see Section~\ref{sec:feti}).
This is true also for sodium, although we have applied the NLTE corrections of \cite{gratton99},
which take into account departures from LTE conditions driven by over-recombination
\citep{bergemann14}\footnote{By adopting the NLTE corrections
of \citet{lind11}, the differential behaviour between AGB and RGB
stars remains the same.}. In any case, we have verified that the
same results are obtained if the Na, Al and Mg abundances are computed
with respect to FeII or H. In agreement with what commonly observed in
Galactic GCs, we find that the Mg abundance is constant within the
uncertainties, while O, Na, and Al show significant (several tenths of
a dex) star-to-star variations in the RGB sample \citep[see also][]{yong14b}.
As shown in Figures~\ref{ONa} and \ref{Al},
the observed star-to-star variations are organized in the same
correlations observed for GCs.  In particular, oxygen and sodium are
anti-correlated, independently of using FeI or FeII for the
computation of the sodium abundance ratio (Figure~\ref{ONa}), while
aluminum and sodium are positively correlated and [AlI/FeI] shows a
$\sim 1$ dex spread for fixed magnesium (Figure~\ref{Al}).
Very interestingly, instead, all abundance ratios are constant for the AGB
sample, with values mainly consistent with those commonly associated
to the FG.

% FIGURE 6
%
%
%

% FIGURE 7
%

The Na-O anti-correlation derived from our RGB sample is qualitatively
compatible with that measured by \citet{yong14b}, who found two groups
of stars well-separated both in [Na/Fe] and [O/Fe]. We note that the
oxygen abundances quoted by \citet{yong14b} are larger than ours, with
an average offset of +0.5 dex for the O-rich stars.  The origin of
this discrepancy can be ascribable to different factors (like atomic
data, telluric correction, etc.), but it is beyond the aims of this
paper. A good agreement with the results of \citet{yong14b} is found
also for the [Al/Fe] and [Mg/Fe] distributions.

The derived Na-O anti-correlation of M62 is more extended than those
observed in most Galactic GCs.  Two discrete groups of stars can be
recognized, a first one with [O/Fe]$\sim$+0.2/+0.3 dex and
[Na/Fe]$\sim$+0.1 dex, and a second group with [O/Fe]$<$0.0 dex (only
upper limits) and [Na/Fe] at $\sim$+0.5 dex.  In particular, the
sub-solar O-poor component \citep[the so-called ``extreme
population''; see ][]{carretta10c} is quite prominent in M62,
while these stars are usually rare, observed only in some massive
systems, as NGC2808 \citep{carretta09c}, M54
\citep{carretta10b}, and $\omega$ Centauri \citep{johnson10}.
We also find a significant lack of ``intermediate population'' stars
\citep[with enhanced Na abundances and mild oxygen depletion][]{carretta10c},
which are instead the dominant component in most GCs.

\begin{figure}[h]
\centering
\includegraphics[trim=0cm 0cm 0cm 0cm,clip=true,scale=0.65,angle=0]{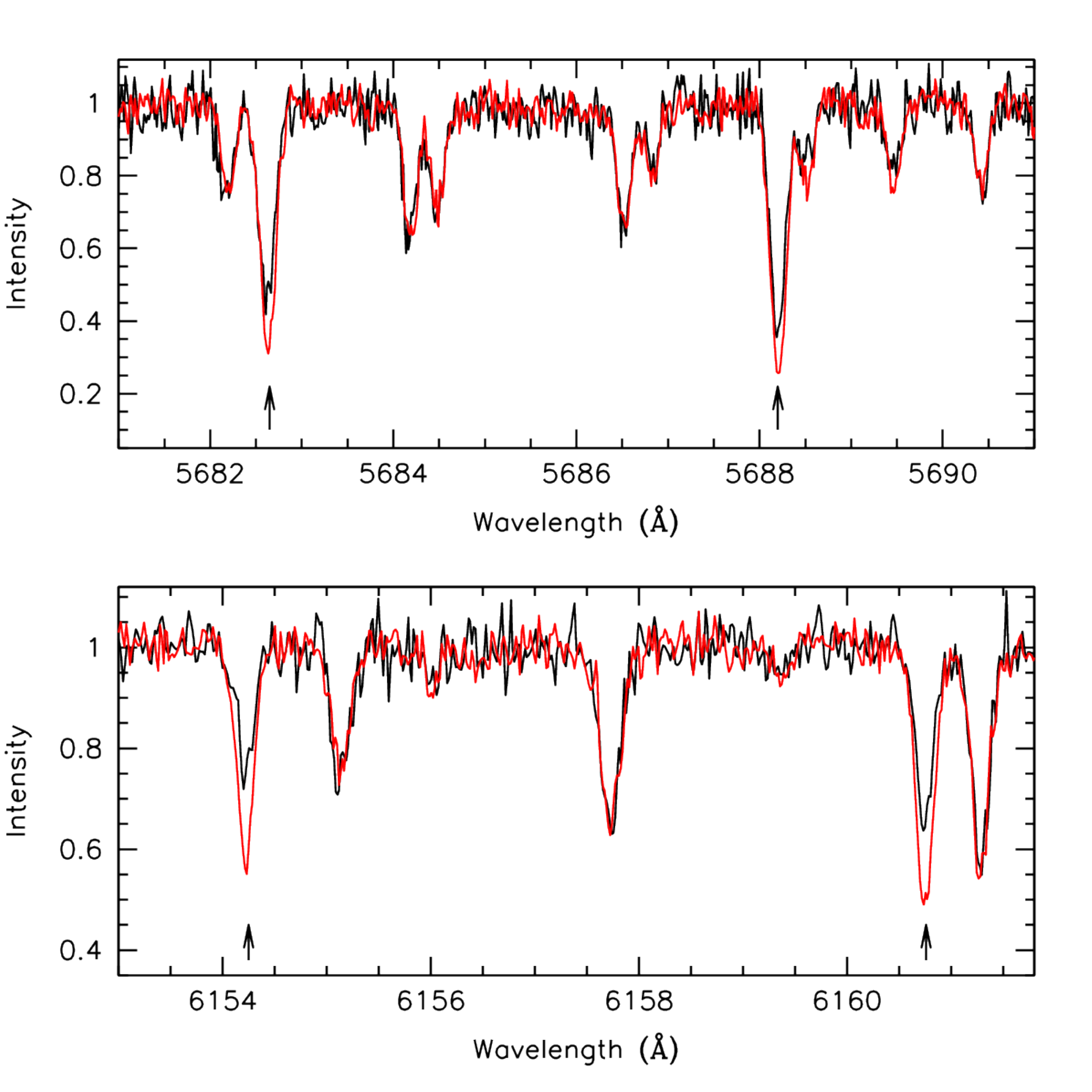}
\caption{Comparison between the spectra of the RGB stars 89 (red line)
and 95 (black line) for the NaI lines at 5682-5688 $\rm\mathring{A}$
(top-panel) and 6154-6160 $\rm\mathring{A}$ (bottom-panel).
The black arrows mark the position of the Na lines.}
\label{spec89}
\end{figure}

We finally note that the RGB star 89 exhibits a Na abundance
[Na/Fe]=+1.08 dex, which is $\sim$0.5 dex larger than that measured
for all the other O-poor stars.  In Figure~\ref{spec89} we compare the
spectrum of star 89 with that of another RGB target (id=95) having
very similar atmospheric parameters and iron abundances (see Table~\ref{tab3_c7}).
As apparent, all lines have the same strengths, with the
notable exception of the two Na doublets, which are significantly
stronger in star 89.

% FIGURE 8
%

To our knowledge, this is one of the most Na-rich giant ever
detected in a genuine GC (see also the comparison with literature
data in Figure~\ref{ONa}), with a [Na/Fe] abundance even higher than
the most Na-rich stars observed in NGC2808 \citep{carretta06} and
NGC4833 \citep{carretta14}, and comparable to a few extremely
Na-rich objects observed in the multi-iron system $\omega$ Centauri
\citep{marino11a}.

%%%%%%%%%%%%%%%%%%%%%%%%%%%%%%%%%%%%%%%%%%%%%%%%%%%%%%%%%% DISCUSSION

\section{Discussion and conclusions}
\label{concl}

The differences in the iron and titanium abundances measured from
neutral and from single-ionized lines in five AGB stars of M62 closely
resemble those found in M5 \citep{ivans01}, 47Tuc \citep{lapenna14}
and NGC 3201 \citep{mucciarelli15}.  These results might be
explained as the consequences of departures from LTE conditions,
affecting the neutral species, while leaving unaltered the ionized
lines.  The final effect is a systematic underestimate of the chemical
abundances if measured from neutral features.  Interestingly, the
findings in M5, 47Tuc and NGC 3201 seem to suggest that this effect
concerns most, but not all, AGB stars, while it is essentially
negligible for RGB targets.  This is inconsistent with the available
NLTE calculations \citep[e.g.][]{lind12,bergemann12}, which predict
the same corrections for stars with similar parameters. Moreover, the
results recently obtained in M22 show that the situation is even more
complex. In M22, in fact, neutral iron abundances systematically lower
than [FeII/H] have been measured also for RGB stars
\citep{mucciarelli15_m22}.  However, the [FeI/H]-[FeII/H] difference
in M22 clearly correlates with the abundance of $s$-process elements
(that show intrinsic star-to-star variations unusual for GCs),
suggesting that this could be a distinct, peculiar case.
All these results seem to suggest that we are still missing some crucial
detail about the behaviour of chemical abundances in the case of
departures from LTE conditions, and/or that other ingredients (as
full 3D, spherical hydro calculations, and the inclusion of
chromospheric components) should be properly taken into account in
modeling the atmospheres of these stars.

For all the studied stars we have also determined the abundances of O,
Na, Al and Mg from the neutral lines available.  As shown in
Figures~\ref{ONa} and \ref{Al}, our sample of RGB stars shows the
typical behaviors observed in all massive GCs, with large and
mutually correlated dispersions of O, Na and Al (and with one
of the most Na-rich giant ever observed in a GC: RGB star 89, with
[Na/Fe]=$+1.08$ dex). Instead, the light-element abundances of AGB
stars are essentially constant and clumped at the low-end of the Na
and Al values of the RGB sample.

If the (still unclear) NLTE effects impacting the FeI and TiI
abundances of the AGB targets significantly weaken the
minority species lines (as it seems reasonable to assume), also the
measured abundances of sodium, aluminum and magnesium could be
underestimated for these objects (even when referred to the neutral
iron abundance, FeI). Thus, although the observed
star-to-star variation of Na in most GCs is often a factor of 5-10
larger than the suspected NLTE effects on Fe and Ti, we caution that
it could be risky to derive firm conclusions about a lack of Na-rich
AGB stars on the basis of the sodium abundance alone until these
effects are properly understood and quantified (of course, the same
holds for any other light-element potentially affected by NLTE
effects, especially if the star-to-star variations of this element
are intrinsically small). In fact, a lack of Na-rich AGB stars
could be either real, or just due to a bias induced by NLTE effects.
A solid evidence, instead, is obtained if the result is based on
elemental species (like the oxygen abundance derived from the
forbidden line considered here) that are virtually unaffected by NLTE
effects.  Hence, Figure~\ref{ONa}, showing that the oxygen abundances of
all AGB stars are larger than those expected for the SG population and
measured, in fact, for a sub-sample of RGB giants, convincingly
indicates that none of the AGB targets studied in M62 is compatible
with the SG of the cluster.

Does this mean that the SG stars in M62 did not experience the AGB
phase (as it has been suggested for NGC6752 by \citealp{campbell13})?
To answer this question we note that, although variable from cluster
to cluster, the typical percentages of FG and SG stars in Galactic GCs
are 30\% and 70\%, respectively \citep[e.g.][]{carretta13b,bastian15b}. On this
basis, we should have observed 4 second generation AGB stars in our
sample. In alternative, from Figures~\ref{ONa} and \ref{Al} we see that
6 out of 13 (46\%) RGB stars likely belong to the SG and, we could
have therefore expected 2-3 AGB stars in the same group, at odds with what observed.
On the other hand, a deficiency of CN-strong
(second generation) AGB stars in several GCs is known since the
pioneering work of \citet{norris81} and it has been recently found
to be most severe in GCs with the bluest HB morphology
\citep[see, e.g.,][and references therein]{gratton10b}.
While M62 has indeed a very extended HB, it shows no deficiency of AGB stars.
In fact, by using ACS and WFC3 HST archive data acquired in the $m_{\rm F390W}$
and $m_{\rm F658N}$ filters, we counted the number of AGB and HB
stars (86 and 640, respectively) in M62, finding that their ratio
(the so-called $R_2$ parameter; \citealp{caputo89}) is $R_2\simeq
0.13$. This value is in very good agreement with the theoretical
predictions based on the ratio between the AGB and the HB
evolutionary timescales \citep[e.g.][]{cassisi01}.
Hence, our observations show that all the sampled AGB stars belong to the FG,
but we cannot exclude that some SG object is present along the AGB of M62.

Clearly, if a complete lack of SG AGB stars is confirmed by future
studies in M62, NGC 6752, M13 \citep[see e.g.][]{sneden00b,johnson12} or any other GC,
this will represent a new challenge for the formation and evolution models of these
stellar systems (as already discussed, e.g., by \citealp{charbonnel13}
and \citealp{cassisi14}).

%%%%%%%%%%%%%%%%%%%%%%%%%%%%%%%%%% TABLES %%%%%%%%%%%%%%%%%%%%%%%%%%%%%%%%%%%%

\begin{landscape}
\begin{deluxetable}{rccccccrc}
\tiny
\tablecolumns{9}
\tablewidth{0pt}
\tablecaption{Photometric properties and radial velocities of the RGB and AGB sample}
\tablehead{\colhead{ID} & \colhead{R.A.} & \colhead{Decl.} & \colhead{U} & \colhead{V}
& \colhead{U$_{0}$} & \colhead{V$_{0}$} & \colhead{RV} & \colhead{Type} \\
& (J2000) & (J2000) & \colhead{(mag)} & \colhead{(mag)}
& \colhead{(mag)} & \colhead{(mag)} & \colhead{(km s$^{-1}$)} & \colhead{}}
\startdata
%\hline
% & & & & & & \\
%\hline
  50  &  255.2961736  &  --30.1122536  &  17.458  &  14.061  &  14.558  &  12.184  &  --95.41 $\pm$ 0.06  &  R  \\
  54  &  255.2968276  &  --30.1110148  &  17.465  &  14.149  &  14.462  &  12.206  &  --68.41 $\pm$ 0.04  &  R  \\
  76  &  255.3016326  &  --30.0879873  &  17.578  &  14.416  &  14.734  &  12.576  &  --69.34 $\pm$ 0.06  &  R  \\
  82  &  255.2908040  &  --30.1230200  &  17.378  &  14.478  &  14.649  &  12.712  &  --56.67 $\pm$ 0.04  &  R  \\
  89  &  255.3053120  &  --30.1235390  &  17.694  &  14.584  &  14.759  &  12.685  &  --68.78 $\pm$ 0.04  &  R  \\
  95  &  255.2683150  &  --30.1061800  &  17.689  &  14.624  &  14.821  &  12.768  &  --85.92 $\pm$ 0.06  &  R  \\
  97  &  255.2746210  &  --30.1078150  &  17.551  &  14.632  &  14.703  &  12.789  &  --92.22 $\pm$ 0.06  &  R  \\
 104  &  255.2990264  &  --30.1195799  &  17.502  &  14.680  &  14.861  &  12.971  &  --81.19 $\pm$ 0.04  &  R  \\
 118  &  255.2953240  &  --30.1054710  &  17.584  &  14.771  &  14.883  &  13.023  &  --90.41 $\pm$ 0.06  &  R  \\
 127  &  255.3064600  &  --30.0967810  &  17.775  &  14.895  &  15.020  &  13.112  &  --55.24 $\pm$ 0.06  &  R  \\
 133  &  255.3025803  &  --30.1265560  &  17.819  &  14.939  &  14.903  &  13.052  &  --90.70 $\pm$ 0.05  &  R  \\
 145  &  255.2959190  &  --30.1263240  &  17.831  &  15.041  &  15.031  &  13.229  &  --63.56 $\pm$ 0.05  &  R  \\
 157  &  255.2998135  &  --30.0934941  &  17.720  &  15.174  &  14.988  &  13.406  &  --74.04 $\pm$ 0.05  &  R  \\
\hline
  79  &  255.3060883  &  --30.1031433  &  17.335  &  14.430  &  14.443  &  12.558  & --109.85 $\pm$ 0.06  &  A  \\
  96  &  255.2885360  &  --30.1173880  &  17.345  &  14.629  &  14.558  &  12.826  &  --81.49 $\pm$ 0.07  &  A  \\
 116  &  255.2778880  &  --30.1205350  &  17.130  &  14.764  &  14.348  &  12.963  &  --87.57 $\pm$ 0.09  &  A  \\
 128  &  255.2980470  &  --30.1078870  &  17.248  &  14.895  &  14.501  &  13.117  &  --72.72 $\pm$ 0.08  &  A  \\
 135  &  255.2914560  &  --30.1287900  &  17.416  &  14.952  &  14.613  &  13.138  &  --60.08 $\pm$ 0.07  &  A  \\
 158  &  255.3017290  &  --30.1013070  &  17.361  &  15.180  &  14.647  &  13.424  &  --53.49 $\pm$ 0.07  &  A  \\
\enddata
\tablecomments{Identification number, coordinates, $U$, $V$, $U_{0}$ and $V_{0}$
magnitudes, heliocentric radial velocity, and type of star (R=RGB, A=AGB).}
\label{tab1_c7}
\end{deluxetable}
\end{landscape}

\begin{deluxetable}{ccrcc}
\tiny
\tablecolumns{5}
\tablewidth{0pt}
\tablecaption{Wavelength, element, oscillator strength, excitation potential, and reference source of adopted lines.}
\tablehead{\colhead{Wavelength} & \colhead{El.} & \colhead{log $gf$} & \colhead{E.P.} & \colhead{Ref.} \\
\colhead{($\mathring{\rm A}$)} & & & \colhead{(eV)} & }
\startdata
 4962.572  &  FeI  &  --1.182  &  4.178  &  \cite{fuhr06}   \\
 4967.897  &  FeI  &  --0.534  &  4.191  &  K   \\
 4969.917  &  FeI  &  --0.710  &  4.217  &  \cite{fuhr88}   \\
 4982.499  &  FeI  &    0.164  &  4.103  &  K   \\
 4983.250  &  FeI  &  --0.111  &  4.154  &  K   \\
 4985.547  &  FeI  &  --1.331  &  2.865  &  \cite{fuhr06}   \\
 4950.106  &  FeI  &  --1.670  &  3.417  &  \cite{fuhr88}   \\
 4962.572  &  FeI  &  --1.182  &  4.178  &  \cite{fuhr06}   \\
 4967.897  &  FeI  &  --0.534  &  4.191  &  K   \\
 4969.917  &  FeI  &  --0.710  &  4.217  &  \cite{fuhr88}   \\
\enddata
\tablecomments{
K = Oscillator strengths (OS) from the R.L.Kurucz on-line database of observed and 
predicted atomic transitions (see http://kurucz.harvard.edu),
NIST = OS from NIST database (see http://www.nist.gov/pml/data/asd.cfm)
S = OS from solar analysis by F. Castelli (see http://wwwuser.oats.inaf.it/castelli/linelists.html).
For AlI lines we derived astrophysical oscillator strengths (labeled as S*) by using the solar flux spectra of
\cite{neckel84} and the model atmosphere for the Sun computed by
F. Castelli\footnote{http://wwwuser.oats.inaf.it/castelli/sun/ap00t5777g44377k1odfnew.dat}
adopting the solar abundances of \cite{grevesse98}.
The entire Table is available in the on-line version, a portion
is shown here for guidance about its form and content.
}
\label{tab2_c7}
\end{deluxetable}

\begin{landscape}
\begin{deluxetable}{rcccrcrcrcrc}
\tiny
\tablecolumns{12}
\tablewidth{0pt}
\tablecaption{Atmospheric parameters, iron and titanium abundances of the measured RGB and AGB stars.}
\tablehead{\colhead{ID} & \colhead{$T_{\rm eff}$} & \colhead{log g}
&  \colhead{$v_{\rm turb}$}
& \colhead{[FeI/H]} & \colhead{n$_{\rm FeI}$} & \colhead{[FeII/H]} & \colhead{n$_{\rm FeII}$} 
& \colhead{[TiI/H]} & \colhead{n$_{\rm TiI}$} & \colhead{[TiII/H]} & \colhead{n$_{\rm TiII}$} \\
& \colhead{(K)} & \colhead{(dex)} &  \colhead{(km s$^{-1}$)} & \colhead{(dex)} & & \colhead{(dex)} & 
& \colhead{(dex)} & & \colhead{(dex)} & }
\startdata
%\hline
% & & & & & & & & & & & \\
%\hline
  50  &  4225  &  0.85  &  1.30  &  --1.13 $\pm$ 0.01  &  128  &  --1.13 $\pm$ 0.02  &  12  &  --0.91 $\pm$ 0.01  &  58  & --0.95 $\pm$ 0.05  &  12  \\
  54  &  4215  &  0.85  &  1.40  &  --1.17 $\pm$ 0.01  &  130  &  --1.14 $\pm$ 0.01  &   7  &  --0.99 $\pm$ 0.01  &  63  & --1.06 $\pm$ 0.03  &  14  \\
  76  &  4375  &  1.15  &  1.35  &  --1.05 $\pm$ 0.01  &  106  &  --1.00 $\pm$ 0.03  &   7  &  --0.87 $\pm$ 0.02  &  37  & --0.88 $\pm$ 0.05  &   6  \\
  82  &  4295  &  1.15  &  1.30  &  --1.06 $\pm$ 0.01  &  104  &  --1.02 $\pm$ 0.03  &  11  &  --0.89 $\pm$ 0.01  &  45  & --0.92 $\pm$ 0.06  &   7  \\
  89  &  4355  &  1.15  &  1.50  &  --1.07 $\pm$ 0.01  &  127  &  --1.08 $\pm$ 0.03  &  10  &  --0.78 $\pm$ 0.01  &  62  & --0.93 $\pm$ 0.03  &  14  \\
  95  &  4365  &  1.20  &  1.45  &  --1.07 $\pm$ 0.01  &  134  &  --1.10 $\pm$ 0.02  &  11  &  --0.89 $\pm$ 0.01  &  58  & --0.93 $\pm$ 0.04  &  15  \\
  97  &  4425  &  1.25  &  1.40  &  --1.01 $\pm$ 0.01  &  142  &  --1.02 $\pm$ 0.02  &  12  &  --0.82 $\pm$ 0.01  &  50  & --0.96 $\pm$ 0.05  &  14  \\
 104  &  4325  &  1.20  &  1.30  &  --1.11 $\pm$ 0.01  &  108  &  --1.00 $\pm$ 0.03  &   7  &  --0.94 $\pm$ 0.01  &  40  & --0.90 $\pm$ 0.05  &   7  \\
 118  &  4450  &  1.35  &  1.40  &  --1.05 $\pm$ 0.01  &  140  &  --1.03 $\pm$ 0.01  &   8  &  --0.81 $\pm$ 0.01  &  56  & --0.88 $\pm$ 0.03  &  13  \\
 127  &  4425  &  1.35  &  1.35  &  --1.06 $\pm$ 0.01  &  102  &  --0.95 $\pm$ 0.02  &  10  &  --0.90 $\pm$ 0.02  &  57  & --0.90 $\pm$ 0.05  &  15  \\
 133  &  4450  &  1.35  &  1.40  &  --1.10 $\pm$ 0.01  &  142  &  --1.05 $\pm$ 0.01  &   9  &  --0.89 $\pm$ 0.01  &  57  & --0.91 $\pm$ 0.04  &  15  \\
 145  &  4475  &  1.45  &  1.30  &  --1.06 $\pm$ 0.01  &  146  &  --0.98 $\pm$ 0.02  &  10  &  --0.88 $\pm$ 0.01  &  47  & --0.92 $\pm$ 0.04  &  13  \\
 157  &  4545  &  1.55  &  1.45  &  --1.04 $\pm$ 0.01  &  136  &  --1.01 $\pm$ 0.02  &  10  &  --0.83 $\pm$ 0.01  &  52  & --0.84 $\pm$ 0.05  &  13  \\
\hline
  79  &  4415  &  1.00  &  1.55  &  --1.19 $\pm$ 0.01  &  131  &  --1.08 $\pm$ 0.01  &   8  &  --1.08 $\pm$ 0.01  &  48  & --1.03 $\pm$ 0.04  &  15  \\
  96  &  4450  &  1.15  &  1.50  &  --1.10 $\pm$ 0.01  &  130  &  --1.13 $\pm$ 0.03  &  11  &  --0.71 $\pm$ 0.03  &  33  & --0.88 $\pm$ 0.07  &   9  \\
 116  &  4760  &  1.35  &  1.80  &  --1.24 $\pm$ 0.01  &  134  &  --1.13 $\pm$ 0.03  &   9  &  --1.12 $\pm$ 0.02  &  27  & --0.99 $\pm$ 0.04  &  13  \\
 128  &  4760  &  1.45  &  1.60  &  --1.21 $\pm$ 0.01  &  138  &  --1.10 $\pm$ 0.04  &  12  &  --1.11 $\pm$ 0.02  &  33  & --0.95 $\pm$ 0.04  &  14  \\
 135  &  4635  &  1.40  &  1.55  &  --1.14 $\pm$ 0.01  &  128  &  --0.98 $\pm$ 0.03  &  10  &  --1.08 $\pm$ 0.02  &  33  & --0.88 $\pm$ 0.04  &  13  \\
 158  &  4840  &  1.60  &  1.65  &  --1.19 $\pm$ 0.01  &  142  &  --1.01 $\pm$ 0.02  &  11  &  --1.10 $\pm$ 0.02  &  27  & --0.92 $\pm$ 0.05  &  14  \\
%\hline
\enddata
\tablecomments{Identification number, spectroscopic temperature and
  photometric gravities, microturbulent velocities, iron and titanium
  abundances with internal uncertainty and number of used lines, as
  measured from neutral and single-ionized lines. For all the stars a
  global metallicity of [M/H]$ = -1.0$ dex has been assumed for the
  model atmosphere. The adopted solar values are from \cite{grevesse98}.}
\label{tab3_c7}
\end{deluxetable}
\end{landscape}

\begin{landscape}
\begin{deluxetable}{rcrrccccccccc}
\tiny
\tablecolumns{13}
\tablewidth{0pt}
\tablecaption{OI, NaI, MgI, AlI, TiI and TiII abundances of the RGB and AGB sample}
\tablehead{\colhead{ID} & \colhead{[OI/FeII]} & \colhead{[NaI/FeI]$_{\rm LTE}$} & \colhead{[NaI/FeI]$_{\rm NLTE}$}
& \colhead{n$_{\rm Na}$} & \colhead{[MgI/FeI]} & \colhead{n$_{\rm Mg}$} & \colhead{[AlI/FeI]} & \colhead{n$_{\rm Al}$}
& \colhead{[TiI/FeI]} & \colhead{n$_{\rm TiI}$} & \colhead{[TiII/FeII]} & \colhead{n$_{\rm TiII}$}\\
& \colhead{(dex)} & \colhead{(dex)} & \colhead{(dex)} & & \colhead{(dex)} & & \colhead{(dex)} & &
\colhead{(dex)} & & \colhead{(dex)} & }
\startdata
%\hline
% & & & & & & & & & & & & & \\
%\hline
  50  &  0.39 $\pm$ 0.05  & --0.01 $\pm$ 0.12  &   0.11 $\pm$ 0.09  &  4  &  0.47 $\pm$ 0.03  &  3  &  0.28 $\pm$ 0.01  &  2  &  0.22 $\pm$ 0.02  &  58  &  0.18 $\pm$ 0.05  &  12  \\
  54  &  0.35 $\pm$ 0.04  &   0.17 $\pm$ 0.10  &   0.30 $\pm$ 0.06  &  4  &  0.51 $\pm$ 0.02  &  3  &  0.28 $\pm$ 0.00  &  2  &  0.18 $\pm$ 0.01  &  63  &  0.08 $\pm$ 0.03  &  14  \\
  76  &       $<$ --0.16  &   0.34 $\pm$ 0.11  &   0.42 $\pm$ 0.07  &  4  &  0.37 $\pm$ 0.02  &  3  &  0.66 $\pm$ 0.02  &  2  &  0.18 $\pm$ 0.02  &  37  &  0.12 $\pm$ 0.05  &   6  \\
  82  &  0.31 $\pm$ 0.07  & --0.05 $\pm$ 0.08  &   0.05 $\pm$ 0.04  &  4  &  0.40 $\pm$ 0.03  &  3  &  0.20 $\pm$ 0.02  &  2  &  0.17 $\pm$ 0.02  &  45  &  0.09 $\pm$ 0.07  &   7  \\
  89  &       $<$ --0.28  &   1.04 $\pm$ 0.06  &   1.08 $\pm$ 0.02  &  4  &  0.46 $\pm$ 0.02  &  3  &  0.78 $\pm$ 0.04  &  2  &  0.29 $\pm$ 0.02  &  62  &  0.15 $\pm$ 0.04  &  14  \\
  95  &       $<$ --0.36  &   0.31 $\pm$ 0.17  &   0.40 $\pm$ 0.16  &  4  &  0.33 $\pm$ 0.05  &  3  &  1.19 $\pm$ 0.06  &  2  &  0.18 $\pm$ 0.02  &  58  &  0.17 $\pm$ 0.05  &  15  \\
  97  &       $<$ --0.34  &   0.41 $\pm$ 0.14  &   0.47 $\pm$ 0.09  &  4  &  0.35 $\pm$ 0.06  &  2  &  1.08 $\pm$ 0.04  &  2  &  0.19 $\pm$ 0.02  &  50  &  0.06 $\pm$ 0.05  &  14  \\
 104  &  0.25 $\pm$ 0.08  &   0.01 $\pm$ 0.11  &   0.11 $\pm$ 0.07  &  4  &  0.43 $\pm$ 0.06  &  3  &  0.29 $\pm$ 0.05  &  1  &  0.17 $\pm$ 0.02  &  40  &  0.10 $\pm$ 0.06  &   7  \\
 118  &  0.39 $\pm$ 0.05  & --0.03 $\pm$ 0.10  &   0.05 $\pm$ 0.07  &  4  &  0.40 $\pm$ 0.06  &  2  &  0.29 $\pm$ 0.03  &  2  &  0.24 $\pm$ 0.02  &  56  &  0.15 $\pm$ 0.03  &  13  \\
 127  &  0.05 $\pm$ 0.09  &   0.08 $\pm$ 0.11  &   0.17 $\pm$ 0.07  &  4  &  0.39 $\pm$ 0.02  &  3  &  0.44 $\pm$ 0.04  &  1  &  0.16 $\pm$ 0.02  &  57  &  0.05 $\pm$ 0.05  &  15  \\
 133  &       $<$ --0.17  &   0.40 $\pm$ 0.10  &   0.46 $\pm$ 0.06  &  4  &  0.34 $\pm$ 0.05  &  2  &  1.12 $\pm$ 0.06  &  2  &  0.21 $\pm$ 0.01  &  57  &  0.14 $\pm$ 0.04  &  15  \\
 145  &  0.13 $\pm$ 0.07  & --0.03 $\pm$ 0.08  &   0.05 $\pm$ 0.06  &  4  &  0.38 $\pm$ 0.01  &  3  &  0.19 $\pm$ 0.04  &  2  &  0.18 $\pm$ 0.01  &  47  &  0.07 $\pm$ 0.04  &  13  \\
 157  &       $<$ --0.40  &   0.55 $\pm$ 0.10  &   0.59 $\pm$ 0.06  &  4  &  0.36 $\pm$ 0.02  &  3  &  0.96 $\pm$ 0.03  &  2  &  0.21 $\pm$ 0.01  &  52  &  0.17 $\pm$ 0.05  &  13  \\
\hline
  79  &  0.37 $\pm$ 0.05  & --0.20 $\pm$ 0.06  & --0.04 $\pm$ 0.04  &  4  &  0.58 $\pm$ 0.08  &  2  &  0.20 $\pm$ 0.07  &  1  &  0.11 $\pm$ 0.01  &  48  &  0.05 $\pm$ 0.04  &  15  \\
  96  &  0.35 $\pm$ 0.06  & --0.03 $\pm$ 0.08  &   0.09 $\pm$ 0.04  &  4  &  0.43 $\pm$ 0.05  &  3  &  0.33 $\pm$ 0.06  &  1  &  0.39 $\pm$ 0.04  &  33  &  0.25 $\pm$ 0.08  &   9  \\
 116  &  0.17 $\pm$ 0.07  &   0.13 $\pm$ 0.06  &   0.23 $\pm$ 0.03  &  4  &  0.38 $\pm$ 0.04  &  2  &  0.46 $\pm$ 0.05  &  1  &  0.12 $\pm$ 0.02  &  27  &  0.14 $\pm$ 0.05  &  13  \\
 128  &  0.19 $\pm$ 0.06  & --0.06 $\pm$ 0.10  &   0.04 $\pm$ 0.09  &  4  &  0.50 $\pm$ 0.09  &  3  &  0.44 $\pm$ 0.07  &  1  &  0.10 $\pm$ 0.03  &  33  &  0.15 $\pm$ 0.06  &  14  \\
 135  &  0.30 $\pm$ 0.07  & --0.20 $\pm$ 0.04  & --0.08 $\pm$ 0.03  &  4  &  0.47 $\pm$ 0.03  &  3  &  0.28 $\pm$ 0.07  &  1  &  0.06 $\pm$ 0.02  &  33  &  0.11 $\pm$ 0.05  &  13  \\
 158  &  0.19 $\pm$ 0.05  &   0.13 $\pm$ 0.06  &   0.20 $\pm$ 0.03  &  3  &  0.48 $\pm$ 0.08  &  3  &  0.46 $\pm$ 0.05  &  1  &  0.09 $\pm$ 0.02  &  27  &  0.09 $\pm$ 0.05  &  14  \\
\hline
\enddata
\tablecomments{The oxygen abundance has been derived from the 6300.3$\rm\mathring{A}$ [OI] line,
the abundances of sodium have been reported without and with NLTE corrections computed following
\cite{gratton99}.
The reference solar values are taken from \cite{caffau11} for the oxygen,
from \cite{grevesse98} for the other species.}
\label{tab4_c7}
\end{deluxetable}
\end{landscape}

%\begin{landscape}
\begin{deluxetable}{ccccc}
\tiny
\tablecolumns{5}
\tablewidth{0pt}
\tablecaption{Abundance uncertainties due to the atmospheric parameters for the stars 157 and 158.}
\tablehead{\colhead{Species} & \colhead{Global} & \colhead{$\delta T_{\rm eff}$}
& \colhead{$\delta \log g$} & \colhead{$\delta v_{\rm turb}$} \\
\colhead{} & \colhead{Uncertainty} & \colhead{$\pm 50 K$}
& \colhead{$\pm 0.1$} & \colhead{$\pm 0.1 km s^{-1}$} \\
& \colhead{(dex)} & \colhead{(dex)} & \colhead{(dex)} & \colhead{(dex)}}
\startdata
%\hline
% & & & & \\
%\hline
& & 157 (RGB) & & \\
\hline
 FeI  &  $\pm$0.07  &  $\pm$0.04  &  $\pm$0.00  &  $\mp$0.06  \\
 FeII &  $\pm$0.08  &  $\mp$0.05  &  $\pm$0.05  &  $\mp$0.04  \\
 OI   &  $\pm$0.04  &  $\pm$0.01  &  $\pm$0.03  &  $\mp$0.02  \\
 NaI  &  $\pm$0.05  &  $\pm$0.04  &  $\mp$0.01  &  $\mp$0.02  \\
 MgI  &  $\pm$0.04  &  $\pm$0.03  &  $\pm$0.00  &  $\mp$0.03  \\
 AlI  &  $\pm$0.04  &  $\pm$0.04  &  $\pm$0.00  &  $\mp$0.02  \\
 TiI  &  $\pm$0.09  &  $\pm$0.08  &  $\pm$0.00  &  $\mp$0.03  \\
 TiII &  $\pm$0.05  &  $\mp$0.02  &  $\pm$0.04  &  $\mp$0.03  \\
\hline
& & 158 (AGB) & & \\
\hline
 FeI  &  $\pm$0.07  &  $\pm$0.06  &  $\pm$0.00  &  $\mp$0.04  \\
 FeII &  $\pm$0.07  &  $\mp$0.03  &  $\pm$0.05  &  $\mp$0.03  \\
 OI   &  $\pm$0.05  &  $\pm$0.02  &  $\pm$0.04  &  $\mp$0.02  \\
 NaI  &  $\pm$0.04  &  $\pm$0.04  &  $\pm$0.00  &  $\mp$0.01  \\
 MgI  &  $\pm$0.03  &  $\pm$0.03  &  $\pm$0.00  &  $\mp$0.01  \\
 AlI  &  $\pm$0.03  &  $\pm$0.03  &  $\pm$0.00  &  $\mp$0.00  \\
 TiI  &  $\pm$0.08  &  $\pm$0.08  &  $\pm$0.00  &  $\mp$0.01  \\
 TiII &  $\pm$0.06  &  $\mp$0.01  &  $\pm$0.05  &  $\mp$0.03  \\
\hline
\enddata
\tablecomments{The second column shows the global uncertainty
  calculated by adding in quadrature the single uncertainties.  The
  other columns list the uncertainties obtained by varying only one
  parameter at a time, while keeping the others fixed.}
\label{tab5_c7}
\end{deluxetable}
%\end{landscape}

%

%% file: c8/ms_8.tex
% CHAPTER 8

\chapter{Weighing Stars: the Identification of an Evolved Blue Straggler Star in the Globular Cluster 47Tucanae}

\label{c8}

{\bf Ferraro et al. 2015, accepted by ApJ}

{\it Globular clusters are known to host peculiar objects, named Blue
Straggler Stars (BSSs), significantly heavier than the normal stellar
population. While these stars can be easily identified during their
core hydrogen-burning phase, they are photometrically
indistinguishable from their low-mass sisters in advanced stages of
the subsequent evolution. A clear-cut identification of these objects
would require the direct measurement of the stellar mass, which is a
very difficult task in Astrophysics. We used the detailed comparison
between chemical abundances derived from neutral and from ionized
spectral lines as a powerful stellar {\it ``weighing device''} to
measure stellar mass and to identify an evolved BSS in the globular
cluster 47 Tucanae.  In particular, high-resolution spectra of three
bright stars located slightly above the level of the ``canonical''
horizontal branch sequence in the color-magnitude diagram of 47
Tucanae, in a region where evolved BSSs are expected to lie, have been
obtained with UVES-FLAMES at the ESO Very Large Telescope.  The
measurements of iron and titanium abundances performed separately from
neutral and ionized lines reveal that two targets have stellar
parameters fully consistent with those expected for low-mass
post-horizontal branch objects, while for the other target the
elemental ionization balance is obtained only by assuming a mass of
$\sim$ 1.4 $M_\odot$, which is significantly larger than the main
sequence turn-off mass of the cluster ($\sim$ 0.85 $M_\odot$). The
comparison with theoretical stellar tracks suggests that this blue
straggler descendant is possibly experiencing its core helium-burning
phase. The large applicability of the proposed method to most of the
globular clusters in our Galaxy opens the possibility to initiate
systematic searches for evolved BSSs, thus giving access to still
unexplored phases of their evolution and providing deep insights even
into their (still unclear) formation scenarios.}

%%%%%%%%%%%%%%%%%%%%%%%%%%%%%%%%%%%%%%%%%%%%%%%%%%%%%%%%%% OBSERV AND RV

\section{Observations and membership}
\label{obs}

\begin{figure}[h]
\centering
\includegraphics[trim=0cm 0cm 0cm 0cm,clip=true,scale=0.60,angle=0]{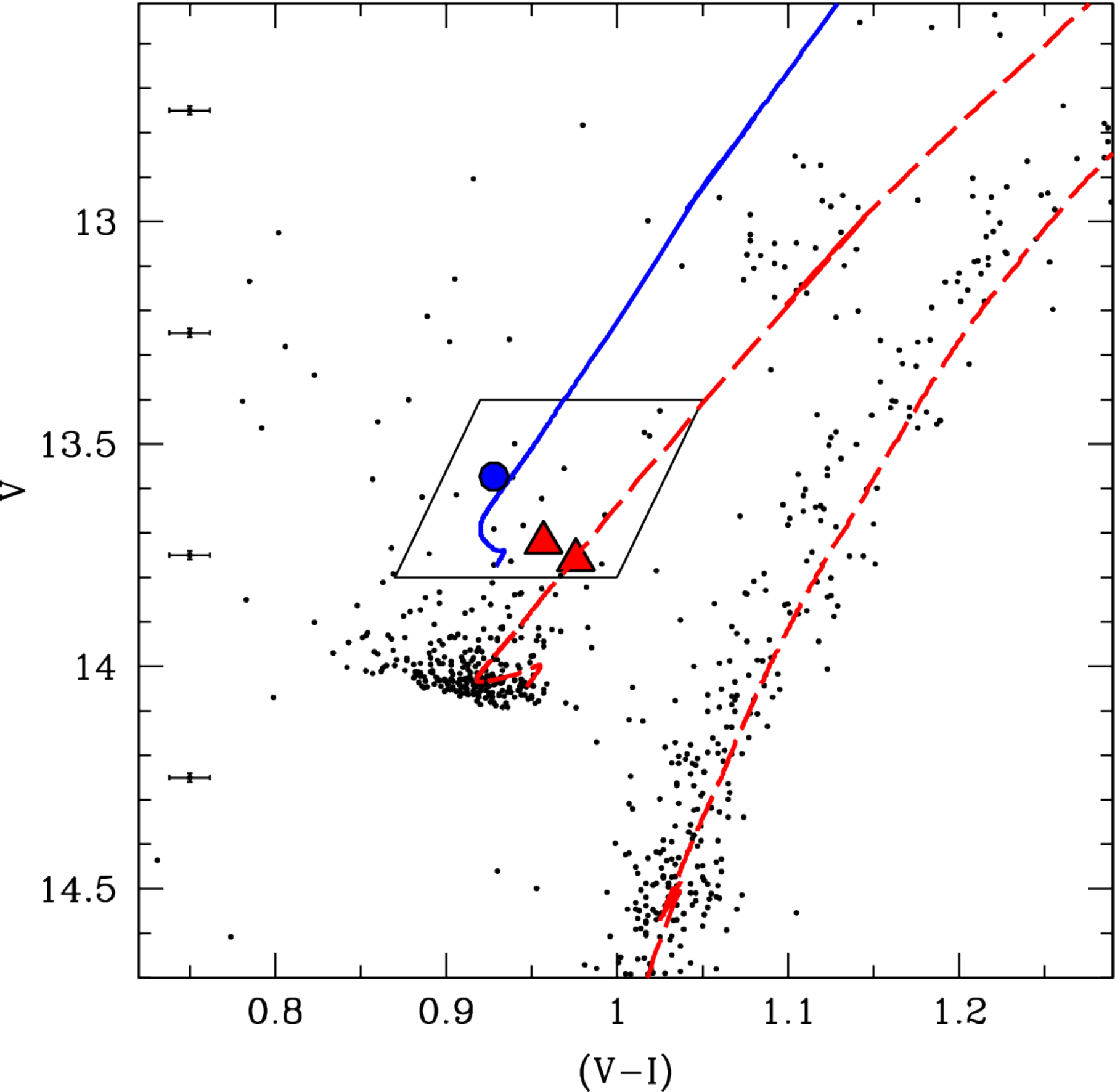}
\caption{Magnified portion of the $(V, V-I)$ CMD of 47 Tuc around the
horizontal branch.  The large blue circle marks the position of
E-BSS1. The position of the two reference targets (bHB1 and bHB2)
are marked with red triangles. For reference, only stars within
$50\arcsec$ from the center are plotted (small black dots). Error
bars are marked at different magnitude levels. The red dashed line
corresponds to the evolutionary track \citep{pietrinferni06} of a
single star of 0.9 $M_\odot$ well reproducing the cluster main
evolutionary sequences.  The evolutionary track, from the HB to the
AGB, for a star with a MS mass of 1.5 $M_\odot$ is also shown (blue
solid line). Because of mass loss during the RGB, the
stellar mass at the HB level for this evolutionary track is $\sim$
1.4 $M_\odot$, well in agreement with the spectroscopic estimate for
E-BSS1. The marked box delimitates the region where evolved BSSs
experiencing the HB stage are expected to lie: the blue and red
boundaries of the box are approximately set by the tracks
corresponding to MS turn-off masses of 1.8 $M_\odot$ and 0.9 $M_\odot$, respectively.}
\label{cmd}
\end{figure}

In the context of the ESO Large Programme 193.D-0232 (PI: Ferraro)
aimed at studying the internal kinematics of Galactic globular
clusters, we have secured UVES-FLAMES \citep{pasquini00}
high-resolution spectra of three stars in 47 Tuc.  The targets
(hereafter named bHB1, bHB2 and E-BSS1).  have been selected in a
regionof the CMD slightly brighter than the red clump (see Figure~\ref{cmd}),
where evolved BSSs experiencing the core helium burning
process are expected to lie (see also \citealp{beccari06}).  All the
targets lie within a distance of $\sim 132\arcsec$ from the cluster
center, corresponding to 4.5 $r_c$ or 0.6 $r_{\rm hm}$
($r_c=29\arcsec$ and $r_{\rm hm}=213\arcsec$ being, respectively, the
core and half-mass radii of 47 Tuc; \citealp{miocchi13}).
Figure~\ref{cmd} shows the ($V, V-I$) CMD obtained from the HST-ACS
photometric catalog of \citet{sarajedini07}, with the target selection box
marked.  The color and magnitude of bHB2 (which is located beyond the
ACS field of view) are from ground-based wide field data
\citep{ferraro04b} homogenized to the Johnson-Cousin photometric
system. The coordinates, $V$ band magnitude, $V-I$ color, and distance
from the center of each target are listed in Table~\ref{tab1_c8}.

% FIGURE 1
%

The target spectra have been acquired with the grating 580 Red Arm
CD\#3, which provides a spectral resolution $R\sim40000$ between
4800$\rm\mathring{A}$\ and 6800$\rm\mathring{A}$. All the spectra have
been reduced by using the dedicated ESO pipeline, performing bias
subtraction, flat-fielding, wavelength calibration and order
merging. During the observations, a number of fibers were allocated on
empty regions to sample the sky background, which has then been
subtracted from the target spectra. The total exposure time is $\sim$
30 min for each star, providing a signal-to-noise ratio (S/N) $\geq$
50 per pixel.

The three stars are all cluster members, as assessed by their radial
velocity (see Table~\ref{tab1_c8}) and the systemic velocity and velocity
dispersion of 47 Tuc ($V_r = -17.6$ km s$^{-1}$ and $\sigma$ = 11.5 km
s$^{-1}$, respectively, from \citealp{lapenna14}; see also
\citealp{carretta04,alvesbrito05,ferraro06,koch08,lane10,
gratton13,cordero14,thygesen14,johnson15}).
The radial velocities have been determined with the code DAOSPEC
\citep{stetson08}, by measuring the position of up to 300 metallic
lines. The final uncertainty was obtained by dividing the dispersion
of the velocities by the square root of the number of lines used.

To further check the possible contamination by field stars, we
extracted the distribution of radial velocities and metallicities of a
sample of about 1700 field objects from the the Besan\c{c}on Galactic
model \citep{robin03}. We found that no field stars are present in the
CMD region corresponding to the position of the targets and with
radial velocities and metallicities similar to those of 47 Tuc.  We
can also safely exclude a contamination from stars belonging to the
Small Magellanic Cloud (SMC), since the brightest SMC objects (at the
RGB tip) are located at much fainter magnitudes ($V \sim
16.5$ mag) and have quite different radial velocities (between $+50$ e
$+250$ km s$^{-1}$; \citealp{harris96}).

%%%%%%%%%%%%%%%%%%%%%%%%%%%%%%%%%%%%%%%%%%%%%%%%%%%%%%%%%% ANALYSIS

\section{Chemical analysis}
\label{analysis}

The chemical analysis has been performed following the same approach
already used in \citet{lapenna14}. The equivalent width (EW) and the
relative uncertainty have been measured with DAOSPEC, iteratively
launched by the 4DAO\footnote{http://www.cosmic-lab.eu/4dao/4dao.php}
code \citep{mucciarelli13b}. Abundances were obtained with the
code GALA\footnote{http://www.cosmic-lab.eu/gala/gala.php}
\citep{mucciarelli13}, by matching the measured and theoretical
EWs, and adopting the ATLAS9 model atmospheres and the solar values of
\citet{grevesse98}.  We have used only the lines with reduced EWs,
defined as log($EW/\lambda$ (with $\lambda$ being the wavelength),
ranging between $-5.6$ and $-4.5$ to avoid saturated or too weak
lines, and we discarded those with EW uncertainties larger than
20\%. The computation of the final iron abundances has been performed
by using up to 127 FeI lines and 9 FeII lines. To derive the
abundances of titanium we exploited 20 TiI lines and 11 TiII lines.

The microturbulence velocity (see Table~\ref{tab2_c8}) has been optimized
spectroscopically by requiring that no trends exist between the
abundance derived from FeI lines and the reduced EWs.  To determine
the stellar surface gravity (log~$g$) an estimate of the stellar mass
and radius is needed.  The latter is obtained from the
Stefan-Boltzmann equation once the surface temperature (see Table~\ref{tab1_c8})
has been determined from the $(V-I)_0$ color-temperature
relation of \citet{alonso99}, after transforming the Johnson-Cousin
$(V-I)_0$ color into the Johnson system following appropriate
transformations \citep{bessell79}.  \emph{As for the stellar mass,
which value is expected for the three targets?} Due to the mass-loss
occurring along the RGB, stars evolving on the HB are
expected to be less massive than MS turn-off stars by $\sim$ 0.1-0.15
$M_\odot$ \citep{renzinifusi88,origlia07,origlia10,origlia14}.
Recently, \cite{gratton10b} derived the mass distribution of HB stars
in several globular clusters, obtaining values between $\sim$ 0.6 and
0.7 $M_\odot$ in the case of 47 Tuc. These values are 0.1-0.2
$M_\odot$ lower than the turn-off mass of the best-fit isochrone
(0.85 $M_\odot$), in full agreement with the expected amount of
mass-loss during the RGB.
Adopting a mass of 0.6 $M_\odot$ and the photometric measure of the effective temperature,
we obtained surface gravities of $\sim 2.0$ and we derived the FeI and
FeII abundances of the three targets (see Table~\ref{tab2_c8}).

For all the targets we found values of [FeI/H[ in agreement with the
mean metallicity of 47 Tuc ([Fe/H]$=-0.83, \sigma = 0.03$ dex;
\citealp{lapenna14}). This suggests that the target stars are not
affected by departures from local thermal equilibrium
(LTE)\footnote{In fact, departures from LTE conditions affect the
minority species, leading to an under-estimate of [FeI/H], while
they have no impact on the abundances obtained from the dominant
species, as single ionized iron lines \citep[see also][]{ivans01,mashonkina11,mucciarelli15}}
and the abundance derived from FeI lines is a reliable measure of
the iron content of the stars.
However, only for two objects (namely bHB1 and bHB2)
the value of the iron abundance obtained from the ionized lines
(FeII) agrees, within 0.01 dex, with that derived from FeI, while
it is sensibly ($\sim 0.2$ dex) lower for E-BSS1. This is the
opposite of what is predicted and observed in the case of
departures from LTE conditions, while it could be explained as an
effect of surface gravity and, hence, of stellar mass.  In fact
the absorption lines of ionized elements are sensitive to changes
in surface gravity, while neutral lines are not. Taking this into
account, we evaluated the effect of increasing the stellar mass by
re-performing the spectral analysis for different values of the
surface gravity.

% FIGURE 2
%
\begin{figure}[h]
\centering
\includegraphics[trim=0cm 0cm 0cm 0cm,clip=true,scale=0.70,angle=0]{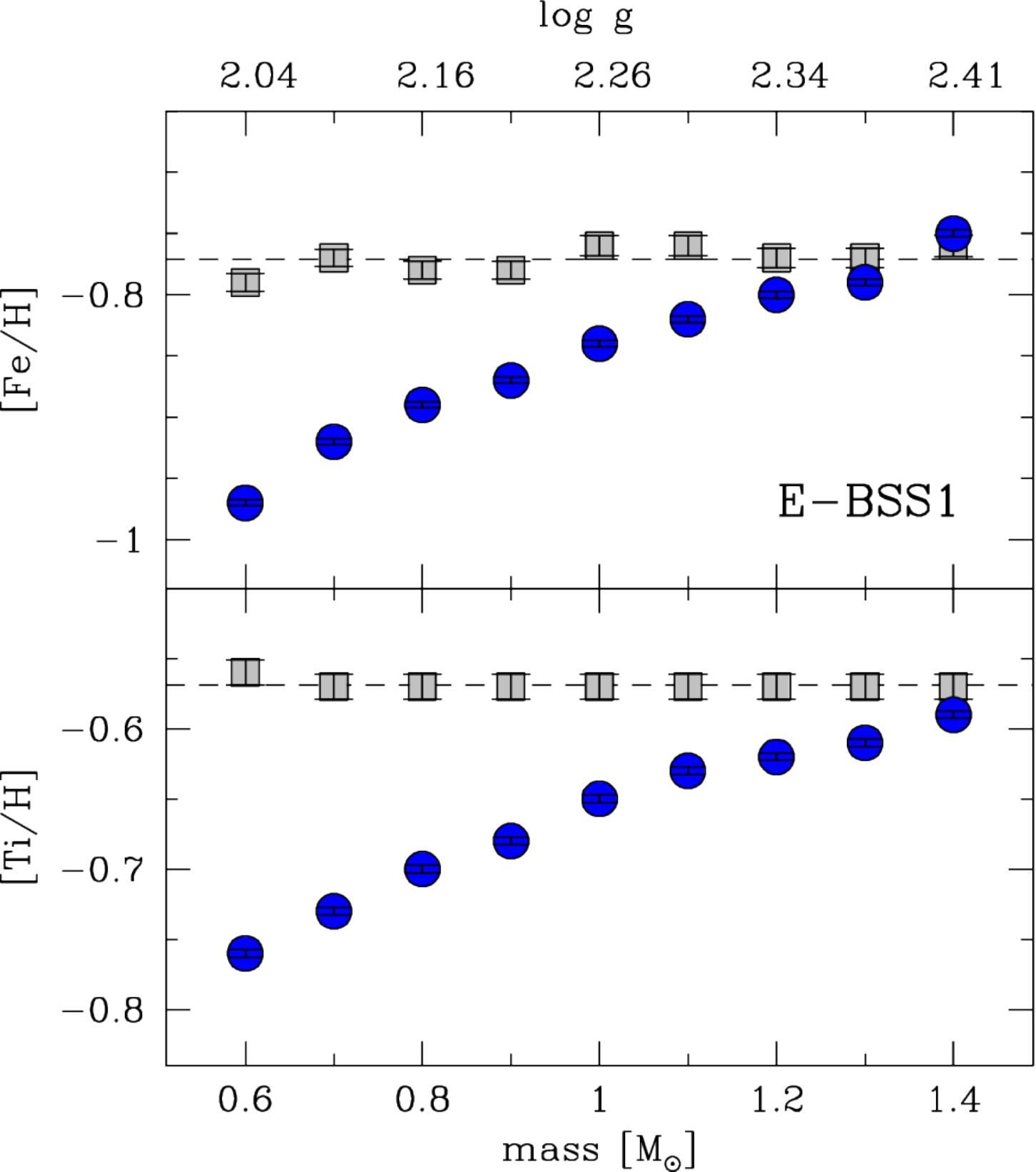}
\caption{\emph{Top panel:} Iron abundance of E-BSS1 derived from FeI
lines (grey squares) and FeII lines (blue circles), as a function of
the assumed stellar mass. Errors in the derived abundances are
smaller than the symbol sizes. The dashed line marks the average FeI
abundance (well corresponding to the metallicity of 47 Tuc:
[Fe/H]$=-0.83$ dex; e.g., \citealp{lapenna14}). In the top axis of
the panel, the logarithmic values of the stellar surface gravity
corresponding to the various adopted masses are labeled.
\emph{Bottom panel:} Same as the top panel, but for the Ti abundance,
as derived from TiI lines (grey squares) and TiII lines (blue circles).}
\label{dfeti1}
\end{figure}

\begin{figure}[h]
\centering
\includegraphics[trim=0cm 0cm 0cm 0cm,clip=true,scale=0.70,angle=0]{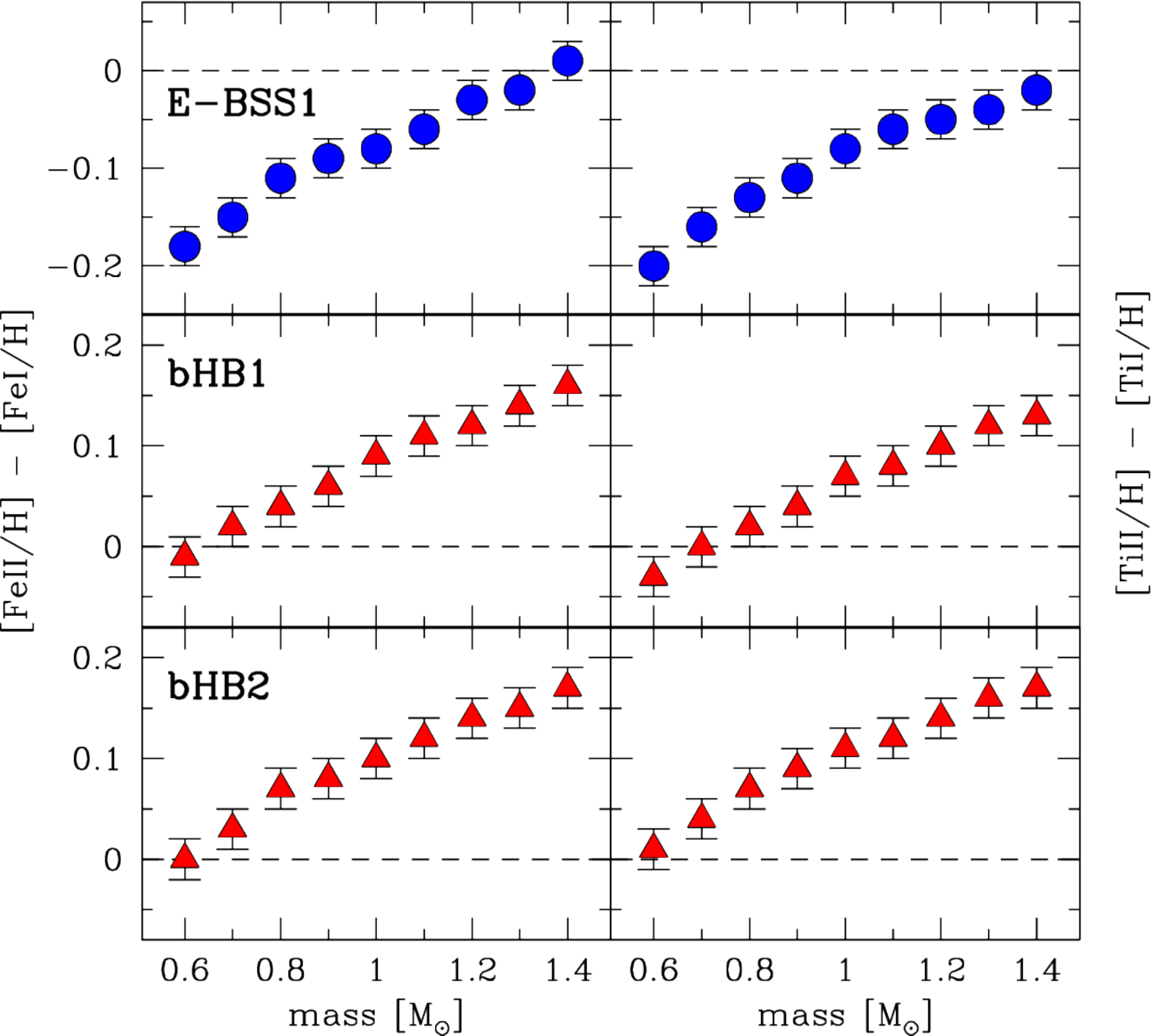}
\caption{\emph{Top panel:} Difference between the iron abundances
derived from ionized lines and that obtained from neutral lines, as
a function of the assumed stellar mass (left-hand panel) for E-BSS1.
The same, but for the titanium abundances is shown in the right-hand panels.
\emph{Mid panel:} As in the top panel, but for target bHB1.
\emph{Bottom panel:} As in the top panel, but for target bHB2.}
\label{dfeti23}
\end{figure}

In Table~\ref{tab2_c8} we list the values of [FeI/H] and [FeII/H]
obtained by varying the star mass in steps of 0.1 $M_\odot$ while
keeping the effective temperature fixed.
The upper panel of Figure~\ref{dfeti1} shows the resulting behavior.
As expected, in all cases the FeI abundance remains essentially unaltered
(and consistent with the cluster metallicity),
while [FeII/H] systematically increases for increasing mass (gravity).
The behavior of the difference between
FeII and FeI abundances as a function of the adopted stellar mass is
plotted in the left panels of Figure~\ref{dfeti23} for the three targets.
Clearly, while for two stars (bHB1 and bHB2) a good agreement between
the FeI and FeII abundances is reached at 0.6 $M_\odot$, for E-BSS1 a
significantly larger stellar mass (1.3-1.4 $M_\odot$, corresponding to
a gravity log~$g$ = 2.4 dex) is needed. Thus, a mass larger than twice
the mass expected for a canonical post-HB cluster star is needed in
order to reconcile the FeI and FeII abundances of E-BSS1.  Conversely,
the difference [FeII/H]-[FeI/H] for the other two targets tends to
diverge for increasing stellar mass (see Figure~\ref{dfeti23}),
indicating that the adopted values of the surface gravity become
progressively unreasonable.

As a double check, the same procedure has been performed on the
titanium lines, since this is one of the few other elements providing
large numbers of both neutral and single ionized lines. Also in this
case, the same abundance of TiI and TiII is reached, within the
errors, only if a mass of 1.4 $M_\odot$ is adopted for E-BSS1, while
the best agreement is reached at 0.6-0.7 $M_\odot$ for bHB1 and bHB2
(see the right panels of Figure~\ref{dfeti23}). This fully confirms the
results obtained from the iron abundance analysis, pointing out that
EBSS-1 is an object significantly more massive than the others.

% FIGURE 3
%

\begin{figure}[h]
\centering
\includegraphics[trim=0cm 0cm 0cm 0cm,clip=true,scale=0.70,angle=0]{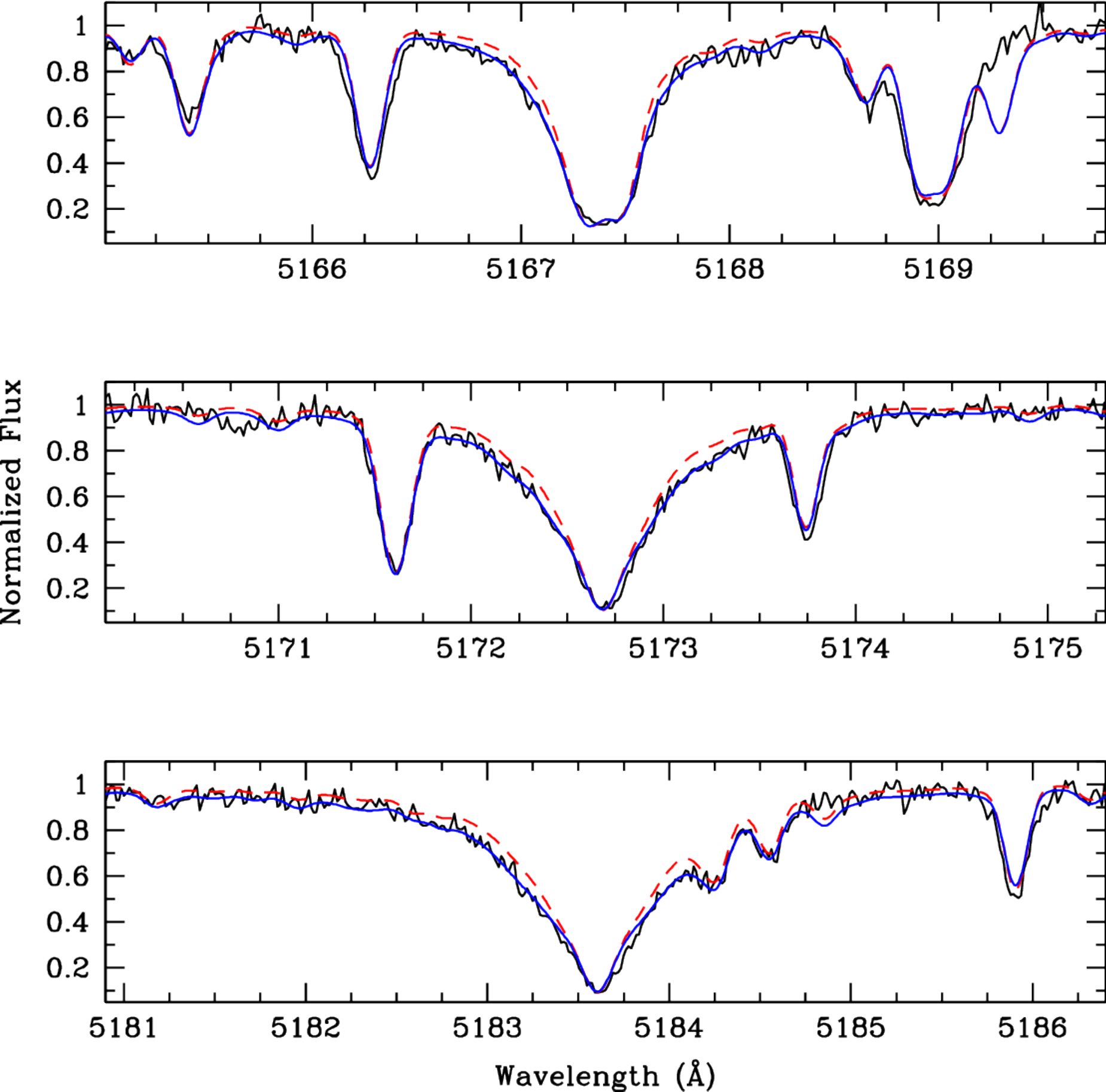}
\caption{Comparison between the observed spectrum (solid black line)
and two synthetic spectra for the MgI lines at 5167.3, 5172.6 and
5183.6$\rm\mathring{A}$. The synthetic spectra have been computed by
adopting $T_{\rm eff} = 5013$ K and $v_{\rm turb} = 1.20$ km
s$^{-1}$ and by adopting two different values of the surface
gravity: log~$g = 2.03$ corresponding to a stellar mass of 0.6
$M_\odot$ (dashed red line) and log~$g = 2.40$ corresponding to 1.4
$M_\odot$ (solid blue line). Clearly, the spectrum obtained for a
0.6 $M_\odot$ star is unable to reproduce the observations, while
the observed wings of the MgI b triplet are very well matched under
the assumption of a 1.4 $M_\odot$ stellar mass.}
\label{mglines}
\end{figure}

Additional support comes from the inspection of another feature that
is known to be sensitive to the stellar surface gravity: the wings of
the MgI b triplet at 5167.3, 5172.6 and 5183.6$\rm\mathring{A}$. In
Figure~\ref{mglines} we show a comparison between the observed spectrum of EBSS-1
and two synthetic spectra computed by assuming the atmospheric
parameters listed in Table~\ref{tab1_c8} and the measured Mg abundance;
only the surface gravity has been varied: we adopted log~$g$ = 2.03 and
log~$g$ = 2.40 dex (corresponding to 0.6 and 1.4 $M_\odot$
respectively). Clearly, the synthetic spectrum computed for log~$g$ =
2.03 dex fails to fit the wings of the MgI b triplet, while that
computed assuming a 1.4 $M_\odot$ stellar mass closely reproduces the
observed spectrum.  All these findings point out that E-BSS1 is an
object significantly more massive than the other targets and canonical
cluster stars.

% FIGURE 4
%

\subsection{Uncertanties}
\label{errors}

It is worth noticing that, in doing the comparison between the
abundances derived from the ionized and the neutral species, we
performed a differential analysis, with the most critical parameters
(as temperature, microturbulence and gravity) set to the same
value. Hence only internal errors, due to the quality of the spectra
and the number of the absorption lines used, need to be considered,
while potential external sources of errors can be neglected.

The global uncertainty on the difference between ionized and neutral
chemical abundances has been determined by taking into account the
effect of atmospheric parameter variations and the covariance terms
due to their correlations \citep{cayrel04}. We estimate that the
global effect of varying the temperature by 40 K (corresponding to an
error in color of the order of 0.015 mag) produces a variation of 0.01
dex on the abundance difference. By adding in quadrature this term
with the uncertainties due to the EW measurements (which are of the
order of 0.01 dex for both the abundance ratios), we estimate a total
error of about 0.02 dex on the derived abundance differences.

We emphasize that the only way to make the FeII abundance in agreement
with that of FeI, while simultaneously complying with the other
constraints, is to assume a large mass (gravity) for E-BSS1. In fact,
departures from LTE conditions would affect the neutral species
(yielding to an under-estimate of [FeI/H]), leaving unaltered the
abundances obtained from single ionized lines
\citep{ivans01,mashonkina11,mucciarelli15}. The micro-turbulence
velocity has a negligible impact on the derived abundances, and its
effect is the same (both in terms of absolute value and direction) on
the abundances derived from neutral and from ionized lines: hence, it
cannot help reconciling the value of [FeII/H] with that of
[FeI/H]. Finally, if for E-BSS1 we assume a mass sensibly lower than
1.4 $M_\odot$ (for instance 0.6-0.8 $M_\odot$, as appropriate for HB
and giant stars in 47 Tuc), FeII would agree with FeI only if the
effective temperature is lowered by $\sim 130$ K.  However, this
solution is not acceptable, because it implies a non-zero slope
between the iron abundance and the excitation potential. Moreover,
such a low value of $T_{\rm eff}$ corresponds to a photometric color
that is completely inconsistent with the observed one (note the
internal accuracy of the HST photometry for such a bright object is
less than 0.01 mag). Thus, the spectra inevitably lead to the
conclusion that E-BSS1 is significantly more massive than the other
stars.

%%%%%%%%%%%%%%%%%%%%%%%%%%%%%%%%%%%%%%%%%%%%%%%%%%%%%%%%%% DISCUSSION

\section{Discussion}
\label{discussion}

Indeed the mass derived for E-BSS1 (1.4 $M_\odot$) is by far too large
for a genuine HB cluster star evolving toward the AGB (the MS turn-off
mass in 47 Tuc  is 0.85 $M_\odot$; see Section~\ref{analysis}).
Moreover, the mass values that we obtained for
targets bHB1 and bHB2 from our analysis turn out to be fully in
agreement with the values (0.6-0.7 $M_\odot$) recently estimated for
typical HB stars in 47 Tuc \citep{gratton10b}. Notably, these are also
the values that we obtained for targets bHB1 and bHB2 from our
analysis. These results clearly demonstrate that the detailed
comparison between neutral and ionized chemical abundances is a
powerful {\it weighing device} able to reliably determine stellar
masses in a self-consistent and differential way (this is quite
relevant since it gets rid of any possible zero-point offset among
different methods).

If E-BSS1 were coeval to the other low-mass stars populating the
cluster, such a massive object would have already evolved into a white
dwarf several Gyr ago.  The only possibility is that it formed more
recently, through a mass-enhancement process: it could therefore be
the descendant of a BSS.  As any other star, BSSs are then expected to
evolve along the various post-MS phases. Indeed, E-BSS1 is located in
the region of the CMD where evolved BSSs experiencing the core helium
burning process are expected to lie. In fact, the collisional models
of \citet{sills09} show that, during the core helium burning stage,
the progeny of collisional BSSs should populate the CMD slightly
blueward of the RGB, between 0.2 and 1 mag brighter than
the ``canonical'' HB level of the host cluster\footnote{Unfortunately
no specific tracks for the post-MS evolution of mass-transfer BSSs
in globular clusters are available at the moment.  However the
models specifically built for the open cluster M67 \citep{tian06}
and the globular cluster M30 \citep{xin15} suggest that, after
mass-transfer, BSSs behave largely as normal single stars of
comparable mass.  Hence they are also expected to populate the same
region of the CMD.}.  Overall, the post-MS evolution of a
collisional product in the CMD is very similar to that of a normal
star of the same mass, the former being just a few tens of degree
hotter (see Figure 6 of \citealp{sills09} for the model of a $\sim$ 1.4
$M_\odot$ star).  In the following analysis, we therefore adopted
``normal'' evolutionary tracks. In Figure~\ref{cmd} we have
superimposed\footnote{We assumed a distance of 4.45 kpc and a color
excess $E(B-V)=0.04$ \citep[see][]{harris96}.} to the CMD of 47 Tuc
the evolutionary track \citep[from][]{pietrinferni06} of a 1.5
$M_\odot$ star.  We note that EBSS-1 lies in the a region very close to
the red clump level of this track, corresponding to an effective
temperature $T_{\rm eff} \simeq 5011$ K and a surface gravity log~$g
\sim 2.47$ dex.  These values are in very good agreement with those
derived from the chemical analysis of this object.  It seems therefore
perfectly reasonable to identify our star with an evolved BSS that is
currently experiencing its red clump (core helium burning) phase
before ascending the AGB.

{\it How rare evolved BSSs are?} The number of evolved BSSs observable
along the HB stage is predicted to be small even in a massive cluster
like 47 Tuc. An estimate can be derived by combining the theoretical
ratio between the characteristic MS and HB lifetimes, with the
observed number of BSSs.  The total BSS population of 47 Tuc likely
counts $\sim 200$ objects.  In fact, observational surveys
\citep{ferraro04b} sampling only on the brightest portion of the
population counted 110 BSSs over the entire cluster extension, with
approximately $40\%$ of the population being segregated within 2 core
radii ($r<50\arcsec$) from the center, which is the distance where
E-BSS1 is located.  However deep UV observations of the very central
region of the cluster, sampling the entire BSS sequence
\citep{ferraro01}, demonstrated that the total population could be 1.7
times larger. Based on these numbers, it is reasonable to expect
$\sim80$ BSSs within $r<50\arcsec$.

On the other hand, evolutionary tracks of collisional BSSs
\citep{sills09} show that the HB lifetime is approximately constant
($\sim 10^8$ years, similar to the HB duration for low-mass single
stars), regardless of the original stellar masses at the time of the
collision.  Instead the MS lifetime of collisional BSSs can change by
3 orders of magnitude, being $40-70\%$ smaller than that of a normal
single star with similar mass. Thus the predicted number of BSSs in
the HB evolutionary stage sensibly depends on the mass of the BSS
progenitor. For instance, the average value of the ratio between the MS
and the HB lifetimes is $t_{\rm MS}/t_{\rm HB}=17.7$, while for a 1.4
$M_\odot$ BSS originated from the collision of a 0.8+0.6 $M_\odot$
pair, we expect $t_{\rm MS}=0.82$ Gyr and $t_{\rm HB}=0.096$ Gyr, thus
yielding $t_{\rm MS}/t_{\rm HB}=8.5$.  This value is more than two
times smaller than that predicted for a 1.4 $M_\odot$ single star
($t_{\rm MS}/t_{\rm HB}=19.3$; \citealp{sills09}).  By considering
these two values as reasonable extremes for the ratio $t_{\rm MS}/t_{\rm HB}$,
we expect to observe 4-8 evolved BSSs experiencing the helium burning phase out
of a total population of 80 BSSs (in the MS stage).

Note that in the same region of the cluster ($r<50\arcsec$), several
hundreds genuine low-mass stars, with quite similar photometric
properties and experiencing the same evolutionary phase, are
observed. However, because of their larger mass, evolved BSSs in the
HB stage are expected to appear brighter than ``normal'' low-mass HB
stars and to lie along the path of low-mass stars evolving toward the
AGB. For genuine low-mass stars the transition from the HB to the AGB
phase (see the box in Figure~\ref{cmd}) is quite rapid: $\sim 3.5$ million
years, corresponding to approximately $4\%$ of the time they spent in
the HB phase.  Based on these considerations and the fact that within
$50\arcsec$ from the cluster center we count 270 objects in the HB
clump, we would have expected to observe $\sim 11$ stars within the
box drawn in Figure~\ref{cmd}. Instead 20 stars are counted. According to
these estimates, roughly half of the stars within $50\arcsec$ from the
cluster center and lying in the selection box could be evolved BSSs.
Hence beside E-BSS1, 8 additional stars in the box could be evolved
BSSs. This is in very good agreement with the prediction above, based
on the number of BSSs observed on the MS. It is also consistent with
previous evidence \citep{beccari06} showing that the radial
distribution of supra-HB stars in 47 Tuc is anomalously segregated in
the center, as expected if a significant fraction of them is made of
objects heavier than the average, sunk to the bottom of the potential
well because of the cluster dynamical evolution \citep{ferraro12}.

%%%%%%%%%%%%%%%%%%%%%%%%%%%%%%%%%%%%%%%%%%%%%%%%%%%%%%%%%% CONCLUSIONS

\section{Summary and conclusions}
\label{conclusions}

In this work we have performed the chemical analysis of three stars
observed between the HB and the AGB regions in the CMD of the Galactic
globular cluster 47 Tuc.  By using high-resolution spectra acquired at
the Very Large Telescope, we have used the difference between iron and
titanium abundances derived from neutral and ionized lines as a {\it
weighing machine} to derive the stellar mass.  This provided
convincing evidence that one target (E-BSS1) is significantly more
massive ($\sim$ 1.4 $M_\odot$) than normal cluster stars, while the
other two targets (bHB1 and bHB2) have masses of 0.6-0.7 $M_\odot$,
perfectly consistent with the theoretical expectations. These results
clearly demonstrate that the detailed comparison between neutral and
ionized chemical abundances is a powerful {\it weighing device} able
to reliably determine stellar masses in a self-consistent and
differential way. The presence of such a high-mass star in that region
of the CMD strongly suggests that it is an evolved descendants of a
BSS, caught during its core He-burning phase. Interestingly, the ratio
between the characteristic MS and HB evolutionary times and the number
of BSSs observed in 47 Tuc suggest that a few other evolved BSSs
should populate the same region of the CMD.

The identification of evolved BSSs is crucial in the context of BSS
formation and evolution models, and the proposed {\it weighing device}
can provide a major advance in this field by efficiently pinpointing
(heavy) evolved BSSs into the dominant and photometrically
indistinguishable population of genuine (low-mass) stars. The
collection of complete samples of these objects in globular clusters
(where BSSs and their descendants are expected to be numerous enough)
allows the determination of meaningful population ratios from which
the characteristic evolutionary time-scales can be empirically constrained.
Moreover, since evolved BSSs are 20 times more luminous
than their progenitors, detailed spectroscopic follow-up studies are
largely facilitated and open the possibility to even go back to the
formation channel. In fact, a few characterizing features impressed by
the formation process could be still observable in such advanced
stages of the evolution. One of the most solid predictions of the
mass-transfer scenario is that mass-transfer BSSs should be bound in a
binary system with a compact (degenerate) companion star (the peeled
core of the donor, probably a helium white dwarf). This was recently
confirmed in an open cluster \citep{gosnell14}. Thanks to the high
luminosity of evolved BSSs, spectroscopic follow-up observations would
make such a prediction easily testable through the measurement of
periodic radial velocity variations. Since no companion is expected in
the collisional scenario (which ends up with the merger of the two
progenitors), this kind of studies is particularly important in
globular clusters, where both formation channels are expected to be
active but their relative efficiency is still unknown. Moreover,
detailed spectroscopic follow-ups providing the entire chemical
pattern of evolved BSSs represent an additional and highly fruitful
route toward the full characterization of their evolution. In fact,
significant depletion of chemical species like carbon and oxygen has
been observed in a sub-sample of BSSs in 47 Tuc and it has been
interpreted as the chemical signature of the mass-transfer origin of
these objects \citep{ferraro06}. However it is still unknown whether
this signature is transient and on which time-scales. Hence, any
additional information obtained from more advanced stages of the
evolution can provide new clues about the degree of mixing experienced
by these stars. Indeed, after the detection of E-BSS1, the proposed
{\it weighing device} promises to boost the identifications of evolved
BSSs, thus providing unprecedented constraints to the theoretical
modeling of these exotica and opening a new perspective on the
comprehension of their evolutionary paths and formation processes.

%%%%%%%%%%%%%%%%%%%%%%%%%%%%%%%%%%% TABLES %%%%%%%%%%%%%%%%%%%%%%%%%%%%%%%%%%%

\begin{landscape}
\begin{deluxetable}{cccccccrr}
\tiny
\tablecolumns{9}
\tablewidth{0pt}
\tablecaption{Observational parameters of the three targets}
\tablehead{\colhead{Name} & \colhead{ID} & \colhead{R.A.}
& \colhead{Dec} & \colhead{$V$} & \colhead{$(V-I)$} & \colhead{$r$} & \colhead{T$_{\rm eff}$} & \colhead{V$_r$} \\
 & & \colhead{(J2000)} & \colhead{(J2000)} &  &  & \colhead{(arcsec)} & \colhead{(K)} & \colhead{(km s$^{-1}$)}}
\startdata
%\hline
 & & & & & & & & \\
%\hline
 EBSS-1&  1090214  &  6.0601164  & $-72.0726528$  &  13.573  &  0.928  &  50.6 & 5013 & $-24.3 \pm 0.05$  \\
 bHB1  &  1109049  &  6.0001040  & $-72.0720222$  &  13.761  &  0.976  &  42.1 & 4896 & $-7.8 \pm 0.06$  \\
 bHB2  &  2625792  &  5.9343037  & $-72.1054776$  &  13.722  &  0.957  & 132.2 & 4940 & $-11.7 \pm 0.06$  \\
\enddata
\tablecomments{Coordinates, $V$ band magnitude, $(V-I)$ color,
  distance from the center, effective temperature and radial velocity
  of the three target stars.  The cluster center used to compute the
  distance from the center is from \citet{miocchi13}.}
\label{tab1_c8}
\end{deluxetable}
\end{landscape}

\begin{landscape}
\begin{deluxetable}{ccccccccccc}
\footnotesize
\tablecolumns{11}
\tablewidth{0pt}
\tablecaption{Abundance ratios of Fe and Ti obtained by  adopting different  stellar mass (gravity) values}
\tablehead{\colhead{Mass} & \colhead{log~$g^{phot}$} & \colhead{$v_{turb}^{spec}$}
& \colhead{[FeI/H]} & \colhead{n(FeI)} & \colhead{[FeII/H]} & \colhead{n(FeII)}
& \colhead{[TiI/H]} & \colhead{n(TiI)} & \colhead{[TiII/H]} & \colhead{n(TiII)} \\
($M_{\odot}$) & (dex) & \colhead{(km s$^{-1}$)} & \colhead{(dex)} &
& \colhead{(dex)} & & \colhead{(dex)} & & \colhead{(dex)} & }
\startdata
%\hline
 & & & & & E-BSS1 & & & & & \\
\hline
 0.60  &  2.03  &  1.30  &  --0.79 $\pm$ 0.01  &  122  &  --0.97 $\pm$ 0.01  &  9  &  --0.56 $\pm$ 0.01  &  21  &  --0.76 $\pm$ 0.05  &  11  \\
 0.70  &  2.10  &  1.25  &  --0.77 $\pm$ 0.01  &  122  &  --0.92 $\pm$ 0.01  &  9  &  --0.57 $\pm$ 0.01  &  20  &  --0.73 $\pm$ 0.05  &  11  \\
 0.80  &  2.15  &  1.25  &  --0.78 $\pm$ 0.01  &  121  &  --0.90 $\pm$ 0.01  &  9  &  --0.57 $\pm$ 0.01  &  20  &  --0.71 $\pm$ 0.05  &  11  \\
 0.90  &  2.21  &  1.25  &  --0.78 $\pm$ 0.01  &  121  &  --0.87 $\pm$ 0.01  &  9  &  --0.57 $\pm$ 0.01  &  20  &  --0.68 $\pm$ 0.05  &  11  \\
 1.00  &  2.25  &  1.20  &  --0.76 $\pm$ 0.01  &  123  &  --0.84 $\pm$ 0.01  &  9  &  --0.57 $\pm$ 0.01  &  20  &  --0.66 $\pm$ 0.05  &  11  \\
 1.10  &  2.30  &  1.20  &  --0.76 $\pm$ 0.01  &  123  &  --0.82 $\pm$ 0.01  &  9  &  --0.57 $\pm$ 0.01  &  21  &  --0.63 $\pm$ 0.05  &  11  \\
 1.20  &  2.33  &  1.20  &  --0.77 $\pm$ 0.01  &  122  &  --0.80 $\pm$ 0.01  &  9  &  --0.57 $\pm$ 0.01  &  21  &  --0.62 $\pm$ 0.05  &  11  \\
 1.30  &  2.37  &  1.20  &  --0.77 $\pm$ 0.01  &  125  &  --0.79 $\pm$ 0.01  &  9  &  --0.57 $\pm$ 0.01  &  21  &  --0.61 $\pm$ 0.05  &  11  \\
 1.40  &  2.40  &  1.15  &  --0.76 $\pm$ 0.01  &  127  &  --0.76 $\pm$ 0.01  &  9  &  --0.57 $\pm$ 0.01  &  20  &  --0.59 $\pm$ 0.05  &  11  \\
\hline
 & & & & & & & & & & \\
 & & & & & bHB1 & & & & & \\
\hline
 0.60  &  2.06  &  1.35  &  --0.84 $\pm$ 0.01  &  129  &  --0.85 $\pm$ 0.01  & 12  &  --0.66 $\pm$ 0.01  &  26  &  --0.69 $\pm$ 0.05  &  14  \\
 0.70  &  2.13  &  1.35  &  --0.84 $\pm$ 0.01  &  128  &  --0.82 $\pm$ 0.01  & 12  &  --0.66 $\pm$ 0.01  &  26  &  --0.66 $\pm$ 0.05  &  14  \\
 0.80  &  2.18  &  1.35  &  --0.84 $\pm$ 0.01  &  128  &  --0.80 $\pm$ 0.01  & 12  &  --0.66 $\pm$ 0.01  &  26  &  --0.64 $\pm$ 0.05  &  14  \\
 0.90  &  2.23  &  1.35  &  --0.84 $\pm$ 0.01  &  127  &  --0.78 $\pm$ 0.01  & 12  &  --0.66 $\pm$ 0.01  &  26  &  --0.62 $\pm$ 0.05  &  14  \\
 1.00  &  2.28  &  1.35  &  --0.84 $\pm$ 0.01  &  126  &  --0.75 $\pm$ 0.01  & 12  &  --0.66 $\pm$ 0.01  &  26  &  --0.59 $\pm$ 0.05  &  14  \\
 1.10  &  2.32  &  1.35  &  --0.84 $\pm$ 0.01  &  126  &  --0.73 $\pm$ 0.01  & 12  &  --0.66 $\pm$ 0.01  &  26  &  --0.58 $\pm$ 0.05  &  14  \\
 1.20  &  2.36  &  1.35  &  --0.84 $\pm$ 0.01  &  126  &  --0.72 $\pm$ 0.01  & 12  &  --0.67 $\pm$ 0.01  &  26  &  --0.57 $\pm$ 0.05  &  14  \\
 1.30  &  2.39  &  1.30  &  --0.83 $\pm$ 0.01  &  126  &  --0.69 $\pm$ 0.01  & 12  &  --0.67 $\pm$ 0.01  &  26  &  --0.55 $\pm$ 0.05  &  14  \\
 1.40  &  2.43  &  1.30  &  --0.83 $\pm$ 0.01  &  127  &  --0.67 $\pm$ 0.01  & 12  &  --0.67 $\pm$ 0.01  &  26  &  --0.54 $\pm$ 0.05  &  14  \\
\hline
 & & & & & & & & & & \\
 & & & & & & & & & & \\
 & & & & & & & & & & \\
 & & & & & & & & & & \\
 & & & & & & & & & & \\
 & & & & & & & & & & \\
 & & & & & & & & & & \\
 & & & & & & & & & & \\
 & & & & & & & & & & \\
 & & & & & & & & & & \\
 & & & & & bHB2 & & & & & \\
\hline
 0.60  &  2.06  &  1.20  &  --0.81 $\pm$ 0.01  &  124  &  --0.81 $\pm$ 0.02  & 10  &  --0.68 $\pm$ 0.01  &  22  &  --0.67 $\pm$ 0.05  &  13  \\
 0.70  &  2.13  &  1.20  &  --0.81 $\pm$ 0.01  &  128  &  --0.78 $\pm$ 0.02  & 10  &  --0.68 $\pm$ 0.01  &  22  &  --0.64 $\pm$ 0.05  &  13  \\
 0.80  &  2.19  &  1.20  &  --0.82 $\pm$ 0.01  &  129  &  --0.75 $\pm$ 0.02  & 10  &  --0.68 $\pm$ 0.01  &  22  &  --0.61 $\pm$ 0.05  &  13  \\
 0.90  &  2.24  &  1.15  &  --0.80 $\pm$ 0.01  &  128  &  --0.72 $\pm$ 0.02  & 11  &  --0.68 $\pm$ 0.01  &  22  &  --0.59 $\pm$ 0.05  &  13  \\
 1.00  &  2.29  &  1.15  &  --0.78 $\pm$ 0.01  &  127  &  --0.68 $\pm$ 0.02  & 11  &  --0.68 $\pm$ 0.01  &  22  &  --0.57 $\pm$ 0.05  &  13  \\
 1.10  &  2.33  &  1.15  &  --0.80 $\pm$ 0.01  &  125  &  --0.68 $\pm$ 0.02  & 11  &  --0.68 $\pm$ 0.01  &  22  &  --0.56 $\pm$ 0.05  &  13  \\
 1.20  &  2.37  &  1.10  &  --0.79 $\pm$ 0.01  &  125  &  --0.65 $\pm$ 0.02  & 11  &  --0.68 $\pm$ 0.01  &  22  &  --0.54 $\pm$ 0.05  &  13  \\
 1.30  &  2.40  &  1.10  &  --0.78 $\pm$ 0.01  &  124  &  --0.63 $\pm$ 0.02  & 11  &  --0.69 $\pm$ 0.01  &  22  &  --0.53 $\pm$ 0.05  &  13  \\
 1.40  &  2.43  &  1.10  &  --0.79 $\pm$ 0.01  &  124  &  --0.62 $\pm$ 0.02  & 11  &  --0.69 $\pm$ 0.01  &  22  &  --0.52 $\pm$ 0.05  &  13  \\
\enddata
\tablecomments{
Iron and titanium abundance ratios obtained for the program stars by
adopting the mass values listed in the first column.  During the
analysis the effective temperature (listed in Table~\ref{tab1_c8}) and
surface gravity (column 2) have been kept fixed, while the
microturbulent velocity (column 3) has been spectroscopically
optimized. The abundances obtained from neutral and ionized lines, and
the number of lines used are listed in columns 4--11 (see labels).
We adopted the solar reference values of \cite{grevesse98}.}
\label{tab2_c8}
\end{deluxetable}
\end{landscape}

%% file: concl/conclusions.tex
% CONCLUSIONS

\chapter*{Conclusions}

The PhD work presented in this manuscript is aimed at clarifying one of
the least studied (yet highly significant) phases of stellar evolution:
the asymptotic giant branch (AGB).
As discussed in the previous chapters, the AGB evolutionary stage is poorly studied in the literature,
especially from the point of view of high-resolution spectroscopy.
However, it may provide important information about exotic stellar species
\citep[like Blue Straggler Stars; see][]{beccari06}
and the evolutionary history of multiple-populations \citep{campbell13} in the parent cluster.
The thesis work is based on the analysis of a high-quality, high-resolution spectra acquired
at the Very Large Telescope (ESO) and at the MPG/ESO-2.2m telescope.
New unexpected results about the chemical composition of AGB stars have been found,
with important consequences for our understanding of this evolutionary phase and
recent claims of significant iron spread in a few GCs.
The main result is the discovery of a previously unknown mechanism affecting
the neutral species of some chemical elements in the atmosphere of most AGB stars: because of it,
the abundances derived from neutral lines are systematically underestimated,
while those measured from ionized lines remain unaffected.
Such a behaviour exactly corresponds to what expected in the case of
non-local thermodynamic equilibrium (NLTE) in the star atmosphere.
However, the observed effect is larger than predicted by the current NLTE models,
thus demonstrating either that these models are not adequate enough,
or that some more complex mechanism is occurring in AGB star atmospheres.
With this caveat in mind, we refer to this effect as ``NLTE effect''.
It affects most (but not all) AGB stars in all the investigated GCs and
it is particularly evident for iron and titanium elements,
which provide the largest number of both neutral and ionized lines.
In the case of M22, also some red giant branch (RGB) stars turn out
to suffer from the same ``problem''.
A deep understanding of the detected phenomenon is crucial for a proper determination
of the chemical abundance patterns and enrichment history of GCs.
In detail, the main results of the thesis can be summarized as follows.

\begin{itemize}

\item (1) {\it The discovery of NLTE effects in an unsuspected metallicity regime} -
The first evidence of the ``NLTE effect'' has been obtained from the
proper spectroscopic analysis of a sample of 24 AGB stars in 47Tucanae,
observed with FEROS at the MPG/ESO-2.2m telescope.
We analyzed neutral and ionized iron lines separately, finding that, for most of the targets (20 out of 24),
the former provide iron abundances systematically lower (by $-0.1$ dex on average) that
those obtained from ionized species and from a sample of RGB stars.
The importance of this result is that, at odds with previous findings \citep{ivans01},
NLTE effects have been observed for the first time in metal-rich ([Fe/H] = $-0.8$ dex) stars,
a metallicity regime where models predict that NLTE should be negligible.

\item (2) {\it The intrinsic iron spread claimed in a few GCs seems to be spurious} - 
The case of NGC3201: The high-resolution spectroscopic analysis of 21 giant stars
in NGC3201 presented in this work, demonstrated that, contrary to RGB stars of similar luminosity,
AGB stars show a clear discrepancy between the abundances derived from neutral and from single-ionized iron lines,
confirming the behaviour observed in 47Tucanae.
Hence, the iron spread claimed by \cite{simmerer13}
turned out to be spurious, and due to (neglected) NLTE effects.
In fact, no iron spread is obtained by considering the iron abundance only from (the unaffected) ionized lines. 
The case of M22: M22 is another GC for which a significant iron spread ($\Delta$[Fe/H] = $0.2$ dex)
has been claimed \citep{marino09,marino11b}.
Our analysis of a sample of 17 giant stars used by Marino et al. demonstrates that
important NLTE effects are present in the majority of AGB stars.
However, in the case of M22, we have observed marginal NLTE effects also among RGB stars,
suggesting that the situation is even more complex.
In any case, the iron abundances derived from ionized lines,
both for RGB and for AGB stars, show no intrinsic spread. 

Note that the presence of any intrinsic iron spread necessarily implies that
the high-velocity supernova ejecta were retained within the system potential well,
thus requiring a much larger mass than currently observed.
Hence, our findings show that, in the cases of both NGC3201 and M22,
no exotic scenario requiring that the clusters were significantly more massive at their birth is needed.
The observed unimodal [FeII/H] distribution rules out the possibility that
these systems are the remnants of now disrupted dwarf galaxies,
with a crucial impact on the current interpretation of GC formation and evolutionary history.
In a more general context, our findings indicate that any claim of intrinsic iron spread
in GCs should be always confirmed with an analysis based on FeII lines and photometric gravities.
If the abundance spread is real, it should be detected also when FeII lines and photometric log~$g$ are adopted,
since FeII lines are the most reliable indicators of the iron abundance. 

\item (3) {\it Additional clues from the AGB population of M62} -
The high-resolution spectroscopic analysis of a sample of 19 giant stars in M62 has revealed,
for the first time, that the same NLTE mechanism affects also the titanium lines.
Moreover, the abundances derived for light elements, like oxygen, sodium and aluminum,
have shown that all the studied AGB stars are compatible with the first stellar generation
(made of O-rich, Na-poor, and Al-poor stars), while the RGB component includes both first and second generation stars.
This result is quite puzzling and closely resembles the case of NGC6752 in which \citep{campbell13}
recently claimed the lack of second generation component in the AGB phase. 

\item (4) {\it Ionization balance as a powerful weighing machine for stars} -
We used the ionization balance between chemical abundances derived from neutral and
from ionized elements to define a powerful weighing device.
In fact, the abundance of a given chemical element obtained from ionized atoms
is sensitively dependent on the stellar gravity,
while such a dependence is negligible for neutral spectral lines of the same element.
Hence, by forcing the former to match the latter, the surface gravity of the star can be estimated.
We used this approach to identify an anomalously heavy (1.4 M$_{\odot}$) star in 47Tucanae.
Because of its position in the Colour-Magnitude Diagram,
this star is probably a Blue Straggler caught during its helium burning phase.

These results clearly demonstrate that the proposed weighting device is able
to reliably determine stellar masses in a self-consistent and differential way:
this is crucial, since it gets rid of any possible zero-point offset among different methods.
The large applicability of the proposed method to most of the GCs in our Galaxy opens
the possibility to initiate systematic searches for evolved BSSs,
thus giving access to still unexplored phases of their evolution by
efficiently pinpointing (heavy) evolved-BSSs into the dominant and
photometrically indistinguishable population of genuine (low-mass) stars.
The collection of complete samples of these objects in globular clusters
(where BSSs and their descendants are expected to be numerous enough)
finally opens the possibility to determine the characteristic BSS evolutionary time-scales,
thus providing crucial constraints to the formation and evolution models of these exotica.

\item (5) {\it ``Side-products''} -
A side product of this thesis work is the characterization of the performances
of new-generation spectrographs, as GIANO at TNG and KMOS at VLT.
In particular, in the case of KMOS the work was aimed at the validation of the quality
of the radial velocities obtainable with this instrument,
in the context of a Large Program designed to determine the next generation
of velocity dispersion and rotation profiles for a representative sample of Galactic GCs.
In the case of GIANO, the work purpose was twofold: (1) to determine position and intensity of OH sky lines
for wavelength calibration and rest-frame reference; (2) to determine the overall continuum airglow emission in the H-band,
which is important for the design of faint-object infrared spectrographs.

\end{itemize}

%% file: ap1/ap_1.tex
% APPENDIX 1

\appendix

\chapter{Radial Velocities from VLT-KMOS Spectra of Giant Stars in the Globular Cluster NGC6388}

\label{a1}

{\bf Published in Lapenna et al. 2015, ApJ, 798, 23}

{\it We present new radial velocity measurements for 82 stars,
members of the Galactic globular cluster NGC6388, obtained from
ESO-VLT KMOS spectra acquired during the instrument Science
Verification. The accuracy of the wavelength calibration is
discussed and a number of tests of the KMOS response are
presented. The cluster systemic velocity obtained
($81.3\pm 1.5$ km s$^{-1}$) is in very good agreement with previous
determinations. While a hint of ordered rotation is found between
$9\arcsec$ and $20\arcsec$ from the cluster centre, where the
distribution of radial velocities is clearly bimodal, more data are
needed before drawing any firm conclusions. The acquired sample of
radial velocities has been also used to determine the cluster velocity
dispersion profile between $\sim 9\arcsec$ and $70\arcsec$,
supplementing previous measurements at $r<2\arcsec$ and
$r>60\arcsec$ obtained with ESO-SINFONI and ESO-FLAMES spectroscopy,
respectively. The new portion of the velocity dispersion profile
nicely matches the previous ones, better defining the knee of the distribution.
The present work clearly shows the effectiveness of a deployable Integral
Field Unit in measuring the radial velocities of individual stars for determining the
velocity dispersion profile of Galactic globular clusters.
It represents the pilot project for an ongoing
large program with KMOS and FLAMES at the ESO-VLT, aimed at
determining the next generation of velocity dispersion and rotation
profiles for a representative sample of globular clusters.}

%%%%%%%%%%%%%%%%%%%%%%%%%%%%%%%%%%%%%%%%%%%%%%%%%%%%%%%%%%% INTRO

\section{Introduction}

Galactic globular clusters (GCs) are massive ($10^4-10^6~M_\odot$)
stellar aggregates, where the two-body relaxation
time-scale is shorter than the age \citep[e.g.,][]{binney87}.
For this reason, they have been traditionally assumed to be quasi-relaxed,
non rotating systems, characterized by spherical symmetry and
orbital isotropy.  Hence spherical, isotropic and non-rotating
models, with a truncated distribution function close to a Maxwellian
\citep{king66,wilson75} are commonly used to fit the observed
surface brightness or density profiles, and to estimate the main GC
structural parameters, like the core and half-mass radii, the
concentration parameter and even the total mass
\citep[e.g.][]{harris96,mclaughlin05}.

However, recent theoretical results indicate that these systems may
have not attained complete energy equipartition \citep{trenti13} and,
depending on the degree of dynamical evolution suffered and the effect
of an external tidal field, they may still preserve some
characteristic kinematical feature \citep{vesperini14}.
In particular, non zero angular momentum has been recognized to affect
the entire dynamical evolution of star clusters \citep{einsel99}, and
central rotation might still be present in GCs hosting an intermediate
mass ($10^2-10^4~M_\odot$) black hole \citep[IMBH;][]{fiestas10}.
Moreover, it is well known that the density profile alone is not
sufficient to fully characterize a gravitational system, and the
information about internal dynamics is also necessary
\citep[e.g.][and references therein]{binney87,meylan97}.

For instance, a star density profile with a shallow cusp deviating from a
King (1966) model and a  velocity dispersion (VD) profile 
with a keplerian central behavior are predicted in the presence of
an IMBH \citep[e.g.][]{baumgardt05,miocchi07}.
Despite its importance, the kinematical properties of Galactic GCs are
still poorly explored from the observational point of view, although
the number of dedicated studies aimed at building their VD and
rotation profiles has significantly increased in the last years
\citep[see, e.g., ][and references therein]{anderson10, noyola10,
lane10b, bellazzini12, mcnamara12, lut13, fabricius14, kacharov14}.
In this context, interesting insights on specific dynamical processes
occurring in the central regions of some clusters have been obtained by
using ``exotic'' stellar populations, like millisecond pulsars and
blue straggler stars \citep[see][]{ferraro03b,fe09_m30,ferraro12}. However
proper VD and rotation profiles, especially in their innermost regions
where the presence of the long-searched IMBHs is expected to leave
characteristic signatures (as a VD cusp and systemic rotation;
\citealp{baumgardt05,miocchi07,einsel99}), are still badly
constrained. This is due to the observationally difficulties,
affecting both proper motion studies and the investigations of the
velocity line-of-sight component.

As for the latter, while determining the line-of-sight rotation curve
and VD profile in external galaxies is relatively simple (requiring
the measurement, respectively, of the Doppler shift and the broadening
of spectral lines in integrated-light spectra), it is much less
straightforward in resolved stellar populations as Galactic GCs.  In
these systems, in fact, the dominant contribution of a few bright
stars may artificially broaden the spectral lines, making the
resulting value a non-representative measure of the true VD of the
underlying stellar population \citep[this is commonly refereed as
``shot noise bias''; e.g.][]{dubath97,noyola10,lut11}. The
alternative approach is to measure the dispersion about the mean of
the radial velocities of statistically significant samples of
individual stars.  Clearly, this methodology is not prone to the
shot-noise bias, provided that the individual stars are well resolved,
sufficiently isolated and bright enough to be negligibly contaminated
by the unresolved stellar background.

The latter approach is becoming increasingly feasible thanks to the
current generation of adaptive-optics (AO) assisted spectrograph
with an integral field unit (IFU), and the improved data analysis techniques
(e.g. \citealp{lanzoni13}, hereafter L13; \citealp{kamann13}),
as clearly demonstrated by the case of NGC~6388.
The VD profile of this cluster derived from the line broadening of
integrated-light spectra shows a steep cusp with a central value of
23-25 km s$^{-1}$, which is best-fitted by assuming that an IMBH of
$2\times 10^4~M_\odot$ is hidden in the system \citep{lut11}.
Instead, if the radial velocities of individual stars are used, a
completely different result is found.  By using SINFONI, an
AO-assisted IFU spectrograph at the ESO-VLT, L13 measured the radial
velocity of 52 individual stars in the innermost $2\arcsec$ of the
cluster, finding a flat VD profile with a central value of only 13-14
km s$^{-1}$, which is well reproduced by no IMBH or, at most, a BH of
$\sim 2000~M_\odot$ (L13; see also \citealp{lanzoni07}).  As discussed in
detail in L13 (see their Sect. 4.1 and their Fig. 12), the integrated
light spectra measured in the innermost part of the cluster are
dominated by the contribution of two bright stars having opposite
radial velocities with respect to the systemic one, despite the
explicit effort by \citet{lut11} to correct for this. This produces a
spuriously large line broadening and a consequent overestimate of the
central VD value.

The results obtained in NGC 6388 clearly demonstrate the feasibility
of the individual radial velocity diagnostics and show that this is
indeed the safest way to measure the stellar VD in Galactic GCs. To
identify other multi-object facilities suitable for this kind of
approach, we took advantage of the new $K$-band Multi Object
Spectrograph \citep[KMOS;][]{sharples10}, recently commissioned at the
ESO-VLT.  During the instrument Science Verification (SV) run, under
proposal 60.A-9448(A) (PI: Lanzoni), we used KMOS multiple pointings
to investigate the region within $\sim 9\arcsec$ and $70\arcsec$ from
the center of NGC 6388. The results obtained from these observations
are the subject of the present paper, and they prompted us to
successfully apply for an ESO Large Program (193.D-0232, PI: Ferraro)
aimed at constructing a new generation of VD profiles for a
representative sample of Galactic GCs.

In Section \ref{obs} we describe the observations and data reduction
procedures.  Section \ref{KA} is devoted to the description of the
kinematic analysis, including a number of tests about the performances
of KMOS (Sect. \ref{test}), the discussion of the determination of the
radial velocities of individual stars (Sect. \ref{RVC}), and the
presentation of the derived VD profile (Sect. \ref{VD}). Discussion
and conclusions are presented in Section \ref{disc}.

%%%%%%%%%%%%%%%%%%%%%%%%%%%%%%%%%%%%%%%%%%%%%%%%%%%%%%%%%%% OBSERVATIONS

\section{Observations and data reduction}
\label{obs}

\begin{figure}[h]
\centering
\includegraphics[scale=0.6]{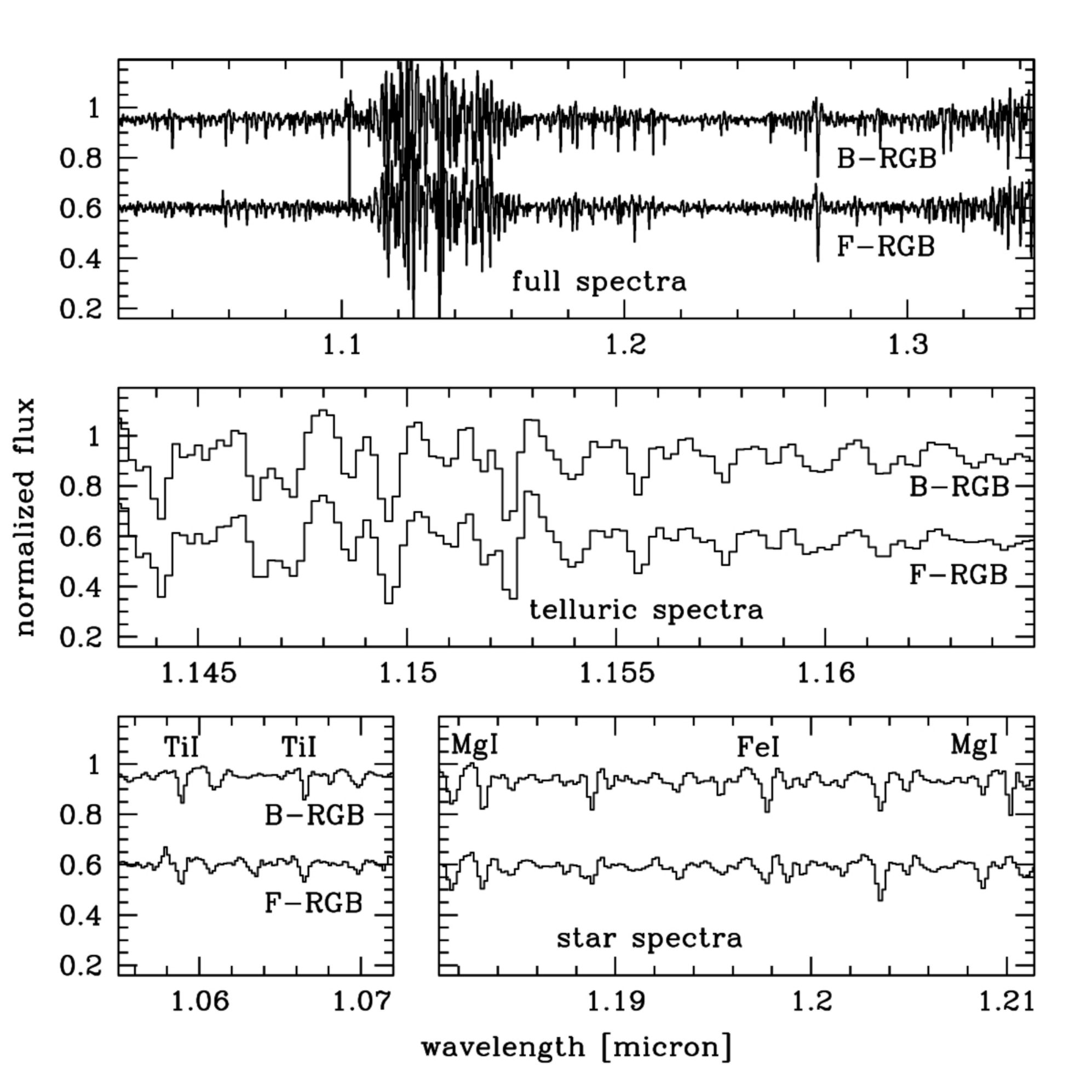}
\caption{An example of the observed KMOS YJ spectra of two giant
stars: a bright B-RGB (star \#245899, $J$ = 10.24 mag, top spectra) and a
faint F-RGB (star \#43138, $J$ = 12.90 mag, bottom spectra).
Top panel: observed spectra. Middle panel: zoomed spectra around 1.15
$\mu$m, including a few isolated telluric lines. Bottom panel:
zoomed spectra around 1.05 and 1.2 $\mu$m, including a few
isolated stellar features of interest.}
\label{spec}
\end{figure}

KMOS is a second generation spectrograph equipped with 24 IFUs that
can be allocated within a $7.2\arcmin$ diameter field of view.
Each IFU covers a projected area on the sky of about
$2.8\arcsec\times2.8\arcsec$, and it is sampled by an array of
14$\times$14 spatial pixels (hereafter spaxels) with an angular size
of $0.2\arcsec$ each.  The 24 IFUs are managed by three identical
spectrographs, each one handling 8 IFUs (1-8, 9-16 and 17-24,
respectively).  At the time of the observations discussed here, IFUs
\#13 and \#16 were not usable.  KMOS is equipped with four gratings
providing a maximum spectral resolution R between $\sim$ 3200 and 4200
over the 0.8-2.5 $\mu$m wavelength range.  We have used the YJ grating
and observed in the 1.00-1.35 $\mu$m spectral range at a resolution
R$\approx$3400, corresponding to a sampling of about 1.75 $\rm
\mathring{A}$ pixel$^{-1}$, i.e. $\sim$ 46 km s$^{-1}$ pixel$^{-1}$ at
1.15 $\mu$m.  This instrumental setup is especially effective in
simultaneously measuring a number of reference telluric lines in the
spectra of giant stars, for an accurate calibration of the radial
velocity, despite the relatively low spectral resolution.  An example
of the observed spectra is shown in Figure~\ref{spec}, with a zoom
around 1.15 $\mu$m to show some isolated telluric lines, and around
1.06 and 1.20 $\mu$m to show some stellar features of interest.

% FIGURE 1
%

The data presented here have been acquired during the KMOS SV, with
four different pointings on NGC6388.  The total on-source integration
time for each pointing was 3-5min and it has been obtained with three
sub-exposures of 60-100s each, dithered by $0.2\arcsec$ for optimal
flat-field correction. The typical signal-to-noise ratio (SNR) of the
observed spectra is $\gtrsim$ 50.  We used the ``nod to sky'' KMOS
observing mode and nodded the telescope to an off-set sky field at
$\approx 6\arcmin$ North of the cluster center, for a proper
background subtraction.

The spectroscopic targets have been selected from near-IR data
acquired with SOFI at the ESO-NTT \citep{valenti07}, based on the star
position in the color-magnitude diagrams (CMD) and the radial
distribution within the cluster.

We selected targets with $J < 14$ mag (in order to always have SNR $> 50$)
and sufficiently isolated, without stars brighter than 15 mag within $1\arcsec$ from
their center. We then used ACS-HST data in the $V$ and $I$ bands,
from \citet{lanzoni07b}, \citet{sarajedini07} and \citet{dalessandro08}, to
identify additional stars not present in the SOFI catalog.

% FIGURE 2
%
\begin{figure}[h]
\centering
\includegraphics[scale=0.8,angle=270]{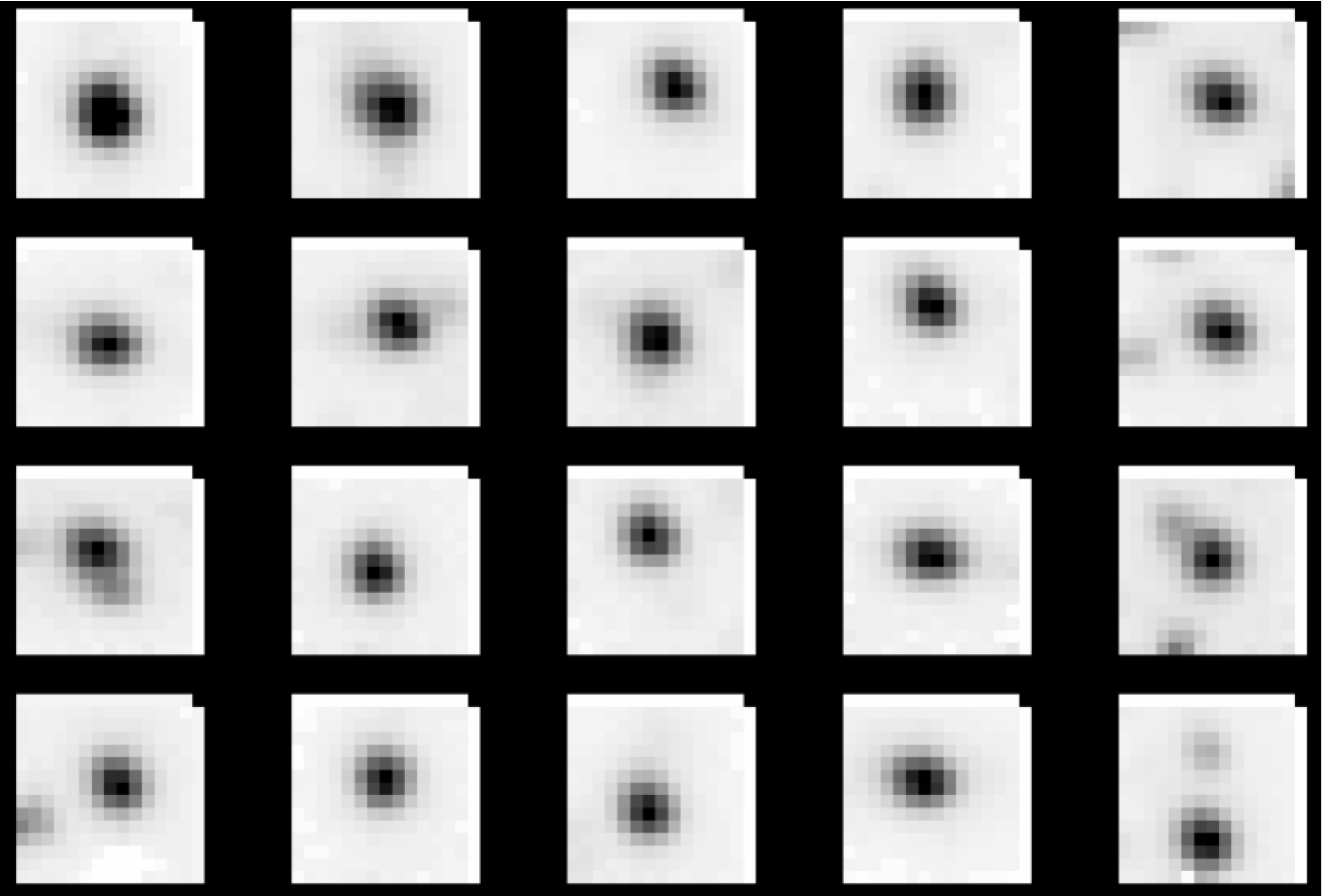}
\caption{Reconstructed images of 20 IFUs obtained during the
first pointing.  In some cases, other stars with a sufficient
spectral SNR and not contaminated by the main target can be
recovered from the same IFU.}
\label{recifu}
\end{figure}

The raw data have been reduced using the KMOS pipeline version 1.2.6,
which performs background subtraction, flat field correction and
wavelength calibration of the 2D spectra.  The 1D spectra have been
extracted manually by visually inspecting each IFU and selecting the
spectrum from to the brightest spaxel in correspondence of each target
star centroid, in order to minimize the effects of possible residual
contamination by nearby stars and/or by the unresolved stellar
background. An example of the reconstructed images of the
stars observed during the first pointing is shown in Figure~\ref{recifu}.

% FIGURE 3
%
\begin{figure}[h]
\centering
\includegraphics[scale=0.6]{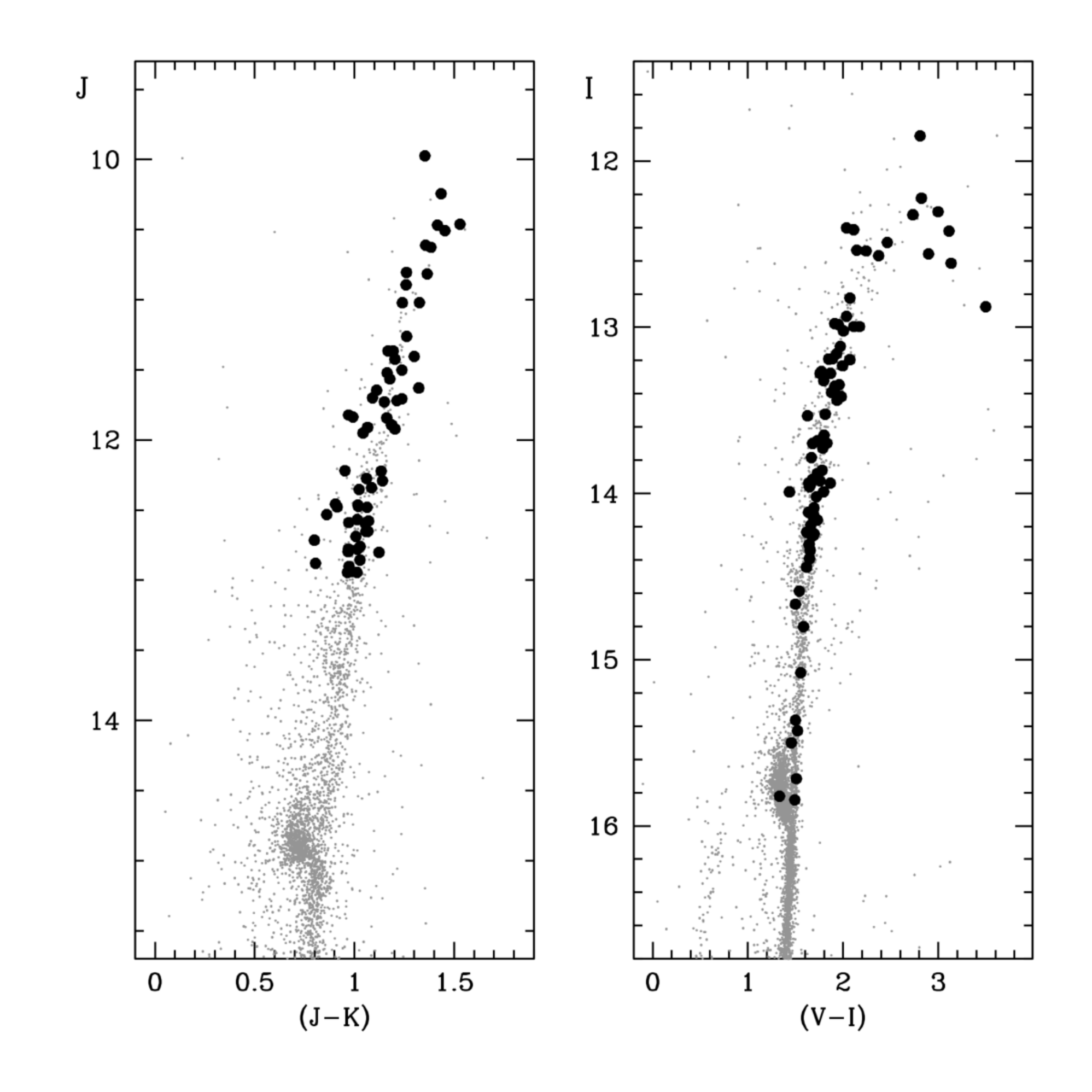}
\caption{The $(J,J-K)$ and $(I,V-I)$ color-magnitude diagrams (left
and right panels, respectively) of NGC6388, with highlighted the
KMOS targets.}
\label{cmd}
\end{figure}

\begin{figure}[h]
\centering
\includegraphics[scale=0.6]{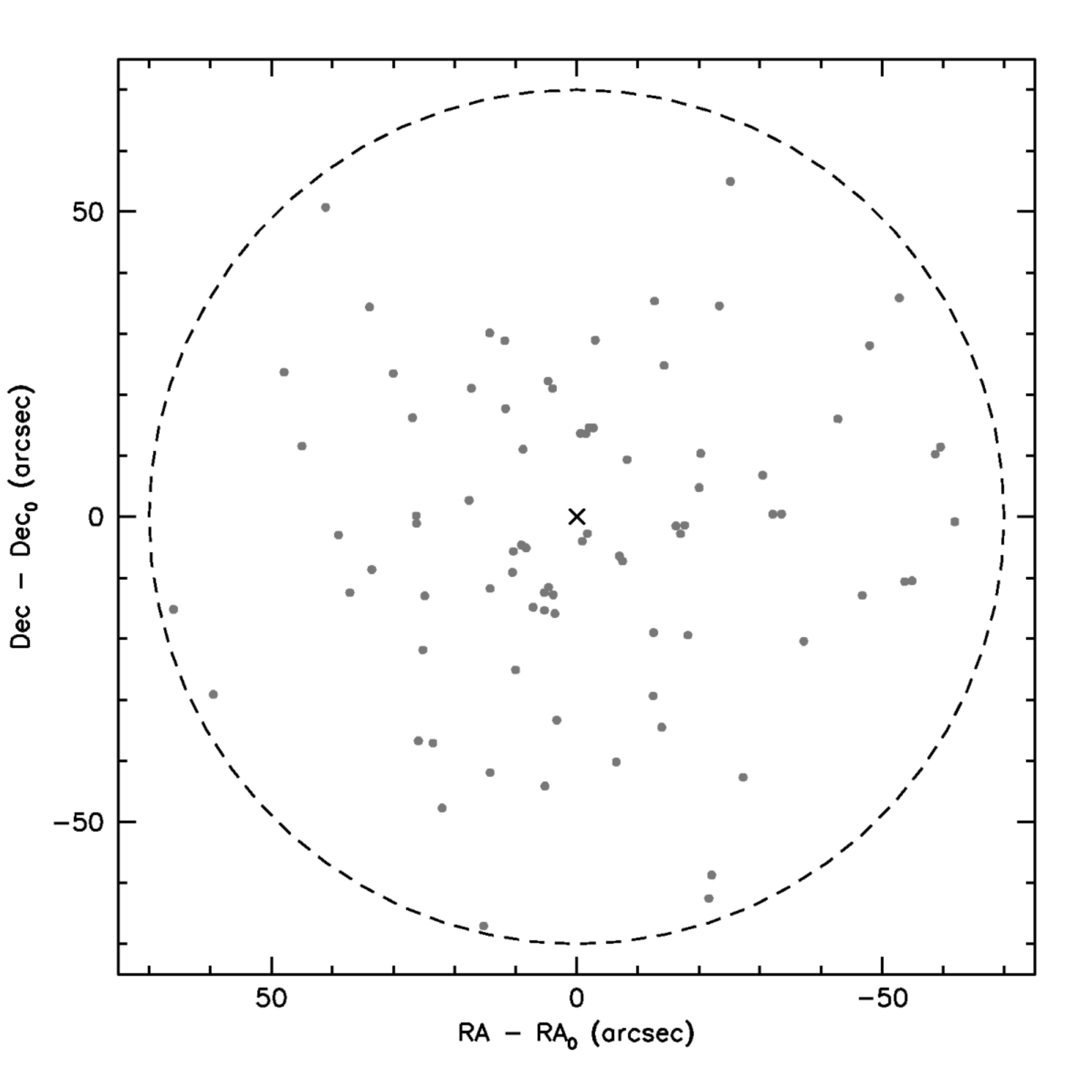}
\caption{The position of the observed target (gray circles) in the central region of NGC6388.
The black cross marks the center of the cluster as derived in \cite{lanzoni07b}
while the black annulus marks a region of $70 \arcsec$ of radius.}
\label{map}
\end{figure}

We measured a total of 82 giant stars located within $\sim 70\arcsec$
from the center of NGC6388. Figure~\ref{cmd} shows the position of the
targets in the the $(I,V-I)$ and $(J,J-K)$ CMDs, while
Figure~\ref{map} displays their location in the RA and Dec plane.
Identification number, coordinates and magnitudes
of each target are listed in Table~\ref{tabifutot} (the complete
version of the table is available in electronic form). Twelve stars
have been observed twice for cross-checking measurements from
different pointings/exposures.  Seven stars are in common with the
FLAMES-VLT radial velocity sample of \citeauthor{lut13} (2013, hereafter L13).
In a few cases, within a single KMOS IFU we could extract the spectra of more
than one star and measure their radial velocity (see Figure~\ref{recifu}).

% FIGURE 4
%
%
%

%%%%%%%%%%%%%%%%%%%%%%%%%%%%%%%%%%%%%%%%%%%%%%%%%%%%%%%%%%%%%%%%%%%%%%%%

\section{Kinematic analysis}
\label{KA}

To accurately measure the radial velocities of the observed targets we
made use of cross-correlation techniques with template spectra (in
particular we used the IRAF task FXCOR).  As telluric template we used
a high resolution spectrum of the Earth's telluric feature
\footnote{Retrieved from \url{http://www.eso.org/sci/facilities/paranal/decommissioned/isaac/tools/spectroscopic$\_$standards.html}.},
convolved at the KMOS YJ grating resolution.
As stellar templates we used synthetic spectra computed with the TURBOSPECTRUM code
\citep{alvarez98, pletz12}, optimized for cool giants.
We used a set of average templates with photospheric parameters representative
of those of the observed stars and [Fe/H] = --0.5 dex,
the metallicity of NGC6388 (see \citealt{harris96}, 2010 edition).
For a given star, we also checked the impact of using a different template
with varying the temperature by $\Delta T_{eff}$ $\pm$ 500 K and
the gravity by $\Delta log~g$ $\pm$ 0.5 dex and we verified that it 
has a negligible effect on the final radial velocity measurements ($<$ 1 km s$^{-1}$).

%%%%%%%%%%%%%%%%%%%%%%%%%%%%%%%%%%%%%%%%%%%%%%%%%%%%%%%%%%%%%%%%%%%%%%%%

\subsection{Accuracy of the wavelength calibration}
\label{test}

With the purpose of quantifying the ultimate accuracy of the radial
velocity measurements of individual giant stars in crowded fields, we
performed a number of tests aimed at checking the reliability and
repeatability of the wavelength calibration of each KMOS IFU.  Since
KMOS is mounted at a Nasmyth focus and rotates, some flexures are
expected, with impact on the overall spatial and especially spectral
accuracy of the reconstructed 2D spectra.

The KMOS Data Reduction Software (DRS) pipeline allows to take
calibration exposures at several rotator angles and to choose the
frames with the rotator angle closest to the one of the input science
frame and eventually interpolate.  The KMOS DRS pipeline has also the
option of refining the wavelength solution by means of the observed OH
lines. We reduced the spectra by selecting all these options to obtain
the best possible accuracy in the spectral calibration.

% FIGURE 5
%
\begin{figure}[h]
\centering
\includegraphics[scale=0.35]{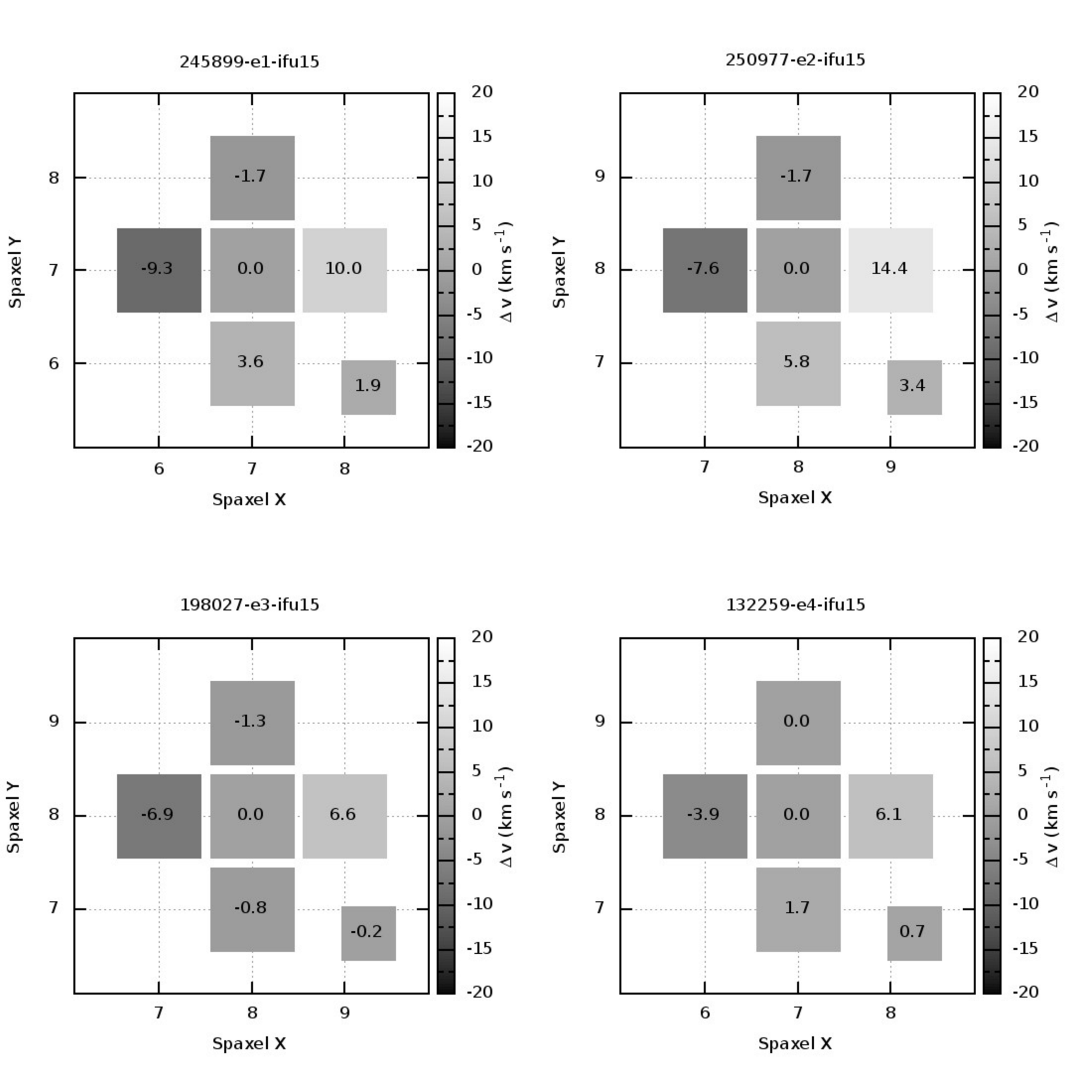}
\caption{The spaxel cross-shaped matrixes used in the wavelength
calibration test. The matrixes are centered on the brightest spaxel
of stars \#245899, \#198027, \#250977 and \#132259 observed with the
IFU \#15 during pointings 1, 2, 3, and 4, respectively.  The number
marked in each spaxel refers to the velocity shift (in km s$^{-1}$)
with respect to the central spaxel, as measured by cross-correlating
telluric lines.  The small square in the bottom-right corner of each
matrix marks the velocity shift with respect to the central spaxel,
as obtained directly from cross-correlating the combined (from the
five spaxels in the cross) spectrum.}
\label{ifu15tell}
\end{figure}

However, residual velocity shifts in different spaxels of a given IFU,
as well as in different IFUs, are still possible.  In order to measure
these residual shifts, we selected the spectral region between 1.14
and 1.16 $\mu$m (see Figure~\ref{spec}) containing telluric lines
only, and we cross-correlated the observed spectra from five different
spaxels in a given IFU with the telluric template.  The five spaxels
are the ones where the star centroid is located (having the highest
signal) and the four surrounding (cross-shape) spaxels.  We then
computed the residual wavelength/velocity shifts of the four
surrounding spaxels with respect to the central one used as
reference. As an example, Figure~\ref{ifu15tell} shows the results for
IFU \#15: four different stars (\#245899, \#250977, \#198027 and
\#132259) have been observed in four pointings.  The measured zero
point shifts are normally well within $\pm$10 km s$^{-1}$,
corresponding to 1/4 of a pixel at the spectral resolution of the KMOS
YJ band\footnote{At this resolution, one pixel corresponds to $\sim 46$ km s$^{-1}$.},
with average values of a few km s$^{-1}$ and corresponding dispersions within 10 km s$^{-1}$
(see Table~\ref{tabifu15}).  For each IFU, we finally combined the spectra
from the five spaxels, by using the IRAF task SCOMBINE, and we
measured the radial velocity in the resulting combined one.  The
obtained values (see Figure~\ref{ifu15tell}) are fully consistent with
the average values from individual spaxels (see Table~\ref{tabifu15}).

% FIGURE 6
%
\begin{figure}[h]
\centering
\includegraphics[scale=0.35]{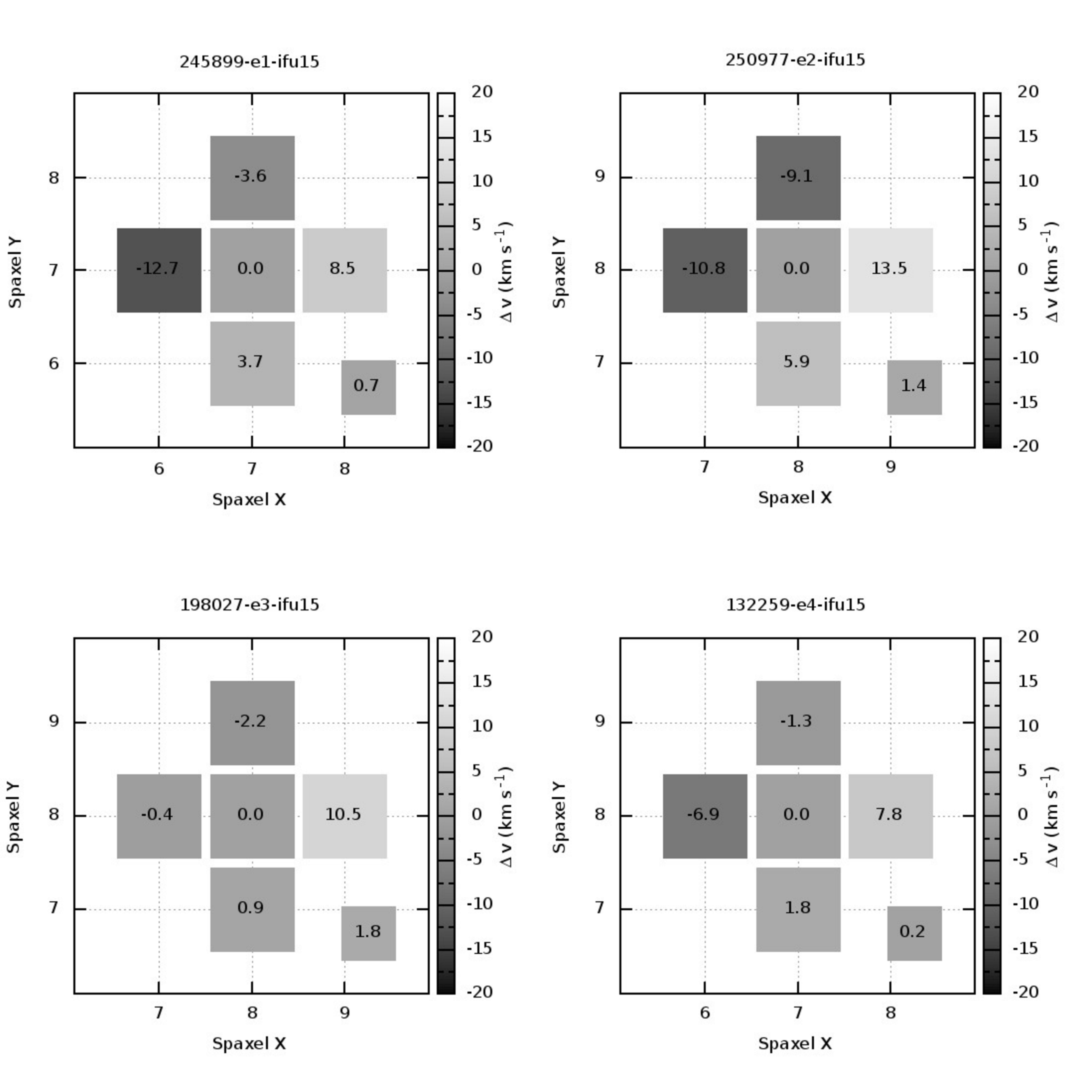}
\caption{As in Figure~\ref{ifu15tell}, but for the velocity shifts
measured by cross-correlating stellar lines.}
\label{ifu15stel}
\end{figure}

For the same four stars observed with IFU \# 15, we also used five,
isolated stellar lines in two spectral regions centered at 1.06 and
1.20 $\mu$m (see Figure~\ref{spec}) to compute the velocity shifts.
Also in this case, for each star we extracted the spectrum of the
spaxel with the highest signal and the spectra of the surrounding
(cross-shape) spaxels.  We cross correlated them with suitable
synthetic spectra having photospheric parameters as those of the
target stars.  The resulting velocity shifts with respect to the
central reference spaxel have been plotted in Figure~\ref{ifu15stel},
while the average values are listed in Table~\ref{tabifu15}.  The
inferred average values and dispersions are fully consistent (at
better than 1/10 of a pixel) with those obtained measuring the
telluric lines.  For each star, we finally combined the spectra from
the five spaxels as done in the first test and we measured the radial
velocity in the resulting spectra.  The obtained values (see
Figure~\ref{ifu15tell}) are fully consistent with the average values
from individual spaxels (see Table~\ref{tabifu15}), as well as with
the shifts measured with the telluric lines.

We repeated the same tests by using other stars observed by different
IFUs in different pointings.
As an example, Table~\ref{tabifu15} also reports the average shifts for other three stars,
namely $\#$124271, $\#$325164 and $\#$216954 observed by the IFUs $\#$20, $\#$11 and $\#$2 and
collected during the pointings \#1, \#2 and \#4, respectively.
We found values very similar to those derived for the IFU $\#$15,
thus ensuring that the overall wavelength
calibration provided by the KMOS DRS pipeline is normally accurate and
stable in time at a level of a fraction (on average, within 1/10) of a
pixel, both within each IFU and among different IFUs.

% FIGURE 7
%
\begin{figure}[h]
\centering
\includegraphics[scale=0.7]{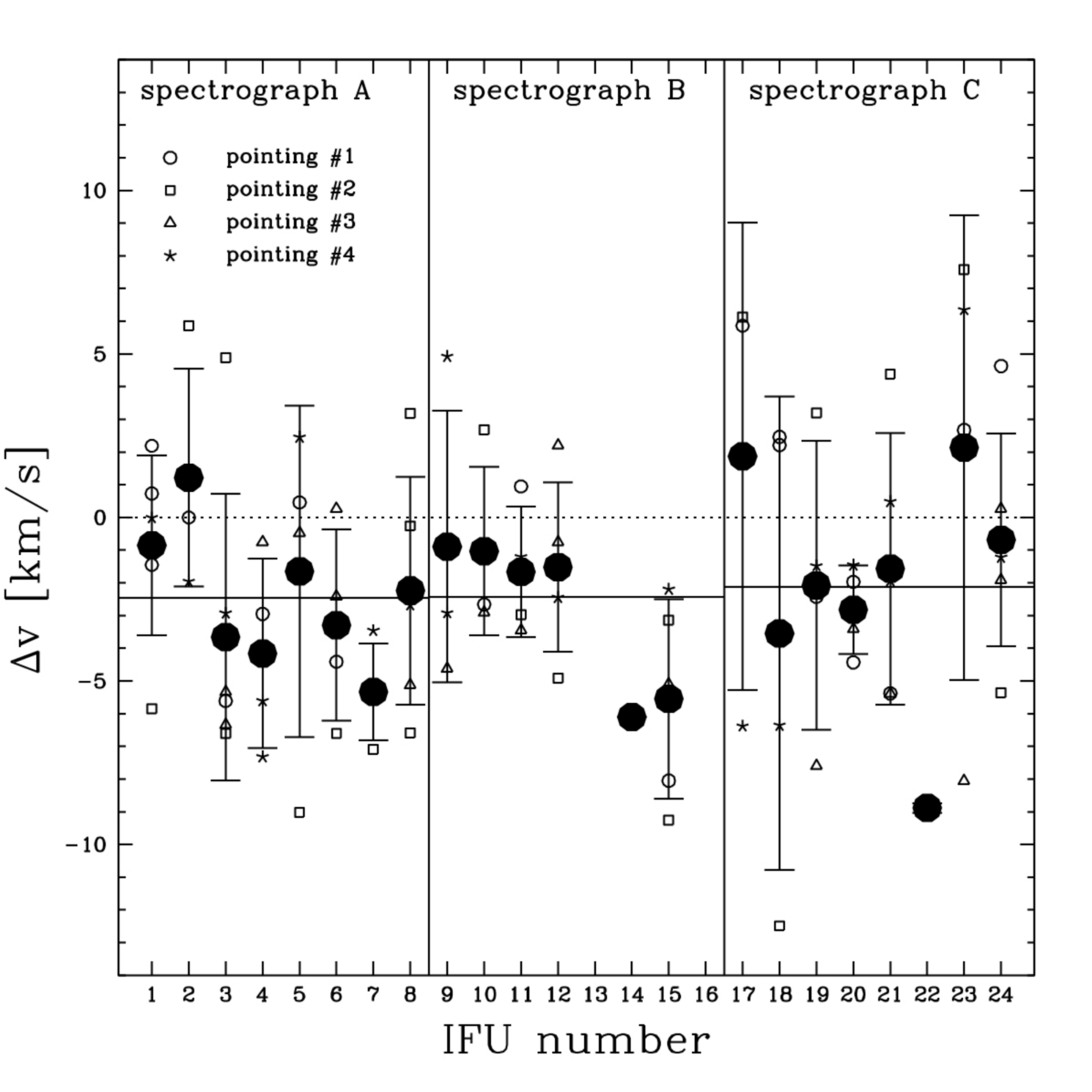}
\caption{The telluric velocity shifts (in units of km s$^{-1}$) of the
spectra extracted from the spaxel with highest signal
for all the target stars observed with the KMOS IFUs (small symbols).
The shifts have been computed by using the telluric template as reference.
The large dots mark the average values and the 1$\sigma$ dispersion as
measured for each IFU, while the horizontal, continuum lines mark
the average values for each of the three spectrographs.
IFUs \#13 and \#16 were not usable during those observations.}
\label{ifutot}
\end{figure}

The tests performed so far, by using both telluric and stellar lines, 
have demonstrated that the velocity shifts between the spectra extracted from the spaxel 
with the highest signal and those obtained by combining the spectra from
the cross-shape spaxels are fully consistent one to each other,
thus we decided to use the spectra of the brightest spaxel only,
which is definitely dominated by the target light.
Hence, as a final wavelength calibration check, we selected the
spectrum corresponding to the spaxel with the highest signal in each
observed star. We then cross-correlated this spectrum with the
telluric template as reference, and computed the residual velocity
shift. Figure~\ref{ifutot} shows the results for the active KMOS
IFUs. We find that for a given IFU, the residual velocity shifts as
measured in different stars observed during different
pointings/exposures are normally consistent to each other with an
average dispersion of 3.4 km s$^{-1}$.  Such a dispersion is
relatively small, taking into account that the four exposures on
NGC6388 were obtained with KMOS at very different rotation angles
with respect to the Nasmyth axis (261$\rm ^o$, 188$\rm ^o$, 326$\rm
^o$, and 97$\rm ^o$ in pointing \#1, \#2, \#3, and \#4, respectively),
indicating that the KMOS optimized calibration procedure is effective
in correcting the effects of spectral flexures.

Since the 24 IFUs of KMOS are managed by three separate spectrographs,
one can also compute the mean shift of each spectrograph, by averaging
the mean shifts from IFUs \#1 to \#8, \#9 to \#16, and \#17 to \#24,
respectively.  We find very similar residual velocity shifts, of
$\approx$-2 km s$^{-1}$ and dispersion of 3-5 km s$^{-1}$.

\subsection{Radial velocity measurements}
\label{RVC}

The tests described in Section~\ref{test} indicate that the wavelength
calibration provided by the KMOS pipeline is well suited for
kinematic studies of extragalactic sources.  However, for
precise radial velocity measurements of individual stars, it is
necessary to refine the calibration, by correcting each spectrum for
the corresponding residual velocity shift, as inferred from the
telluric lines. 

Once corrected for such a residual shift, the radial velocity of each
star was computed by cross-correlating the observed spectra with
suitable synthetic ones.  We finally applied the heliocentric
correction by using the IRAF task RVCORRECT.
The final radial velocity errors have been computed from
the dispersion of the velocities derived from each line divided by
the number of lines used (that is $eV_{r}$ = $\sigma / \sqrt{N_{lines}}$).
The average uncertainty in the velocity estimates is 2.9 km s$^{-1}$.

% FIGURE 8
%
\begin{figure}[h]
\centering
\includegraphics[scale=0.7]{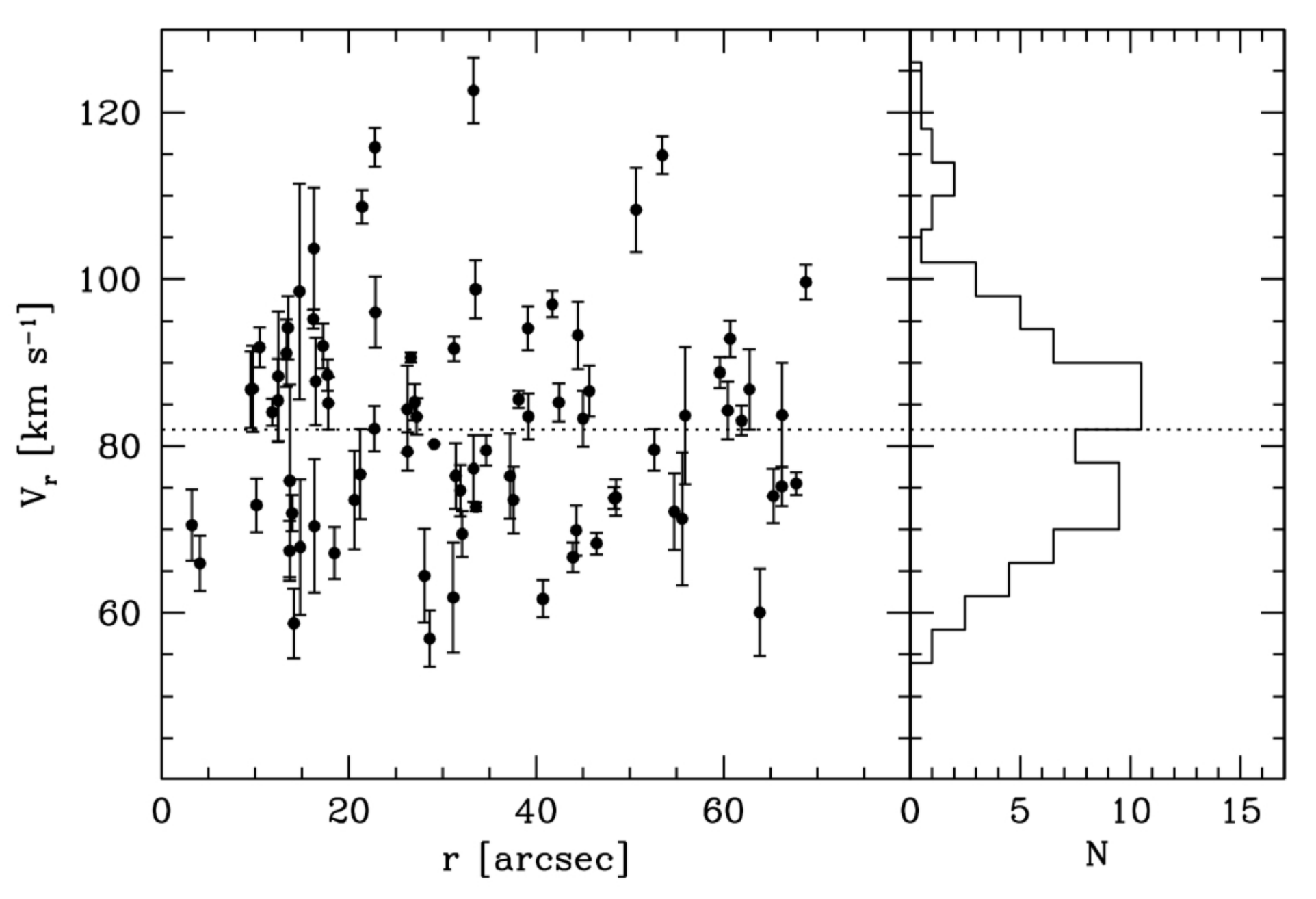}
\caption{Radial velocities as a function of the distance from the
cluster center (left panel) and histogram of their distribution
(right panel), for the 82 stars of NGC6388 observed with KMOS.
The dotted line marks the systemic velocity of the
cluster (82 km s$^{-1}$, from L13).}
\label{rvtot}
\end{figure}

Figure~\ref{rvtot} shows the inferred radial velocities as a function
of the radial distance from the cluster center and the histogram of their distribution.
Table~\ref{tabifutot} lists the radial velocity values and
corresponding errors, as well as the KMOS IFU and pointing reference
numbers.  For the twelve stars observed twice, the average value of
the measured radial velocities has been adopted. For these stars the measured radial
velocities are in excellent agreement, with an average difference of
$\sim 1$ km s$^{-1}$ between two different exposures and a dispersion
of 3.6 km s$^{-1}$. 

Only seven stars have been found in common with the FLAMES sample
of L13, and the inferred radial velocities from the KMOS spectra turn out to be
in good agreement with the FLAMES ones. In fact by applying a $2-\sigma$ 
rejection criterion, an average difference of
$\langle \Delta \rm V_{KMOS-FLAMES} \rangle$ = --0.2 km s$^{-1}$
($\sigma$ = 2.2 km s$^{-1}$) is found.

We computed the systemic velocity of the KMOS sample by conservatively
using all stars with radial velocities between 60 and 105 km s$^{-1}$,
as done in L13 for the SINFONI and FLAMES samples.
In this velocity range, 75 stars are counted, representing 91$\%$ of the entire KMOS sample.
We found $81.3 \pm 1.5$ km s$^{-1}$, in very good agreement with the value of $82.0 \pm
0.5$ km s$^{-1}$ found by L13 and indicating that all samples are
properly aligned on the same radial velocity scale.

\subsection{Line-of-sight rotation and velocity dispersion profiles}
\label{VD}

To compute the projected rotation and VD profiles from the measured
radial velocities of individual stars we adopted the same approach
described in L13. All the 82 KMOS targets have been considered as
cluster members, since they all have radial velocities between 50 and
130 km s$^{-1}$, which has been adopted as cluster membership
criterion in L13.

To study the possible presence of a rotation signal, we restricted the
analysis to the sample of targets providing a symmetric coverage of
the surveyed area, namely 52 stars located between $9\arcsec$ and
$40\arcsec$ from the centre. For further increasing the sample size,
we took into account 6 additional stars in the same radial range from
the L13 data-set.  We then used the method described in \citeauthor{bellazzini12}
(2012, and references therein; see also L13). No significant rotation
signal has been found from this sample.  Interestingly, however, the
distribution of radial velocities for stars within $20\arcsec$ from
the cluster centre is clearly bimodal (see Figure~\ref{rvtot}), thus
suggesting the possible presence of ordered rotation. Unfortunately
only 23 stars have been measured within this radial range and more
data are needed before drawing any firm conclusion about 
ordered rotation in the central regions of NGC6388 (see also L13).

% FIGURE 9
%
\begin{figure}[h]
\centering
\includegraphics[scale=0.6]{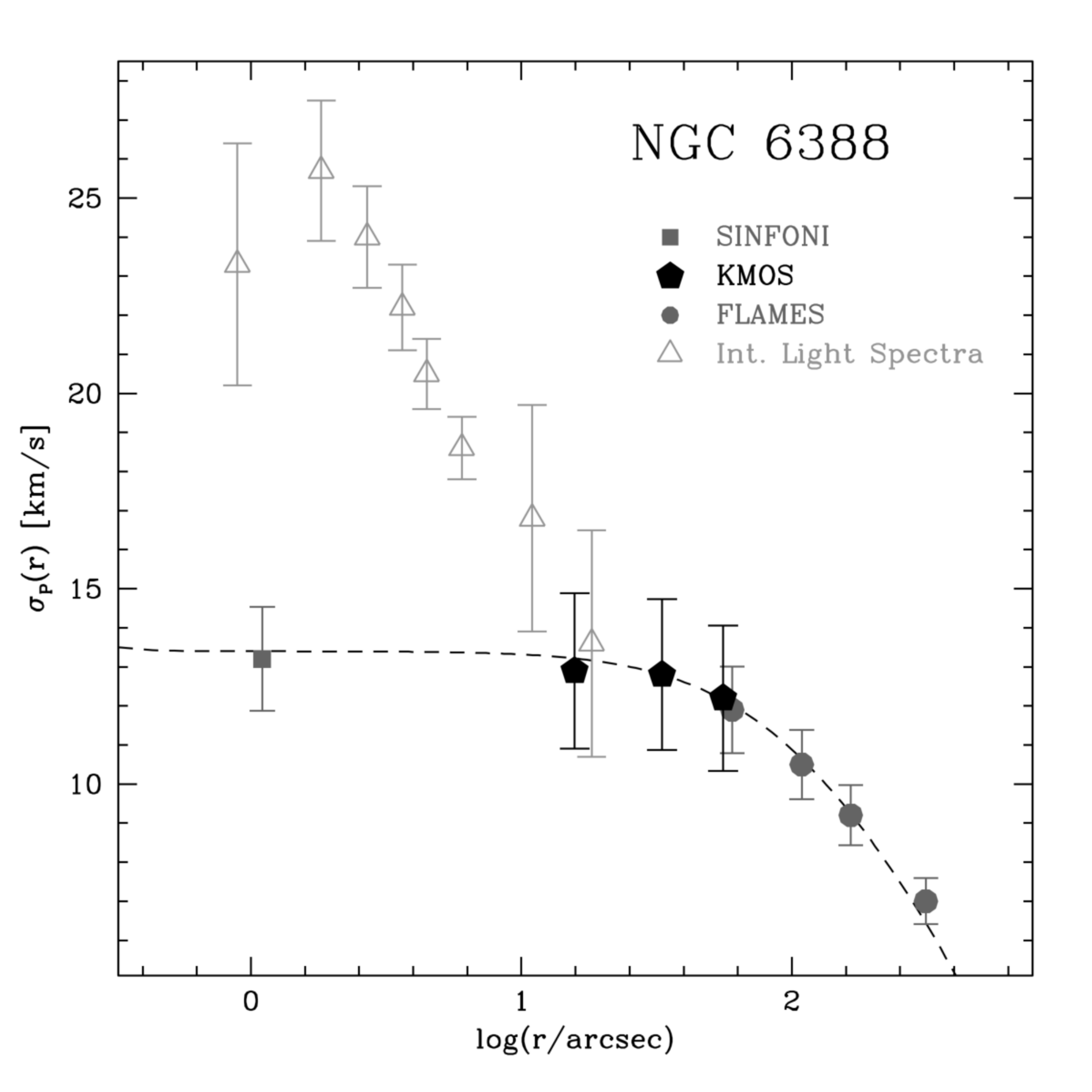}
\caption{Line-of-sight velocity dispersion profile of NGC6388
computed from the radial velocities of individual stars, as measured
with KMOS (black pentagons, this paper), SINFONI and FLAMES (dark
grey squares and circles, respectively; from L13). The dashed line
correspond to the self-consistent King model plotted in Figure 13 of
L13. The velocity dispersion profile obtained from integrated-light
spectra \citep{lut11} is also shown for comparison (light grey empty
triangles).}
\label{vdisp}
\end{figure}

To compute the VD profile we used the entire KMOS sample (but the two
innermost targets at $r<5\arcsec$ have been conservatively excluded)
and divided the surveyed area in three radial bins,
each containing approximately the same number of
stars: namely $9\arcsec\le r\le 23\arcsec$ (29 stars), $23\arcsec\le
r\le 43\arcsec$ (26 stars), and $43\arcsec\le r\le 70\arcsec$ (25
stars).  The values obtained are $12.9 \pm 2.0$ km s$^{-1}$, at an
average distance of $16\arcsec$, $12.8 \pm 1.9$ km s$^{-1}$ at
$r=33\arcsec$, and $12.2\pm 1.9$ km s$^{-1}$ at $r=56\arcsec$.
The errors have been estimated by following \citealt{pryor93}.
The corresponding profile is plotted in Figure~\ref{vdisp}.

\section{Discussion and conclusions}
\label{disc}

The velocity dispersion values obtained with KMOS are presented in Figure~\ref{vdisp}.
We have included for comparison the measurements obtained with SINFONI
and FLAMES from L13 and those derived from integrated-light spectra by \citet{lut11}.
We note that the outermost point of the KMOS VD profile well matches the innermost
FLAMES measure of L13. At the same time, the innermost point of the
KMOS profile is also consistent with the most external value of \citet{lut11}.
Overall, the three new KMOS measurements allow us to sample the velocity
dispersion profile in the spatial region between 9$\arcsec$ and 70$\arcsec$,
and better define the knee of the distribution around 40$\arcsec$ from the
cluster center.

Unfortunately, both crowding and mechanical constraints did not allow
us to allocate more than 1-2 KMOS IFUs per pointing in the very
central region, i.e. at $r\le 9\arcsec$.
Hence, given the limited amount of observing time
during the SV run, only a few stars have been measured in the
innermost region, preventing us to compute a precise VD value closer
to the center.

% FIGURE 10
%
\begin{figure}[h]
\centering
\includegraphics[scale=0.6]{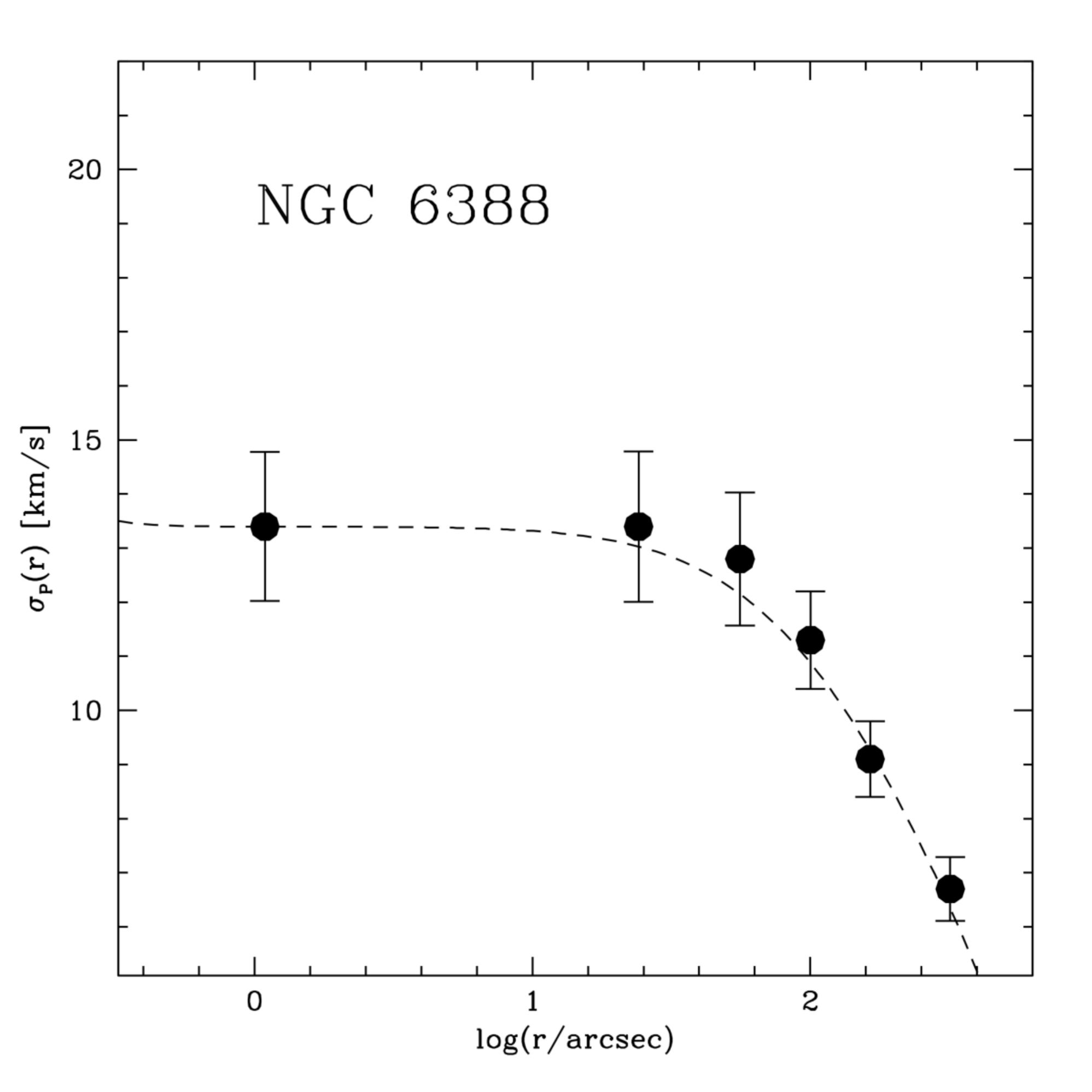}
\caption{Final velocity dispersion profile of NGC6388, obtained from
the combined sample of individual radial velocities, as measured
from SINFONI, KMOS and FLAMES spectra. The dashed line is as in
Figure~\ref{vdisp}.}
\label{vdisp_final}
\end{figure}

The final velocity dispersion profile of NGC6388, obtained from the combination of the
entire sample (namely, SINFONI, KMOS and FLAMES spectra) is presented
in Table~\ref{final_vd} and shown in Figure~\ref{vdisp_final}.

The results presented here demonstrate the effectiveness of an IFU
facility to perform multi-object spectroscopy of individual stars even
in dense stellar systems and at a modest spectral resolution.  The
KMOS deployable IFU is especially useful in studying the internal
kinematics of GCs for a number of reasons: 1) it allows to sample
stars over a rather large and tunable field of view, according to the
cluster central density and extension; 2) it allows to measure
individual stars and their surroundings, avoiding slit losses and
best-accounting for possible blending effects due to crowding and
unresolved stellar background; 3) it covers a rather wide spectral
range in a single exposure, to simultaneously record stellar features
and telluric lines, and measure accurate radial velocities even at a
spectral resolution R~$\approx$3000.

%%%%%%%%%%%%%%%%%%%%%%%%%%%%%%%%%%%%%%%%%%%%%%%%%%%%%%%%%%%%%%%%%%%%%%%% TABLES

\begin{landscape}
\begin{deluxetable}{cccccccccrc}
\tablewidth{0pt}
\tablecaption{Individual giant stars in NGC6388 observed with KMOS.}
\tablewidth{0pt} 
\tablehead{ 
\colhead{star \#} & 
\colhead{RA (2000)}&
\colhead{Dec (2000)}&
\colhead{$V$}&
\colhead{$I$}&
\colhead{$J$}&
\colhead{$K$}&
\colhead{$V_r$}&
\colhead{$eV_r$}&
\colhead{IFU \#}&
\colhead{pointing} 
}
\startdata
 28786 & 264.0777424 & -44.7539232 &  15.12 &  13.32 &  11.91 &  10.84 &  99.7 &  2.1 & 18 &  4 \\ 
 43138 & 264.0633117 & -44.7526742 &  15.85 &  14.19 &  12.90 &  11.93 &  75.2 &  2.4 & 23 &  1 \\ 
 43163 & 264.0631292 & -44.7516067 &  15.64 &  13.86 &  12.48 &  11.42 &  86.8 &  4.8 &  4,19 &  2,3 \\ 
 78741 & 264.0976030 & -44.7395056 &  14.94 &  12.57 &  10.81 &   9.54 &  75.5 &  1.4 &  1,11 &  2,3 \\ 
 81865 & 264.0950571 & -44.7433792 &  15.94 &  14.25 &  12.78 &  11.81 &  83.7 &  6.2 & 10 &  3 \\ 
 93646 & 264.0870297 & -44.7361283 &  15.40 &  13.42 &  11.92 &  10.72 &  94.1 &  2.6 &  6 &  1 \\ 
 94946 & 264.0863051 & -44.7387495 &  15.09 &  13.12 &  11.57 &  10.39 &  83.5 &  2.8 &  5 &  1 \\ 
 96560 & 264.0849040 & -44.7377027 &  14.52 &  12.41 &  10.89 &   9.63 &  79.5 &  1.8 &  2 &  2 \\ 
100296 & 264.0711060 & -44.7360611 &  15.80 &  13.94 &  11.63 &  10.31 &  70.5 &  4.3 &  3 &  3 \\ 
101667 & 264.0820265 & -44.7355902 &  16.16 &  14.67 &   0.00 &   0.00 &  84.4 &  5.2 &  3 &  2 \\ 
\enddata
\tablecomments{Identification number, coordinates, optical and NIR
magnitudes, radial velocities ($V_r$) and errors ($eV_r$) in km
s$^{-1}$, KMOS IFU and pointing numbers (two values are marked for the twelve
targets observed twice). The full table is available in the online
version of the paper.}
\label{tabifutot}
\end{deluxetable}
\end{landscape}

\begin{deluxetable}{crrcc}
\tablewidth{0pt}
\tablecaption{Results of the wavelength calibration tests.}
\tablewidth{0pt}
\tablehead{ 
\colhead{star \#} & 
\colhead{$\langle \Delta v_{cross}^{telluric}\rangle$} & 
\colhead{$\langle\Delta v_{cross}^{stellar}\rangle$} &
\colhead{IFU \#} &
\colhead{pointing}
}
\startdata
%\multicolumn{14}{l}{}\\
245899 & $+0.7 (8.2)$ & $-1.0 (9.2)$ & 15 & 1\\
250977 & $-0.6 (5.5)$ & $+2.2 (5.7)$ & 15 & 2\\
198027 & $+2.7 (9.5)$ & $-0.1 (11.8)$ & 15 & 3\\
132259 & $+1.0 (4.1)$ & $+0.3 (7.1)$ & 15 & 4\\
 & & & &\\
124271 & $+0.3 (8.2)$ & $+0.7 (9.7)$ & 20 & 1\\ % #83 %
325164 & $+0.4 (8.2)$ & $+1.3 (11.5)$ & 11 & 2\\ % #18 %
216954 & $-0.1 (5.7)$ & $-0.5 (7.8)$ & 2 & 4\\ % #8 %
\enddata
\tablecomments{Average velocity shifts and dispersion (in bracket)
among different spaxels distributed in cross-shaped matrixes, with
respect to the reference central spaxel.  Velocity in km s$^{-1}$.}
\label{tabifu15}
\end{deluxetable}

\begin{deluxetable}{rrrcrr}
\tablecaption{Velocity dispersion profile of NGC6388}
\tablewidth{0pt} 
\tablehead{ 
\colhead{$r_i$} & 
\colhead{$r_e$} &
\colhead{$r_m$} &
\colhead{$N_\star$} &
\colhead{$\sigma_P$} &
\colhead{$e_{\sigma_P}$}}
\startdata
  0.2 &   1.9 &   1.1 & 51 & 13.40 & 1.38 \\ 
  9.0 &  40.0 &  24.1 & 58 & 13.40 & 1.39 \\ 
 40.0 &  75.0 &  55.9 & 57 & 12.80 & 1.23 \\ 
 75.0 & 130.0 & 100.2 & 81 & 11.30 & 0.90 \\ 
130.0 & 210.0 & 164.2 & 84 &  9.10 & 0.70 \\ 
210.0 & 609.0 & 318.8 & 67 &  6.70 & 0.59 \\ 
\enddata
\tablecomments{The final profile has been obtained from the combined
sample of SINFONI, KMOS and FLAMES spectra. The three first columns
give the internal, external and mean radii (in arcseconds) of each
considered radial bin ($r_m$ is computed as the average distance from
the centre of all the stars belonging to the bin), $N_\star$ is the
number of star in the bin, $\sigma_P$ and $e_{\sigma_P}$ are the
velocity dispersion and its rms error (in km s$^{-1}$), respectively.}
\label{final_vd}
\end{deluxetable}

%% file: ap2/ap_2.tex
% APPENDIX 2

\chapter{Lines and Continuum Sky Emission in the Near Infrared: 
Observational Constraints from Deep High Spectral Resolution Spectra 
with GIANO-TNG}

\label{a2}

\newcommand{\ETACOLD}{\eta_{rc}}
\newcommand{\ETAHOT}{\eta_{rh}}
\newcommand{\TCOLD}{T_{rc}}
\newcommand{\THOT}{T_{rh}}

\setcounter{table}{2}

{\bf Published in Oliva et al. 2015, A$\&$A, 581, A47\\}

{\it Aim - Determining the intensity of lines and continuum airglow emission in the 
H-band is important for the design of faint-object infrared
spectrographs. Existing spectra at low/medium resolution cannot
disentangle the true sky-continuum from instrumental effects (e.g.
diffuse light in the wings of strong lines).
We aim to obtain, for the first time, a high resolution
infrared spectrum deep enough to set significant constraints on the 
continuum emission between the lines in the H-band.

Methods - During the second commissioning run of the GIANO high-resolution 
infrared spectrograph at the La Palma Observatory,
we pointed the instrument directly to the sky and obtained 
a deep spectrum that extends from 0.97 to 2.4 $\mu$m.

Results - The spectrum shows about 1500 emission lines, a factor of two more than in
previous works. Of these, 80\% are identified as 
OH transitions; half of these are from highly excited molecules (hot-OH 
component) that are not included in the
OH airglow emission models normally used for astronomical applications.
 The other lines are attributable to O$_2$ or unidentified.
Several of the faint lines are in spectral regions that 
were previously believed to be free of line emission.
The continuum in the H-band is marginally detected
at a level of about 300 photons/m$^2$/s/arcsec$^2$/$\mu$m,
equivalent to 20.1 AB-mag/arcsec$^2$.
The observed spectrum and the list of observed sky-lines are published in electronic
format.

Conclusions - Our measurements indicate that the sky continuum in the H-band 
could be even darker than previously believed. However, the myriad
of airglow emission lines severely limits the spectral ranges where very
low background can be effectively achieved with low/medium resolution 
spectrographs.
We identify a few spectral bands that could still remain quite dark
at the resolving power foreseen for VLT-MOONS (R$\simeq$6,600).
}

%%%%%%%%%%%%%%%%%%%%%%%%%%%%%%%%%%%%%%%%%%%%%%%%%%%%%%%%%%% INTRO

\section{Introduction}

The sky emission spectrum at infrared wavelengths and up to 1.8 $\mu$m
(Y, J, H bands) is dominated by lines (airglow) emitted by OH and
O$_2$ molecules; see e.g. \cite{Sharma}. 
These lines are intrinsically very narrow and, when observed at a high enough spectral resolution,
they occupy only a small fraction of the spectrum.
Therefore, by filtering the lines out, one
could in principle decrease the sky background by orders of magnitudes,
down to the level set by the sky continuum emission in between the lines.
This apparently simple idea, often reported as "OH sky-suppression",
has fostered a long and active field of research; see e.g. \cite{Oliva92},
\cite{Maihara93}, \cite{Herbst94}, \cite{Content96}, \cite{Ennico98},
\cite{cuby00}, \cite{Rousselot00}, \cite{Iwamuro01}, \cite{blandhawthorn04},
\cite{Iwamuro06}, \cite{ellis12}, \cite{Trinh13}.
However, in spite of the intense work devoted to measuring and modelling
the properties of the sky spectrum, it is still
not clear what is the real level of the sky continuum in between the
airglow lines in the H-band (1.5-1.8 $\mu$m).

A detailed study of the infrared sky continuum emission 
was recently reported by \cite{Sullivan}. Using
spectra at a resolving power R=6,000 they were able
to correct the spectra for all instrumental effects and derive accurate
measurements of the sky continuum at wavelengths shorter than 1.3 $\mu$m 
(Y, J bands). However, they could not obtain precise results
in the H-band (1.5-1.8 $\mu$m) because the sky continuum is well below the
light diffused in the instrumental wings of the airglow lines. 
This problem was already noted in earlier works.
In particular, \cite{blandhawthorn04} claimed that the continuum level
between the OH lines could be as low as the zodiacal light level and
much lower than that measurable with classical (i.e. not properly 
OH suppressed) spectrographs.
This claim was later retracted by \cite{ellis12} after 
measuring the interline continuum with an optimised
OH-suppression device based on a Bragg fibre grating. 
\cite{Trinh13} subsequently attempted to model
the interline continuum based on spectral models and measurements that did 
not reach the depth and completeness of the data presented in this paper.

The net - and somewhat surprising - result is that so far nobody has been 
able to improve the earliest measurements of \cite{Maihara93} who reported a continuum emission
of 590 photons/m$^2$/s/arcsec$^2$/$\mu$m measured at 1.665 $\mu$m
(equivalent to 19.4 AB-mag/arcsec$^2$)
using a spectrometer with resolving power R=17,000,
equipped with one of the first-generation 256$^2$ HgCdTe infrared detectors

A proper understanding of the line and continuum emission from the sky
is of fundamental importance when designing new infrared spectrographs
optimised for observations of very faint targets. A representative case
is that of MOONS, the multi-objects optical and near infrared spectrometer
for the VLT, see \cite{cirasuolo11,cirasuolo14}.
This instrument includes an arm covering the H-band at a resolving power
R$\simeq$6,600. The requirements on instrumental background and stray light 
strongly depend on the sky continuum one assumes, see \cite{licausi} for details.

In a previous work \citep{paper1} we presented,
for the first time, observations of the infrared sky spectrum
at high spectral resolution and covering a very wide spectral range.
The spectrum revealed 750 emission lines, many of these never reported
before. However, the data were not deep enough to provide significant
constraints on the continuum emission in between the lines.

Here we present and discuss new measurements taken with GIANO during the
second commissioning run at Telescopio Nazionale Galileo (TNG). In 
Section~\ref{observations} we briefly describe the instrument, the measurements,
and the data reduction. In Sections~\ref{results} and \ref{discussion}
we present and discuss the results.

  \begin{figure}[h]
  \includegraphics[width=\hsize]{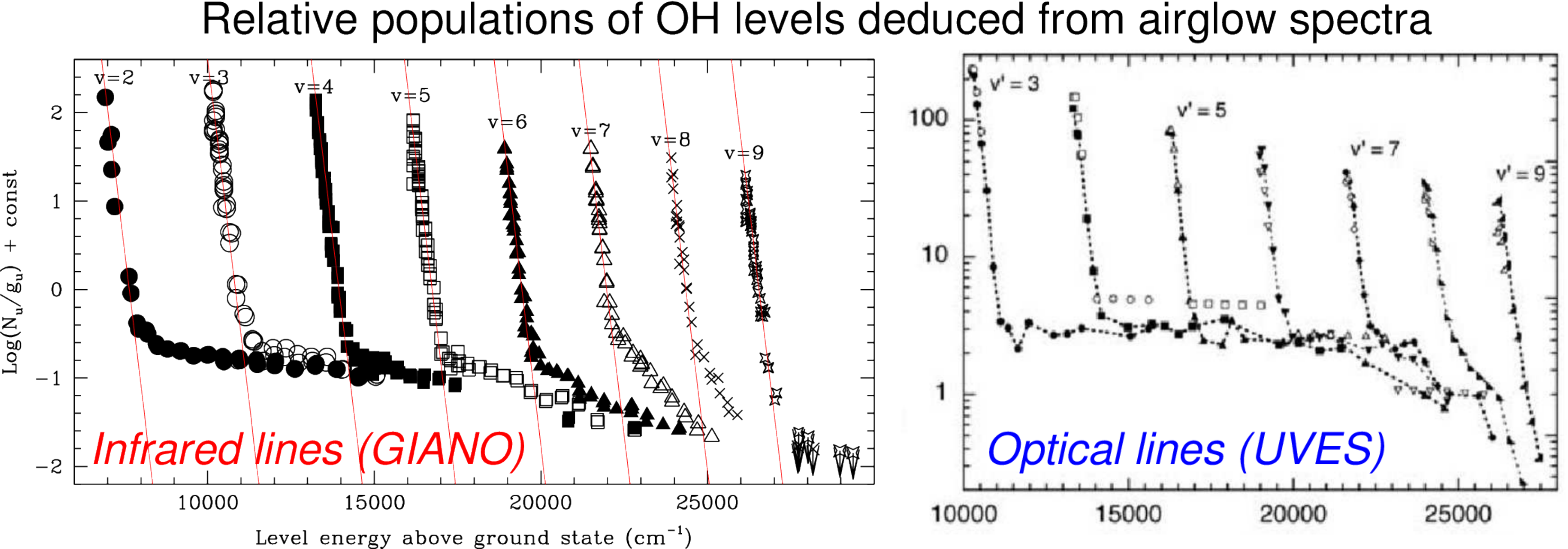}
      \caption{
        Derived column densities of the OH levels plotted against the
        energy of the levels above the ground-state of the molecule.
        The left panel shows the values derived from the infrared
        lines discussed here while the right hand panel -- reproduced
        under permission from
        Figure~16 of \cite{Cosby};
        \copyright\ Canadian Science Publishing or its licensors
        -- summarizes the results based
        on optical (UVES) spectra.
        The steep straight lines in the left panel
        show the distribution predicted by standard
        models with rotational levels thermalised at 200~K.
        The quasi-flat tails reveal the hot-OH component, see
        Section~\ref{OH_hot} for details.
        }
        \label{fig_OH_levels_single}
  \end{figure}
  \begin{figure}[h]
  \includegraphics[width=\hsize]{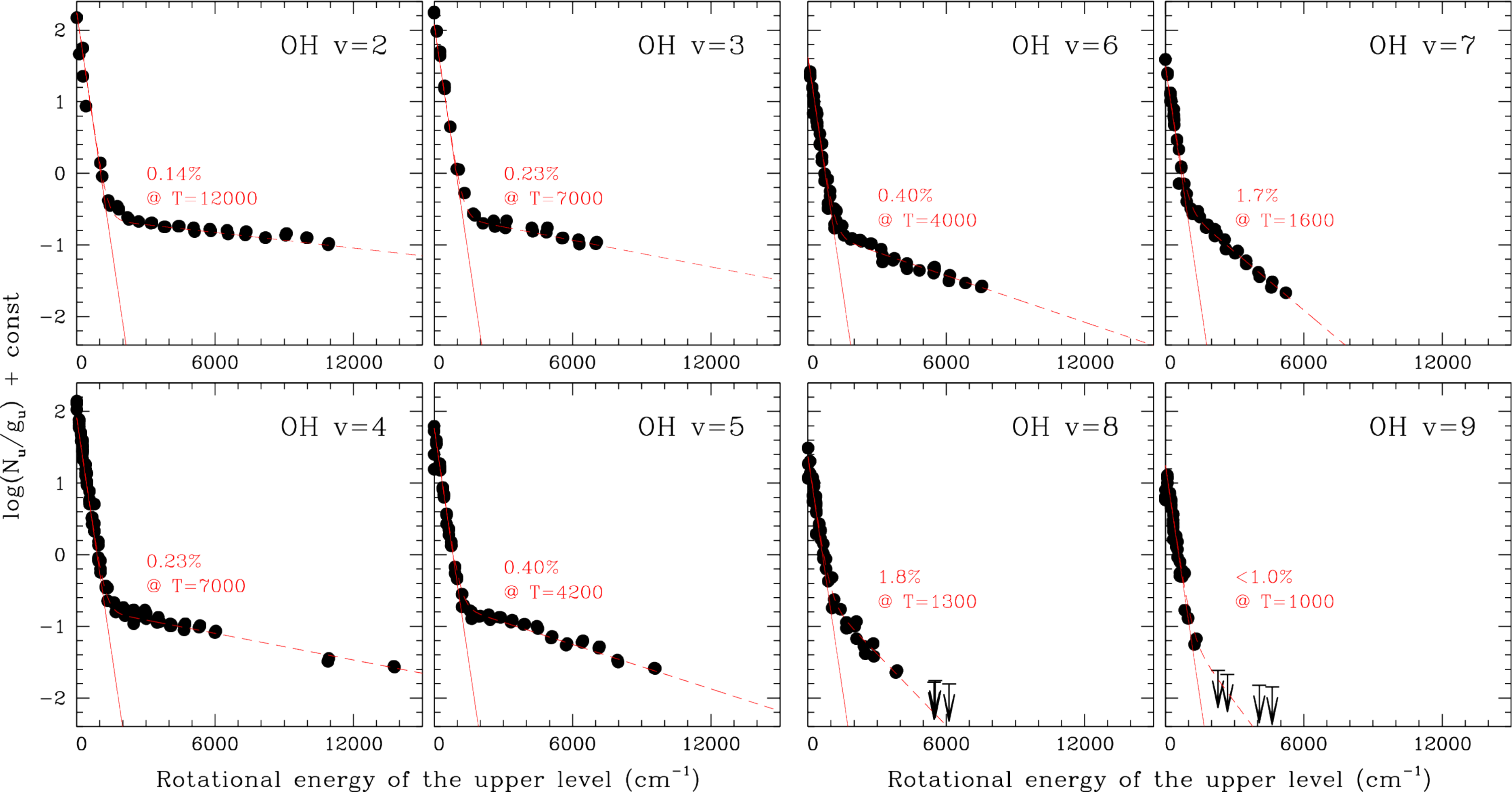}
      \caption{
        Same as Figure~\ref{fig_OH_levels_single} but with separate
        panels for each vibrational level. The straight solid lines
        represent the cold-OH component while the dashed curves show the
        distribution obtained adding a fraction of hot-OH molecules.
        The numerical fraction and rotational temperature of the
        hot-OH molecules is reported within each panel.
        See text, Section~\ref{OH_hot} for details.}
         \label{fig_OH_levels_details}
   \end{figure}

\section{Observations and spectral analysis}
\label{observations}

GIANO is a cross-dispersed cryogenic spectrometer that simultaneously 
covers the spectral range from 0.97~$\mu$m to 2.4~$\mu$m with a
maximum resolving power of R$\simeq$50,000 for a 2-pixels slit.
The main disperser is a commercial R2 echelle grating
with 23.2 lines/mm that works on a $\oslash$100 mm collimated beam.
Cross dispersion is performed by prisms (one made of fused silica and two made of ZnSe)
that work in double pass. The prisms
cross-disperse the light both before and after it is dispersed by
the echelle gratings;
this setup produces a curvature of the images of the spectral orders.
The detector is a HgCdTe Hawaii-II-PACE with 2048$^2$ pixels. Its control
system is extremely stable with a remarkably low read-out noise \citep[see][]{oliva_spie2012_2}.
More technical details on the instrument can be found in 
\cite{oliva_spie2012_1} and references therein.

GIANO was designed and built for direct light feeding from the TNG 
3.5~m telescope.
Unfortunately, the focal station originally reserved 
was not available when GIANO was commissioned. Therefore we were
forced to position the spectrograph on the rotating building and develop
a complex light-feed system using a pair of IR-transmitting ZBLAN fibres 
with two separate opto-mechanical interfaces. 
The first interface is used to feed the telescope light into
the fibres; it also includes the guiding camera and the calibration
unit. The second interface re-images the light from the fibres onto
the cryogenic slit; see \cite{tozzi14} for more details. 

The overall performances of GIANO are negatively affected by 
the complexity of the interfaces and by problems intrinsic to the
fibres: the efficiency has been lowered by almost a factor of 3 and the
spectra are affected by modal-noise, especially at longer wavelengths.
Consequently, the observations of the sky taken during normal
operations are not appropriate to reveal the faintest airglow lines and
the continuum emission in between. To overcome this problem we took 
advantage of the early part of a test night (September 3, 2014) to secure
a direct spectrum of the sky with GIANO. We moved the spectrograph to the 
main entrance of the TNG and arranged a simple pre-slit system (a lens
doublet and two flat mirrors) to feed the cryogenic slit with the light from
a sky-area in the ESE quadrant at a zenith distance of about 25 degrees.
The half moon was in the SSW quadrant at a zenith distance of 50 degrees.

We integrated the sky for two hours using the 3-pixels slit that is normally 
used in combination with the fibre interface. This yields a resolving
power R$\simeq$32,000.
For calibration we took several long series 
of darks interspaced with flat frames. This strategy was chosen because 
the sky spectra showed some residuals (persistency) of a flat frame taken 
many hours before. We therefore re-created different pseudo-darks with
different levels of persistency and, during the reduction, we selected the 
combination of pseudo-darks that best reproduced the persistency pattern.
The criterion for selection relied on the assumption
that the sky continuum emission is zero in the spectral regions were the 
atmosphere is opaque, i.e. in the 1.37-1.40~$\mu$m range
(orders 54-56 of the GIANO echellogram). In other words,
we took advantage of the fact that the flat and its residual persistency
have similar intensities over the full spectral range, while the spectrum of
any light source above the troposphere (i.e. astronomical
targets and the sky airglow emission) is absorbed by the water bands.

The acquisitions were performed using the standard setup
of the controller, i.e. on chip
integrations of 5 minutes with multiple non-destructive read-outs every
10 seconds. All the read-outs were separately stored. The "ramped-frames"
were constructed later-on using the algorithm described in 
\cite{oliva_spie2012_2} that, besides applying the standard 
Fowler sampling, it also minimises the effects of 
reset-anomaly and cosmic rays.

The 1D spectra were extracted by summing 20 pixels along the slit.
Wavelength calibration was performed using U-Ne lamp frames taken
after the series of darks. The spectrometer is stable to $<$0.1 pixels
(i.e. $\Delta\lambda/\lambda\!<\!10^{-6}$ r.m.s.).
The wavelengths of the uranium lines were taken from \cite{redman2011},
while for neon we used the table available
on the NIST website\footnote{physics.nist.gov/PhysRefData/ASD/lines$\_$form.html}.
The resulting wavelength accuracy was about 0.07~\AA\ r.m.s. for lines in 
the H-band.

The flat exposures were used to determine and correct the variation of
instrumental efficiency within each order. An approximate flux calibration was
performed by assuming that the relative efficiencies of the orders are
the same as when observing standard stars through the fibre-interface 
and the TNG telescope. This is a very reasonable assumption within the
relatively narrow wavelength range covered by the H-band. However, it may
cause systematic errors (up to 0.3 dex) in the relative fluxes of lines 
with very different wavelengths.
Absolute flux calibration was roughly estimated by imposing that the flux
of the OH [4-0]Q1(1.5) line at 1.5833 $\mu$m is 270 photons/m$^2$/s/arcsec$^2$;
i.e. the typical value measured during normal observing nights.

\section{The sky lines and continuum emission}
\label{results}
A total of about 1500 airglow lines were detected in the spectrum.
Compared to Paper~1, we have doubled the number of emission features measured.
In the following we separately discuss the OH lines, the other emission
features and the continuum emission in the Y, J, and H bands.

\subsection{OH lines and the hot-OH component}
\label{OH_hot}

Table~1 (available only in electronic format) lists
the lines identified as OH transitions.
For each $\Lambda$-doublet we give the wavelengths (in vacuum)
and the total observed flux of the doublet, normalised to the brightest
transition. 
For the fluxes we assumed that the two components of each
doublet have equal
intensities, i.e. that the '$e$' and '$f$' sub-levels are in thermal
equilibrium; this is appropriate for the density and temperature of the
mesosphere.
The listed wavelengths are derived from the
newest OH molecular constants by \cite{Bernath}.
These include highly excited rotational states and
allowed us to identify OH lines from rotational levels as high
as J=22.5, thus adding important constraints on the hot component 
of OH emission. 
This component was already reported by \cite{Cosby} and in Paper~1.
It is not included in any of the
models of OH airglow emission normally used for astronomical applications. 
These assume that the OH molecules have
a very high vibrational temperature ($T_{vib}\simeq 9000$~K) and a much lower
rotational temperature ($T_{rot}\simeq 200$~K).
In other words they assume that the gas density is high enough
to make collisional transitions between rotational states much faster
than radiative de-excitations. This brings the rotational temperature
to values similar to the kinetic temperature of the gas.
The net result is that all the lines from levels with rotational quantum 
number J$>$8.5 are normally predicted to be extremely faint and totally
negligible. 
The number of lines that are missed by standard models can be directly
visualised in Figure~\ref{fig_OH_levels_single} that plots the column densities
of the upper levels of the measured lines as a function 
of the excitation energy of the levels. The steep lines show the distribution
expected for a single gas component with rotational levels thermalised
at T=200~K. 
The points in the quasi-flat tails represent emission lines
from hot molecules that are not thermalised. 
According to \cite{Cosby}, this hot component is related to low density clouds at higher altitudes.
Here the gas density is lower than the critical density of the rotational
levels and, therefore, the population of the levels remain similar to that
set at the moment the OH molecule is formed.

In order to provide a practical tool to predict the intensities of all OH
lines we have fitted the observed level distribution with a mixture
of two components. The first is the standard model (cold-OH), while the 
second (hot-OH) has a rotational temperature that is empirically determined from the observed
values. Each vibrational state must be separately fit to obtain a good matching.
 
This simple model works as follows: let $N_u$ (cm$^{-2}$) 
be the column density of a given state ($v,J,F$) of the
OH molecule. This quantity is related to the excitation
temperatures by the standard Boltzmann equations, i.e.
$$ {N_u\over g_u N_{OH} } = e^{-E_v/k T_v} \ \ 
 \left[ \ETACOLD {e^{-E_{J,F}/k\TCOLD}\over U(T_v,\TCOLD)} 
 + \ETAHOT {e^{-E_{J,F}/k\THOT}\over U(T_v,\THOT)} \right] \eqno{(1)} $$
where $g_u$ is statistical weight of the level, $N_{OH}$
is the total column density of OH molecules, $E_v$ is 
the vibrational
energy of the level, $T_v$ is the vibrational temperature, $E_{J,F}$ is 
the rotational energy of the level, $\TCOLD$ is the rotational temperature 
of the cold component, $\ETACOLD$ is the fraction of cold molecules,
$\THOT$ is the rotational temperature of the hot
component, $\ETAHOT$ is the fraction of hot molecules
 and $U(T_v,T_r)$ is the partition function.
The photon-flux of a given transition arising from the same level is given by 
$$ I_{ul} = N_u \cdot A_{ul} \eqno{(2)} $$
where $A_{ul}$ (s$^{-1}$) is the transition probability.
The points in Figures~\ref{fig_OH_levels_single}, \ref{fig_OH_levels_details}
are computed from Eq.~(2) using the observed line intensities together with 
the molecular parameters of \cite{Bernath} and the transition probabilities 
of \cite{vanderloo}.
The steep straight lines in the left panel of Figure~\ref{fig_OH_levels_single} 
plot the function defined in
Eq.~(1) for $T_v$=9000~K, $\TCOLD$=200~K and $\ETACOLD$=1 (i.e. only cold-OH).
The same function is displayed in Figure~\ref{fig_OH_levels_details}
where the dashed curves show the results obtained adding a hot-OH component
with parameters ($\ETAHOT$,$\THOT$) adjusted for each vibrational level; the
values of the parameters are indicated in each panel.

The hot-OH component is most prominent in the lowest vibrational 
state (v=2) and becomes progressively weaker and cooler going to higher
vibrational states; it virtually disappears at v=9.

\subsection{ O$_2$ and unidentified lines }
\label{sect_O2}

The lines that cannot be associated with OH transitions are listed in
Table~2 (this table is available only in electronic format).
For the identification of the O$_2$ lines we used 
the HITRAN database \citep{hitran}. 
Most of the identified transitions were already reported in Paper~1.
A comparison between the two spectra shows that the intensity ratio 
between O$_2$ and OH lines has varied by almost a factor of 2 between
the two epochs. This is not surprising: the Oxygen lines are known
to vary by large factors even on timescales of hours. In our case
the variation can be used to select those features that follow the
time-behaviour of the O$_2$ lines. These lines are identified as
``O2?'' (i.e. probably O$_2$) in Table~2.

\begin{figure}[h]
   \includegraphics[width=\hsize]{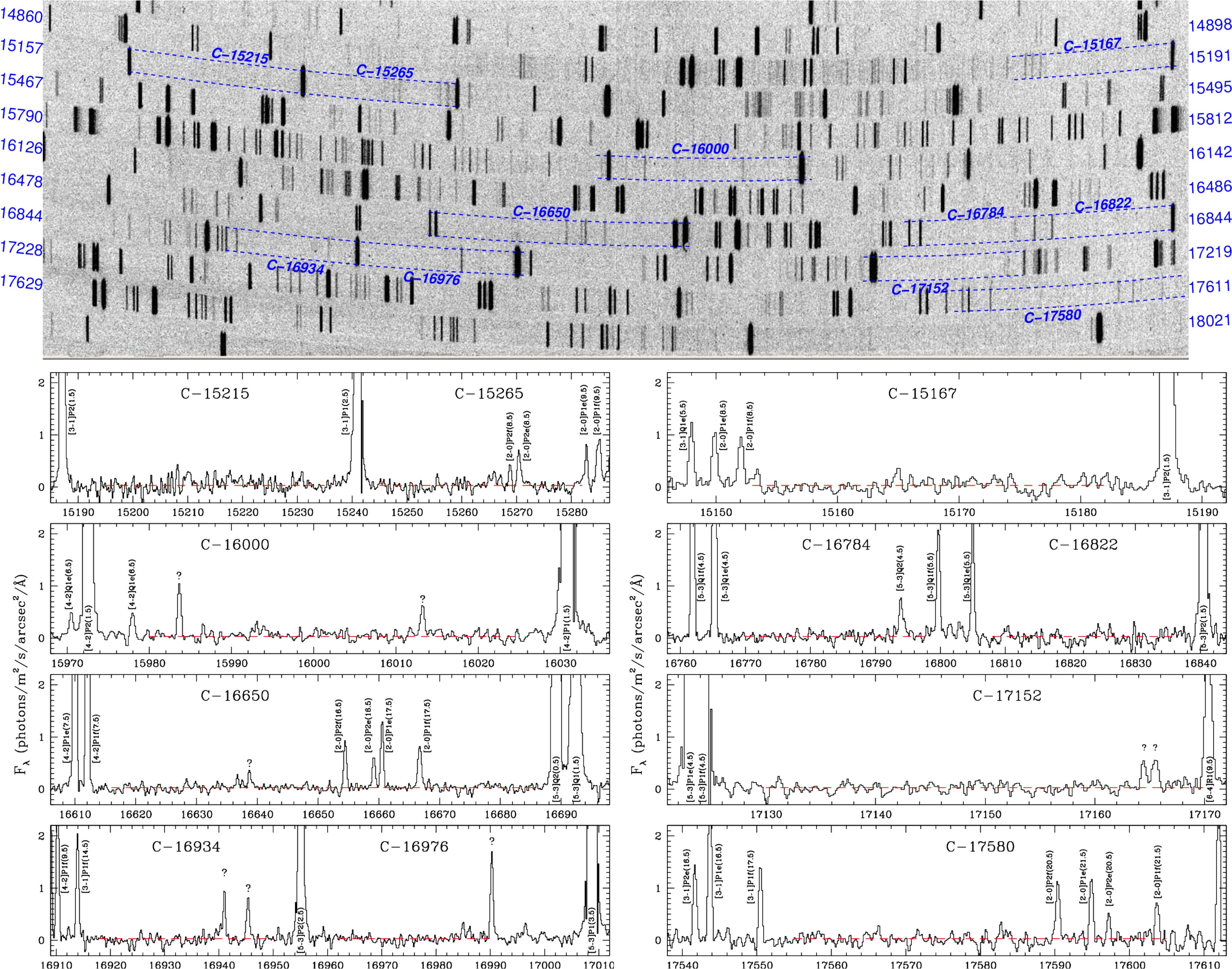}
      \caption{
           Upper panel: GIANO echelle spectrum of the H atmospheric band.
           Lower panels: extracted spectra in regions relatively free of
           line emission. The horizontal dashed lines show the level of
           300 photons/m$^2$/s/arcsec$^2$/$\mu$m
           (equivalent to 20.1 AB-mag/arcsec$^2$).
              }
         \label{fig_nice_H}
   \end{figure}

The remaining features are not identified. Of these 34 lines
are closely spaced doublets with equal intensities. A representative
example are the lines at $\lambda\lambda$17164.5, 17165.5~\AA\ visible in
the lower-right panel of Figure~\ref{fig_nice_H}. Several of these features were
already detected in Paper~1.

They are very similar to other $\Lambda$-split OH doublets
detected in our spectra. However, their wavelengths do not correspond
to any OH transition with $J_u\le40.5$ and $v_u\le10$.
The possibility that these doublets are produced by OH isotopologues
(e.g. $^{18}$OH) should be investigated, but is beyond the aims of
this paper.

   \begin{figure}[h]
   \includegraphics[width=\hsize]{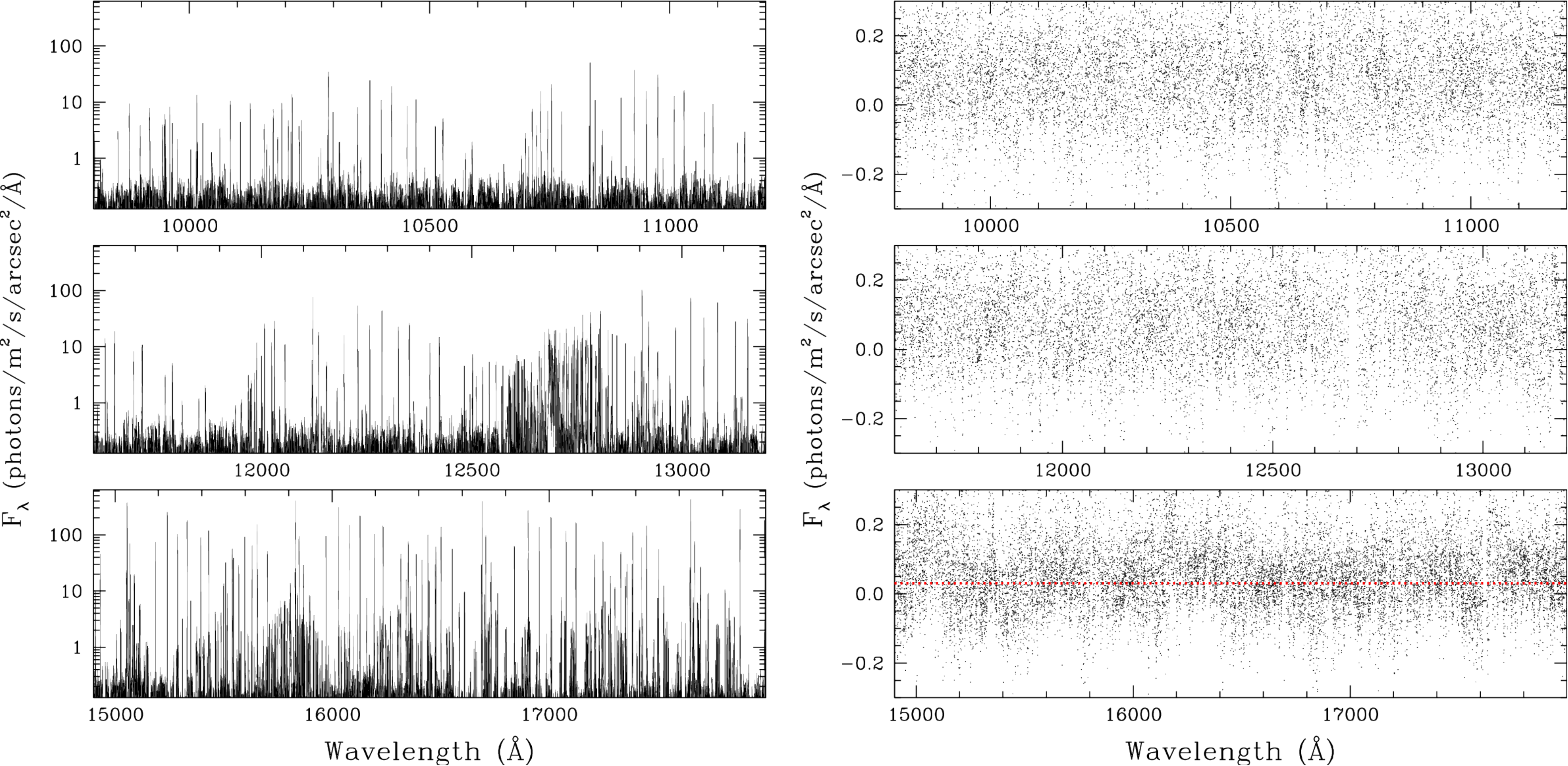}
      \caption{
        Left-hand panel: overview of the GIANO spectrum of the airglow.
        Right-hand panel: downscale to low flux levels. The spectral
        measurements are shown as separate dots to avoid confusion.
        The horizontal dashed line in the lowest panel shows the level of
           300 photons/m$^2$/s/arcsec$^2$/$\mu$m
           (equivalent to 20.1 AB-mag/arcsec$^2$).
        }
        \label{fig_show_all_and_zoom}
   \end{figure}

\begin{table}
 \centering
 \caption{Spectral bands with low contamination from lines.}
 \label{table_selected_bands}
 \begin{tabular}{|l|c|c|c|}
 \hline\hline
  Band & $\lambda$-range (\AA) &
    $\Delta\lambda/\lambda$ &
    Lines Flux(1) \\
 \hline
  C-15167    & 15153 -- 15183 &  0.0020 & 0.6 (210\ ;\ 20.5) \\
  C-15215    & 15195 -- 15235 &  0.0026 & --        \\
  C-15265    & 15245 -- 15285 &  0.0026 & 1.6 (400\ ;\ 19.8) \\
  C-16000    & 15980 -- 16020 &  0.0025 & 1.2 (320\ ;\ 20.0) \\
  C-16650(2) & 16620 -- 16680 &  0.0036 & 2.3 (380\ ;\ 19.9) \\
  C-16784    & 16770 -- 16798 &  0.0017 & 0.5 (170\ ;\ 20.7) \\
  C-16822    & 16808 -- 16836 &  0.0017 & --        \\
  C-16934    & 16918 -- 16950 &  0.0019 & 1.3 (390\ ;\ 19.8) \\
  C-16976    & 16962 -- 16990 &  0.0016 & 1.0 (360\ ;\ 19.9) \\
  C-17152    & 17134 -- 17170 &  0.0021 & 0.7 (190\ ;\ 20.6) \\
  C-17580    & 17555 -- 17605 &  0.0028 & 2.2 (430\ ;\ 19.7) \\
 \hline\hline
 \end{tabular}
 {\\
 \begin{flushleft}
  {
   (1) First entry is the lines flux in photons/m$^2$/s/arcsec$^2$. Numbers in
   brackets are the equivalent continuum flux (i.e. the line flux averaged
   over the band-width) in photons/m$^2$/s/arcsec$^2$/$\mu$m
   and in AB-mag/arcsec$^2$.
  }\\
  {
   (2) Region used by \cite{Maihara93}
   to define the sky-continuum
  }
 \end{flushleft}
  }
\end{table}

\subsection{The sky continuum emission}
\label{results_cont}

Within the H-band (1.5--1.8 $\mu$m) we detected:
\begin{itemize}
\item 514 lines of OH, half of which are produced by the 
hot-OH component described in Section~\ref{OH_hot};
\item 41 lines of O$_2$, including two broad and prominent band-heads;
\item 79 unidentified features.
\end{itemize}
Finding spectral regions free of emission features and far from bright airglow
lines is already difficult in our spectra. It becomes
virtually impossible at the lower resolving powers foreseen for MOONS
(R$\simeq$6,600) and other faint-object IR spectrometers.
In Figure~\ref{fig_nice_H}
we show the observed 2D echellogram of GIANO and the extracted 1D spectra of
selected regions with relatively low contamination from lines. Their
main parameters are
listed in Table~\ref{table_selected_bands}. They were selected
with the following criteria:
\begin{itemize}
\item The width of the band must correspond to at least 10 resolution 
elements of MOONS (i.e. $\Delta\lambda/\lambda\!>\!1/660$)
\item The band must include only faint lines whose total flux,
averaged over the band-width, is less than 500 
photons/m$^2$/s/arcsec$^2$/$\mu$m (equivalent to 19.6 AB-mag/arcsec$^2$).
\end{itemize}
The broadest band is C-16650. It coincides with the region used by \cite{Maihara93}
to measure an average sky-continuum of
590 photons/m$^2$/s/arcsec$^2$/$\mu$m (equivalent to 19.4 AB-mag/arcsec$^2$).
We find that about 65\% of this
flux can be ascribed to five emission features (4 lines from hot-OH
and one unidentified, see Figure~\ref{fig_nice_H}) that lie close to
the centre of this band. 
Taken at face value, this would imply that the
true continuum is $\simeq$200 photons/m$^2$/s/arcsec$^2$/$\mu$m
(equivalent to 20.6 AB-mag/arcsec$^2$).
However, this number is affected by large uncertainties intrinsic to the
procedure used to extract/average the continuum level from the spectrum and
to variations of the sky lines between different epochs. Indeed, to
reach a more reliable conclusion one would
should re-analyse the raw data of \cite{Maihara93} 
and correct them for the contribution of the sky-emission lines before
computing the continuum level.

We attempted to measure the sky continuum emission using
the extracted GIANO spectrum. This spectrum is shown in 
Figure~\ref{fig_show_all_and_zoom} and listed in 
Table~4 (available only in electronic format). 
The H-band has enough S/N ratio to show a faint continuum
of about 300 photons/m$^2$/s/arcsec$^2$/$\mu$m (equivalent to 20.1
AB-mag/arcsec$^2$); this level is shown as
a dashed line in the figure. 
It corresponds to 5 e$^-$/pixel/hr at the GIANO detector.
A formal computation of noise (i.e. including read-out, dark-current
and photon statistics) yields a convincing 5$\sigma$ detection
once the spectrum is re-sampled to a resolving power of R=5,000.
The contribution by systematic errors is more difficult to estimate. 
On the one hand, the procedure used to subtract detector dark and persistency 
(see Section~\ref{observations}) has
correctly produced a zero continuum in the bands where the
atmosphere is opaque (the uppermost order in the 2D frame of 
Figure~\ref{fig_nice_H}).
On the other hand, however, we cannot
exclude that second order effects have left some residual instrumental
artifacts in the H-band. 
An analysis of the dark frames affected by persistency indicates that
second order effects tend to increase the residuals, rather than 
over-subtracting the residual continuum level in the H-band.
Therefore, we are reasonably confident that the true sky-continuum
cannot be larger than the observed value. 

In the Y and J bands our spectra have a lower S/N ratio
because the efficiency of the GIANO detector drops at shorter wavelengths. 
The measured upper limits correspond to about 19 AB-mag/arcsec$^2$ and
are compatible with the measurements by \cite{Sullivan}.
In general, the Y and J bands are much less contaminated by line emission 
and the higher resolving power of GIANO is no longer
needed to find spectral regions that properly sample the sky continuum.

\section{Discussion and conclusions}
\label{discussion}

We took advantage of the second commissioning of the GIANO high-resolution
infrared spectrograph at La Palma Observatory
to point the instrument directly to the sky.
This yielded a sky spectrum much deeper than those collected through the
fibre-interface to the TNG telescope and published in \cite{paper1}.
The spectrum extends from 0.97 to 2.4 $\mu$m and includes the whole
Y, J, and H-bands. 

The spectrum shows about 1500 emission lines, a factor of two more than in
previous works. Of these, 80\% are identified as
OH transitions while the others are attributable to O$_2$ or unidentified.
Roughly half of the OH lines arise from highly excited rotational
states, presumably associated with lower density clouds at higher altitudes.
We derive physical parameters useful to model this hot-OH component that
as yet has never been included in the airglow models used by astronomers.

Several of the faint lines are in spectral regions that
were previously believed to be free of lines emission.
The continuum in the H-band is marginally detected
at a level of about 300 photons/m$^2$/s/arcsec$^2$/$\mu$m
equivalent to 20.1 AB-mag/arcsec$^2$.
In spite of the very low sky-continuum level, the myriad of airglow 
emission lines in the H-band severely limits 
the spectral ranges that can be
properly exploited for deep observations of faint objects
with low/medium resolution spectrographs.
We have identified a few spectral bands that could still remain quite dark
at the resolving power foreseen for the faint-object spectrograph
VLT-MOONS (R=6,600).

The spectrum and the updated lists of observed infrared sky-lines are published in electronic
format.

%% file: acknowl/acknowledgements.tex
% ACKNOWLEDGEMENTS

\chapter*{Acknowledgements}

My PhD research project would not have been possible without the help of many people.
I wish to thank in particular Davide Massari, Emanuele Dalessandro and Paolo Donati.
I have also spent a wonderful and fruitful period at Royal Observatory of Edinburgh
made possible by Michele Cirasuolo.
I want to thank also my friends, in Bologna and in my hometown Potenza,
for their support and encouragement.
This work is dedicated to all my family, a special thought goes to
the memory of my uncle and my cousin.

{\color{white}All the passion of these years, now, survives in Q.}